\documentclass[twocolumn,rmp,aps,amssymb]{revtex4}

\usepackage{epsf}
\usepackage{psfrag}

\usepackage{graphicx}
\usepackage{dcolumn}
\usepackage{bm}

\usepackage{graphicx}
\usepackage{amssymb}
\usepackage{times}
\usepackage{amsmath}
\usepackage{color}

\input{Qcircuit}

\renewcommand{\Qcircuit}[1][0em]{\xymatrix @*=<#1>}

\newcommand{\node}[2][]{{\begin{array}{c} \ _{#1}\  \\ {#2} \\ \
\end{array}}\drop\frm{o} }
\newcommand{\link}[2]{\ar @{-} [#1,#2]}

\newcommand{\proj}[1]{\ket{#1}\bra{#1}}

\definecolor{Blue}{rgb}{0,0,1}

\begin{document}

\title{Gaussian Quantum Information}

\author{Christian Weedbrook}   \affiliation{Center for Quantum Information and Quantum
Control, Department of Electrical and Computer Engineering and Department of Physics, University of Toronto, Toronto, M5S 3G4, Canada} \affiliation{Research Laboratory of Electronics, Massachusetts Institute of Technology, Cambridge MA 02139, USA}

\author{Stefano Pirandola} \affiliation{Department of Computer Science,
University of York, Deramore Lane, York YO10 5GH, United Kingdom}

\author{Ra\'ul Garc\'ia-Patr\'on}\affiliation{Research Laboratory of Electronics, Massachusetts Institute of Technology, Cambridge MA 02139, USA} \affiliation{Max-Planck-Institut f\"ur Quantenoptik, Hans-Kopfermann-Strasse 1, Garching, D-85748, Germany}

\author{Nicolas J.~Cerf}\affiliation{Quantum Information and Communication,
Ecole Polytechnique, Universit\'e Libre de Bruxelles, 1050 Brussels, Belgium} \affiliation{Research Laboratory of Electronics, Massachusetts Institute of Technology, Cambridge MA 02139, USA}

\author{Timothy C.~Ralph}\affiliation{Centre for Quantum Computation and Communication Technology, Department of Physics, University
of Queensland, Brisbane, Queensland 4072, Australia}

\author{Jeffrey H.~Shapiro}\affiliation{Research Laboratory of Electronics, Massachusetts Institute of Technology, Cambridge MA 02139, USA}

\author{Seth Lloyd}\affiliation{Research Laboratory of Electronics, Massachusetts Institute of Technology, Cambridge MA 02139, USA}\affiliation{Department of Mechanical Engineering, Massachusetts Institute of Technology, Cambridge MA 02139, USA}

\date{\today}

\begin{abstract}

The science of quantum information has arisen over the last two decades centered on the manipulation of individual
quanta of information, known as quantum bits or qubits. Quantum computers, quantum cryptography and quantum teleportation
are among the most celebrated ideas that have emerged from this new field. It was realized later on
that using continuous-variable quantum information carriers, instead of qubits, constitutes an extremely powerful alternative
approach to quantum information processing. This review focuses on continuous-variable quantum information processes
that rely on any combination of Gaussian states, Gaussian operations, and Gaussian measurements. Interestingly, such a
restriction to the Gaussian realm comes with various benefits,  since on the theoretical side, simple analytical  tools are available and, on the experimental side, optical components
effecting Gaussian processes are readily available in the laboratory. Yet, Gaussian quantum information processing opens the way to a wide
variety of tasks and applications, including quantum communication, quantum
cryptography, quantum computation, quantum teleportation, and
quantum state and channel discrimination. This review reports on
the state of the art in this field, ranging from the basic theoretical tools and landmark
experimental realizations to the most recent successful developments.

\end{abstract}


\maketitle

\tableofcontents

\section{Introduction}

Quantum mechanics is the branch of physics that studies how the
universe behaves at its smallest and most fundamental level.
Quantum computers and quantum communication systems transform and
transmit information using systems such as atoms and photons whose
behavior is intrinsically quantum mechanical.   As the size of
components of computers and the number of photons used to transmit
information has pressed downwards to the quantum regime, the
study of quantum information processing has potential practical relevance.
Moreover, the strange and counterintuitive features of quantum
mechanics translate into novel methods
for information processing that have no classical analogue.
Over the past two decades, a detailed theory of quantum information processing
has developed, and prototype quantum computers and quantum communication
systems have been constructed and tested experimentally.
Simple quantum algorithms have been performed, and a wide variety
of quantum communication protocols have been demonstrated, including
quantum teleportation and quantum cryptography.

Quantum information comes in two forms, \textit{discrete} and \textit{continuous}.
The best-known example of discrete quantum information is the quantum
bit or `qubit', a quantum system with two distinguishable states.
Examples of quantum systems that can be used to register a qubit
are spin $1/2$ particles such as electrons and many nuclear spins,
the two lowest energy states of semiconductor quantum dots or
quantized superconducting circuits, and the two polarization states
of a single photon.  The best-known example of continuous quantum
information~\cite{Bra04,Cerf2007a,Andersen2010,Braunstein2003} is the quantized harmonic oscillator, which can be
described by continuous variables such as position and momentum (an alternative description is the discrete but infinite-dimensional
representation in terms of energy states).  Examples of continuous-variable
quantum systems include quantized modes of bosonic systems such as
the different degrees of freedom of the electromagnetic field, vibrational modes of solids, atomic ensembles, nuclear spins in a quantum dot, Josephson junctions, and
Bose-Einstein condensates.  Because they supply the quantum
description of the propagating electromagnetic field, continuous-variable quantum systems are particularly relevant for quantum
communication and quantum-limited
techniques for sensing, detection and imaging. Similarly, atomic or solid-state based encoding of
continuous-variable systems can be used to perform quantum computation. Bosonic systems are not only useful in the physical modeling of qubit-based quantum computation, e.g., the quantized vibrational modes of ions embody the medium of communication between qubits in
ion-trap quantum computers, but also allows for new approaches to quantum computation.

\subsection{Gaussian quantum information processing}

This review reports on the state of the art of quantum information processing
using continuous variables.  The primary tools for analyzing
continuous-variable quantum information processing are Gaussian
states and Gaussian transformations.  Gaussian states are continuous-variable states that have a representation in terms of Gaussian
functions, and Gaussian transformations are those that take
Gaussian states to Gaussian states.  In addition to offering an
easy description in terms of Gaussian functions, Gaussian states
and transformations are of great practical relevance.
The ground state and thermal states of bosonic systems are Gaussian,
as are states created from such states by linear amplification
and loss.   Frequently, nonlinear operations
can be approximated to a high degree of accuracy by Gaussian
transformations.  For example, squeezing is a process that
decreases the variance of one continuous variable (position
or electric field, for example) while increasing the variance
of the conjugate variable (momentum or magnetic field).
Linear squeezing is Gaussian, and
nonlinear squeezing can typically be approximated to
first order by a linear, Gaussian process.  Moreover,
any transformation of a continuous-variable state can
be built up by Gaussian processes together with a repeated
application of a single nonlinear process such as photodetection.

In reviewing the basic facts of Gaussian quantum information processing~\cite{Braunstein2003,Bra04,Ferraro2005,Eisert2003}
and in reporting recent developments, we have attempted to present
results in a way that is accessible to two communities.  Members
of the quantum optics and atomic physics communities are very familiar with the basic
aspects of Gaussian quantum states and transformations, but may
be less acquainted with the application of Gaussian techniques
to quantum computation, quantum cryptography, and quantum
communication.  Members of the quantum information community
are familiar with quantum information processing techniques
such as quantum teleportation, quantum algorithms, and quantum
error correction, but may have less experience in the continuous-variable
versions of these protocols, which exhibit a range of features
that do not arise in their discrete versions.

The review is self-contained in the sense that study of
the introductory material should suffice to follow the detailed
derivations of more advanced methods of Gaussian quantum
information processing presented in the body of the paper.
Finally, the review supplies a comprehensive set of references
both to the foundations of the field of Gaussian quantum
information processing, and to recent developments.

\subsection{Outline of review}

The large subject matter and page length limit means that this review will take a mostly theoretical approach to Gaussian quantum information. In particular, we focus on \textit{optical} Gaussian protocols as they are the natural choice of medium for a lot of the protocols presented in this review. However, we do make mention of Gaussian atomic ensemble protocols~\cite{Hammerer2010} due to the close correspondence between continuous variables for light and atomic ensembles. Furthermore, experiments (both optical and atomic) will be mentioned and cited where appropriate. We also note that fermionic Gaussian states have also been studied in the literature (e.g., see \cite{Eisert2010,DiVicenzo2005,Bravyi2005}) but are outside the scope of this review. We limit our discussion of entanglement, quantum teleportation, quantum cloning, and quantum dense coding as these have all been discussed in detail previously, e.g., see \textcite{Bra04}. On the other hand, we give a detailed account of bosonic quantum channels, continuous-variable quantum cryptography and quantum computation.

We begin in Sec.~\ref{sec: elements of GQIP theory} by
introducing the fundamental theoretical concepts of Gaussian quantum
information. This includes Gaussian states and their phase-space representations
and symplectic structure, along with Gaussian unitaries,
which are the simplest quantum operations transforming Gaussian
states into Gaussian states. We then give examples
of both Gaussian states and Gaussian unitaries.  Multimode Gaussian states are discussed next using powerful techniques based on the manipulation of the second-order
statistical moments. The quantum
entanglement of bipartite Gaussian states is presented with the various measures associated with
it. We end this section by introducing the basic models of measurement, such as homodyne detection,
heterodyne detection and direct detection.

In Sec.~\ref{chapter: distinguishability of gaussian states} we begin to go more deeply into Gaussian quantum information processing via the process of distinguishing
between Gaussian states.  We present general
bounds and measures of distinguishability, and discuss
specific models for discriminating between optical coherent states. In Sec.~\ref{Chapter basic quantum protocols} we introduce basic Gaussian quantum information processing
protocols including quantum teleportation and quantum cloning.

In Sec.~\ref{chapter bosonic gaussian channels} we review bosonic communication channels which is one of the fundamental areas of research in quantum information. Bosonic channels include communication by electromagnetic waves (e.g., radio waves, microwaves,
and visible light), with Gaussian quantum channels being the most important example. These channels represent the standard model of noise in many quantum information protocols as well as being a good approximation to current optical telecommunication schemes. We begin by first reviewing the general formalisms and chief properties of bosonic channels, and specifically, those of Gaussian channels. This naturally leads to the study of the important class of one-mode Gaussian channels. The established notions of Gaussian channel capacity, both the classical and quantum versions, are presented next. Next up is entanglement-assisted classical capacity with quantum dense coding being a well-known example. This is followed by the concepts of entanglement distribution over noisy Gaussian channels and secret-key capacities. Finally, we end with the discrimination of quantum channels and the protocols of quantum illumination and quantum reading.

The state of the art in the burgeoning field of continuous-variable quantum cryptography is presented in Sec.~\ref{chap:QKD}. We begin by introducing how a generic quantum cryptographic scheme works followed by examples of the most commonly studied protocols. We then consider aspects of their security including what it means to be secure along with the main types of eavesdropping attacks. We continue with the practical situation of finite-size keys and the optimality and full characterization of collective Gaussian attacks before deriving the secret-key rates for the aforementioned protocols. We conclude with a discussion on the future avenues of research in continuous-variable quantum cryptography.

In Sec.~\ref{chapter computation} we review the most recent of the continuous-variable quantum information protocols, namely, quantum computation using continuous-variable cluster states. We begin by listing the most commonly used continuous-variable gates and by discussing the Lloyd-Braunstein criterion which provides the necessary and sufficient conditions for gates to form a universal set. The basic idea of one-way quantum computation using continuous variables is discussed next with teleportation providing an elegant way of understanding how computations can be achieved using only measurements. A common and convenient way of describing cluster states, via graph states and the nullifier formalism, is presented. Next, we consider the practical situations of Gaussian computational errors along with the various optical implementations of Gaussian cluster states. This leads into a discussion on how to incorporate universal quantum computation and quantum error correction into the framework of continuous-variable cluster state quantum computation. We end by introducing two quantum computational algorithms and provide directions for future research. In Sec.~\ref{chapter conclusion}, we offer perspectives and concluding remarks.

\subsection{Further readings}

For additional readings, perhaps the first place to start for an overview of continuous-variable quantum information is the well-known review article by \textcite{Bra04}. Furthermore, there is also the recent review by \textcite{Andersen2010} as well as two edited books on the subject by \textcite{Braunstein2003} and \textcite{Cerf2007a}. On Gaussian quantum information specifically there is the review article by \textcite{Wang2007} and the lecture notes of \textcite{Ferraro2005}. For a quantum optics~\cite{Gerry2005,Leonhardt2010,Bachor2004} approach to quantum information see the textbooks by \textcite{Walls1995}, \textcite{Kok10} and \textcite{Furusawa2011} (who provide a joint theoretical and experimental point of view). Whilst, the current state of the art of continuous variables using atomic ensembles can be found in the review article of \textcite{Hammerer2010}. For a detailed treatment of Gaussian systems see the textbook by \textcite{Holevo2011}. An overview of Gaussian entanglement is presented in the review of \textcite{Eisert2003}. For an elementary introduction to Gaussian quantum channels, see \textcite{Eisert2007}, and for continuous-variable quantum cryptography, see the reviews of \textcite{Scarani2009} and \textcite{Cerf2007}. Cluster state quantum computation using continuous variables is treated in the books of \textcite{Kok10} and \textcite{Furusawa2011}.

\subsection{Comment on notation}

Throughout this review, the variance of the vacuum noise is normalized to $1$. Such a normalization is commonly and conveniently thought of as setting Planck's constant $\hbar$ to a particular value, in our case $\hbar=2$. Currently, in continuous-variable quantum information there is no general consensus about the value of the variance of the vacuum, with common choices being either $1/4~(\hbar=1/2),1/2~(\hbar=1)$ or $1$. This is important to point out to the reader who, when referring to the many references in this review, should be aware that different papers use different choices of normalization. Another nomenclature issue in the literature is the notation used to define the quadrature operators (and the corresponding eigenvalues). Here we define the `position' quadrature by $\hat{q}$ and the `momentum' quadrature by $\hat{p}$. In this review, the logarithm ($\log$) can be taken to be base 2 for bits or ln for nats.
$\bold{I}$ represents the identity matrix which may be $2 \times 2$ (for one mode) or
$2N \times 2N$ (for arbitrary N modes). The correct dimensions, if not specified,
can be deducted from the context. Since we deal with continuous variables, we can have both discrete
and continuous ensembles of states, measurement operators, etcetera.
In order to keep the notation as simple as possible, in some parts
we consider discrete ensembles. It is understood that the extension
to continuous ensembles involves the replacement of sums by integrals. Finally, integrals are taken from $-\infty$ to $+\infty$ unless otherwise stated.

\section{Elements of Gaussian Quantum Information Theory}\label{sec: elements of GQIP theory}

\subsection{Bosonic systems in a nutshell}

A quantum system is called a \textit{continuous-variable
system} when it has an infinite-dimensional Hilbert space described by observables with continuous eigenspectra. The prototype of a continuous-variable system is
represented by $N$ bosonic modes, corresponding to $N$ quantized radiation modes of the electromagnetic field, i.e., $N$ quantum harmonic oscillators. In general, $N$ bosonic modes are associated with a
tensor-product Hilbert space $\mathcal{H}^{\otimes N}=\otimes_{k=1}%
^{N}\mathcal{H}_{k}$ and corresponding $N$ pairs of bosonic field operators
$\{\hat{a}_{k},\hat{a}_{k}^{\dagger}\}_{k=1}^{N}$, which are called the
annihilation and creation operators, respectively. These operators can be
arranged in a vectorial operator $\mathbf{\hat{b}}:=(\hat{a}_{1},\hat{a}%
_{1}^{\dagger},\cdots,\hat{a}_{N},\hat{a}_{N}^{\dagger})^{T}$, which must
satisfy the bosonic commutation relations
\begin{equation}
\lbrack\hat{b}_{i},\hat{b}_{j}]=\Omega_{ij}~,~(i,j=1,\cdots,2N) \label{BCR}%
\end{equation}
where $\Omega_{ij}$ is the generic element of the $2N\times2N$\ matrix%
\begin{equation}
\boldsymbol{\Omega}:=\bigoplus\limits_{k=1}^{N}\boldsymbol{\omega}=\left(
\begin{array}
[c]{ccc}%
\boldsymbol{\omega} &  & \\
& \ddots & \\
&  & \boldsymbol{\omega}%
\end{array}
\right)  ~~\boldsymbol{\omega}:=\left(
\begin{array}
[c]{cc}%
0 & 1\\
-1 & 0
\end{array}
\right)  ~, \label{Symplectic_Form}%
\end{equation}
known as the symplectic form. The Hilbert
space of this system is separable and infinite-dimensional. This is because
the single-mode Hilbert space $\mathcal{H}$ is spanned by a countable basis
$\{\left\vert n\right\rangle \}_{n=0}^{\infty}$, called the Fock or number state basis,
which is composed by the eigenstates of the number operator $\hat{n}:=\hat
{a}^{\dagger}\hat{a}$, i.e., $\hat{n}\left\vert n\right\rangle =n\left\vert
n\right\rangle $. Over these states the action of the bosonic operators is
well-defined, being determined by the bosonic commutation relations. In
particular, we have
\begin{equation}
\hat{a}\left\vert 0\right\rangle =0,~~~\hat{a}\left\vert n\right\rangle
=\sqrt{n}\left\vert n-1\right\rangle ~~\text{(for}~n\geq1\text{),}%
\end{equation}
and
\begin{equation}
\hat{a}^{\dagger}\left\vert n\right\rangle =\sqrt{n+1}\left\vert
n+1\right\rangle ~~\text{(for}~n\geq0\text{).}%
\end{equation}

Besides the bosonic field operators, the bosonic system may be described by
another kind of field operators. These are the quadrature field operators
$\{\hat{q}_{k},\hat{p}_{k}\}_{k=1}^{N}$, formally arranged in the vector
\begin{equation}
\mathbf{\hat{x}}:=(\hat{q}_{1},\hat{p}_{1},\ldots,\hat{q}_{N},\hat{p}_{N}%
)^{T}~. \label{vector_quadratures}%
\end{equation}
These operators derive from the cartesian decomposition of the bosonic field
operators, i.e., $\hat{a}_{k}:=\frac{1}{2}(\hat{q}_{k}+i\hat{p}_{k})$
or equivalently,
\begin{equation}
\hat{q}_{k}:=\hat{a}_{k}+\hat{a}_{k}^{\dagger},~\hat{p}_{k}:=i(\hat{a}%
_{k}^{\dagger}-\hat{a}_{k}). \label{Carte2}%
\end{equation}
The quadrature field operators represent dimensionless canonical observables
of the system and act like the position and momentum operators of the quantum
harmonic oscillator. In fact, they satisfy the canonical commutation relations
in natural units ($\hbar=2$)
\begin{equation}
\lbrack\hat{x}_{i},\hat{x}_{j}]=2i\Omega_{ij}, \label{CCR}%
\end{equation}
which are easily derivable from the bosonic commutator relations of
Eq.~(\ref{BCR}). In the following, we will use both kinds of field operators,
the bosonic field operators and the quadrature field operators.

Now it is important to note that the quadrature operators are observables with
continuous eigenspectra. In fact the two quadrature operators $\hat{q}$ (position) and $\hat{p}$
(momentum) have eigenstates\footnote{Strictly speaking, $\ket{q}$ and $\ket{p}$ are \textit{improper} eigenstates since they
are nonnormalizable, thus lying outside the Hilbert space. Correspondingly, $q$ and $p$ are \textit{improper} eigenvalues. In the remainder we take this mathematical subtlety for granted.}
\begin{equation}
\hat{q}\left\vert q\right\rangle =q\left\vert q\right\rangle ,~\hat
{p}\left\vert p\right\rangle =p\left\vert p\right\rangle ,
\end{equation}
with continuous eigenvalues $q\in\mathbb{R}$ and $p\in\mathbb{R}$. The two
eigensets $\{\left\vert q\right\rangle \}_{q\in\mathbb{R}}$ and $\{\left\vert
p\right\rangle \}_{p\in\mathbb{R}}$\ identify two bases which are connected by
a Fourier transform%
\begin{equation}\label{eq: basis states relation via fourier }
\left\vert q\right\rangle =\frac{1}{2\sqrt{\pi}}\int dpe^{-iqp/2}\left\vert
p\right\rangle ,~\left\vert p\right\rangle =\frac{1}{2\sqrt{\pi}}\int
dq e^{iqp/2}\left\vert q\right\rangle .
\end{equation}
In general, for the $N$-mode Hilbert space we can write%
\begin{equation}
\mathbf{\hat{x}}^{T}\left\vert \mathbf{x}\right\rangle =\mathbf{x}%
^{T}\left\vert \mathbf{x}\right\rangle , \label{quad_EIGEN}%
\end{equation}
with $\mathbf{x}\in\mathbb{R}^{2N}$ and $\left\vert \mathbf{x}\right\rangle
:=\left(  \left\vert x_{1}\right\rangle ,\ldots,\left\vert x_{2N}\right\rangle
\right)  ^{T}$. Here the quadrature eigenvalues $\mathbf{x}$\ can be used as
continuous variables to describe the entire bosonic system. This is possible
by introducing the notion of phase-space representation.

\subsubsection{Phase-space representation and Gaussian states}

All the physical information about the $N$-mode bosonic system is contained in
its quantum state. This is represented by a density operator $\hat{\rho}$, which is
a trace-one positive operator acting on the corresponding Hilbert space, i.e.,
$\hat{\rho}:\mathcal{H}^{\otimes N}\rightarrow\mathcal{H}^{\otimes N}$. We denote by
$\mathcal{D}(\mathcal{H}^{\otimes N})$ the space of the density operators,
also called the state space. Whenever $\hat{\rho}$
is a projector ($\hat{\rho}^{2}=\hat{\rho}$) we say that $\hat{\rho}$ is pure and the state can
be represented as $\hat{\rho}=\left\vert \varphi\right\rangle \left\langle
\varphi\right\vert $ where $\left\vert \varphi\right\rangle \in\mathcal{H}%
^{\otimes N}$. Now it is important to note that any density operator has an
equivalent representation in terms of a quasi-probability distribution (Wigner
function) defined over a real symplectic space (phase space). In fact, let us
introduce the Weyl operator%
\begin{equation}
D(\boldsymbol{\xi}):=\exp(i\mathbf{\hat{x}}^{T}\boldsymbol{\Omega\xi}),
\label{Weyl operator}%
\end{equation}
where $\boldsymbol{\xi}\in\mathbb{R}^{2N}$. Then, an arbitrary $\hat{\rho}$ is
equivalent to a Wigner characteristic function
\begin{equation}
\chi(\boldsymbol{\xi})=\mathrm{Tr}\left[  \hat{\rho} D(\boldsymbol{\xi})\right]  ,
\label{CH_function}%
\end{equation}
and, via Fourier transform, to a Wigner function%
\begin{equation}
W(\mathbf{x})=\int\limits_{\mathbb{R}^{2N}}\frac{d^{2N}\boldsymbol{\xi}}%
{(2\pi)^{2N}}~\exp\left(  -i\mathbf{x}^{T}\boldsymbol{\Omega\xi}\right)
\chi(\boldsymbol{\xi}), \label{Wig_function}%
\end{equation}
which is normalized to one but generally non-positive (quasi-probability
distribution). In Eq.~(\ref{Wig_function}) the continuous variables
$\mathbf{x}\in\mathbb{R}^{2N}$ are the eigenvalues of quadratures operators
$\mathbf{\hat{x}}$. These variables span a real symplectic space
$\mathcal{K}:=(\mathbb{R}^{2N},\mathbf{\Omega})$ which is called the
\textit{phase space}. Thus, an arbitrary quantum
state $\hat{\rho}$ of a $N$-mode bosonic system is equivalent to a Wigner function
$W(\mathbf{x})$\ defined over a $2N$-dimensional phase space $\mathcal{K}$.

The most relevant quantities that characterize the Wigner representations
($\chi$ or $W$) are the statistical moments of the quantum state. In
particular, the first moment is called the displacement
vector or, simply, the mean value
\begin{equation}
\mathbf{\bar{x}}:=\left\langle \mathbf{\hat{x}}\right\rangle =\mathrm{Tr}%
(\mathbf{\hat{x}}\hat{\rho}), \label{Displacement_1mom}%
\end{equation}
and the second moment is called the covariance
matrix $\mathbf{V}$, whose arbitrary element is
defined by%
\begin{equation}
V_{ij}:=\tfrac{1}{2}\left\langle \left\{  \Delta\hat{x}_{i},\Delta\hat{x}%
_{j}\right\}  \right\rangle , \label{CM_definition}%
\end{equation}
where $\Delta\hat{x}_{i}:=\hat{x}_{i}-\langle\hat{x}_{i}\rangle$ and $\{,\}$
is the anti-commutator. In particular, the diagonal elements of the covariance matrix provide
the variances of the quadrature operators, i.e.,%
\begin{equation}
V_{ii}=V(\hat{x}_{i}),
\end{equation}
where $V(\hat{x}_{i})=\langle(\Delta\hat{x}_{i})^{2}\rangle=\langle\hat{x}_{i}^{2}%
\rangle-\langle\hat{x}_{i}\rangle^{2}$. The covariance matrix is a $2N\times2N$, real and
symmetric matrix which must satisfy the uncertainty principle \cite{Simon1994}%
\begin{equation}
\mathbf{V}+i\mathbf{\Omega}\geq0, \label{HeisCM}%
\end{equation}
directly coming from the commutation relations of Eq.~(\ref{CCR}), and
implying the positive definiteness $\mathbf{V}>0$. From
the diagonal terms in Eq.~(\ref{HeisCM}), one can easily derive the usual
Heisenberg relation for position and momentum
\begin{equation}
V(\hat{q}_{k})V(\hat{p}_{k})\geq1. \label{Heis_usual}%
\end{equation}
For a particular class of states the first two moments are sufficient for a
complete characterization, i.e., we can write $\hat{\rho}=\hat{\rho}(\mathbf{\bar{x}%
},\mathbf{V})$. This is the case of the \textit{Gaussian
states}~\cite{Holevo1975,Holevo2011}. By definition, these are bosonic states whose
Wigner representation ($\chi$ or $W$) is Gaussian, i.e.,%
\begin{align}
\chi(\boldsymbol{\xi})  &  =\exp\left[  -\frac{1}{2}\boldsymbol{\xi}%
^{T}\left(  \boldsymbol{\Omega}\mathbf{V}\boldsymbol{\Omega}^{T}\right)
\boldsymbol{\xi}-i\left(  \boldsymbol{\Omega}\mathbf{\bar{x}}\right)
^{T}\boldsymbol{\xi}\right]  ,\\
W(\mathbf{x})  &  =\frac{\exp\left[  -\frac{1}{2}(\mathbf{x}-\mathbf{\bar{x}%
})^{T}\mathbf{V}^{-1}(\mathbf{x}-\mathbf{\bar{x}})\right]  }{(2\pi)^{N}%
\sqrt{\det\mathbf{V}}}.
\end{align}
It is interesting to note that a pure state is Gaussian, if and only if,
its Wigner function is non-negative~\cite{Hudson1974,Soto1983,Mandilara2009}.

\subsubsection{Gaussian unitaries}\label{sec: gaussian unitaries}

Since Gaussian states are easy to characterize, it turns out that a large
class of transformations acting on these states are easy to describe too. In
general, a quantum state undergoes a transformation called a
quantum operation~\cite{Nielsen2000}. This is a linear map
$\mathcal{E}:\hat{\rho}\rightarrow\mathcal{E}(\hat{\rho})$ which is completely positive
and trace-decreasing, i.e., $0\leq\mathrm{Tr}\left[  \mathcal{E}(\hat{\rho})\right]
\leq1$. A quantum operation is then called a quantum
channel when it is trace-preserving, i.e., $\mathrm{Tr}%
\left[  \mathcal{E}(\hat{\rho})\right]  =1$. The simplest quantum channels are the
ones which are reversible. These are represented by unitary transformations
$U^{-1}=U^{\dagger}$, which transform a state according to the rule
$\hat{\rho}\rightarrow U\hat{\rho} U^{\dagger}$ or, simply, $\left\vert \varphi
\right\rangle \rightarrow U\left\vert \varphi\right\rangle $, if the state is pure.

Now we say that a quantum operation is Gaussian when it transforms Gaussian
states into Gaussian states. Clearly, this definition applies to the specific
cases of quantum channels and unitary transformations. Thus, Gaussian channels
(unitaries) are those channels which preserve the Gaussian
character of a quantum state.
%
Gaussian unitaries are generated via $U=\exp(-i\hat{H}/2)$ from Hamiltonians
$\hat{H}$ which are second-order polynomials in the field operators. In terms of
the annihilation and creation operators $\mathbf{\hat{a}}:=(\hat{a}_{1},\cdots,\hat{a}_{N})^{T}$
and $\mathbf{\hat{a}}^{\dagger}:=(\hat{a}_{1}^{\dagger},\cdots,\hat{a}%
_{N}^{\dagger})$ this means that%
\begin{equation}
\hat{H}=i(\mathbf{\hat{a}}^{\dagger}\boldsymbol{\alpha}+\mathbf{\hat{a}%
}^{\dagger}\mathbf{F\hat{a}+\hat{a}}^{\dagger}\mathbf{G\mathbf{\hat{a}}%
}^{\dagger T})+h.c.~, \label{Bilinear_Hamiltonian}%
\end{equation}
where $\boldsymbol{\alpha}\in\mathbb{C}^{N}$ and $\mathbf{F},\mathbf{G}$ are
$N\times N$ complex matrices and $h.c.$ stands for `Hermitian conjugate'. In the Heisenberg picture, this kind of unitary
corresponds to a linear unitary Bogoliubov transformation
\begin{equation}
\mathbf{\hat{a}}\rightarrow U^{\dagger}\mathbf{\hat{a}}U=\mathbf{A\hat{a}%
}+\mathbf{B\hat{a}}^{\dagger}+\boldsymbol{\alpha}~,
\end{equation}
where the $N\times N$ complex matrices $\mathbf{A}$ and $\mathbf{B}$ satisfy
$\mathbf{AB}^{T}=\mathbf{BA}^{T}$ and $\mathbf{AA}^{\dagger}=\mathbf{BB}%
^{\dagger}+\mathbf{I}$ with $\mathbf{I}$ being the identity matrix. In terms of the quadrature
operators, a Gaussian unitary is more simply described by an affine map
\begin{equation}
(\mathbf{S},\mathbf{d}):\mathbf{\hat{x}}\rightarrow\mathbf{S\hat{x}+d},
\label{LUBT}%
\end{equation}
where $\mathbf{d}\in\mathbb{R}^{2N}$ and $\mathbf{S}$ is a $2N\times2N$ real
matrix. This transformation must preserve the commutation relations of
Eq.~(\ref{CCR}), which happens when the matrix $\mathbf{S}$ is
\textit{symplectic}, i.e.,%
\begin{equation}
\mathbf{S\Omega S}^{T}=\mathbf{\Omega}~. \label{Sympl_cond}%
\end{equation}
Clearly the eigenvalues $\mathbf{x}$ of the quadrature operators
$\mathbf{\hat{x}}$ must follow the same rule, i.e., $(\mathbf{S}%
,\mathbf{d}):\mathbf{x}\rightarrow\mathbf{Sx+d}$. Thus, an arbitrary Gaussian
unitary is equivalent to an affine symplectic map $(\mathbf{S},\mathbf{d})$
acting on the phase space, and can be denoted by $U_{\mathbf{S},\mathbf{d}}$.
In particular, we can always write $U_{\mathbf{S},\mathbf{d}}=D(\mathbf{d}%
)U_{\mathbf{S}}$, where the canonical
unitary $U_{\mathbf{S}}$\ corresponds to a linear
symplectic map $\mathbf{x}\rightarrow\mathbf{Sx}$, and the Weyl operator
$D(\mathbf{d})$ to a phase-space translation $\mathbf{x}\rightarrow
\mathbf{x+d} $. Finally, in terms of the statistical moments, $\mathbf{\bar
{x}}$ and $\mathbf{V}$, the action of a Gaussian unitary $U_{\mathbf{S}%
,\mathbf{d}}$ is characterized by the following transformations%
\begin{equation}
\mathbf{\bar{x}}\rightarrow\mathbf{S\bar{x}+d~,~V}\longrightarrow
\mathbf{SVS}^{T}~. \label{CM_congruence}%
\end{equation}
Thus the action of a Gaussian unitary $U_{\mathbf{S},\mathbf{d}}$ over a
Gaussian state $\hat{\rho}(\mathbf{\bar{x}},\mathbf{V})$ is completely characterized
by Eq.~(\ref{CM_congruence}).

\subsection{Examples of Gaussian states and Gaussian unitaries}\label{sec: examples of gaussian states}

Here we introduce some elementary Gaussian states that play a major role in
continuous-variable quantum information. We also introduce
the simplest and most common Gaussian unitaries and discuss their connection
with basic Gaussian states. In these examples we first consider one and
then two bosonic modes with the general case (arbitrary $N$) discussed in
Sec.~\ref{Symplectic_SECTION}.

\subsubsection{Vacuum states and thermal states}

The most important Gaussian state is the one with zero photons ($\bar{n}=0$),
i.e., the \textit{vacuum state} $\ket{0}$. This is also the eigenstate with zero eigenvalue
of the annihilation operator ($\hat{a}\left\vert 0\right\rangle =0 $). The covariance matrix of the vacuum is just the identity, which means that position and
momentum operators have noise-variances equal to one, i.e., $V(\hat{q}%
)=V(\hat{p})=1$. According to Eq.~(\ref{Heis_usual}), this is the minimum
variance which is reachable symmetrically by position and momentum. It is also
known as \textit{vacuum noise} or \textit{quantum shot-noise}.

As we will soon see, every Gaussian state can be decomposed into \textit{thermal states}. From this point of view, a thermal state can be thought of as the most fundamental Gaussian state. By definition, we call thermal a bosonic
state which maximizes the von Neumann entropy
\begin{equation}\label{eq: von neumann entropy}
S:=-\mathrm{Tr}(\hat{\rho}\ln\hat{\rho})~,
\end{equation}
for fixed energy $\mathrm{Tr}(\hat{\rho}\hat{a}^{\dagger}\hat{a})=\bar{n}$, where
$\bar{n}\geq0$ is the mean number of photons in the bosonic mode. Explicitly,
its number-state representation is given by%
\begin{equation}
\hat{\rho}^{th}(\bar{n})=\sum_{n=0}^{+\infty}\frac{\bar{n}^{n}}{(\bar{n}+1)^{n+1}%
}\left\vert n\right\rangle \left\langle n\right\vert ~. \label{Thermal_state}%
\end{equation}
One can easily check that its Wigner function is Gaussian, with zero mean and
covariance matrix $\mathbf{V}=(2\bar{n}+1)\mathbf{I}$ where $\mathbf{I}$ is the $2\times2$
identity matrix.

\subsubsection{Displacement and coherent states}

The first Gaussian unitary we introduce is the \textit{displacement operator},
which is just the complex version of the Weyl operator. The displacement
operator is generated by a linear Hamiltonian and is defined by%
\begin{equation}
D(\alpha):=\mathrm{exp}(\alpha\hat{a}^{\dagger}-\alpha^{\ast}\hat{a})~,
\end{equation}
where $\alpha=(q+ip)/2$ is the complex
amplitude. In the Heisenberg picture, the annihilation operator is transformed by the linear unitary Bogoliubov transformation $\hat{a}\rightarrow\hat{a}+\alpha$, and the
quadrature operators $\mathbf{\hat{x}}=(\hat{q},\hat{p})^{T}$ by the
translation $\mathbf{\hat{x}}\rightarrow\mathbf{\hat{x}}+\mathbf{d}_{\alpha}$,
where $\mathbf{d}_{\alpha}=(q,p)^{T}$. By displacing the vacuum state, we generate coherent states $\ket{\alpha}=D(\alpha)\ket{0}$. They have
the same covariance matrix of the vacuum ($\mathbf{V}=\mathbf{I}$) but different mean values
($\mathbf{\bar{x}}=\mathbf{d}_{\alpha}$). Coherent states are the eigenstates
of the annihilation operator $\hat{a}\ket{\alpha}=\alpha\ket{\alpha}$ and can
be expanded in number states as
\begin{equation}
\ket{\alpha}=\mathrm{exp}\left(  -\tfrac{1}{2}|\alpha|^{2}\right)  \sum
_{n=0}^{\infty}\frac{\alpha^{n}}{\sqrt{n!}}\ket{n}~.
\end{equation}
Furthermore, they form an overcomplete basis, since they are
non-orthogonal. In fact, given two coherent states $\ket{\alpha}$ and
$\ket{\beta}$, the modulus squared of their overlap is given by%
\begin{equation}
|\langle\beta\ket{\alpha}|^{2}=\exp(-|\beta-\alpha|^{2})~.
\label{eq: overlap coherent state}%
\end{equation}

\subsubsection{One-mode squeezing and squeezed states}

When we pump a nonlinear crystal with a bright laser, some of the pump photons
with frequency $2\omega$ are split into pairs of photons with frequency
$\omega$. Whenever the matching conditions for a degenerate optical parametric
amplifier (OPA) are satisfied \cite{Walls1995}, the outgoing mode is ideally
composed of a superposition of even number states ($\ket{2n}$). The interaction
Hamiltonian must then contain a $\hat{a}^{\dagger2}$ term to generate pairs of
photons and a term $\hat{a}^{2}$ to ensure hermiticity. The corresponding
Gaussian unitary is the one-mode squeezing
operator, which is defined as
\begin{equation}
S(r):=\mathrm{exp}[r(\hat{a}^{2}-\hat{a}^{\dagger2})/2]~,
\end{equation}
where $r\in\mathbb{R}$ is called the squeezing
parameter. In the Heisenberg picture, the annihilation operator is transformed by the linear unitary Bogoliubov transformation $\hat{a}\rightarrow(\cosh r)\hat{a}-(\sinh
r)\hat{a}^{\dagger}$ and the quadrature operators $\mathbf{\hat{x}}=(\hat
{q},\hat{p})^{T}$ by the symplectic map $\mathbf{\hat{x}}\rightarrow
\mathbf{S}(r)\mathbf{\hat{x}}$, where%
\begin{equation}
\mathbf{S}(r):=\left(
\begin{array}
[c]{cc}%
e^{-r} & 0\\
0 & e^{r}%
\end{array}
\right)  ~. \label{CM_1mode_Squeeze}%
\end{equation}
Applying the squeezing operator to the vacuum we generate a
\textit{squeezed vacuum state}~\cite{Yuen1976},
\begin{equation}
\ket{0,r}=\frac{1}{\sqrt{\cosh r}}\sum_{n=0}^{\infty}\frac{\sqrt{(2n)!}}%
{2^{n}n!}\tanh r^{n}\ket{2n}.
\end{equation}
Its covariance matrix is given
by $\mathbf{V=S}(r)\mathbf{S}(r)^{T}=\mathbf{S}(2r)$ which has different quadrature noise-variances, i.e., one variance is
squeezed below the quantum shot-noise, whilst the other is anti-squeezed above
it.

\subsubsection{Phase rotation}

The phase is a crucial element of the wave behavior of the electromagnetic
field with no physical meaning for a single mode on its own. In
continuous-variable systems the phase is usually defined with respect to a
local oscillator, i.e., a mode-matched classical beam. Applying a phase shift
on a given mode is done by increasing the optical path length of the beam
compared to the local oscillator. For instance, this can be done by adding a
transparent material of a tailored depth and with a higher refractive index
than vacuum. The phase rotation operator is generated by the free propagation Hamiltonian
$\hat{H}=2\theta\hat{a}^{\dagger}\hat{a}$, so that it is defined by $R(\theta
)=\exp(-i\theta\hat{a}^{\dagger}\hat{a})$. In the Heisenberg picture, it
corresponds to the simple linear unitary Bogoliubov transformation $\hat{a}\rightarrow e^{i\theta}\hat{a}$ for the
annihilation operator. Correspondingly, the quadratures are transformed via the
symplectic map $\mathbf{\hat{x}}\rightarrow\mathbf{R}(\theta)\mathbf{\hat{x}}%
$, where
\begin{equation}
\mathbf{R}(\theta):=\left(
\begin{array}
[c]{cc}%
\cos\theta & \sin\theta\\
-\sin\theta & \cos\theta
\end{array}
\right)  ~,
\end{equation}
is a proper rotation with angle $\theta$.

\subsubsection{General one-mode Gaussian states}

Using the singular value decomposition, one can show that any $2\times2$
symplectic matrix can be decomposed as $\mathbf{S}=\mathbf{R}(\theta
)\mathbf{S}(r)\mathbf{R}(\phi)$. This means that any one-mode Gaussian unitary
can be expressed as $D(\mathbf{d})U_{\mathbf{S}}$ where $U_{\mathbf{S}%
}=R(\theta)S(r)R(\phi)$. By applying this unitary to a thermal state
$\hat{\rho}_{th}(\bar{n})$, the result is a Gaussian state with mean $\mathbf{d}$
and covariance matrix
\begin{equation}
\mathbf{V}=(2\bar{n}+1)\mathbf{R}(\theta)\mathbf{S}(2r)\mathbf{R}(\theta
)^{T}~. \label{V_gen_1mode}%
\end{equation}
This is the most general one-mode Gaussian state. This result can be
generalized to arbitrary $N$ bosonic modes as we will see in
Sec.~\ref{Symplectic_SECTION}. Now, by setting $\bar{n}=0$ in
Eq.~(\ref{V_gen_1mode}), we achieve the covariance matrix of the most general one-mode pure
Gaussian state. This corresponds to a rotated and displaced squeezed state
$\left\vert \alpha,\theta,r\right\rangle =D(\alpha)R(\theta)S(r)\left\vert
0\right\rangle $.

\subsubsection{Beam splitter}

In the case of two bosonic modes one of the most important Gaussian unitaries is
the \textit{beam splitter transformation}, which is the simplest example of an
interferometer. This transformation is defined by
\begin{equation}\label{eq: beam splitter}
B(\theta)=\mathrm{exp}[\theta(\hat{a}^{\dagger}\hat{b}-\hat{a}\hat{b}%
^{\dagger})]~,
\end{equation}
where $\hat{a}$ and $\hat{b}$ are the annihilation operators of the two modes,
and $\theta$ which determines the transmissivity of the beam splitter $\tau=\mathrm{{cos}%
^{2}\theta}\in\lbrack0,1]$. The beam splitter is called balanced when $\tau=1/2$. In the Heisenberg picture, the
annihilation operators are transformed via the linear unitary Bogoliubov transformation
\begin{equation}
\left(
\begin{array}
[c]{c}%
\hat{a}\\
\hat{b}%
\end{array}
\right)  \rightarrow\left(
\begin{array}
[c]{cc}%
\sqrt{1-\tau} & \sqrt{\tau}\\
-\sqrt{\tau} & \sqrt{1-\tau}%
\end{array}
\right)  \left(
\begin{array}
[c]{c}%
\hat{a}\\
\hat{b}%
\end{array}
\right)  ~, \label{BS_anni}%
\end{equation}
and the quadrature operators $\mathbf{\hat{x}}:=(\hat{q}_{a},\hat{p}_{a}%
,\hat{q}_{b},\hat{p}_{b})^{T}$\ are transformed via the symplectic map
\begin{equation}\label{eq: beam splitter symplectic}
\mathbf{\hat{x}}\rightarrow\mathbf{B}(\tau)\mathbf{\hat{x}}~,~\mathbf{B(}%
\tau \mathbf{)}:=\left(
\begin{array}
[c]{cc}%
\sqrt{1-\tau}\mathbf{I} & \sqrt{\tau}\mathbf{I}\\
-\sqrt{\tau}\mathbf{I} & \sqrt{1-\tau}\mathbf{I}%
\end{array}
\right)  ~.
\end{equation}

\subsubsection{Two-mode squeezing and EPR states}

Pumping a nonlinear crystal in the non-degenerate OPA regime, we generate
pairs of photons in two different modes, known as the signal and the idler. This
process is described by an interaction Hamiltonian which contains the bilinear
term $\hat{a}^{\dagger}\hat{b}^{\dagger}$. The corresponding Gaussian unitary
is known as the \textit{two-mode squeezing operator} and is defined as
\begin{equation}
S_{2}(r)=\exp\left[  r(\hat{a}\hat{b}-\hat{a}^{\dagger}\hat{b}^{\dagger
})/2\right]  ~,
\end{equation}
where
$r$ quantifies the two-mode squeezing \cite{Bra04}. In the Heisenberg
picture, the quadratures $\mathbf{\hat{x}}:=(\hat{q}_{a},\hat{p}_{a},\hat
{q}_{b},\hat{p}_{b})^{T}$ undergo the symplectic map
\begin{equation}\label{eq: two mode sqz symplectic}
\mathbf{\hat{x}}\rightarrow\mathbf{S_{2}(}r\mathbf{)\hat{x}}~,~\mathbf{S_{2}%
(}r\mathbf{)}=\left(
\begin{array}
[c]{cc}%
\mathrm{cosh}r\mathrm{~}\mathbf{I} & \mathrm{sinh}r\mathrm{~}\mathbf{Z}\\
\mathrm{sinh}r{\mathrm{~}}\mathbf{Z} & \mathrm{cosh}r{\mathrm{~}}\mathbf{I}%
\end{array}
\right)  ~,
\end{equation}
where $\mathbf{I}$ is the identity matrix and $\mathbf{Z:}=\mathrm{diag}%
(1,-1)$. By applying $S_{2}(r)$ to a couple of vacua, we obtain the \textit{two-mode
squeezed vacuum state}, also known as an \textit{Einstein-Podolski-Rosen (EPR)
state} $\hat{\rho}^{epr}(r)=\left\vert r\right\rangle \left\langle r\right\vert
_{epr}$, where
\begin{equation}
\ket{r}_{epr}=\sqrt{1-\lambda^{2}}\sum_{n=0}^{\infty}({-\lambda})^{n}%
\ket{n}_a\ket{n}_b~,
\end{equation}
where $\lambda=\mathrm{tanhr}\in\lbrack0,1]$.
This is a Gaussian state with zero mean and covariance matrix
\begin{equation}
\mathbf{V}_{epr}=\left(
\begin{array}
[c]{cc}%
\nu\mathbf{I} & \sqrt{\nu^{2}-1}\mathbf{Z}\\
\sqrt{\nu^{2}-1}\mathbf{Z} & \nu\mathbf{I}%
\end{array}
\right):=\mathbf{V}_{epr}(\nu)  ~, \label{eq:EPR_CM}%
\end{equation}
where $\nu=\textrm{cosh}2r$ quantifies the noise-variance in the
quandratures (afterwards, we also use the notation $\left\vert \nu
\right\rangle _{epr}$). Using Eq.~(\ref{eq:EPR_CM}) one can easily check that%
\begin{equation}
V(\hat{q}_{-})=V(\hat{p}_{+})=e^{-2r}~,
\end{equation}
where $\hat{q}_{-}:=(\hat{q}_{a}-\hat{q}_{b})/\sqrt2$ and $\hat{p}_{+}:=(\hat{p}%
_{a}+\hat{p}_{b})/\sqrt2$. Note that for $r=0$, the EPR state corresponds to two vacua
and the previous variances are equal to 1, corresponding to the
quantum shot-noise. For every two-mode squeezing $r>0$, we have $V(\hat{q}%
_{-})=V(\hat{p}_{+})<1$, meaning that the correlations between the quadratures
of the two systems beat the quantum shot-noise. These correlations are known
as \textit{EPR correlations} and they imply the presence of bipartite
entanglement. In the limit of $r \rightarrow \infty$ we have an ideal EPR state with perfect correlations: $\hat{q}_a=\hat{q}_b$ and $\hat{p}_a=-\hat{p}_b$. Clearly, EPR correlations can also exist in the symmetric case for $\hat{q}_{+}$ and $\hat{p}_{-}$ using the replacement $\mathbf{Z}\rightarrow
-\mathbf{Z}$ in Eq.~(\ref{eq:EPR_CM}).

The EPR state is the most commonly used Gaussian
entangled state and has maximally-entangled
quadratures, given its average photon number. Besides the use of a non-degenerate parametric amplifier,
an alternative way to generate the EPR state is by combining two appropriately
rotated squeezed vacuum states (outputs of two degenerate OPAs) on a balanced
beam splitter~\cite{Bra04,Furusawa1998}. This passive generation
of entanglement from squeezing has been generalized by \textcite{Wolf2003}. When one considers Gaussian atomic processing, the same state can also be created using two atomic (macroscopic) objects as shown by \textcite{Julsgaard2001}. Finally, let us note the important
relation between the EPR state and the thermal state. By tracing out one of the
two modes of the EPR state, e.g., mode $b$, we get $\mathrm{Tr}_{b}[\hat{\rho}^{epr}(r)]=\hat{\rho}
_{a}^{th}(\bar{n})$, where $\bar{n}=\mathrm{sinh}^2r$. Thus, the
surviving mode is described by a thermal state, whose mean photon number is
related to the two-mode squeezing. Because of this, we also say that the
EPR\ state is the purification of the thermal state.

\subsection{Symplectic analysis for multimode Gaussian
States\label{Symplectic_SECTION}}

In this section we discuss the most powerful approach to studying Gaussian states
of multimode bosonic systems. This is based on the analysis and manipulation
of the second-order statistical moments, and its central tools are Williamson's theorem and the Euler decomposition.

\subsubsection{Thermal decomposition of Gaussian states\label{WilliamSection}}

According to Williamson's theorem, every positive-definite real matrix of even
dimension can be put in diagonal form by a symplectic transformation
\cite{Williamson1936}. In particular, this theorem can be applied to covariance matrices. Given an arbitrary $N$-mode covariance matrix $\mathbf{V}$, there exists a symplectic matrix
$\mathbf{S}$ such that%
\begin{equation}
\mathbf{V}=\mathbf{S\mathbf{V}^{\oplus}\mathbf{S}}^{T}~,~\mathbf{\mathbf{V}%
^{\oplus}:}=\bigoplus\limits_{k=1}^{N}\nu_{k}\mathbf{I}~, \label{Diag_SymplCM}%
\end{equation}
where the diagonal matrix $\mathbf{V}^{\oplus}$ is called the
Williamson form of $\mathbf{V}$, and the
$N$ positive quantities $\nu_{k}$ are called the symplectic
eigenvalues of $\mathbf{V}$. Here the symplectic spectrum $\{\nu_{k}\}_{k=1}^{N}$ can be
easily computed as the standard eigenspectrum of the matrix $|i\mathbf{\Omega
V}|$, where the modulus must be understood in the operatorial
sense. In fact, the matrix $i\mathbf{\Omega V}$ is Hermitian and is
therefore diagonalizable by a unitary transformation. Then, by taking the
modulus of its $2N$ real eigenvalues, one gets the $N$ symplectic eigenvalues
of $\mathbf{V}$. The symplectic spectrum is very important since it
provides powerful ways to express the fundamental properties of the
corresponding quantum state. For example, the uncertainty principle of
Eq.~(\ref{HeisCM}) is equivalent to
\begin{equation}
\mathbf{V}>0,~\mathbf{\mathbf{V}^{\oplus}}\geq\mathbf{I}~. \label{Heis_WILL}%
\end{equation}
In other words, a quantum covariance matrix must be positive definite and its symplectic eignevalues must satisfy $\nu_k \geq1$. Then, the von Neumann entropy $S(\hat{\rho})$ of a Gaussian state $\hat{\rho}$ can be
written as~\cite{Holevo1999}
\begin{equation}
S(\hat{\rho})=\sum\limits_{k=1}^{N}g(\nu_{k}), \label{VN_Gauss}%
\end{equation}
where
\begin{equation}
g(x):= \Big(\frac{x+1}{2}\Big) \log\Big(\frac{x+1}{2}\Big)-\Big(\frac{x-1}{2}\Big)\log\Big(\frac{x-1}{2}\Big).
\label{g_explicit}%
\end{equation}
In the space of density operators, the symplectic decomposition of
Eq.~(\ref{Diag_SymplCM}) corresponds to a thermal
decomposition for Gaussian states. In fact, let us consider
a zero-mean Gaussian state $\hat{\rho}(\mathbf{0},\mathbf{V})$. Because of
Eq.~(\ref{Diag_SymplCM}), there exists a canonical unitary $U_{\mathbf{S}}$
such that $\hat{\rho}(\mathbf{0,V})=U_{\mathbf{S}}\hat{\rho}(\mathbf{0,V}^{\oplus
})U_{\mathbf{S}}^{\dagger}$, where%
\begin{equation}
\hat{\rho}(\mathbf{0,V}^{\oplus})=\bigotimes\limits_{k=1}^{N}\hat{\rho}^{th}\left(
\tfrac{\nu_{k}-1}{2}\right)  \label{Thermal_product}%
\end{equation}
is a tensor-product of one-mode thermal states whose photon numbers are
provided by the symplectic spectrum $\{\nu_{k}\}$ of the original state. In
general, for an arbitrary Gaussian state $\hat{\rho}(\mathbf{\bar{x}},\mathbf{V})$
we can write the thermal decomposition%
\begin{equation}
\hat{\rho}(\mathbf{\bar{x}},\mathbf{V})=D(\mathbf{\bar{x}})U_{\mathbf{S}}\left[
\hat{\rho}(\mathbf{0,V}^{\oplus})\right]  U_{\mathbf{S}}^{\dagger}D(\mathbf{\bar{x}%
})^{\dagger}~. \label{Thermal_product_2}%
\end{equation}

Using the thermal decomposition of Eq.~(\ref{Thermal_product_2}) and the fact
that thermal states are purified by EPR states, we can derive a very simple
formula for the purification of an arbitrary Gaussian state~\cite{Holevo2001}. In fact, let us
denote by $A$ a system of $N$ modes described by a Gaussian state $\hat{\rho}
_{A}(\mathbf{\bar{x}},\mathbf{V})$, and introduce an additional reference
system $R$ of $N$ modes. Then, we have $\hat{\rho}_{A}(\mathbf{\bar{x}}%
,\mathbf{V})=\mathrm{Tr}_{R}[\hat{\rho}_{AR}(\mathbf{\bar{x}}^{\prime}%
,\mathbf{V}^{\prime})]$, where $\hat{\rho}_{AR}$ is a pure Gaussian state for the
composite system $AR$, having mean$\ \mathbf{\bar{x}}^{\prime}%
=\mathbf{(\mathbf{\bar{x},0})}^{T}$ and covariance matrix
\begin{equation}
\mathbf{V}^{\prime}=\left[
\begin{array}
[c]{cc}%
\mathbf{V} & \mathbf{SC}\\
\mathbf{C}^{T}\mathbf{S}^{T} & \mathbf{V}^{\oplus}%
\end{array}
\right]  ~,~\mathbf{C}:=\bigoplus_{k=1}^{N}\sqrt{\nu_{k}^{2}-1}\mathbf{Z}~.
\end{equation}

\subsubsection{Euler decomposition of canonical unitaries}\label{Euler Decomposition of Canonical Unitaries}

The canonical unitary $U_{\mathbf{S}}$ in Eq.~(\ref{Thermal_product_2}) can be
suitably decomposed using the Euler decomposition~\cite{Arvind1995}, alternatively known as the Bloch-Messiah reduction~\cite{Bra05}. First of
all, let us distinguish between active and passive canonical unitaries. By
definition, a canonical unitary $U_{\mathbf{S}}$\ is called passive (active)
if it is photon number preserving (non-preserving). A passive $U_{\mathbf{S}}$
corresponds to a symplectic matrix $\mathbf{S}$ which preserves the trace of
the covariance matrix, i.e., $\mathrm{Tr}(\mathbf{SVS}^{T})=\mathrm{Tr}(\mathbf{V})$ for any $\mathbf{V}$. This
happens when the symplectic matrix $\mathbf{S}$\ is orthogonal, i.e.,
$\mathbf{S}^{T}\mathbf{=S}^{-1}$. Passive canonical unitaries describe
multiport interferometers, e.g., the beam splitter in the case of two modes.
By contrast, active canonical unitaries correspond to symplectic matrices
which are not trace-preserving and, therefore, cannot be orthogonal. This is
the case of the one-mode squeezing matrix of Eq.~(\ref{CM_1mode_Squeeze}).
Arbitrary symplectic matrices contain both the previous elements. In fact,
every symplectic matrix $\mathbf{S}$ can be written as%
\begin{equation}
\mathbf{S}=\mathbf{K}\left[  \bigoplus\limits_{k=1}^{N}\mathbf{S}%
(r_{k})\right]  \mathbf{L}, \label{Euler}%
\end{equation}
where $\mathbf{K},\mathbf{L}$ are symplectic and orthogonal, while
$\mathbf{S}(r_{1})$, $\cdots$, $\mathbf{S}(r_{N})$ is a set of one-mode
squeezing matrices. Direct sums in phase space correspond to tensor products
in the state space. As a result, every canonical unitary $U_{\mathbf{S}}$\ can
be decomposed as
\begin{equation}
U_{\mathbf{S}}=U_{\mathbf{K}}\left[  \bigotimes_{k=1}^{N}S(r_{k})\right]
U_{\mathbf{L}}~, \label{Euler_unitary}%
\end{equation}
i.e., a multiport interferometer ($U_{\mathbf{L}}$), followed by a parallel
set of $N$ one-mode squeezers ($\otimes_{k}S(r_{k})$), followed by another
passive transformation ($U_{\mathbf{K}}$). Combining the thermal decomposition
of Eq.~(\ref{Thermal_product_2}) with the Euler decomposition of
Eq.~(\ref{Euler_unitary}), we see that an arbitrary multimode Gaussian state
$\hat{\rho}(\mathbf{\bar{x}},\mathbf{V})$ can be realized by preparing $N$ thermal
states $\hat{\rho}(\mathbf{0,V}^{\oplus})$, applying multimode interferometers
and one-mode squeezers according to Eq.~(\ref{Euler_unitary}), and finally
displacing them by $\mathbf{\bar{x}}$.

\subsubsection{Two-mode Gaussian states\label{Two_modes_SEC}}

Gaussian states of two bosonic modes ($N=2$) represent a remarkable case. They
are characterized by simple analytical formulas and represent the simplest
states for studying properties like quantum entanglement. Given a two-mode
Gaussian state $\hat{\rho}(\mathbf{\bar{x}},\mathbf{V})$, let us write its covariance matrix in the
block form
\begin{equation}
\mathbf{V}=\left(
\begin{array}
[c]{cc}%
\mathbf{A} & \mathbf{C}\\
\mathbf{C}^{T} & \mathbf{B}%
\end{array}
\right)  ~, \label{CM_2modes}%
\end{equation}
where $\mathbf{A}=\mathbf{A}^{T}$, $\mathbf{B}=\mathbf{B}^{T}$ and
$\mathbf{C}$ are $2\times2$ real matrices. Then, the Williamson form is simply
$\mathbf{V}^{\oplus}=(\nu_{-}\mathbf{I})\oplus(\nu_{+}\mathbf{I})$, where
symplectic spectrum $\{\nu_{-},\nu_{+}\}$ is provided by%
\begin{equation}
\nu_{\pm}=\sqrt{\frac{\Delta \pm\sqrt{\Delta^{2}%
-4\det\mathbf{V}}}{2}}~, \label{symple}%
\end{equation}
with $\Delta:=\det\mathbf{A}+\det\mathbf{B}+2\det\mathbf{C}$ and $\det$ is the determinant~\cite{Serafini2004}. In this case the uncertainty principle is equivalent to the bona-fide conditions~\cite{Serafini2006,Pirandola2009}
\begin{align}
\mathbf{V}>0, \hspace{1mm} \det\mathbf{V}\geq1 \hspace{2mm} \rm{and} \hspace{2mm} \Delta \leq 1+\det\mathbf{V}.
\end{align}
An important class of two-mode Gaussian states has covariance matrix in the
standard form~\cite{Duan2000,Simon2000}
\begin{equation}
\mathbf{V}=\left(
\begin{array}
[c]{cc}%
a\mathbf{I} & \mathbf{C}\\
\mathbf{C} & b\mathbf{I}%
\end{array}
\right)  ~,~\mathbf{C}=\left(
\begin{array}
[c]{cc}%
c_{1} & 0\\
0 & c_{2}%
\end{array}
\right)  ~,\label{V_simple}%
\end{equation}
where $a,b,c_{1},c_{2}\in\mathbb{R}$ must satisfy the previous bona-fide
conditions. In particular, for $c_{1}=-c_{2}:=c\geq0$, the symplectic
eigenvalues are simply $\nu_{\pm}=[\sqrt{y}\pm\left(  b-a\right)  ]/2$, where
$y:=(a+b)^{2}-4c^{2}$. In this case, we can also derive the matrix $\mathbf{S}$ realizing the symplectic decomposition $\mathbf{V}%
=\mathbf{S\mathbf{V}^{\oplus}S}^{T}$. This is given by the formula%
\begin{equation}
\mathbf{S}=\left(
\begin{array}
[c]{cc}%
\omega_{+}\mathbf{I} & \omega_{-}\mathbf{Z}\\
\omega_{-}\mathbf{Z} & \omega_{+}\mathbf{I}%
\end{array}
\right)  ,~\omega_{\pm}:=\sqrt{\frac{a+b\pm\sqrt{y}}{2\sqrt{y}}}~.
\label{Symplectic_expression}%
\end{equation}

\subsection{Entanglement in bipartite Gaussian states}

Entanglement is one of the most important properties of quantum mechanics,
being central in most quantum information protocols. To begin with let us consider two bosonic systems, $A$ with $N$ modes
and $B$ with $M$ modes, having Hilbert spaces $\mathcal{H}_{A}$ and
$\mathcal{H}_{B}$, respectively. The global bipartite system $A+B$ has a
Hilbert space $\mathcal{H}=\mathcal{H}_{A}\otimes\mathcal{H}_{B}$. By
definition, a quantum state $\hat{\rho}\in\mathcal{D}(\mathcal{H})$ is said to be
separable if it can be written as convex
combination of product states, i.e.,
\begin{equation}
\hat{\rho}=\sum\limits_{i}p_{i}~\hat{\rho}_{i}^{A}\otimes\hat{\rho}_{i}^{B},~~\hat{\rho}_{i}^{A(B)}%
\in\mathcal{D}(\mathcal{H}_{A(B)}), \label{separable_state_Definition}%
\end{equation}
where $p_{i}\geq0$ and $\sum_{i}p_{i}=1$. Note that the index can also be continuous. In such a case, the previous sum becomes an integral and the
probabilities are replaced by a probability density function. Physically, Eq.~(\ref{separable_state_Definition}) means that a
separable state can be prepared via local (quantum) operations
and classical communications (LOCCs). By definition, a state is called entangled when it is not separable, i.e., the
correlations between $A$ and $B$ are so strong that they cannot be created by any
strategy based on LOCCs. In entanglement theory there are two central questions to answer: ``Is the state entangled?",
and if the answer is yes, then ``how much entanglement does it have?". In what follows we review how we can answer
those two questions for Gaussian states.

\subsubsection{Separability}

As first shown by \textcite{Horodecki1996} and \textcite{Peres1996}, a key-tool for
studying separability is the partial transposition, i.e., the
transposition with respect to one of the two subsystems, e.g., system $B$. In
fact, if a quantum state $\hat{\rho}$ is separable, then its partial transpose
$\hat{\rho}^{T_{B}}$ is a valid density operator and in particular positive, i.e., $\hat{\rho}^{T_{B}}\geq0$.
Thus, the positivity of the partial
transpose represents a necessary condition for separability. On the other hand,
the non-positivity of the partial transpose
represents a sufficient condition for entanglement. Note that, in general, the positivity of the partial transpose is not a sufficient condition for
separability, since there exist entangled states with positive partial transpose. These states are bound
entangled meaning that their entanglement cannot be distilled into maximally entangled states~\cite{Horodecki1998,Horodecki2009}.

The partial transposition operation corresponds to a local time reversal~\cite{Horodecki1998}. For bosonic
systems the quadratures $\mathbf{\hat{x}}$ of the bipartite system $A+B$ undergo the transformation $\mathbf{\hat{x}}\rightarrow(\mathbf{I}%
_{A}\oplus\mathbf{T}_{B})\mathbf{\hat{x}}$, where $\mathbf{I}_{A}$ is the
$N$-mode identity matrix while $\mathbf{T}_{B}:=\oplus_{k=1}^{M}\mathbf{Z}$~\cite{Simon2000}.
Let us consider an arbitrary Gaussian state $\hat{\rho}(\mathbf{\bar{x}}%
,\mathbf{V})$ of the bipartite system $A+B$, also known as an $N\times M$
bipartite Gaussian state. Under the partial transposition operation, its covariance matrix is transformed via
the congruence
\begin{equation}
\mathbf{V\rightarrow(\mathbf{I}_{A}\oplus\mathbf{T}_{B})V(\mathbf{I}_{A}%
\oplus\mathbf{T}_{B}):=\tilde{V}}~. \label{V_tilde}%
\end{equation}
where the partially-transposed matrix $\mathbf{\tilde{V}}$ is positive definite.
If the state is separable, then
$\mathbf{\tilde{V}}$ satisfies
the uncertainty principle, i.e., $\mathbf{\tilde{V}}+i\mathbf{\Omega}\geq0$. Since $\mathbf{\tilde{V}}>0$,
this is equivalent to check the condition $\mathbf{\tilde{V}}^{\oplus}\geq\mathbf{I}$, where
$\mathbf{\tilde{V}}^{\oplus}$\ is the Williamson form of $\mathbf{\tilde{V}}$. This is
also equivalent to check $\tilde{\nu}_{-}\geq1$, where
$\tilde{\nu}_{-}$ is the minimum eigenvalue in the symplectic spectrum
$\{\tilde{\nu}_{k}\}$ of $\mathbf{\tilde{V}}$.

The satisfaction (violation) of
the condition $\tilde{\nu}_{-}\geq1$ corresponds to having
the positivity (non-positivity) of the partially
transposed Gaussian state. In some restricted situations, this positivity is
equivalent to separability. This happens for $1\times M$ Gaussian states
\cite{Werner2001}, and for a particular class of $N\times M$ Gaussian states
which are called bisymmetric~\cite{Serafini2005}. In general, the equivalence is not true, as shown already for $2\times2$ Gaussian states by \textcite{Werner2001}.
Finally, note that the partial transposition is not the only way to study separability. In
\cite{Duan2000} the authors constructed an inseparability criterion, generalizing the
EPR correlations, which gives a sufficient condition
for entanglement (also necessary for 1x1 Gaussian states). Two other useful techniques exist to fully characterize the separability of bipartite Gaussian states. The first
uses nonlinear maps as shown by~\textcite{Giedke2001b}, where the second reduces the separability
problem to a semi-definite program~\cite{Hyllus2006}.

\subsubsection{Entanglement measures}\label{sec: measures of entanglement}

In the case of pure $N\times M$ Gaussian states $\left\vert \varphi
\right\rangle $, the entanglement is provided by the
\textit{entropy of entanglement} $E_{V}(\left\vert \varphi\right\rangle )$.
This is defined as the von Neumann entropy of the reduced states $\hat{\rho}
_{A,B}=\mathrm{Tr}_{B,A}(\left\vert \varphi\right\rangle \left\langle
\varphi\right\vert )$, i.e., $E_{V}(\left\vert \varphi\right\rangle
)=S(\hat{\rho}_{A})=S(\hat{\rho}_{B})$ \cite{Bennett1996}, which can be easily calcuated
using Eq.~(\ref{VN_Gauss}).
The entropy of entanglement gives the amount of entangled qubits (measured in e-bits) that can be extracted from
the state together with the amount of entanglement needed to generate the state, i.e., distillation and
generation being reversible for pure states (in the asymptotic limit) \cite{Nielsen2000}.
Any bipartite pure Gaussian state can be mapped, using local Gaussian unitaries,
into a tensor product of EPR states of covariance matrix $\bigoplus_k \mathbf{V}_{epr}(\nu_k)$~\cite{Botero2003,Holevo2001}.
A LOCC mapping a Gaussian pure state to another one exists if and only if $ \nu_k \geq \nu_{k}'$ for all k,
where their respective $\nu_k$ and $\nu_k'$ are in descending order~\cite{Giedke2003b}.

Unfortunately, for mixed states we do not have a single definition of measure of entanglement
\cite{Horodecki2009}.
Different candidates exist, each one with its own operational interpretation.
Among the most well known is the \textit{entanglement of formation}~\cite{Bennett1996b},
\begin{equation}
E_{F}(\hat{\rho})=\min_{\{p_{k},\left\vert \varphi_{k}\right\rangle \}}\sum_{k}%
p_{k}E_{V}(\left\vert \varphi_{k}\right\rangle )~,
\end{equation}
where the minimization is taken over all the possible decompositions
$\hat{\rho}=\sum_{k}p_{k}\left\vert \varphi_{k}\right\rangle \left\langle
\varphi_{k}\right\vert $ (the sum becomes an integral for continuous decompositions). In general, this optimization is very difficult to
carry out. In continuous variables, we only know the solution for two-mode
symmetric Gaussian states~\cite{Giedke2003b}. These are two-mode Gaussian
states whose covariance matrix is symmetric under the permutation of the two modes, i.e.,
$\mathbf{A}=\mathbf{B}$ in Eq.~(\ref{CM_2modes}),
where $E_F(\hat{\rho})$ is then a function of $\tilde{\nu}_{-}$. Interestingly the optimal decomposition $\{p_{k},\left\vert \varphi
_{k}\right\rangle \}$ leading to this result is obtained from Gaussian states
$\left\vert \varphi_{k}\right\rangle $. This is conjectured to be true for any the Gaussian
state, i.e., the Gaussian entanglement of
formation $GE_{F}(\hat{\rho})$, defined by the minimization over Gaussian
decompositions satisfies $GE_{F}(\hat{\rho})=E_{F}(\hat{\rho})$~\cite{Wolf2004}.

The distillable entanglement $D(\hat{\rho})$ quantifies the amount of
entanglement that can be distilled
from a given mixed state $\hat{\rho}$ \cite{Horodecki2009}.
It is easy to see that $D(\hat{\rho})\leq E_F(\hat{\rho})$,
otherwise we could generate an infinite amount of entanglement
from finite resources, where for pure states we have
$D(\ket{\psi})=E_F(\ket{\psi})=E_V(\ket{\psi})$.
The entanglement distillation is also hard to calculate, as it
needs an optimization over all possible distillation protocols.
Little is known about $D(\hat{\rho})$ for Gaussian states, except trivial lower-bounds
given by the coherent information~\cite{Devetak2004} and its reverse counterpart~\cite{GarciaPatron2009}.
In \textcite{Giedke2001} it was shown that bipartite Gaussian states are
distillable if and only if they have a non-positive partial transpose.
However, the distillation of mixed
Gaussian states into pure Gaussian states is not possible using only Gaussian LOCC
operations \cite{Eisert2002,Fiurasek2002,Giedke2002}, but can be achieved using
non-Gaussian operations that map Gaussian states into Gaussian states
\cite{Bro03}, as recently demonstrated by \textcite{Takahashi2010}.

The two previous entanglement measures, i.e., $E_F(\hat{\rho})$ and $D(\hat{\rho})$,
are unfortunately very difficult to calculate in full generality.
However, a measure easy to compute is the logarithmic negativity~\cite{Vidal2002}
\begin{equation}
E_N(\hat{\rho})=\log||\hat{\rho}^{T_B}||_1
\end{equation}
which quantifies how much the state fails to
satisfy the positivity of the partial transpose condition.
For Gaussian states it reads
\begin{equation}
 E_N(\hat{\rho})=\sum_kF(\tilde{\nu}^k)
\end{equation}
where $F(x)=-\log(x)$ for $x<1$ and $F(x)=0$ for $x\geq1$ \cite{Vidal2002}.
It was shown to be an entanglement monotone~\cite{Eisert2001,Plenio2005}
and an upperbound of $D(\hat{\rho})$~\cite{Vidal2002}.
The logarithmic negativity of $1\times M$ and $N\times M$ bisymmetric Gaussian states was characterized by
\textcite{Adesso2007} and \textcite{Adesso2004}, respectively. Finally, we briefly mention that although the separability of a quantum state implies zero entanglement, other types of quantum correlations can exist for separable (non-entangled) mixed states. One measure of such correlations is the quantum discord and has recently been extended to Gaussian states~\cite{Adesso2010,Giorda2010}.

\subsection{Measuring Gaussian states}

A quantum measurement is described by a set of operators $\{E_{i}\}$
satisfying the completeness relation $\sum_{i}E_{i}^{\dagger}E_{i}=I$ where $I$ is the identity operator. Given
an input state $\hat{\rho}$, the outcome $i$ is found with probability
$p_{i}=\mathrm{Tr}(\hat{\rho} E_{i}^{\dagger}E_{i})$ and the state is
projected onto $\hat{\rho}_{i}=p_{i}^{-1}E_{i}\hat{\rho} E_{i}^{\dagger}$. If
we are only interested in the outcome of the measurement we can set $\Pi
_{i}:=E_{i}^{\dagger}E_{i}$ and describe the measurement as a positive
operator-valued measure (POVM).
In the case of continuous-variable systems, quantum measurements are often described by
continuous outcomes $i\in\mathbb{R}$, so that $p_i$ becomes a probability density. Here we define a measurement as being Gaussian when its application to Gaussian states
provides outcomes which are Gaussian-distributed. From a practical point of view, any Gaussian measurement can be accomplished using homodyne detection, linear optics (i.e., active and passive Gaussian unitaries), and Gaussian ancilla modes. A general property of a Gaussian measurement is the following: suppose a Gaussian measurement is made on $N$ modes of an $N+M$ Gaussian state where $N,M \geq 1$; then the classical
outcome from the measurement is a Gaussian distribution and the unmeasured $M$ modes
are left in a Gaussian state.

\subsubsection{Homodyne detection}

The most common Gaussian measurement in continuous-variable quantum
information is homodyne detection, consisting of the measurement of
the quadrature $\hat{q}$ (or $\hat{p}$) of a bosonic mode. Its measurement
operators are projectors over the quadrature basis $\left\vert q\right\rangle
\left\langle q\right\vert $ (or $\left\vert p\right\rangle \left\langle
p\right\vert $), i.e., infinitely squeezed states. The corresponding outcome $q$ (or $p$) has
a probability distribution $P(q)$ (or $P(p)$) which is given by the marginal
integral of the Wigner function over the conjugate quadrature, i.e.,
\begin{equation}
P(q)=\int W(q,p)dp, \hspace{2mm} P(p)=\int
W(q,p)dq.\label{eq:homodyneproability}%
\end{equation}
This can be generalized to the situation of partially homodyning a multimode
bosonic system by including the integration over both quadratures of the
non-measured modes. Experimentally a homodyne measurement is implemented by
combining the target quantum mode with a local oscillator into a balanced beam
splitter and measuring the intensity of the outgoing modes using two
photo-detectors. The subtraction of the signal of both photo-detectors gives a
signal proportional to $\hat{q}$~\cite{Bra04}. The $\hat{p}$ quadrature is
measured by applying a $\pi/2$ phase shift to the local oscillator. Corrections due to bandwidth effects or limited local oscillator power have also been addressed~\cite{Braunstein1990,Braunstein1991}. Homodyne detection is also a powerful tool in quantum tomography~\cite{Lvovsky2009a}. For
instance, by using a single homodyne detector, one can experimentally
reconstruct the covariance matrix of two-mode Gaussian states~\cite{Auria2009,Buono2010}. In tandem to well known homodyne measurements on light, homodyne measurements of the atomic Gaussian spin states via a quantum non-demolition measurement by light have also been developed. For example, the work of \textcite{Fernholz2008} demonstrated the quantum tomographic reconstruction of a spin squeezed state of the atomic ensemble.

\subsubsection{Heterodyne detection and Gaussian POVMs}

The quantum theory of heterodyne detection was established by
\textcite{Yuen1980} and is an important example of a Gaussian POVM.
Theoretically, heterodyne detection corresponds to a projection onto coherent states, i.e.,
$E(\alpha):=\pi^{-1/2}\left\vert \alpha\right\rangle \left\langle
\alpha\right\vert $. A heterodyne detector combines the measured bosonic mode
with a vacuum ancillary mode into a balanced beam splitter and homodynes the
quadratures $\hat{q}$ and $\hat{p}$ of the outcome modes. This approach can be generalized to any POVM composed of projectors over
pure Gaussian states. As shown by \textcite{Giedke2002} and
\textcite{Eisert2003}, such measurements can be decomposed into a Gaussian
unitary applied to the input system and extra ancillary (vacuum) modes followed by
homodyne measurements on all the output modes. Finally, a general noisy Gaussian
POVM is modeled as before but with part of the output modes traced out.

\subsubsection{Partial Gaussian measurement}

When processing a quantum system we are usually interested in measuring only
part of it (for example, subsystem $B$ which contains $1$ mode) in order to extract information
and continue processing the remaining part (say, subsystem $A$ with $N$
modes). Let us consider a Gaussian state for the global system $A+B$ where the covariance matrix is in block form similar to Eq.~(\ref{CM_2modes}) (but with $N+1$ modes). Measuring  the
$\hat{q}$ quadrature of subsystem $B$ transforms the
covariance matrix of subsystem $A$ as follows~\cite{Eisert2002,Fiurasek2002}
\begin{equation}
\mathbf{V}=\mathbf{A}-\mathbf{C}(\mathbf{\Pi}\mathbf{B}%
\mathbf{\Pi})^{-1}\mathbf{C}^{T}~, \label{eq: CM under homo}%
\end{equation}
where $\mathbf{\Pi}:=\mathrm{diag}(1,0)$ and $(\mathbf{\Pi}\mathbf{B}\mathbf{\Pi})^{-1}$ is a pseudoinverse since
$\mathbf{\Pi}\mathbf{B}\mathbf{\Pi}$ is singular. In particular, we have
$(\mathbf{\Pi}\mathbf{B}\mathbf{\Pi})^{-1}=B_{11}^{-1}\mathbf{\Pi}$, where
$B_{11}$ is the top-left element of $\mathbf{B}$. Note that the output covariance matrix
does not depend on the specific result of the measurement. This technique can be generalized to model
any partial Gaussian measurement, which consists of appending ancillary modes
to a system, applying a Gaussian unitary, and processing the output modes as
follows: part is homodyned, another part is discarded and the remaining part is
the output system. As an example, we can easily derive the effect on a
multi-mode subsystem $A$ after we heterodyne a single-mode subsystem $B$. By heterodyning the last mode, the first
$N$ modes are still in a Gaussian state, and the output covariance matrix is given by%
\begin{equation}
\mathbf{V}=\mathbf{A}-\mathbf{C}(\mathbf{B}+\mathbf{I}%
)^{-1}\mathbf{C}^{T}~, \label{CM_het_POVM}%
\end{equation}
or, equivalently, $\mathbf{V}=\mathbf{A}-\Theta^{-1}%
\mathbf{C(\boldsymbol{\omega}\mathbf{B}}\boldsymbol{\omega}^{T}%
\mathbf{\mathbf{+I)}C}^{T}$, where $\Theta:=\det\mathbf{B}+\mathrm{Tr}%
\mathbf{B}+1$, and $\mathbf{\boldsymbol{\omega}}$ is defined in
Eq.~(\ref{Symplectic_Form}).

\subsubsection{Counting and detecting photons}

Finally, there are two measurements, that despite being non-Gaussian, play an important
role in certain Gaussian quantum information protocols, e.g., distinguishability of Gaussian states, entanglement distillation and universal quantum computation. The first one is the von Neumann measurement in the
number state basis, i.e., $E_{n}:=\left\vert n\right\rangle \left\langle
n\right\vert $. The second one is the avalanche photo-diode that discriminates
between vacuum $E_{0}=\proj{0}$ and one or more photons $E_{1}%
=I-\proj{0}$. Realistic avalanche photo-diode detectors usually have small
efficiency, i.e., they detect only a small fraction of the impinging photons.
This is modeled theoretically by adding a beam splitter before an ideal
avalanche photo-diode detector, with transmissivity given by the efficiency of
the detector. Recent technological developments allow experimentalists to approach ideal
photon-counting capability for photon numbers of up to five to ten~\cite{Lita2008}.

\section{Distinguishability of Gaussian States}\label{chapter: distinguishability of gaussian states}

The laws of quantum information tell us that in general it is impossible to
perfectly distinguish between two non-orthogonal quantum states
\cite{Fuchs2000,Nielsen2000}. This limitation of quantum measurement theory
\cite{Helstrom1976} is inherent in a number of Gaussian quantum information
protocols including quantum cloning and the security of quantum cryptography.
Closely related to this is \textit{quantum state discrimination} which is
concerned with the distinguishability of quantum states. There are two
commonly used distinguishability techniques~\cite{A.Chefles2000,Bergou2004}:
(1) minimum error state discrimination, and (2) unambiguous state discrimination. In minimum error state discrimination, a number of approaches have been
developed which allows one to (imperfectly) distinguish between quantum states
provided we allow a certain amount of uncertainty or error in our measurement
results. On the other hand, unambiguous state discrimination, is an error-free
discrimination process but relies on the fact that sometimes the observer gets
an inconclusive result \cite{A.Chefles1998,Elk2002}. There also exists an
intermediate discrimination regime which allows for both errors and
inconclusive results~\cite{Che98,Fiurasek2003,Wit10}. Here we discuss
minimum error state discrimination which is more developed than unambiguous state discrimination
in the continuous-variable framework particularly in the case
of Gaussian states.

This section is structured as follows. In Sec.~\ref{sec:MeasDIST} we begin by
introducing some of the basic measures of distinguishability, such as the
Helstrom bound, the quantum Chernoff bound and the quantum fidelity. We give
their formulation for arbitrary quantum states, providing analytical formulas
in the specific case of Gaussian states. Then, in Sec.~\ref{Sec: distinguish coherent states}, we consider the most common
Gaussian discrimination protocol: distinguishing optical coherent
states with minimum error.

\subsection{Measures of distinguishability\label{sec:MeasDIST}}

\subsubsection{Helstrom bound}

Let us suppose that a quantum system is described by an unknown quantum state
$\hat{\rho}$\ which can take two possible forms, $\hat{\rho}_{0}$ or $\hat{\rho}_{1}$, with the
same probability (more generally, the problem can be formulated for
quantum states which are not equiprobable). For discriminating between
$\hat{\rho}_{0}$ and $\hat{\rho}_{1}$, we can apply an arbitrary quantum measurement to
the system. Without loss of generality, we can consider a dichotomic POVM
$\{\Pi_{0},\Pi_{1}:=I-\Pi_{0}\}$ whose outcome $u=0,1$
is a logical bit solving the discrimination. This happens up to an error
probability
\begin{equation}
p_{e}=\frac{p(u=0|\hat{\rho}=\hat{\rho}_{1})+p(u=1|\hat{\rho}=\hat{\rho}_{0})}{2}~,
\end{equation}
where $p(u|\hat{\rho})$ is the conditional probability of getting the outcome $u$
given the state $\hat{\rho}$. Then we ask: what is the minimum error probability we
can achieve by optimizing over the (dichotomic) POVMs? The answer to this
question is provided by the Helstrom bound \cite{Helstrom1976}. Helstrom
showed that an optimal POVM is given by $\Pi_{1}=P(\gamma_{+})$,
which is a projector onto the positive part $\gamma_{+}$ of the non-positive
operator $\gamma:=\hat{\rho}_{0}-\hat{\rho}_{1}$, known as the \textit{Helstrom matrix}.
As a result, the minimum error probability is equal to the
\textit{Helstrom bound}%
\begin{equation}
p_{e,\text{\textit{min}}}=\frac{1}{2}\left[  1-D(\hat{\rho}_{0},\hat{\rho}_{1})\right]  ~,
\label{P_err_Eq2}%
\end{equation}
where
\begin{equation}\label{eq: trace distance}
D(\hat{\rho}_{0},\hat{\rho}_{1}):=\frac{1}{2}\mathrm{Tr}\left\vert \hat{\rho}_{0}-\hat{\rho}_{1}\right\vert = \frac{1}{2} \sum |\lambda_j|,
\end{equation}
is the \textit{trace distance} between the two quantum states
\cite{Nielsen2000}. Here $\sum |\lambda_j|$ is the summation of the absolute values of the eigenvalues of the matrix $\hat{\rho}_{0}-\hat{\rho}_{1}$. In the case of two pure states, i.e., $\hat{\rho}_{0}=\left\vert
\psi_{0}\right\rangle \left\langle \psi_{0}\right\vert $ and $\hat{\rho}
_{1}=\left\vert \psi_{1}\right\rangle \left\langle \psi_{1}\right\vert $, the
Helstrom bound takes the simple form
\begin{equation}
p_{e,\text{\textit{min}}}=\frac{1}{2}\left(  1-\sqrt{1-|\left\langle \psi
_{0}\right\vert \psi_{1}\rangle|^{2}}\right)  ~. \label{eq: helstrom bound}%
\end{equation}

\subsubsection{Quantum Chernoff bound}

In general, deriving an analytical expression for the trace distance is not easy and, therefore,
the Helstrom bound is usually approximated by other distinguishability measures. One of the most recent is the \textit{quantum Chernoff
bound}~\cite{Audenaert2007,Audenaert2008,Calsamiglia2008,Nussbaum2009}. This is an upper bound
$p_{e,\text{\textit{min}}}\leq p_{QC}$, defined by
\begin{equation}
p_{QC}:=\frac{1}{2}\left(  \inf_{0\leq s\leq1}C_{s}\right),~C_{s}%
:=\mathrm{Tr}\left(  \hat{\rho}_{0}^{s}\hat{\rho}_{1}^{1-s}\right). \label{P_QC}%
\end{equation}
Note that the quantum Chernoff bound involves a minimization in s $\in [0,1]$.
In particular, we must use an infimum because of
possible discontinuities of $C_{s}$ at the border $s=0,1$. By ignoring the minimization and setting $s=1/2$, we derive a weaker but
easier-to-compute upper bound. This is known as the \textit{quantum
Bhattacharyya bound} \cite{Pirandola2008b}
\begin{equation}
p_{B}:=\frac{1}{2}\mathrm{Tr}\left(  \sqrt{\hat{\rho}_{0}}\sqrt{\hat{\rho}_{1}}\right)  ~.
\end{equation}

\paragraph{General formula for multimode Gaussian
states\label{sec:QCBformula}\\}

In the case of Gaussian states the quantum Chernoff
bound can be computed from the first two statistical moments.
A first formula, valid for single-mode Gaussian states, was shown by
\textcite{Calsamiglia2008}. Later, \textcite{Pirandola2008b} provided
a general formula for multimode Gaussian states, relating the quantum
Chernoff bound to the symplectic spectra (Williamson forms).
Here we review this general formula. Since it concerns the term $C_{s}$ in Eq.~(\ref{P_QC}), it also applies to the quantum Bhattacharyya bound.

First of all it is useful to introduce the two real functions%
\begin{equation}
G_{s}(x):=2^{s}\left[  \left(  x+1\right)  ^{s}-\left(  x-1\right)
^{s}\right]  ^{-1},
\end{equation}
and%
\begin{equation}
\Lambda_{s}(x):=\frac{\left(  x+1\right)  ^{s}+\left(  x-1\right)  ^{s}%
}{\left(  x+1\right)  ^{s}-\left(  x-1\right)  ^{s}}~,
\end{equation}
which are positive for $x\geq1$ and $s>0$. These functions can be computed
over a Williamson form $\mathbf{V}^{\oplus}$\textbf{\ }via the rule%
\begin{equation}
f(\mathbf{V}^{\oplus})=f\left(  \bigoplus\limits_{k=1}^{N}\nu_{k}%
\mathbf{I}\right)  =\bigoplus\limits_{k=1}^{N}f(\nu_{k})\mathbf{I~.}%
\end{equation}
Using these functions we can state the following result~\cite{Pirandola2008b}.
Let us consider two $N$-mode Gaussian states, $\hat{\rho}_{0}(\mathbf{\bar{x}}%
_{0},\mathbf{V}_{0})$ and $\hat{\rho}_{1}(\mathbf{\bar{x}}_{1},\mathbf{V}_{1})$,
whose covariance matrices have symplectic decompositions
\begin{equation}
\mathbf{V}_{0}=\mathbf{S}_{0}\mathbf{V}_{0}^{\oplus}\mathbf{S}_{0}%
^{T}~,~\mathbf{V}_{1}=\mathbf{S}_{1}\mathbf{V}_{1}^{\oplus}\mathbf{S}_{1}%
^{T}~\mathbf{.}%
\end{equation}
Then, for every $s\in(0,1)$, we can write the Gaussian formula
\begin{equation}
C_{s}=2^{N}\sqrt{\frac{\det\boldsymbol{\Pi}_{s}}{\det\boldsymbol{\Sigma}_{s}}%
}\exp\left(  -\frac{\mathbf{d}^{T}\boldsymbol{\Sigma}_{s}^{-1}\mathbf{d}}%
{2}\right)  ~,
\end{equation}
where $\mathbf{d}:=\mathbf{\bar{x}}_{0}-\mathbf{\bar{x}}_{1}$ and%
\begin{align}
\boldsymbol{\Pi}_{s}  &  :=G_{s}(\mathbf{V}_{0}^{\oplus})G_{1-s}%
(\mathbf{V}_{1}^{\oplus})~,\\
\boldsymbol{\Sigma}_{s}  &  :=\mathbf{S}_{0}\left[  \Lambda_{s}(\mathbf{V}%
_{0}^{\oplus})\right]  \mathbf{S}_{0}^{T}+\mathbf{S}_{1}\left[  \Lambda
_{1-s}(\mathbf{V}_{1}^{\oplus})\right]  \mathbf{S}_{1}^{T}.
\end{align}
In the previous formula, the matrix $\boldsymbol{\Pi}_{s}$ is diagonal and
very easy to compute, depending only on the symplectic spectra. In particular,
for pure states ($\mathbf{V}_{0}^{\oplus}=\mathbf{V}_{1}^{\oplus}=\mathbf{I}$)
we have $\boldsymbol{\Pi}_{s}=\mathbf{I}$. By contrast, the computation of
$\boldsymbol{\Sigma}_{s}$ is not straightforward due to the explicit presence of the
two symplectic matrices $\mathbf{S}_{0}$ and $\mathbf{S}_{1}$, whose
derivation may need non-trivial calculations in the general case (however see \ref{Two_modes_SEC}\ for two modes). If the
computation of $\mathbf{S}_{0}$ and $\mathbf{S}_{1}$\ is too difficult, one
possibility is to use weaker bounds which depend on the symplectic spectra
only, such as the \textit{Minkowski bound}~\cite{Pirandola2008b}.

\subsubsection{Quantum fidelity}

Further bounds can be constructed using the \textit{quantum fidelity}. In
quantum teleportation and quantum cloning, the fidelity $F$ is a commonly used
measure to compare the input state to the output state. Given two quantum
states, $\hat{\rho}_{0}$ and $\hat{\rho}_{1}$, their fidelity is defined by
\cite{Jozsa1994,Uhlmann1976}
\begin{equation}
F(\hat{\rho}_{0},\hat{\rho}_{1}):=\left[  \mathrm{Tr}\left(  \sqrt{\sqrt{\hat{\rho}_{0}}\hat{\rho}
_{1}\sqrt{\hat{\rho}_{0}}}\right)  \right]  ^{2}, \label{fid_definition}%
\end{equation}
which ranges from zero (for orthogonal states) to one (for identical
states).
In the specific case of two single-mode Gaussian states, $\hat{\rho}_{0}(\bar{x}%
_{0},\mathbf{V}_{0})$ and $\hat{\rho}_{1}(\bar{x}_{1},\mathbf{V}_{1})$, we have
\cite{Scutaru1998,Nha2005,Olivares2006,Holevo1975}
\begin{equation}
F(\hat{\rho}_{0},\hat{\rho}_{1})=\frac{2}{\sqrt{\Delta+\delta}-\sqrt{\delta}}\exp\left[
-\frac{1}{2}\mathbf{d}^{T}(\mathbf{V}_{0}+\mathbf{V}_{1})^{-1}\mathbf{d}%
\right]  , \label{Fidelity_pairGaussian}%
\end{equation}
where $\Delta:=\det(\mathbf{V}_{0}+\mathbf{V}_{1})$, $\delta:=(\det
\mathbf{V}_{0}-1)(\det\mathbf{V}_{1}-1)$ and $\mathbf{d}:=\bar{x}_{1}-\bar
{x}_{0}$. Using the fidelity, we can define the two fidelity bounds
\cite{Fuchs99a}%
\begin{equation}
F_{-}:=\tfrac{1}{2}\left[  1-\sqrt{1-F(\hat{\rho}_{0},\hat{\rho}_{1})}\right]
,~F_{+}:=\tfrac{1}{2}\sqrt{F(\hat{\rho}_{0},\hat{\rho}_{1})}~,
\end{equation}
which provide further estimates for the minimum error probability. In
particular, they satisfy the chain of inequalities%
\begin{equation}
F_{-}\leq p_{e,\text{\textit{min}}}\leq p_{QC}\leq p_{B}\leq F_{+}~.
\label{Bounds_Fidelities}%
\end{equation}

\subsubsection{Multicopy discrimination}

In general, let us assume that we have $M$ copies of the unknown quantum state
$\hat{\rho}$, which again can take the two possible forms, $\hat{\rho}_{0}$ or $\hat{\rho}_{1}$, with
the same probability. In other words, we have the two equiprobable hypotheses
\begin{align}
H_{0}  &  :\hat{\rho}^{\otimes M}=\hat{\rho}_{0}^{\otimes M}:=\overset{M}{\overbrace
{\hat{\rho}_{0}\otimes\cdots\otimes\hat{\rho}_{0}}}~,\\
H_{1}  &  :\hat{\rho}^{\otimes M}=\hat{\rho}_{1}^{\otimes M}:=\underset{M}{\underbrace
{\hat{\rho}_{1}\otimes\cdots\otimes\hat{\rho}_{1}}}~.
\end{align}
The optimal quantum measurement for discriminating the two cases is now a
collective measurement involving all the $M$ copies. This is the same
dichotomic POVM as before, now projecting on the positive part of the Helstrom
matrix $\gamma=\hat{\rho}_{0}^{\otimes M}-\hat{\rho}_{1}^{\otimes M}$. Correspondingly,
the Helstrom bound for the $M$-copy state discrimination takes the form
\begin{equation}
p_{e,\text{\textit{min}}}^{(M)}=\frac{1}{2}\left[  1-D\left(  \hat{\rho}
_{0}^{\otimes M},\hat{\rho}_{1}^{\otimes M}\right)  \right]  ~.
\end{equation}
This quantity is upper bounded by the general
$M$-copy expression of the quantum Chernoff bound, i.e.,%
\begin{equation}
p_{e,\text{\textit{min}}}^{(M)}\leq p_{QC}^{(M)}:=\frac{1}{2}\left(
\inf_{0\leq s\leq1}C_{s}\right)  ^{M}~.
\end{equation}
Interestingly, in the limit of many copies ($M\gg1$), the quantum Chernoff bound
is exponentially tight \cite{Audenaert2007}. This means that, for large $M$,
the two quantities $p_{e,\text{\textit{min}}}^{(M)}$ and $p_{QC}^{(M)}$ decay
exponentially with the same error-rate exponent, i.e.,%
\begin{equation}
p_{e,\text{\textit{min}}}^{(M)}\rightarrow\vartheta\exp(-M\kappa)~,~p_{QC}%
^{(M)}\rightarrow\upsilon\exp(-M\kappa)~,
\end{equation}
where $\vartheta \leq \upsilon$ and $\kappa$ is known as the quantum Chernoff
information~\cite{Calsamiglia2008}. Note that we can also consider other measures of
distinguishability, like the $M$-copy version of the quantum Bhattacharyya
bound
\begin{equation}
p_{QC}^{(M)}\leq p_{B}^{(M)}:=\frac{1}{2}\left[  \mathrm{Tr}\left(  \sqrt
{\hat{\rho}_{0}}\sqrt{\hat{\rho}_{1}}\right)  \right]  ^{M}~.
\end{equation}
However, even though it is easier to compute, it is not exponentially tight in
the general case.

\subsection{Distinguishing optical coherent states\label{Sec: distinguish coherent states}}

The distinguishing of coherent states with minimum error is one of the
fundamental tasks in optical communication theory. For example, we can consider a simple theoretical way of modeling current telecommunication systems by considering weak coherent states to send binary
information which has been encoded via the amplitude or phase modulation of a
laser beam\footnote{More specifically, fiber
communications currently employ in-line optical amplifiers, so the
states that are received are bathed in amplified spontaneous emission
noise and, moreover, received by direct detection.  Future fiber
systems -- in which bandwidth efficiency is being sought -- will go to
coherent detection, but they will use much larger than binary signal
constellations, i.e., quadrature amplitude modulation.  Laser communication from
space will use direct detection and $M$-ary pulse-position modulation
rather than binary modulation.}. Such states have small amplitudes and are largely overlapping (i.e., nonorthogonal) and hence the ability to successfully decode this classical
information is bounded by the minimum error given the Helstrom bound. Note that starting off with
orthogonal states might make more sense, however, if orthogonal
states were to be used their orthogonality is typically lost due to real world
imperfections such as energy dissipation and excess noise on the optical
fibre. By achieving the lowest error possible, the information
transfer rate between the sender and receiver can be maximized. We will now
illustrate a typical protocol involving the distinguishing of coherent states.

Suppose we have a sender, Alice, and a receiver, Bob. Alice prepares one of
two binary coherent states $\hat{\rho}_{0}$ and $\hat{\rho}_{1}$ where one may be encoded
as a logical ``0" and the other a logical ``1", respectively. These two states
form what is known as the \textit{alphabet} of possible states from which
Alice can choose to send and whose contents are also known by Bob.
Furthermore, the probabilities of each state being sent, $p_{0}$ and $p_{1}$,
are also known by Bob. Alice can decide to use either an amplitude modulation
keyed encoding, or a binary phase-shift keyed encoding, given respectively as,
\begin{align}
\ket{0} \hspace{2mm} \text{and} \hspace{2mm} \ket{2 \hspace{0.5mm} \alpha}, \hspace{4mm} \ket{\alpha} \hspace{2mm} \text{and} \hspace{2mm} \ket{-\alpha}.
\end{align}
We note that it is possible to transform between the two encoding schemes by
using a displacement, e.g., by applying the displacement operator $D(\alpha)$ to each of the two binary phase-shift keyed
coherent states we retrieve the amplitude modulation keyed encodings: $D(\alpha) \ket{-\alpha} =
\ket{0}$ and $D(\alpha) \ket{\alpha} = \ket{2 \hspace{0.5mm} \alpha}$. Bob's
goal is to decide with minimum error, which of the two coherent states he
received from Alice (over, for example, a quantum channel with no loss and no noise).
Bob's strategy is based on \textit{quantum hypothesis testing} in which he
devises two hypotheses: $H_{0}$ and $H_{1}$. Here $H_{0}$ corresponds to the
situation where $\hat{\rho}_{0}$ was sent whilst $H_{1}$ corresponds to $\hat{\rho}_{1}$
being sent. As mentioned earlier, the POVM that optimizes this decision
problem is actually a projective or von Neumann measurement, i.e., described
by the two operators $\Pi_{0}$ and $\Pi_{1}$, such that $\Pi_{i} \geq0$ for $i=0,1$ and $\Pi_{0} + \Pi_{1} = I$. Here
the measurement described by the operator $\Pi_{0}$ selects the state
$\hat{\rho}_{0}$ while $\Pi_{1} = I - \Pi_{0}$ selects $\hat{\rho}_{1}$.
The probability of error quantifies the probability in misinterpreting which
state was actually received by Bob and is given by
\begin{align}
p_{e}  &  = p_{0} \hspace{1mm} p(H_{1}|\hat{\rho}_{0}) + p_{1} \hspace{1mm} p(H_{0}|\hat{\rho}_{1}),
\end{align}
where $p(H_{i}|\hat{\rho}_{j})$ is defined as the conditional probability, i.e.,
probability that Bob decided it was hypothesis $H_{i}$ when in fact it was
$\hat{\rho}_{j}$, for $i \neq j$. The conditional probabilities can be written as
\begin{align}
p(H_{1}|\hat{\rho}_{0})    = \mathrm{tr} [\Pi_{1} \hat{\rho}_{0}], \hspace{4mm} p(H_{0}|\hat{\rho}_{1})    = \mathrm{tr} [\Pi_{0} \hat{\rho}_{1}].
\end{align}
Consequently, in the binary phase-shift keyed setting we can write the Helstrom bound as
\begin{align}
\label{eq: BPSK PE}p_{e} = p_{0} \bra{\alpha} \Pi_{1} \ket{\alpha} +
p_{1} \bra{-\alpha} \Pi_{0} \ket{-\alpha},
\end{align}
and for the amplitude modulation
keyed encoding
\begin{align}
\label{eq: AMK PE}p_{e} = p_{0} \bra{0} \Pi_{1} \ket{0} + p_{1}
\bra{2\alpha} \Pi_{0} \ket{2\alpha}.
\end{align}

The optimal type of measurement needed to achieve the Helstrom bound when
distinguishing between two coherent states was shown~\cite{Helstrom1976} to
correspond to a Schrodinger cat-state basis (i.e., a superposition of two
coherent states~\cite{Jeo05}): $\Pi_{0}=\ket{\psi}\bra{\psi}$ with
$\ket{\psi}=c_{0}(\gamma)\ket{0}+c_{1}(\gamma)\ket{\gamma}$ where the actual
weightings ($c_{1}$ and $c_{2}$) depend on the displacement $\gamma$. After
Helstrom introduced his error probability bound in 1968, it was not until 1973,
that two different physical models of implementing the receiver were discovered. The first
construction, by \textcite{Ken73}, involved building a receiver
based on direct detection (or photon counting) that was \textit{near-optimal},
i.e., an error probability that was larger than the optimal Helstrom bound.
However, building on Kennedy's initial proposal, \textcite{Dol73} discovered how
one could achieve the optimal bound using an adaptive feedback process with
photon counting. Over the years other researchers have continued
to make further progress in this area~\cite{Bon93,Osa96,Oli04,Ger04}.
Recently, Kennedy's original idea was improved upon with a receiver that was
much simpler to implement than Dolinar's (although still near-optimal) but
produced a smaller error probability than Kennedy's. Such a device is called
an optimized displacement receiver~\cite{Tak08,Wit08}. However, the simplest
possible receiver to implement is the conventional homodyne receiver, a common
element in optical communication which is also near-optimal outperforming the Kennedy receiver, albeit only for small
coherent amplitudes. We will now review each of these receivers in more detail.

\subsubsection{Kennedy receiver}

\textcite{Ken73} gave the first practical realization of a receiver with an error probability twice that of the Helstrom bound.
The Kennedy receiver distinguishes between the alphabet $\ket{\alpha}$ and
$\ket{-\alpha}$ by first displacing each of the coherent states by $\alpha$,
i.e., $\ket{-\alpha} \rightarrow\ket{0}$ and $\ket{\alpha} \rightarrow
\ket{2\alpha}$. Bob then measures the number of incoming photons between the
times $t=0$ and $t=T$ using direct photon counting, represented by the
operators
\begin{align}
\label{Eq: KR projectors}\Pi_{1} = \ket{0}\bra{0} \hspace{2mm}
\text{and} \hspace{2mm} \Pi_{2} = I -\ket{0}\bra{0}.
\end{align}
If the number of photons detected during this time is zero then $\ket{0}$ is
chosen (as the vacuum contains no photons) otherwise it is assumed to have
been $\ket{2 \alpha}$. Hence, the Kennedy receiver always
chooses $\ket{0}$ correctly (ignoring experimental imperfections), where the
error in the decision results from the vacuum fluctuations in $\ket{2 \alpha}$
(as any coherent state has some finite overlap with the vacuum state). Using
Eq.~(\ref{eq: AMK PE}), where from now on we use the least classical
probability situation of $p_{1}=p_{2}=1/2$, the error probability is given by
$p_{e}^{k} = \frac{1}{2} \bra{2 \alpha} \Pi_{1} \ket{2 \alpha}$
which is equal to
\begin{align}
p_{e}^{k} = \frac{1}{2} \exp(-4 |\alpha|^{2}).
\end{align}
where we have used Eq.~(\ref{eq: overlap coherent state}). The above error
bound is sometimes known as the \textit{shot-noise error}.

\subsubsection{Dolinar receiver}

\textcite{Dol73} built upon the results of Kennedy by constructing a physical scheme
that saturates the Helstrom bound. Using Eq.~(\ref{eq: helstrom bound}) with
Eq.~(\ref{eq: overlap coherent state}), the Helstrom bound for two pure
coherent states $\ket{\alpha}$ and $\ket{-\alpha}$ is given by
\[
p_{e,min}=\frac{1}{2}(1-\sqrt{1-\exp(-4|\alpha|^{2})}).
\]
This is the lowest possible error in distinguishing between two pure coherent
states. Dolinar's scheme combined photon counting with real-time
quantum feedback. Here the incoming coherent signal is combined on a beam
splitter with a local oscillator whose amplitude is causally dependent on the
number of photons detected in the signal beam. Such an adaptive process is
continually repeated throughout the duration of the signal length where a
decision is made based on the parity of the final number of photons
detected~\cite{Helstrom1976,Ger04,Tak05}. Many years after Dolinar's proposal,
other approaches, such as using a highly nonlinear unitary
operation~\cite{Sas96} or fast feedforward~\cite{Tak05}, have also achieved
the Helstrom bound by approximating the required Schrodinger cat state
measurement basis (the actual creation of such a basis is experimentally very
difficult~\cite{Ourjoumtsev2007}). However, an experimental implementation of Dolinar's original
approach was recently demonstrated in a proof-of-principle
experiment~\cite{Coo07}.

\subsubsection{Homodyne receiver}

As its name suggests the homodyne receiver uses a homodyne detector to
distinguish between the coherent states $\ket{\alpha}$ and $\ket{-\alpha}$.
Such a setup is considered the simplest setup possible and unlike the other
receivers relies only on Gaussian operations. The POVMs for the homodyne
receiver are modeled by the projectors
\begin{align}
\label{eq: homodyne projectors}\Pi_{1} = \int_{0}^{\infty} dx \ket{x}
\bra{x} \hspace{3mm} \text{and} \hspace{3mm} \Pi_{2} = I -
\Pi_{1},
\end{align}
where a positive (negative) outcome is obtained identifying $\ket{\alpha}$
($\ket{-\alpha}$). It was proven by \textcite{Tak08} that the simple homodyne
detector is optimal among all available Gaussian measurements.
In fact,
for weak coherent states (amplitudes $|\alpha|^{2} <0.4$), the homodyne
receiver is near-optimal and has a lower error probability than the Kennedy
receiver. Such a regime corresponds to various quantum communication protocols
as well as deep space communication. Using Eq.~(\ref{eq: BPSK PE}) with the
projectors from Eq.~(\ref{eq: homodyne projectors}) and the fact that
$|\bra{-\alpha} x\rangle|^{2} = \pi^{-1/2} \exp[-(x+|\alpha|/2)^{2}]$, the
error probability for the homodyne receiver is given
by~\cite{Oli04,Tak08}
\begin{align}
p_{e}^{h} = \frac{1}{2} \Big(1- \mathrm{erf}\Big[|\alpha|/2\Big]\Big),
\end{align}
where $\mathrm{erf}[\cdot]$ is the error function. This limit is known as the
\textit{homodyne limit}.

\subsubsection{Optimized displacement receiver}

The optimized displacement receiver~\cite{Tak08} is a modification of
the Kennedy receiver where instead of displacing $\ket{\alpha}$ and
$\ket{-\alpha}$ by $\alpha$, both are now displaced by an optimized value
$\beta$, where $\alpha,\beta\in\mathbb{R}$. This displacement $D(\beta)$ is
based on optimizing both terms in the error probability of
Eq.~(\ref{eq: BPSK PE}). When considering the Kennedy receiver, only the $p_{1}
\bra{-\alpha} \Pi_{1} \ket{-\alpha}$ term is minimized. However, the
optimized displacement receiver, is based on optimizing the sum of the two
probabilities as a function of the displacement $\beta$. The signal states
$\ket{\pm \alpha}$ are now displaced by $\beta$ according to
\begin{align}
\label{eq: ODR new coherent states}\ket{\pm \alpha} \rightarrow
\ket{\pm \sqrt{\tau} \alpha + \beta},
\end{align}
for a transmission $\tau$ in the limit of $\tau\rightarrow1$. As with the Kennedy
receiver photon detection is used to detect the incoming states and is
described by the projectors given in Eq.~(\ref{Eq: KR projectors}). Using
Eqs.~(\ref{eq: BPSK PE}) and (\ref{eq: overlap coherent state}) but with the
coherent states now given by Eq.~(\ref{eq: ODR new coherent states}), the
error probability can be expressed as
\begin{align}
p_{e}^{\beta} = \frac{1}{2} - \exp[-(\tau |\alpha|^{2} + |\beta|^{2})] \sinh(2
\sqrt{\tau} \alpha\beta).
\end{align}
The optimized displacement receiver outperforms both the homodyne receiver and
the Kennedy receiver for all values of $\alpha$. It is interesting to note that such a receiver
has applications in quantum cryptography where it has been shown to increase the  secret-key rates
of certain protocols~\cite{Wit10,Wit10b}. Furthermore, by including squeezing with the displacement, an improvement in the performance of the receiver can be achieved~\cite{Tak08}. The optimized displacement receiver has also been demonstrated experimentally~\cite{Tsujino2011,Wit08}.

To summarize, in terms of performance, the hierarchy for the above mentioned
receivers is the following: (1) Dolinar receiver, (2) optimized displacement
receiver, (3) Kennedy receiver and (4) homodyne receiver. Again, out of the ones mentioned, the Dolinar
receiver is the only one that is optimal.
Furthermore, the Kennedy receiver has a lower error probability than the
homodyne receiver for most values of amplitude. Finally, we point out that our discussions of binary receivers (photon counters) presumes unity quantum efficiency with no dark noise or thermal noise, and hence paints an ideal theoretical comparison
between all of the mentioned receivers.

\section{Examples of Gaussian Quantum Protocols}\label{Chapter basic quantum protocols}

\subsection{Quantum teleportation and variants}

Quantum teleportation is one of the most beautiful protocols in quantum
information. Originally developed for qubits~\cite{Bennett1993}, it was
later extended to continuous-variable systems~\cite{Vaidman1994,Braunstein1998,Ral98},
where coherent states are teleported via the
EPR correlations shared by two distant parties. It has also been demonstrated experimentally~\cite{Furusawa1998,Bowen2003,Zhang2003}. Here we review the quantum
teleportation protocol for Gaussian states using the formalism of
\cite{Chizhov2002,Fiurasek2002b,Pirandola2006b}.

Two parties, say Alice and Bob, possess two
modes, $a$ and $b$, prepared in a zero-mean Gaussian state $\hat{\rho}
(\mathbf{0},\mathbf{V})$ whose covariance matrix $\mathbf{V}$ can be written in the
($\mathbf{A,B,C}$)-block form of Eq.~(\ref{CM_2modes}). This state can be seen
as a virtual channel that Alice can exploit
to transfer an input state to Bob. In principle the input state can be completely arbitrary. In practical applications she will typically pick her state from some previously agreed alphabet. Consider the case in which she wishes to transfer a Gaussian state $\hat{\rho}_{in}(\mathbf{\bar{x}}_{in}%
,\mathbf{V}_{in})$, with fixed covariance matrix $\mathbf{V}_{in}$ but unknown mean
$\mathbf{\bar{x}}_{in}$ (chosen from a Gaussian distribution), from her input mode $in$ to Bob. To accomplish this task, Alice must destroy her
state $\hat{\rho}_{in}$ by combining modes $in$ and $a$ in a joint Gaussian measurement,
known as a \textit{Bell measurement}, where Alice mixes
$in$ and $a$ in a balanced beam splitter and homodynes the output modes,
$-$ and $+$ by measuring $\hat{q}_{-}$ and $\hat{p}_{+}$,
respectively. The outcome of the measurement $\gamma:=(q_{-}+ip_{+})/2$ is
then communicated to Bob via a standard telecom line. Once he receives this
information, Bob can reconstruct Alice's input state by applying a
displacement $D(\gamma)$ on his mode $b$,
which outputs a Gaussian state $\hat{\rho}_{out}\simeq\rho_{in}$. The performance of the protocol is expressed by the teleportation fidelity $F
$. This is the fidelity between the input and the output states averaged over all the
outcomes of the Bell measurement. Assuming pure Gaussian states as input, one
has \cite{Fiurasek2002b}%
\begin{equation}
F=\frac{2}{\sqrt{\det\boldsymbol{\Gamma}}},~\boldsymbol{\Gamma}:=2\mathbf{V}%
_{in}+\mathbf{ZAZ}+\mathbf{B}-\mathbf{ZC}-\mathbf{C}^{T}\mathbf{Z}^{T}.
\end{equation}
where again $\mathbf{Z:}=\mathrm{diag}%
(1,-1)$. This formula can be generalized to virtual channels $\hat{\rho}(\mathbf{\bar{x}%
},\mathbf{V})$ with arbitrary mean $\mathbf{\bar{x}}=(\bar{q}_{a},\bar{p}%
_{a},\bar{q}_{b},\bar{p}_{b})^{T}$. This is possible if Bob performs the
modified displacement $D(\gamma+\tilde{\gamma})$, where $\tilde{\gamma
}:=[(\bar{q}_{b}-\bar{q}_{a})-i(\bar{p}_{b}+\bar{p}_{a})]/2\sqrt{2}$
\cite{Pirandola2006b}.

In order to be truly quantum, the teleportation must have a fidelity above a
classical threshold $F_{class}$. This value corresponds to the classical
protocol where Alice measures her states, communicates the results to Bob who,
in turn, reconstructs the states from the classical information. In general, a
necessary condition for having $F>F_{class}$ is the presence of entanglement
in the virtual channel. For bosonic systems, this is usually assured by the
presence of EPR\ correlations. For instance, let us consider the case where
the input states are coherent states chosen from a broad Gaussian distribution and the virtual channel is an EPR state
$\hat{\rho}^{epr}(r)$. In this case, the teleportation fidelity is simply given by~\cite{Furusawa1998,Adesso2005,Mari2008}
\begin{equation}
F=(1+\tilde{\nu}_{-})^{-1}~,~~\tilde{\nu}_{-}=\exp(-2\left\vert r\right\vert
)~. \label{Fid_Ent_relation}%
\end{equation}
Here the presence of EPR correlations ($r>0$) guarantees the presence of
entanglement ($\tilde{\nu}_{-}<1$) and, correspondingly, one has $F>1/2$,
i.e., the fidelity beats the classical threshold for coherent states
\cite{Braunstein2001,Hammerer2005}. A more stringent threshold for teleportation is to require that the quantum
correlations between the input field and the teleported field are retained~\cite{Ral98}. In turn this implies that the teleported field is the best copy of the input allowed by the no-cloning bound~\cite{Grosshans2001}. At unity gain this requires that $\tilde{\nu}_-<1/2$ and corresponds to a coherent state fidelity $F>2/3$ as was first demonstrated by \textcite{Takei2005}.

In continuous variables, the protocol of quantum teleportation has been
extended in several ways, including number-phase teleportation~\cite{Milburn1999}, all-optical teleportation~\cite{Ralph1999a}, quantum teleportation networks
\cite{Loock2000b}, teleportation of single photon states~\cite{Ide2001,Ralph2001}, quantum telecloning~\cite{Loock2001}, quantum gate teleportation~\cite{Bartlett2003}, assisted quantum
teleportation~\cite{Pirandola2005}, quantum teleportation games
\cite{Pirandola2005b}, and teleportation channels~\cite{Wolf2007b}. One of the most important variants of the protocol is
the teleportation of entanglement also known as \textit{entanglement
swapping}~\cite{Loock1999,Jia2004,Takei2005}. Here Alice and Bob possess two entangled states, $\hat{\rho}_{aa^{\prime}}$\ and
$\hat{\rho}_{bb^{\prime}}$, respectively. Alice keeps mode $a$ and sends mode
$a^{\prime}$ to a Bell measurement, while Bob keeps mode $b$ and sends
$b^{\prime}$. Once $a^{\prime}$ and $b^{\prime}$ are measured and the
outcome communicated, Alice and Bob will share an output state $\hat{\rho}_{ab}$,
where $a$ and $b$ are entangled. For simplicity, let us suppose that Alice's
and Bob's initial states are EPR\ states, i.e., $\hat{\rho}_{aa^{\prime}}%
=\hat{\rho}_{bb^{\prime}}=\hat{\rho}^{epr}(r)$. Using the input-output relations given in \textcite{Pirandola2006}, one can easily check that the output Gaussian state
$\hat{\rho}_{ab}$ has logarithmic negativity $E_{N}(\hat{\rho}_{ab})=\ln\cosh(2r)$,
corresponding to entanglement for every $r>0$. By generalizing to two-mode entanglement, polarization entanglement can be swapped and shown to still violate a Bell inequality for $r>0$~\cite{Polkinghorne1999}.

Teleportation and entanglement swapping are protocols which may involve
bosonic systems of different nature. For example, in \textcite{Sherson2006} a
quantum state was teleported from an optical mode onto a macroscopic
object consisting of an atomic ensemble of about $10^{12}$ Caesium atoms.
Theoretically, this kind of result can also be realized by using radiation
pressure. In fact, by impinging a strong monochromatic laser beam onto a highly
reflecting mirror, it is possible to generate a scattering process where an
optical mode becomes entangled with an acoustic (massive) mode excited over
the surface of the mirror. Exploiting this hybrid entanglement, the
teleportation from an optical to an acoustic mode is possible in principle
\cite{Mancini2003,Pirandola2003}, as well as the generation of entanglement
between two acoustic modes by means of entanglement swapping
\cite{Pirandola2006}.

\subsection{Quantum cloning}

Following the seminal works of \textcite{Woo82} and \textcite{Dieks82}, it is well known that a quantum transformation that outputs two perfect copies of an arbitrary input state
$\ket{\psi}$ is precluded by the laws of quantum mechanics. This is the content of the celebrated {\it quantum no-cloning theorem}. More precisely, perfect cloning is possible, if and only if, the input state is drawn from a set of orthogonal states. Then, a simple von Neumann measurement enables the perfect discrimination of the states~(see Sec.~\ref{chapter: distinguishability of gaussian states}), which in turn enables the preparation of exact copies of the measured state. In contrast,
if the input state is drawn from a set of non-orthogonal states, perfect cloning is impossible. A notable
example of this are coherent states which cannot be perfectly distinguished nor cloned as a result of Eq.~(\ref{eq: overlap coherent state}).
Interestingly, although perfect cloning is forbidden, one can devise approximate {\it cloning machines}, which produce imperfect copies of the original state. The concept of a cloning machine was introduced by \textcite{Buzek96}, where the cloning machine produced two identical and optimal clones
of an arbitrary single qubit. Their work launched a whole new field
of investigation~\cite{GisinCloning05,CerfCloning06}.
Cloning machines are intimately related to quantum cryptography (see Sec.~\ref{chap:QKD}) as they usually constitute
the optimal attack against a given protocol, so that finding the best cloning machine is crucial
to address the security of a quantum cryptographic protocol~\cite{Cerf2007}.

The extension of quantum cloning to continuous-variable systems was first carried out
 by~\textcite{Cerf2000} and \textcite{Lindblad2000}, where a Gaussian cloning machine was shown to produce two noisy
copies of an arbitrary coherent state (where the figure of merit here is the single-clone excess noise
variance). The input mode, described by the quadratures $(\hat{q}_{in},\hat{p}_{in})$, is transformed
into two noisy clones $(\hat{q}_{1(2)},\hat{p}_{1(2)})$ according to
\begin{eqnarray}
\hat{q}_{1(2)}=\hat{q}_{in}+\hat{N}_{q1(2)}, \hspace{3mm}
\hat{p}_{1(2)}=\hat{p}_{in}+\hat{N}_{p1(2)},
\end{eqnarray}
where $\hat{N}_{q1(2)}$ and $\hat{N}_{p1(2)}$ stand for the added noise operators
on the output mode 1 (2). We may impose $\langle \hat{N}_{q1(2)} \rangle = \langle \hat{N}_{p1(2)} \rangle =0 $, so that the mean values of the output quadratures coincide with those of the original state. It is the variance of the
added noise operators which translates the cloning imperfection: a generalized uncertainty relation
for the added noise operators can be derived \cite{Cerf2000,Cerf-chapter},
\begin{eqnarray}\label{ineqcloning}
\Delta\hat{N}_{q1}\Delta\hat{N}_{p2} \geq 1, \hspace{2mm}
\Delta\hat{N}_{p1}\Delta\hat{N}_{q2}\geq 1,
\end{eqnarray}
which is saturated (i.e., lower bounded) by this cloning machine. The above inequalities clearly imply
that it is impossible to have two clones with simultaneously vanishing noise in the two canonically conjugate quadratures. This can be straightforwardly linked to the impossibility of simultaneously measuring perfectly the two canonically conjugate quadratures of the input mode: if we measure $\hat{q}$
on the first clone and $\hat{p}$ on the second clone, the cloning machine actually produces
the exact amount of noise that is necessary to prevent this procedure from beating the optimal
(heterodyne) measurement~\cite{Lindblad2000}.

The Gaussian cloning machine was first derived in the quantum circuit language \cite{Cerf2000},
which may, for example, be useful for the cloning of light states onto atomic ensembles \cite{Fiurasek2004}. However, an optical version was later developed by  \textcite{Braunstein2001b} and \textcite{Fiurasek2001},  which is better suited for our purposes here. The cloning machine can be realized with a linear phase-insensitive amplifier of intensity gain two, followed by a balanced beam splitter. The two clones are then found in the two output ports of the beam splitter, while an anti-clone is found in the idler output
of the amplifier. The anti-clone is defined as an imperfect version of the phase-conjugate $\ket{\alpha^*}$
of the input state $\ket{\alpha}$, where $\alpha = (q+ip)/2$ and $\alpha^* = (q-ip)/2$. The symplectic transformation on the quadrature operators $\mathbf{\hat x}= (\hat{q}_1,\hat{p}_1,\hat{q}_2,\hat{p}_2,\hat{q}_3,\hat{p}_3)^T$ of the three input modes reads
\begin{equation}
\mathbf{\hat x} \to \mathbf{C} \mathbf{\hat x} , \qquad
\mathbf{C} = ( \mathbf{B} \oplus \mathbf{I} ) ( \mathbf{I} \oplus \mathbf{S}_2 )
\end{equation}
where $\mathbf{B}$ is the symplectic map of a beam splitter with transmittance $\tau=1/2$
as defined in Eq.~(\ref{eq: beam splitter symplectic}), $\mathbf{S}_2$ is the symplectic map of a two-mode squeezer
with intensity gain $\cosh^2 r=2$ as defined in Eq.~(\ref{eq: two mode sqz symplectic}), and $\mathbf{I}$ is a
$2\times 2$ identity matrix. The input mode of the cloner is the signal mode of the amplifier (mode 2),
while the idler mode of the amplifier (mode 3) and the second input mode of the beam splitter (mode 1)
are both prepared in the vacuum state. At the output, modes 1 and 2 carry the two clones,
while mode 3 carries the anti-clone. By reordering the three $\hat{q}$ quadratures before
the three $\hat{p}$ quadratures, we can express the cloning symplectic map as
\begin{equation}
\mathbf{C} =
\left(
\begin{array}{ccc}
2^{-1/2} & 1& 2^{-1/2} \\
-2^{-1/2} & 1& 2^{-1/2} \\
0 & 1&  2^{1/2}
\end{array}
\right)
\oplus
\left(
\begin{array}{ccc}
2^{-1/2} & 1& -2^{-1/2} \\
-2^{-1/2} & 1& -2^{-1/2} \\
0 & -1&  2^{1/2}
\end{array}
\right)
\label{eq-cloning}
\end{equation}
The second columns of the $\hat{q}$ and $\hat{p}$ blocks immediately imply that the two clones are centered
on the input state $(\hat{q}_2,\hat{p}_2)$, while the anti-clone is centered on the phase conjugate
of the input state $(\hat{q}_2,-\hat{p}_2)$.
We can also check that the covariance matrix of the output modes can be expressed as
\begin{equation}
\mathbf{V}_1 = \mathbf{V}_2 = \mathbf{V}_{in} +  \mathbf{I} , \qquad
\mathbf{V}_3 = \mathbf{Z} \mathbf{V}_{in} \mathbf{Z} +  2 \mathbf{I}
\end{equation}
where $\mathbf{V}_{in}$ is the covariance matrix of the input mode (mode 2) and again $\mathbf{Z}= {\rm diag}(1,-1)$.
Thus, the two clones suffer exactly one unit of additional shot-noise, while
the anti-clone suffers two shot-noise units. This can be expressed in terms of the
cloning fidelities of Eq.~(\ref{Fidelity_pairGaussian}). The fidelity of each of the clones
is given by $F=2/3$, regardless of which coherent state is cloned. The anti-clone is noisier, and characterized by a fidelity of $F=1/2$. Note that this latter fidelity is precisely that
of an optimal joint measurement of the two conjugate quadratures~\cite{Arthurs1965},
so that optimal (imperfect) phase conjugation can be classically achieved by heterodyning the state and preparing its phase-conjugate~\cite{Cerf2001b}.

A variant of this optical cloner was demonstrated experimentally by \textcite{Andersen2005},
where the amplifier was replaced by a feed forward optical scheme which only requires
linear optical components and homodyne detection~\cite{Lam1997}. A fraction of the signal beam
is tapped off and measured using heterodyne detection. The outcomes of this measurement are then used
to apply an appropriate displacement to the remaining part of the signal beam. This setup
demonstrates near optimal quantum noise limited performances, and can also be adapted
to produce a phase-conjugate output~\cite{Josse2006}. This 1-to-2 Gaussian cloner can be straightforwardly extended to a more general setting,
where $M$ identical clones are produced from $N$ identical replica of an unknown
coherent state with a fidelity $F = MN/(MN+M-N)$~\cite{Cerf2000a}.
More generally, one can add $N'$ replica of the phase-conjugate state at the input
and produce $M'=M+N'-N$ anti-clones~\cite{Cerf2001a}. In this more elaborate scheme,
the signal mode carries all inputs and clones, while the idler mode
carries all phase-conjugate inputs and anti-clones. Interestingly, for a fixed total number of inputs
$N+N'$ the clones have a higher fidelity if $N'>0$, a property which holds regardless of $M$
and even survives at the limit of a measurement $M\to \infty$. So the cloning or measurement
performances are enhanced by phase-conjugate inputs. For example,
the precision of measuring the quadratures of two phase-conjugate states
$\ket{\alpha}\ket{\alpha^*}$ is as high as that achieved when measuring four identical states
$\ket{\alpha}^{\otimes 4}$ though half of the mean energy is needed,
as experimentally demonstrated by \textcite{Niset2007}. Furthermore, the cloning of phase-conjugate
coherent states was suggested by \textcite{Chen2007} and also suggested, as well as demonstrated, by \textcite{Sabuncu2007} using the linear cloner of \textcite{Andersen2005}.

Gaussian cloners have also been theoretically devised in an asymmetric setting, where
the clones have different fidelities~\cite{Fiurasek2001}. The way to achieve asymmetry
 is to use an additional beam splitter that deflects a fraction of the input beam
before entering the signal mode of the amplifier. This deflected beam bypasses the amplifier
and feeds the vacuum input port of the beam splitter that yields the two clones. By tuning the
transmittance of the beam splitters, one can generate the entire family of cloners saturating
Eq.~(\ref{ineqcloning}). This idea can also be generalized to define the optimal asymmetric
cloner producing $M$ different clones~\cite{Fiurasek2007}. Other research into Gaussian quantum cloning includes, the relationship of the no-cloning limit to the quality of continuous-variable teleportation~\cite{Grosshans2001}, the optimal cloning of coherent states with a finite distribution~\cite{Cochrane2004}, the cloning of squeezed and thermal states~\cite{Olivares2006}, and the cloning of both entangled Gaussian states and Gaussian entanglement~\cite{Weedbrook2008}. Finally, it is worth noting that all the Gaussian cloners discussed above are optimal if the
added noise variance is taken as the figure of merit. The Gaussian transformation
of Eq.~(\ref{eq-cloning}) produces clones with the minimum noise variance,
namely one unit of shot-noise. Surprisingly, if the single-clone fidelity is chosen instead
as the figure of merit, the optimal cloner is a non-Gaussian cloner which slightly
outperforms the Gaussian cloner (its fidelity is 2.4~\% higher) for the cloning of Gaussian (coherent) states~\cite{Cerf2005b}.

\section{Bosonic Gaussian Channels}\label{chapter bosonic gaussian channels}

A central topic in quantum information theory is the study of
bosonic channels, or more properly, linear bosonic channels~\cite{Demoen1977,Lindblad2000}. In particular, Gaussian channels represent the standard
model of noise in many quantum communication protocols
\cite{Holevo1999,Holevo2001,Eisert2007}. They describe all those communication
processes where the interaction between the bosonic system carrying the
information and the external decohering environment is governed by a linear
and/or bilinear Hamiltonian. In the simplest scenario, Gaussian channels are memoryless,
meaning that different bosonic systems are affected
independently and identically. This is
the case of the one-mode Gaussian channels, where
each mode sent through the channel is perturbed in this way~\cite{Holevo2001,Holevo2007}.

This section is structured as follows. In Sec.~\ref{sec:general}, we give a
general introduction to bosonic channels and, particularly, Gaussian channels,
together with their main properties. Then, in Sec.~\ref{sec:onechannels}, we
discuss the specific case of one-mode Gaussian channels and their recent full
classification. In Secs.~\ref{sec:ccap} and \ref{sec:qcap} we discuss the standard notions of classical and quantum capacity, respectively, with quantum dense coding and entanglement-assisted classical capacity revealed in Sec.~\ref{sec: quantum dense coding}.
Entanglement distribution and secret key capacities are discussed in Sec.~\ref{sec:escap}. Finally, in Sec.~\ref{sec:qcd}, we consider the problem of Gaussian channel discrimination
and its potential applications.

\subsection{General formalism\label{sec:general}}

\begin{figure}[th]
\vspace{-1.7cm}
\par
\begin{center}
\includegraphics[width=7.5cm]{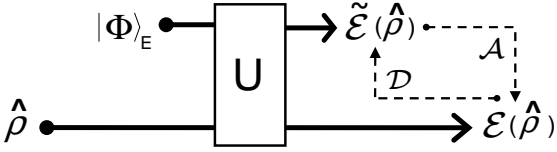}
\end{center}
\par
\vspace{-1.9cm} \caption{Stinespring dilation of a bosonic channel
$\mathcal{E}$. The input state $\hat{\rho}$ interacts unitarily with a pure
state $\left\vert \Phi\right\rangle _{E}$ of the environment, which can be
chosen to be the vacuum. Note that, besides the output $\mathcal{E}(\hat{\rho
})$, there is a complementary output $\mathcal{\tilde{E}}(\hat{\rho})$ for the
environment. In some cases, the two outputs are connected by CPT\ maps (see
text).}%
\label{StinesPIC}%
\end{figure}

Let us consider a multimode bosonic system, with arbitrary $N$ modes, whose
quantum state is described by an arbitrary density operator $\hat{\rho}%
\in\mathcal{D}(\mathcal{H}^{\otimes N})$. Then, an $N$-mode bosonic channel is
a linear map $\mathcal{E}:\hat{\rho}\rightarrow\mathcal{E}(\hat{\rho}%
)\in\mathcal{D}(\mathcal{H}^{\otimes N})$, which must be completely positive
and trace-preserving~(CPT)~\cite{Nielsen2000}. There are several equivalent ways to represent this map, one of the most
useful being the Stinespring dilation \cite{Stinespring1955}. As depicted in
Fig.~\ref{StinesPIC}, a multimode bosonic channel can be represented by a
unitary interaction $U$ between the input state $\hat{\rho}$ and a pure state
$\left\vert \Phi\right\rangle _{E}$ of ancillary $N_{E}$ modes associated with
the environment. Then the output of the channel is given by tracing out the
environment after interaction, i.e.,%
\begin{equation}
\mathcal{E}(\hat{\rho})=\mathrm{Tr}_{E}\left[  U\left(  \hat{\rho}%
\otimes\left\vert \Phi\right\rangle \left\langle \Phi\right\vert _{E}\right)
U^{\dagger}\right]  ~. \label{Sdilation}%
\end{equation}
An important property of the Stinespring dilation is its uniqueness up to
partial isometries \cite{Paulsen2002}. As a result, one can always choose
$\left\vert \Phi\right\rangle _{E}=\left\vert 0\right\rangle _{E}$, where
$\left\vert 0\right\rangle _{E}$ is a multimode vacuum state.

Note that, in the physical representation provided by the Stinespring
dilation, the environment has an output too. In fact, we can consider the
\emph{complementary} bosonic channel $\mathcal{\tilde{E}}:\hat{\rho
}\rightarrow\mathcal{\tilde{E}}(\hat{\rho})$ which is defined by tracing out
the system after interaction. For particular kinds of bosonic channels, the
two outputs $\mathcal{E}(\hat{\rho})$\ and $\mathcal{\tilde{E}}(\hat{\rho})$
are connected by CPT\ maps. This happens when the channel is
\textit{degradable} or \textit{anti-degradable}. By definition, we say that a
bosonic channel $\mathcal{E}$ is \textit{degradable} if there exists a CPT map
$\mathcal{D}$ such that $\mathcal{D}\circ\mathcal{E}=\mathcal{\tilde{E}}$
\cite{Devetak2005b}. This means that the environmental output $\mathcal{\tilde
{E}}(\hat{\rho})$\ can be achieved from the system output $\mathcal{E}%
(\hat{\rho})$ by applying another bosonic channel $\mathcal{D}$. By contrast,
a bosonic channel $\mathcal{E}$ is called \textit{anti-degradable} when there
is a CPT\ map $\mathcal{A}$ such that $\mathcal{A}\circ\mathcal{\tilde{E}%
}=\mathcal{E}$ \cite{Caruso2006b} (see Fig.~\ref{StinesPIC}).

The most important bosonic channels are the Gaussian channels, defined as
those bosonic channels transforming Gaussian states into Gaussian states. An
arbitrary $N$-mode Gaussian channel can be represented by a Gaussian dilation.
This means that the interaction unitary $U$ in Eq.~(\ref{Sdilation}) is
Gaussian and the environmental state $\left\vert \Phi\right\rangle _{E}$ is
pure Gaussian (or, equivalently, the vacuum). Furthermore, we can choose an
environment composed of $N_{E}\leq2N$ modes \cite{Caruso2008,Caruso2010}. The
action of a $N$-mode Gaussian channel over an arbitrary Gaussian state
$\hat{\rho}(\mathbf{\bar{x}},\mathbf{V})$ can be easily expressed in terms of
the first and second statistical moments. In fact, we have \cite{Holevo2001}%
\begin{equation}
\mathbf{\bar{x}}\rightarrow\mathbf{T\bar{x}+d~},~~\mathbf{V}\rightarrow
\mathbf{TVT}^{T}+\mathbf{N}~, \label{Transformation_BGC}%
\end{equation}
where $\mathbf{d}\in\mathbb{R}^{2N}$ is a displacement vector, while
$\mathbf{T}$ and $\mathbf{N}=\mathbf{N}^{T}$ are $2N\times2N$ real matrices,
which must satisfy the complete positivity condition%
\begin{equation}
\mathbf{N}+i\mathbf{\Omega}-i\mathbf{T\Omega T}^{T}\geq0~,
\label{N_K_relation}%
\end{equation}
where $\mathbf{\Omega}$ is defined in Eq.~(\ref{Symplectic_Form}). Note that,
for $\mathbf{N=0}$ and $\mathbf{T}:=\mathbf{S}$ symplectic, the channel
corresponds to a Gaussian unitary $U_{\mathbf{S},\mathbf{d}}$ (see
Sec.~\ref{sec: gaussian unitaries}).

\subsection{One-mode Gaussian channels\label{sec:onechannels}}

The study of one-mode Gaussian channels plays a central role in quantum
information theory, representing one of the standard models to describe the
noisy evolution of one-mode bosonic states. Furthermore, these channels
represent the manifest effect of the most important eavesdropping strategy in
continuous-variable quantum cryptography, known as collective Gaussian
attacks, which will be fully discussed in Sec.~\ref{sec:collectiveGauss}. One
of the central results in the theory of one-mode Gaussian channels is the
Holevo's canonical classification. This result was originally derived by
\textcite{Holevo2007} and then exploited by several authors to study the
degradability and security properties of these channels
\cite{Caruso2006,Pirandola2008c,Pirandola2009b}.

\begin{figure}[th]
\vspace{-1.7cm}
\par
\begin{center}
\includegraphics[width=8cm]{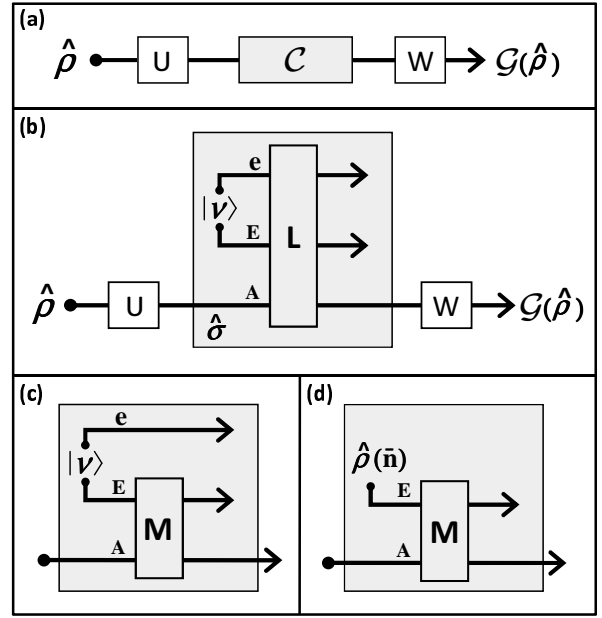}
\end{center}
\par
\vspace{-1.7cm} \caption{(a) A\ generic one-mode Gaussian channel
$\mathcal{G}$ can be represented by a canonical form $\mathcal{C}$\ up to
input and output Gaussian unitaries $U$ and $W$. (b) An arbitrary canonical
form $\mathcal{C}=\mathcal{C}(\tau,r,\bar{n})$ can be dilated to a three-mode
canonical unitary $U_{\mathbf{L}}$ which is described by a class-dependent
symplectic transformation\ $\mathbf{L}=\mathbf{L}(\tau,r)$. This unitary
evolves the input state $\hat{\sigma}$ together with an EPR\ state $\left\vert
\nu\right\rangle $ with noise-variance $\nu=2\bar{n}+1$ and belonging to the
environment. (c) Apart from class $B_{2}$, all the other classes can be
dilated using $\mathbf{L}(\tau,r)=\mathbf{M}(\tau,r)\oplus\mathbf{I}_{e}$.
This means that only one mode $E$ of the EPR state $\left\vert \nu
\right\rangle $ is combined with the input mode $A$. (d) Tracing out mode $e$,
we get a thermal state $\hat{\rho}(\bar{n})$ on mode $E$. Thus the canonical
forms of all the classes but $B_{2}$ can be represented by a single-mode
thermal state interacting with the input state via a two-mode symplectic
transformation $\mathbf{M}$.}%
\label{CanoPIC}%
\end{figure}

An arbitrary one-mode Gaussian channel $\mathcal{G}$\ is fully characterized
by the transformations of Eq.~(\ref{Transformation_BGC}), where now
$\mathbf{d}\in\mathbb{R}^{2}$ and $\mathbf{T},\mathbf{N}$ are $2\times2$ real
matrices, satisfying
\begin{equation}
\mathbf{N}=\mathbf{N}^{T}\geq0,~\det\mathbf{N}\geq\left(  \det\mathbf{T}%
-1\right)  ^{2}~. \label{N_K_1mode}%
\end{equation}
The latter conditions can be derived by specifying Eq.~(\ref{N_K_relation}) to one mode ($\textit{N}=1$). According to \textcite{Holevo2007}, the mathematical structure of a one-mode
Gaussian channel $\mathcal{G}=\mathcal{G}(\mathbf{d},\mathbf{T},\mathbf{N})$
can be greatly simplified. As depicted in Fig.~\ref{CanoPIC}(a), every
$\mathcal{G}$ can be decomposed as%
\begin{equation}
\mathcal{G}(\hat{\rho})=W\left[  \mathcal{C}(U\hat{\rho}U^{\dagger})\right]
W^{\dagger}~,
\end{equation}
where $U$\ and $W$ are Gaussian unitaries, while the map $\mathcal{C}$, which
is called the \textit{canonical form}, is a simplified Gaussian channel
$\mathcal{C}=\mathcal{C}(\mathbf{d}_{c},\mathbf{T}_{c},\mathbf{N}_{c})$ with
$\mathbf{d}_{c}=\mathbf{0}$ and $\mathbf{T}_{c},\mathbf{N}_{c}$ diagonal. The
explicit expressions of $\mathbf{T}_{c}$ and $\mathbf{N}_{c}$ depend on three
quantities which are preserved by the action of the Gaussian unitaries. These invariants are the generalized
\textit{transmissivity }$\tau:=\det\mathbf{T}$ (ranging from $-\infty$\ to
$+\infty$), the \textit{rank} of the channel $r:=\mathrm{min}[$%
rank$(\mathbf{T}),$rank$(\mathbf{N})]$ (with possible values $r=0,1,2$) and
the \textit{thermal number}\emph{\ }$\bar{n}$, which is a non-negative number
defined by%
\begin{equation}
\bar{n}:=\left\{
\begin{array}
[c]{c}%
(\det\mathbf{N})^{1/2}~,~~~\text{for~}~\tau=1~,~~~~~~~~\\
\\
\dfrac{(\det\mathbf{N})^{1/2}}{2\left\vert 1-\tau\right\vert }-\dfrac{1}%
{2}~,~~~\text{for~}~\tau\neq1~.
\end{array}
\right.
\end{equation}
These three invariants $\{\tau,r,\bar{n}\}$ fully characterize the two
matrices $\mathbf{T}_{c}$ and $\mathbf{N}_{c}$, thus identifying a unique
canonical form $\mathcal{C}=\mathcal{C}(\tau,r,\bar{n})$. In particular, the
first two invariants $\{\tau,r\}$ determine the \textit{class} of the form.
The full classification is shown in table~\ref{table1}.
\begin{table}[th]
\centering
\begin{equation}
\begin{tabular}
[c]{c|c||c||c||c|c}%
$\tau$ & $~r~$ & Class & ~~~Form~~ & $\mathbf{T}_{c}$ & $\mathbf{N}_{c}%
$\\\hline
$0$ & $0$ & $A_{1}$ & $\mathcal{C}(0,0,\bar{n})$ & $\mathbf{0}$ & $(2\bar
{n}+1)\mathbf{I}$\\
$0$ & $1$ & $A_{2}$ & $\mathcal{C}(0,1,\bar{n})$ & $\frac{\mathbf{I}%
+\mathbf{Z}}{2}$ & $(2\bar{n}+1)\mathbf{I}$\\
$1$ & $1$ & $B_{1}$ & $\mathcal{C}(1,1,0)$ & $\mathbf{I}$ & $\frac
{\mathbf{I}-\mathbf{Z}}{2}$\\
$1$ & $2$ & $B_{2}$ & $\mathcal{C}(1,2,\bar{n})$ & $\mathbf{I}$ & $\bar
{n}\mathbf{I}$\\
$1$ & $0$ & $B_{2}(Id)$ & $\mathcal{C}(1,0,0)$ & $\mathbf{I}$ & $\mathbf{0}$\\
$(0,1)$ & $2$ & $C(Loss)$ & $\mathcal{C}(\tau,2,\bar{n})$ & $\sqrt{\tau
}\mathbf{I}$ & $(1-\tau)(2\bar{n}+1)\mathbf{I}$\\
$>1$ & $2$ & $C(Amp)$ & $\mathcal{C}(\tau,2,\bar{n})$ & $\sqrt{\tau}%
\mathbf{I}$ & $(\tau-1)(2\bar{n}+1)\mathbf{I}$\\
$<0$ & $2$ & $D$ & $\mathcal{C}(\tau,2,\bar{n})$ & $\sqrt{-\tau}\mathbf{Z}$ &
$(1-\tau)(2\bar{n}+1)\mathbf{I}$%
\end{tabular}
\end{equation}
\caption{The values of $\{\tau,r\}$ in the first two columns specify a
canonical class $A_{1},A_{2},B_{1},B_{2},C$ or $D$ (third column). Within each
class, the possible canonical forms are expressed in the fourth column, where
also the invariant $\bar{n}$ must be considered. The corresponding expressions
of $\mathbf{T}_{c}$ and $\mathbf{N}_{c}$ are shown in the last two columns,
where $\mathbf{Z}:=\mathrm{diag}(1,-1)$, $\mathbf{I}:=\mathrm{diag}(1,1)$ and
$\mathbf{0}$ is the zero matrix.}\label{table1}%
\end{table}

Let us discuss the various classes. Class $A_{1}$ is composed by forms which
are completely depolarizing channels, i.e., replacing input states with
thermal states. Classes $A_{2}$ and $B_{1}$\ are special and involve canonical
forms transforming the quadratures asymmetrically. Class $B_{2}$\ describes
the classical-noise channels, transforming the quadratures as $\mathbf{\hat
{x}}\rightarrow\mathbf{\hat{x}}+\boldsymbol{\xi}$\ where $\boldsymbol{\xi}$ is
Gaussian noise with classical covariance matrix $\bar{n}\mathbf{I}$. This
class collapses to the identity channel for $\bar{n}=0$. Class $C$ describes
canonical forms with transmissivities $0<\tau\neq1$. This class is further
divided in two subclasses: $C$(Loss) for $0<\tau<1$, and $C$(Amp) for $\tau
>1$. Canonical forms in $C$(Loss) are known as \textit{lossy channels}, also
denoted by $\mathcal{L}(\tau,\bar{n}):=\mathcal{C}(0<\tau<1,2,\bar{n})$. These
are the most important ones, representing the basic model to describe
communication lines such as optical fibers. In a lossy channel, the input
signals are attenuated and combined with thermal noise, i.e., we have
$\mathbf{\hat{x}}\rightarrow\sqrt{\tau}\mathbf{\hat{x}}+\sqrt{1-\tau
}\mathbf{\hat{x}}_{th}$, where $\mathbf{\hat{x}}_{th}$ are in a thermal state
with $\bar{n}$\ photons. Canonical forms in $C$(Amp) are known as
\textit{amplifying channels}, denoted by $\mathcal{A}(\tau,\bar{n}%
):=\mathcal{C}(\tau>1,2,\bar{n})$. They describe optical processes, such as
phase-insensitive amplifiers, where the input signals are amplified with the
addition of thermal noise, i.e., $\mathbf{\hat{x}}\rightarrow\sqrt{\tau
}\mathbf{\hat{x}}+\sqrt{\tau-1}\mathbf{\hat{x}}_{th}$. Finally, class $D$ is
associated with negative transmissivities. Its forms can be seen as
complementary outputs of the amplifying channels. 

We can easily construct the Stinespring dilation of all the canonical forms
\cite{Pirandola2008c}. As depicted in Fig.~\ref{CanoPIC}(b), an arbitrary form
$\mathcal{C}(\tau,r,\bar{n})$ can be dilated to a three-mode canonical unitary
$U_{\mathbf{L}}$ corresponding to a $6\times6$ symplectic matrix $\mathbf{L}$.
This unitary transforms the input state $\hat{\sigma}$ (mode $A$) together
with an environmental EPR state $\left\vert \nu\right\rangle $ (modes $E$ and
$e$) of suitable noise-variance $\nu$ [see Eq.~(\ref{eq:EPR_CM})]. In
particular, the symplectic matrix is determined by the class, i.e.,
$\mathbf{L}=\mathbf{L}(\tau,r)$, while the EPR state is determined by the
thermal number, i.e., $\nu=2\bar{n}+1$. Let us analyze $\mathbf{L}(\tau,r)$
for the various classes, starting from $B_{2}$. For null rank, class $B_{2}%
$\ collapses to the identity and we simply have $\mathbf{L}(1,0)=\mathbf{I}$.
However, for full rank the symplectic matrix $\mathbf{L}(1,2)$ does not have a
simple expression \cite{Holevo2007}. If we exclude the class $B_{2}$, the
symplectic matrix $\mathbf{L}$ can always be decomposed as $\mathbf{L}%
(\tau,r)=\mathbf{M}(\tau,r)\oplus\mathbf{I}_{e}$, where $\mathbf{M}$ describes
a two-mode canonical unitary acting on modes $A$\ and $E$, while
$\mathbf{I}_{e}$ is just the identity on mode $e$. As depicted in
Fig.~\ref{CanoPIC}(c), this means that only one mode $E$ of the EPR state
$\left\vert \nu\right\rangle $ is actually combined with the input mode $A$.
Clearly by tracing out the unused EPR mode $e$, we get a thermal state with
$\bar{n}$\ photons on mode $E$, as depicted in Fig.~\ref{CanoPIC}(d). Thus the
canonical forms $\mathcal{C}(\tau,r,\bar{n})$ of all the classes but $B_{2}$
admit a physical representation where a single-mode thermal state $\hat{\rho
}(\bar{n})$ interacts with the input state via a two-mode symplectic
transformation $\mathbf{M}(\tau,r)$. Despite being simpler than the
Stinespring dilation, this unitary dilation involves a mixed environmental
state and, therefore, it is not unique up to partial isometries. The explicit
expressions of $\mathbf{M}(\tau,r)$\ are relatively easy \cite{Caruso2006}.
For classes $A_{1}$, $A_{2}$ and $B_{1}$, we have
\begin{gather}
\mathbf{M}(0,0)=\left(
\begin{array}
[c]{cc}%
\mathbf{0} & \mathbf{I}\\
\mathbf{I} & \mathbf{0}%
\end{array}
\right)  ~,~\mathbf{M}(0,1)=\left(
\begin{array}
[c]{cc}%
\frac{\mathbf{I+Z}}{2} & \mathbf{I}\\
\mathbf{I} & \frac{\mathbf{Z-I}}{2}%
\end{array}
\right)  ~,\\
\mathbf{M}(1,1)=\left(
\begin{array}
[c]{cc}%
\mathbf{I} & \frac{\mathbf{I+Z}}{2}\\
\frac{\mathbf{I-Z}}{2} & \mathbf{-I}%
\end{array}
\right)  ~.
\end{gather}
Then, for classes $C($Loss$)$, $C($Amp$)$ and$\ D$, we have%
\begin{align}
\mathbf{M}(0  &  <\tau<1,2)=\left(
\begin{array}
[c]{cc}%
\sqrt{\tau}\mathbf{I} & \sqrt{1-\tau}\mathbf{I}\\
-\sqrt{1-\tau}\mathbf{I} & \sqrt{\tau}\mathbf{I}%
\end{array}
\right)  ~,\label{BS_canonical}\\
\mathbf{M}(\tau &  >1,2)=\left(
\begin{array}
[c]{cc}%
\sqrt{\tau}\mathbf{I} & \sqrt{\tau-1}\mathbf{Z}\\
\sqrt{\tau-1}\mathbf{Z} & \sqrt{\tau}\mathbf{I}%
\end{array}
\right)  ~,\\
\mathbf{M}(\tau &  <0,2)=\left(
\begin{array}
[c]{cc}%
\sqrt{-\tau}\mathbf{Z} & \sqrt{1-\tau}\mathbf{I}\\
-\sqrt{1-\tau}\mathbf{I} & -\sqrt{-\tau}\mathbf{Z}%
\end{array}
\right)  ~.
\end{align}
Here it is important to note that Eq.~(\ref{BS_canonical}) is just the beam
splitter matrix (cf. Eq.~(\ref{eq: beam splitter symplectic})). This means
that the Stinespring dilation of a lossy channel $\mathcal{L}(\tau,\bar{n})$
is an entangling cloner%
~\cite{Grosshans2003}, i.e., a beam splitter with transmissivity $\tau$ which
combines the input mode with one mode of an environmental EPR\ state
$\left\vert \nu\right\rangle $. Clearly, this implies the well-known physical
representation for the lossy channel where a beam splitter of transmissivity
$\tau$ mixes the input state with a single-mode thermal state $\hat{\rho}%
(\bar{n})$. A particular case of lossy channel is the pure-loss channel
$\mathcal{L}(\tau,0)$ which is given by setting $\bar{n}=0$. In this case the
Stinespring dilation is just a beam splitter mixing the input with the vacuum.

Finally, let us review the degradability properties of the one-mode Gaussian
channels. Since these properties are invariant by unitary equivalence, we have
that a degradable (antidegradable) channel $\mathcal{G}$ corresponds to a
degradable (antidegradable) form $\mathcal{C}$. All the forms $\mathcal{C}%
(\tau,r,\bar{n})$ with transmissivity $\tau\leq1/2$ are antidegradable
\cite{Caruso2006b}. This includes all the forms of classes $A_{1}$, $A_{2}$,
$D$ and part of the forms of class $C$, i.e., lossy channels $\mathcal{L}%
(\tau,\bar{n})$ with $\tau\leq1/2$. By unitary equivalence, this means that
one-mode Gaussian channels with transmissivity $\tau\leq1/2$ are all
antidegradable. For $\tau \geq1/2$ the degradability properties are not so straightforward.
However, we know that pure-loss channels
$\mathcal{L}(\tau,0)$ with $\tau\geq 1/2$ and ideal amplifying channels
$\mathcal{A}(\tau,0)$ are all degradable.

\subsection{Classical capacity of Gaussian channels\label{sec:ccap}}

Shannon proved that sending information through a noisy channel can be
achieved with vanishing error, in the limit of many uses of the channel. He
developed an elegant mathematical theory in order to calculate the ultimate
limits on data transmission rates achievable over a classical communication
channel $\mathcal{N}$, known as the channel capacity~\cite{Shannon1948}. Let
us consider two parties, Alice and Bob, which are connected by an arbitrary
noisy channel $\mathcal{N}$. At the input, Alice draws letters from a random
variable (or alphabet) $A:=\{a,p_{a}\}$, where the letter $a$ occurs with
probability $p_{a}$. The information content of this variable is expressed in
terms of bits per letter and quantified by the Shannon entropy $H(A)=-\sum
_{a}p_{a}\log p_{a}$ (it is understood that when we consider continuous
variables, the probabilities are replaced by probability densities and sums by
integrals.) By drawing many times, Alice generates a random message
$a_{1},a_{2},\cdots$ which is sent to Bob through the noisy channel. As long
as the channel is memoryless, i.e., it does not create correlations between
different letters, Bob's output message can be described by drawings from
another random variable $B:=\{b,p_{b}\}$ correlated to the input one
$B=\mathcal{N}(A)$. On average, the number of bits per letter which are
communicated to Bob is given by the mutual information $I(A:B)=H(B)-H(B|A)$,
where $H(B|A)$ is Shannon entropy of $B$ conditioned on the knowledge of $A$
\cite{Cover2006}. Now, the channel capacity $C(\mathcal{N})$, expressed in
bits per channel use, is given by maximizing the mutual information over all
of Alice's possible inputs
\begin{equation}
C(\mathcal{N})=\max_{A}I(A:B)~. \label{eq:Shannon}%
\end{equation}

It is important to note that many communication channels, such as wired and
wireless telephone channels and satellite links are currently modeled as
classical Gaussian channels. Here the input variable $A$ generates a
continuous signal $a$ with variance $P$ which is transformed to a continuous
output $b=\tau a+\xi$, where $\tau$ is the transmissivity of the channel, and
$\xi$ is drawn from a Gaussian noise-variable of variance $V$. Shannon's
theory gives the capacity
$C(\mathcal{N})=\frac{1}{2}\log\left(1+\tau PV^{-1}\right)$~\cite{Cover2006}. We remark that this
result predicts an infinite communication rate through a noiseless channel $(V=0)$. This counterintuitive result is due to the
lack of limitation to the measurement accuracy in classical physics. This is
no longer true when we consider the actual quantum nature of the physical
systems. In fact, if we encode classical information in the temporal modes
(pulses)\ of the quantized electromagnetic field, then the capacity of the
identity channel is no longer infinite but depends on the input energy. As
shown by \textcite{Yuen1993}, the capacity of the identity channel
$\mathcal{I}$ is given by $C(\mathcal{I})=g(2\bar{m}+1)$, where $g(\cdot)$ is
given in Eq.~(\ref{g_explicit}) and $\bar{m}$ is the mean number of photons
per pulse. Thus, a quantum mechanical treatment of the problem gives a finite
solution for finite energy, showing that quantum mechanics is mandatory in
understanding the ultimate limits of communication.

Since information is fundamentally encoded in a physical system and quantum
mechanics is the most accurate representation of the physical world, it is
therefore natural to ask what are the ultimate limits set by quantum mechanics
to communication? Since the 1980s several groups started studying quantum
encoding and detection over optical channels, modeled as Gaussian quantum
channels \cite{Yuen1980,Shapiro1984,Cav94,Hall1994}. An important milestone
was achieved with the Holevo-Schumacher-Westmoreland (HSW) theorem~\cite{Schumacher1997,Holevo1998}, which laid the basis for a quantum
generalization of Shannon's communication theory. First of all, let us
introduce the notions of quantum ensemble and Holevo bound~\cite{Holevo1973}.
An arbitrary random variable $A=\{a,p_{a}\}$\ can be encoded in a quantum
ensemble (or source) $\mathcal{A}=\{\hat{\rho}_{a},p_{a}\} $, where each
letter $a$\ is associated with a quantum letter-state $\hat{\rho}_{a} $
occurring with probability $p_{a}$. Since quantum states are generally
non-orthogonal, a non-trivial question is the following: what is the maximum
information that we can extract from $\mathcal{A}$ using a quantum
measurement? This quantity is called the accessible information of the
ensemble and is less than or equal to the Holevo bound, defined as
\begin{equation}
\chi(\mathcal{A})=S({\hat{\sigma}}_{A})-\sum_{a}p_{a}S(\hat{\rho}_{a})~,
\label{eq:Holevo_info}%
\end{equation}
where $S(\cdot)$ is the von Neumann entropy and $\hat{\sigma}_{A}=\sum_{a}p_{a}%
\hat{\rho}_{a}$ is the average state of the ensemble (for continuous ensembles, the previous
sums become integrals). Now, the key-result of
the HSW theorem is that the Holevo bound is asymptotically achievable when we
consider a large number of extractions from the source and a collective
quantum measurement. In this limit, the Holevo bound
$\chi(\mathcal{A})$ provides the accessible information per letter-state.

These results can be directly applied to memoryless quantum channels
$\mathcal{M}$.
In this case, the letter-states drawn from a source $\mathcal{A}=\{\hat{\rho
}_{a},p_{a}\}$ are transformed identically and independently by the channel,
i.e., $\hat{\rho}_{a_{1}}\otimes\hat{\rho}_{a_{2}}\cdots\rightarrow
\mathcal{M}(\hat{\rho}_{a_{1}})\otimes\mathcal{M}(\hat{\rho}_{a_{2}})\cdots$
By performing a collective measurement on the output message-state, Bob can
extract an average of $\chi(\mathcal{A},\mathcal{M})$ bits per channel use,
where%
\begin{equation}
\chi(\mathcal{A},\mathcal{M})=S\left[  \mathcal{M}(\hat{\sigma}_{A})\right]
-\sum_{a}p_{a}S[\mathcal{M}(\hat{\rho}_{a})]~. \label{eq:Hole2}%
\end{equation}
Thus the Holevo bound $\chi(\mathcal{A},\mathcal{M})$ gives the optimal
communication rate which is achievable over the memoryless quantum channel
$\mathcal{M}$ for fixed source $\mathcal{A}$. Maximizing this quantity over
all the sources $\mathcal{A}$ we obtain the (single-shot) capacity of the
channel%
\begin{equation}
C^{(1)}(\mathcal{M})=\max_{\mathcal{A}}\chi(\mathcal{A},\mathcal{M})~.
\label{eq:Holevo_capacity}%
\end{equation}
For bosonic systems, where memoryless channels are usually one-mode channels,
the quantity of Eq.~(\ref{eq:Holevo_capacity}) must be constrained by
restricting the maximization over sources with bounded energy $\mathrm{Tr}%
(\hat{\sigma}_{A}\hat{n})\leq \bar{m}$.

Note that we have introduced the notation single-shot in the definition of
Eq.~(\ref{eq:Holevo_capacity}). This is because we are restricting the problem
to single-letter sources which input product states. In general, we can
consider multi-letter sources which input states that are (generally)
entangled between $n$ uses of the channel $\mathcal{M}^{\otimes n}$. Then, we
can define the full capacity of the channel via the regularization
\begin{equation}
C(\mathcal{M})=\lim_{n\rightarrow\infty}\frac{1}{n}C^{(1)}(\mathcal{M}%
^{\otimes n})~. \label{eq:Capacity_Ultimate}%
\end{equation}
For one-mode bosonic channels, the computation of
Eq.~(\ref{eq:Capacity_Ultimate}) involves the maximization over sources which
emit $n$-mode entangled states and satisfying the energy constraint
$\mathrm{Tr}(\hat{\sigma}_{A}\hat{n}^{\otimes n})\leq n \bar{m}$. Now an important
question to ask is if the presence of entanglement can really enhance the rate
of classical communication. In other words, do we have $C(\mathcal{M}%
)>C^{(1)}(\mathcal{M})$? \textcite{Hastings2009} proved the existence of channels for which this is the
case. However, for one-mode bosonic Gaussian channels this is still an
open
question.

A first step in this direction has been the computation of the capacity of a
pure-loss channel $\mathcal{L}_{p}:=\mathcal{L}(\tau,0)$. By exploiting the
sub-additivity of the von Neumann entropy, \textcite{Giovannetti2004a}
obtained an upper-bound for $C(\mathcal{L}_{p})$ coinciding with the
lower-bound reported by \textcite{Holevo1999} and \textcite{Holevo2001}. As a
result, a pure-loss channel $\mathcal{L}_{p}$ of transmissivity $\tau$ has
classical capacity $C(\mathcal{L}_{p})=g(\tau\mu +1 -\tau)$, where $\mu:=2\bar{m}+1$
and $\bar{m}$ is the mean number of photons per input mode. Interestingly, one
can achieve this capacity by sending coherent states modulated with a Gaussian
distribution of variance $V=\mu-1$. At the detection stage, collective
measurements might be necessary. However, this is not the case in the regime
of many photons, where heterodyne detection is sufficient to achieve the capacity.

The model of pure-loss channel $\mathcal{L}_{p}$ can be adopted to describe
broadband communication lines, such as wave guides, where the losses are
independent from the frequency. For a pure-loss channel of this kind which
employs a set of frequencies $\omega_{k}=k\delta\omega$ for integer $k$, one
can derive the capacity $C=\xi\sqrt{\tau P}T$, where $\tau$ is the
transmissivity, $T=2\pi/\delta\omega$ is the transmission time, $P$ is the
average transmitted power, and $\xi$ is a constant
\cite{Yuen1993,Giovannetti2003,Giovannetti2004a}. Another important scenario
is free-space optical communication. Here, transmitter and receiver
communicate through circular apertures of areas $A_{t}$ and $A_{r}$ which are
separated by a distance $L$. Far-field regime corresponds to having a single
spatial mode, which happens when $A_{t}A_{r}\omega^{2}\left(  2\pi cL\right)
^{-2}:=\tau(\omega)\ll1$, where $c$ is the speed of light and $\tau(\omega)$
is the transmissivity of the optimal spatial mode with frequency $\omega$
\cite{Yuen1978}. We have a broadband far-field regime when we use frequencies
up to a critical frequency $\omega_{c}$, such that $\tau(\omega_{c})\ll1$. In
this case, we can compute the capacity
\begin{equation}
C=(\omega_{c}T/2\pi y_{0})\int_{0}^{y_{0}}dx~g[(e^{1/x}-1)^{-1}],
\end{equation}
where $y_{0}$ is a parameter which is connected with the energy constraint
\cite{Giovannetti2004a,Shapiro2005,Guha2008}. Recently, the computation of
this classical capacity has been generalized to the presence of optical
refocusing systems between transmitter and receiver \cite{Lupo2011}.

\subsubsection{Bosonic minimum output entropy conjecture}

Despite a huge research effort in recent years, little
progress has been achieved in the calculation of the classical capacity
of other one-mode Gaussian channels. However, by using Gaussian encodings,
we can easily give lower bounds~\cite{Lupo2011a}. For instance, using a coherent state encoding at the input of a lossy channel
$\mathcal{L}(\tau,\bar{n})$, we can compute the following lower bound for the
capacity
\begin{equation}
C(\mathcal{L})\geq g[\tau\mu+(1-\tau)\nu]-g[\tau+(1-\tau)\nu]:=\underline
{C}(\mathcal{L}),\label{eq:conjecture_capacity}%
\end{equation}
where $\nu:=2\bar{n}+1$, $\mu:=2\bar{m}+1$, and $\bar{m}$ is the mean number
of photons per input mode \cite{Holevo1999}. It is believed that this
lower-bound is tight, i.e., $C(\mathcal{L})=\underline{C}(\mathcal{L})$. This
conjecture is implied by another conjecture, known as the bosonic minimum
output entropy conjecture and stating that the minimum entropy at the output
of a lossy channel is realized by a vacuum state at the input, i.e.,
$S[\mathcal{L}(\left\vert 0\right\rangle \left\langle 0\right\vert )]\leq
S[\mathcal{L}(\hat{\rho})]$ for every $\hat{\rho}$. It seems extremely
reasonable to assume that sending nothing through the channel is the best way
of minimizing the noise (entropy) at its output. However, such a simple
statement is still today without a proof. Using Lagrangian minimization it has
been possible to prove that vacuum is a local minimum of the output entropy
\cite{Giovannetti2004b,Lloyd2009}. In the work of \textcite{Giovannetti2004b}
a simulated annealing optimization suggested that outputs produced by a vacuum
input majorize all the other outputs, and therefore have smaller entropy.
Other studies showed that the output R\'{e}nyi entropy of integer orders
$\geq2$ is minimized by the vacuum input and is also additive
\cite{Giovannetti2004c,Giovannetti2004d}. Unfortunately, the von Neumann
entropy is the R\'{e}nyi entropy of order $1$, which is therefore not covered
by these results. By restricting the input states to Gaussian states it was
proven that vacuum gives the minimal output entropy
\cite{Giovannetti2004b,Serafini2005b,Hiroshima2006}; unfortunately this does
not preclude the possibility of having non-Gaussian input states performing
better. Finally, alternative approaches to the problem were also proposed,
such as proving the \textit{entropy photon-number inequality}~\cite{Guha2008}, which
is a quantum version of the classical \textit{entropy power inequality}~\cite{Cover2006}.

\subsection{Quantum capacity of Gaussian channels\label{sec:qcap}}

\begin{figure}[th]
\vspace{-2.7cm}
\par
\begin{center}
\includegraphics[width=9.5cm]{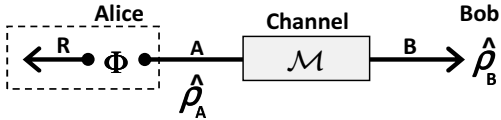}
\end{center}
\par
\vspace{-2.8cm} \caption{Alice's input state $\hat{\rho}_{A}$ is transformed
into Bob's output state $\hat{\rho}_{B}$ by a generic memoryless channel
$\mathcal{M}$. The input state $\hat{\rho}_{A}$ can be purified by introducing
an additional mirror system $R$.}%
\label{CInfoPIC}%
\end{figure}

Quantum channels can be used to transfer not just classical information but
also quantum information. In the typical quantum communication scenario, Alice
aims to transmit quantum states to Bob through a memoryless quantum channel
$\mathcal{M}$. The quantum capacity $Q(\mathcal{M})$ of the channel gives the
number of qubits per channel use that can be reliably transmitted. As shown by
\textcite{Schumacher1996}, a crucial role in the definition of the quantum
capacity is played by the coherent information $J(\mathcal{M},\hat{\rho}_{A}%
)$, which is a function of Alice's input $\hat{\rho}_{A}$ and the channel
$\mathcal{M}$. In order to define this quantity, let us introduce a mirror
system $R$ and the purification $\Phi_{RA}=\left\vert \Phi\right\rangle
\left\langle \Phi\right\vert _{RA}$ of the input state\ $\hat{\rho}%
_{A}=\mathrm{Tr}_{R}(\Phi_{RA})$, as shown in Fig.~\ref{CInfoPIC}.
Then, the coherent information is defined by
\begin{equation}
J(\mathcal{M},\hat{\rho}_{A})=S(\hat{\rho}_{B})-S(\hat{\rho}_{RB})~,
\end{equation}
where $\hat{\rho}_{RB}:=(\mathcal{I}_{R}\otimes\mathcal{M)}(\Phi_{RA})$,
$\mathcal{I}_{R}$ being the identity channel on the mirror system $R$. The
(single-shot) quantum capacity is computed by maximizing over all the input
states
\begin{equation}
Q^{(1)}(\mathcal{M})=\max_{\hat{\rho}_{A}}J(\mathcal{M},\hat{\rho}_{A}).
\end{equation}
Since this quantity is known to be non-additive
\cite{DiVicenzo1998,Smith2007,Smith2008,Smith2011}, the correct definition of quantum
capacity is given by the regularization \cite{Lloyd1997,Devetak2005a,Shor2002}
\begin{equation}
Q(\mathcal{M})=\lim_{n\rightarrow\infty}\frac{1}{n}\max_{\hat{\rho}_{A^{n}}%
}J(\mathcal{M}^{\otimes n},\hat{\rho}_{A^{n}})~, \label{eq:UnEDCap}%
\end{equation}
where the input state $\hat{\rho}_{A^{n}}$ is generally entangled over $n$
uses of the channel $\mathcal{M}^{\otimes n}$. It is important to note that
the coherent information computed over bosonic channels is finite even for
infinite input energy. As a result the quantum capacity of bosonic channels is
still defined as in Eq.~(\ref{eq:UnEDCap}) without the need of energy
constraints. Another important consideration regards degradable and
antidegradable channels. As shown by \textcite{Devetak2005b}, degradable
channels have additive quantum capacity, i.e., $Q(\mathcal{M})=Q^{(1)}%
(\mathcal{M})$. By contrast, antidegradable channels have null quantum
capacity $Q(\mathcal{M})=0$ \cite{Caruso2006b}.

Let us consider the specific case of one-mode Gaussian channels. In this case
a lower bound can be computed by restricting the quantum capacity to a single
use of the channel and pure Gaussian states. Thus, for an arbitrary one-mode
Gaussian channel $\mathcal{G}$ with transmissivity $\tau\neq1$, we can write
the lower-bound \cite{Holevo2001,Pirandola2009b}%
\begin{equation}
Q(\mathcal{G})\geq Q^{(1,g)}(\mathcal{G})=\max\left\{  0,\log\left\vert
\frac{\tau}{1-\tau}\right\vert -g(\nu)\right\}  ~, \label{bound_Q_Gauss}%
\end{equation}
where $\nu:=2\bar{n}+1$ and $\bar{n}$ is the thermal number of the channel.
Clearly this formula applies to all the canonical forms of classes $A_{1}$,
$A_{2}$, $C$ and $D$. There are remarkable cases where the bound in
Eq.~(\ref{bound_Q_Gauss}) is tight. This happens when the one-mode Gaussian
channel is degradable. The proof given in \cite{Wolf2007b} combines the
additivity for degradable channels $Q(\mathcal{G})=Q^{(1)}(\mathcal{G})$ with
the extremality of Gaussian states $Q^{(1)}(\mathcal{G})=Q^{(1,g)}%
(\mathcal{G})$~\cite{Wolf2006}. Important examples of degradable one-mode
Gaussian channels are the ideal amplifying channels $\mathcal{A}(\tau,0)$ and
the pure-loss channels $\mathcal{L}(\tau,0)$ with transmissivity $\tau \geq 1/2$.
Another case, where the previous bound is tight, regards all the one-mode Gaussian
channels with transmissivity $\tau\leq1/2$. These channels are in fact
antidegradable and we have $Q(\mathcal{G})=Q^{(1,g)}(\mathcal{G})=0$. Note that we can also compute a lower bound to the quantum capacity for $\tau=1$ in the case of a canonical form $B_{2}$. This is achieved by using continuous-variable stabilizer codes~\cite{Harrington2001}.

\subsection{Quantum dense coding and entanglement-assisted classical capacity}\label{sec: quantum dense coding}

The classical capacity of a quantum channel can be increased if Alice and Bob
share an entangled state. This effect is known as \textit{quantum dense
coding}. The analysis that reaches this conclusion ignores the cost of
distributing the entanglement. The rationale for doing this is that the
entanglement does not carry any information per se. Originally introduced in
the context of qubits \cite{Ben92}, dense coding was later extended to
continuous variables \cite{Bra99,Ral02,Ban1999}, with a series of experiments in both
settings~\cite{Mat96,Per00, Li02, Miz05}.

The basic setup in continuous variables considers the distribution of
information over an identity channel $\mathcal{I}$ by means of a single
bosonic mode. Here Bob possesses an EPR state of variance $V$. This state can
be generated by combining  a pair of single-mode squeezed states with
orthogonal squeezings into a balanced beam splitter. In particular, the
squeezed quadratures must have variance $V_{sq}=V-\sqrt{V^{2}-1}$. Bob sends
one mode of the EPR state to Alice, while keeping the other mode. To transmit
classical information, Alice modulates both quadratures and sends the mode
back to Bob, with mean number of photons equal to $\bar{m}$. To retrieve
information, Bob detects both received and kept modes by using a Bell
measurement with detector efficiency $\eta\in\lbrack0,1]$. The achievable rate
is given by~\cite{Ral02}
\begin{equation}
R_{dc}(\mathcal{I})=\log\left[  1+\frac{\eta(4\bar{m}-V_{sq}-1/V_{sq}%
+2)}{4(\eta V_{sq}+1-\eta)}\right].  \label{ccdciarb}%
\end{equation}
This rate can exceed the classical capacity of the identity channel
$C(\mathcal{I})$\ at the same fixed average photon number $\bar{m}$ for a
considerable range of values of $V_{sq}$ and $\eta$.

The advantages of quantum dense coding can be extended to an arbitrary
memoryless channel $\mathcal{M}$. This leads to the notion of
entanglement-assisted classical capacity $C_{E}(\mathcal{M})$, which is
defined as the maximum asymptotic rate of reliable bit transmission over a
channel $\mathcal{M}$ assuming the help of unlimited pre-shared entanglement.
As shown by \textcite{Bennett2002} this is equal to%
\begin{equation}
C_{E}(\mathcal{M})=\max_{\hat{\rho}_{A}}I(\mathcal{M},\hat{\rho}_{A}),
\end{equation}
where $I(\mathcal{M},\hat{\rho}_{A})=S(\hat{\rho}_{A})+J(\mathcal{M},\hat
{\rho}_{A})$ is the quantum mutual information associated with the channel
$\mathcal{M}$\ and the input state $\hat{\rho}_{A}$ ($J$ is the coherent
information). For one-mode Gaussian channels, the capacity $C_{E}$ must be
computed under the energy constraint $\mathrm{Tr}(\hat{\rho}_{A}\hat{n}%
)\leq\bar{m}$. In particular, for a pure-loss channel $\mathcal{L}%
_{p}=\mathcal{L}(\tau,0)$, we have (Holevo and Werner, 2001)%
\begin{equation}
C_{E}(\mathcal{L}_{p})=g(\mu)+g(\tau\mu+1-\tau)-g(\lambda_{-})-g(\lambda_{+}),
\end{equation}
where $\mu=2\bar{m}+1$ and $\lambda_{\pm}=D\pm\bar{m}(1-\tau)$, with%
\begin{equation}
D=\{[1+\bar{m}(\tau+1)]^{2}-4\tau\bar{m}(\bar{m}+1)\}^{1/2}~.
\end{equation}
This capacity is achieved by a Gaussian state. For $\tau\rightarrow1$ we have the identity channel and we get $C_{E}%
(\mathcal{I})=2g(\mu)$, which is twice its classical capacity $C(\mathcal{I})$.

\subsection{Entanglement distribution and secret-key
capacities\label{sec:escap}}

Other important tasks that can be achieved in quantum information are the
distribution of entanglement and secret keys over quantum noisy channels.
Given a memoryless channel $\mathcal{M}$, its entanglement distribution
capacity $E(\mathcal{M})$ quantifies the number of entanglement-bits which are
distributed per use of the channel. As shown by \textcite{Barnum00}, this
quantity coincides with the quantum capacity, i.e., $E(\mathcal{M}%
)=Q(\mathcal{M})$. Then, the secret-key capacity $K(\mathcal{M})$\ of the
channel provides the number of secure bits which are distributed per use of
the channel \cite{Devetak2005a}. Since secret bits can be extracted from
entanglement bits, we generally have $K(\mathcal{N})\geq E(\mathcal{M})$.
Using classical communication, Alice and Bob can improve all these
capacities. However, they need feedback classical communication, since the
capacities assisted by forward classical communication, i.e., $K_{\rightarrow
}(\mathcal{M})$, $E_{\rightarrow}(\mathcal{M})$ and $Q_{\rightarrow
}(\mathcal{M})$, coincide with the corresponding \textit{unassisted
}capacities, $K(\mathcal{M})$, $E(\mathcal{M})$ and $Q(\mathcal{M})$
\cite{Barnum00,Devetak2005a}.

Unfortunately, the study of feedback-assisted capacities is a very difficult
task. Alternatively, we can introduce simpler capacities, called reverse
capacities, defined by the maximization over protocols which are assisted by a
single feedback classical communication (known as reverse
protocols). A reverse protocol can be explained considering the purified scenario of Fig.~\ref{CInfoPIC}. Alice sends to Bob a large number of $A$ modes while keeping the $R$ modes. Then Bob applies a quantum operation over all the output $B$ modes and communicates a classical variable to Alice (single final classical communication). Exploiting this information, Alice applies a conditional quantum operation on all the $R$ modes. Thus we have the reverse $(\blacktriangleleft)$ entanglement
distribution capacity $E_{\blacktriangleleft}(\mathcal{M})$ and the reverse
secret-key capacity $K_{\blacktriangleleft}(\mathcal{M})$, which clearly\ must
satisfy $K_{\blacktriangleleft}(\mathcal{M})\geq E_{\blacktriangleleft
}(\mathcal{M})$. Interestingly, these capacities can be lower bounded by a
quantity which is very easy to compute. In fact, as shown by
\textcite{GarciaPatron2009}, we can define the \textit{reverse coherent
information}
\begin{equation}
J_{R}(\mathcal{M},\hat{\rho}_{A})=S(\hat{\rho}_{R})-S(\hat{\rho}_{RB})~.
\end{equation}
This quantity differs from the coherent information $J(\mathcal{M},\hat{\rho
}_{A})$ by the replacement $S(\hat{\rho}_{B})\rightarrow S(\hat{\rho}%
_{R})=S(\hat{\rho}_{A})$. For this reason, we can have $J_{R}(\mathcal{M}%
,\hat{\rho}_{A})>J(\mathcal{M},\hat{\rho}_{A})$ for channels which decrease
entropy, i.e., $S(\hat{\rho}_{A})>S(\hat{\rho}_{B})$. Optimizing the reverse
coherent information over all the inputs, we can define the (one-shot) reverse
coherent information capacity $E_{R}^{(1)}(\mathcal{M})$ and the corresponding
regularization $E_{R}(\mathcal{M})$. Interestingly, this quantity turns out to
be additive for all channels, so that we simply have $E_{R}(\mathcal{M}%
)=E_{R}^{(1)}(\mathcal{M})$. Now the capacity
$E_{R}(\mathcal{M})$ provides a lower bound for the reverse capacities, i.e.,
$K_{\blacktriangleleft}(\mathcal{M})\geq E_{\blacktriangleleft}(\mathcal{M}%
)\geq E_{R}(\mathcal{M})$. The expression of $E_{R}(\mathcal{M})$ can be very
simple. As shown by~\textcite{Pirandola2009b}, an arbitrary one-mode Gaussian
channel\ $\mathcal{G}$ with transmission $\tau\neq1$ has%
\begin{equation}
E_{R}(\mathcal{G})=\max\left\{  0,\log\left\vert \frac{1}{1-\tau}\right\vert
-g(\nu)\right\}  ~,
\end{equation}
where $\nu:=2\bar{n}+1$ and $\bar{n}$ is the thermal number of the channel.
Note that $E_{R}(\mathcal{G})$ can be positive for $\tau\leq1/2$, where the
channel is antidegradable and, therefore, $E(\mathcal{G})=Q(\mathcal{G})=0$.
Thus, despite the fact that the unassisted (forward-assisted) capacities are
zero, the use of a single feedback classical communication is sufficient to
distribute entanglement ($E_{\blacktriangleleft}(\mathcal{G})>0$) and secret
keys ($K_{\blacktriangleleft}(\mathcal{G})>0$). In cryptographic terms,
antidegradibility means that an eavesdropper is able to reconstruct the output
state of Bob. Despite this, Alice and Bob are still able to extract a secret
key from their shared correlations by using a reverse secret-key protocol. This is a
remarkable feature of reverse
reconciliation, further discussed in Chapter~\ref{chap:QKD}.

\subsection{Gaussian channel discrimination and applications\label{sec:qcd}}

\begin{figure}[th]
\vspace{-1.7cm}
\par
\begin{center}
\includegraphics[width=8.5cm]{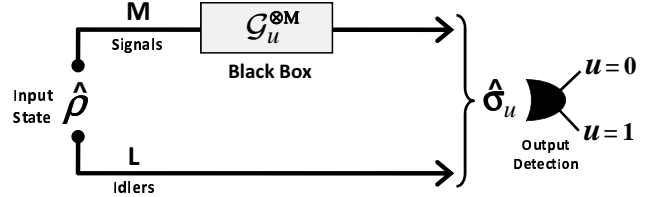}
\end{center}
\par
\vspace{-1.9cm} \caption{Gaussian channel discrimination. The input state
$\hat{\rho}$\ describes $M$ signal modes and $L$\ idler modes. Only the
signals probe the black box which contains one of two possible (one-mode)
Gaussian channels $\mathcal{G}_{0}$\ or $\mathcal{G}_{1}$ (encoding a bit
$u$). At the output, signals and idlers are described by a dichotomic quantum
state $\hat{\sigma}_{u}$\ whose detection gives an estimate of the bit.}%
\label{BoxPIC}%
\end{figure}

The discrimination of quantum channels represents one of the basic problems in
quantum information theory
\cite{Childs2000,Acin2001,Sacchi2005,Wang2006,Chiribella2008,Duan2009,Hayashi2009,Harrow2010,Invernizzi2010}%
. Here we discuss the problem of distinguishing between two Gaussian channels.
Suppose that we have a black box which implements one of two possible
(one-mode)\ Gaussian channels, $\mathcal{G}_{0}$ or $\mathcal{G}_{1}$, with
the same probability, and we want to find out which one it is. In other words,
the box contains an unknown Gaussian channel $\mathcal{G}_{u}$ encoding a
logical bit $u=0,1$ and we want to retrieve the value of this bit. The basic
approach involves probing the box with a one-mode quantum state $\hat{\rho}$
and detecting the corresponding output $\hat{\sigma}_{u}:=\mathcal{G}_{u}(\hat{\rho
})$ by means of a quantum measurement. However, this approach can be readily
generalized. In fact, we can consider multiple access to the box by inputting
$M$ signal modes. Then, we can also consider additional $L$ idler modes, which
are not processed by the box but are directly sent to the output measurement,
as shown in Fig.~\ref{BoxPIC}. Thus, for a given state $\hat{\rho}$ of the
input $M+L$ modes, we have two possible output states, $\hat{\sigma}_{0}$ and
$\hat{\sigma}_{1}$, described by the dichotomic state $\hat{\sigma}_{u}:=(\mathcal{G}%
_{u}^{\otimes M}\otimes\mathcal{I}^{\otimes L})(\hat{\rho})$, where
$\mathcal{G}_{u}^{\otimes M}$ is applied to the signals and the identity
$\mathcal{I}^{\otimes L}$ to the idlers. This output is detected by a
multimode quantum measurement whose outcome estimates the encoded bit. Now,
since $\hat{\sigma}_{0}$ and $\hat{\sigma}_{1}$ are generally non-orthogonal, the bit is
decoded up to an error probability $p_{err}$. Thus, the main goal of the
problem is the minimization of $p_{err}$, which must be done on both input and
output. For fixed input state $\hat{\rho}$, the optimal detection of the
output is already known: this is the Helstrom's dichotomic POVM discussed in
Sec.~\ref{sec:MeasDIST}. However, we do not know which state is optimal at the
input. More precisely, we do not know the optimal input state when we
constrain the signal energy irradiated over the box. Here there are two kinds
of constraints that we can actually consider. The first one is a global energy
constraint, where we restrict the mean total number of photons $m_{tot}$
irradiated over the box. In this case the minimum value of $p_{err}$ can be
non-zero. The second one is a local energy constraint, where we restrict the
mean number of photons $\bar{m}$ per signal mode. In this case, the value of
$p_{err}$ generally goes to zero for $M\rightarrow+\infty$ and the problem is
to achieve the most rapid decaying behavior.\ In both cases finding the
optimal input state for fixed energy is an open problem.

However, we can try to answer related questions: for fixed energy, does
entanglement help? Or more generally: do we need non-classical states for
minimizing $p_{err}$? By definition a state is called classical
(non-classical) when it can (cannot) be written as a probabilistic mixture of coherent states,
i.e., $\hat{\rho}=\int d\boldsymbol{\alpha}P(\boldsymbol{\alpha})\left\vert
\boldsymbol{\alpha}\right\rangle \left\langle \boldsymbol{\alpha}\right\vert
$, where $\left\vert \boldsymbol{\alpha}\right\rangle =\left\vert \alpha
_{1}\right\rangle \otimes\cdots\otimes\left\vert \alpha_{L+M}\right\rangle $
and $P(\boldsymbol{\alpha})$ is a probability density function. Classical
states are always separable and represent the standard sources in today's
optical applications. By contrast, non-classical states (such as number
states, squeezed and entangled states) are only generated in quantum optics
labs. Thus, we can formulate the following question: for fixed signal energy
($m_{tot}$ or $\bar{m}$) and optimal output detection, can we find a
non-classical state which outperforms any classical state in the
discrimination of two Gaussian channels? This basic question has motivated
several theoretical investigations
\cite{Tan2008,Yuen2009,Usha2009,Pirandola2010}. In particular, it has been
answered in two interesting scenarios, with non-trivial implications in quantum technology.

The first scenario is known as \textit{quantum illumination}~\cite{Lloyd2008} with
Gaussian states~\cite{Tan2008}. Here the Gaussian channel
discrimination $\mathcal{G}_{0}\neq\mathcal{G}_{1}$ is related with the
problem of sensing the presence of a low-reflectivity object in a bright
thermal-noise environment. In this case, the black box of Fig.~\ref{BoxPIC}
represents a target region from where the signals are reflected back to the
detector. If the object is absent (bit-value $u=0$) we have a completely
depolarizing channel $\mathcal{C}(0,0,\bar{n})$ which replaces each signal
mode with an environmental mode in a thermal state with $\bar{n}\gg1$ photons.
By contrast, if the object is present (bit-value $u=1$), we have a lossy
channel $\mathcal{L}(\kappa,\bar{n}^{\prime})$ with high loss $\kappa\ll1$ and
high thermal number $\bar{n}^{\prime}:=\bar{n}/(1-\kappa)\gg1$. These channels
are entanglement-breaking, i.e., no entanglement survives at the output. Now,
assuming very few photons per signal mode $\bar{m}\ll1$ (local
constraint), we ask if a non-classical state is able to outperform any
classical state. To this goal, we construct an EPR\ transmitter composed of
$M$ signals and $M$ idlers in a tensor product of EPR\ states, i.e., $\Phi
_{M}:=\hat{\rho}_{11}(r)\otimes\cdots\otimes\hat{\rho}_{MM}(r)$, where
$\hat{\rho}_{ij}(r)$ is an EPR\ state of squeezing $r$ which entangles signal
mode $i$ and idler mode $j$. The corresponding error probability
$p_{err}=p_{\text{EPR}}(M)$\ can be computed using the Gaussian formula of the quantum
Chernoff bound (see Sec.~\ref{sec:QCBformula}). For large $M$, we derive
$p_{\text{EPR}}(M)\simeq\exp(-M\kappa\bar{m}/\bar{n})/2$, which decays to zero
more rapidly than the error probability of any classical state with $M$
signals and arbitrary $L$ idlers. In particular, if we restrict
the classical states to coherent states, then we have an error probability
$p_{coh}(M)\simeq\exp(-M\kappa\bar{m}/4\bar{n})/2$ which is $6$~dB worse than
$p_{\text{EPR}}(M)$~\cite{Tan2008}. Interestingly, the quantum illumination advantage accrues
despite the fact that no entanglement survives at the output. In fact, even if the output
signal-idler correlations are within the classical bounds, there is no
classical input state that can produce a close approximation to this output state. Further studies on quantum illumination of targets have been pursued by \textcite{Shapiro2009} and \textcite{Guha2009}.

The second scenario regards the use of non-classical transmitters
to read data from classical digital memories, such as optical
disks (CDs and DVDs). This is known as \textit{quantum
reading}~\cite{Pirandola2010}. Here the discrimination of Gaussian
channels is associated with the retrieval of information from a
memory cell, modeled as a medium with two possible reflectivities.
This cell is equivalent to the black box of Fig.~\ref{BoxPIC}
where the bit $u=0,1$ specifies two lossy channels,
$\mathcal{L}(\kappa_{0},\bar{n})$ and
$\mathcal{L}(\kappa_{1},\bar{n})$, with the same thermal number
$\bar{n}$ but different losses $\kappa_{0}\neq \kappa_{1}$. For
optical disks, we can consider low noise ($\bar{n}\ll1$) and
$\kappa_{1}$ close to $1$. In these conditions, and irradiating
relatively few photons over the cell $m_{tot}\simeq10$ (global
constraint), we can find an EPR transmitter $\Phi_{M^{\prime}}$,
with small $M^{\prime}$, which is able to outperform any classical
state with any $M$ and $L$. As shown by \textcite{Pirandola2010}, the difference in the readout of
information can be surprising (up to one bit per cell), with
non-trivial implications for the technology of data storage. Follow up studies on the quantum reading of memories have been pursued by various authors~\cite{Nair2011,Bisio2011,Hirota2011,Pirandola2011}.

\section{Quantum Cryptography using Continuous Variables}\label{chap:QKD}

Cryptography is the theory and practice of hiding information
\cite{Menezes1997}. The development of the information age and telecommunications in the last century has made secure communication a must. In the 1970s, public-key cryptography was developed and deals with the tremendous demand for encrypted
data in finance, commerce and government affairs. Public-key cryptography is
based on the concept of one-way functions, i.e., functions which are easy to
compute but extremely hard to invert. As an example, most of the current
internet transactions are secured by the RSA protocol, which is based on the
difficulty of factorizing large numbers~\cite{Rivest1978}. Unfortunately, its
security is not unconditional, being based on the assumption that no efficient
factorization algorithm is known for classical computers. Furthermore, if quantum computers were available today, RSA could be easily
broken by Shor's algorithm~\cite{Shor1997}.

Ideally, it would be desirable to have a completely secure way of communicating, i.e., unconditional security. \textcite{Shannon1949} proved that this is indeed possible
using the one-time pad~\cite{Ver26}. Here two parties, Alice and
Bob, share a pre-established secret key unknown to a potential eavesdropper,
Eve. In this technique, Alice encodes her message by applying a modular
addition between the plaintext bits and an equal amount of random bits from
the secret key. Then, Bob decodes the message by applying the same modular
addition between the ciphertext received from Alice and the secret key. The
main problem of the one-time pad is the secure generation and exchange of the
secret key, which must be at least as long as the message and can only be used once. Distributing very
long one-time pad keys is inconvenient and usually poses a significant
security risk. For this reason, public-key
cryptography is more widely used than the one-time pad.

Quantum cryptography, or quantum key distribution~(QKD) as it is more
accurately known\footnote{Technically, quantum cryptography refers not only to quantum key distribution but also other secrecy tasks such as quantum money, quantum secret and state sharing~\cite{Tyc2002,Lance2004}, quantum bit commitment (albeit with certain constraints~\cite{Magnin2010,Mandilara2011}), and quantum random number generators~\cite{Gabriel2010}. However, it is not uncommon for quantum cryptography and quantum key distribution to be used synonymously in the literature.}, is a quantum technology allowing Alice and Bob to generate
secret keys that can later be used to communicate with theoretically unconditional security.
This is used in conjunction with the one-time pad or another symmetric
cryptographic protocol such as pretty good privacy~\cite{Schneier1995}. The
unconditional security of QKD is guaranteed by the laws of quantum
mechanics~\cite{Gisin2002} and, more precisely, the no-cloning theorem (cf.
Sec.~\ref{sec: quantum dense coding}), which can be understood as a manifestation of the Heisenberg
uncertainty principle. The first QKD protocol was the BB84 protocol~\cite{Bennett1984}. Since then QKD has become one of the
leading fields in quantum information. Despite being a quantum technology, QKD
is not hard to implement experimentally. In fact, the use of telecom
components over normal optical fibers is sufficient to distribute secret keys
with reasonable rates over metropolitan network areas, as recently
demonstrated by the European Union's SECOQC project \cite{SECOQC2007}. Today
QKD can be considered as a mature field \cite{Scarani2009} with several
start-up companies formed around the world.

In this section, we review the continuous-variable version of QKD, whose key elements are the
modulation (encoding) of Gaussian states and Gaussian measurements (decoding), e.g., homodyne and heterodyne detection.
The first continuous-variable QKD protocols were based on a discrete modulation of Gaussian states~\cite{Ralph1999,Hillery2000,Reid2000}. The first
protocol based on a continuous (Gaussian) modulation of Gaussian
states was introduced by \textcite{Cerf2001} and employed squeezed states for the
secret encoding. This idea was readily extended by \textcite{Grosshans2002}
and \textcite{Grosshans2003}, with the design and implementation of the first
continuous-variable QKD protocol based on the Gaussian modulation of coherent
states and homodyne detection. Shortly afterwards, another coherent-state protocol was proposed~\cite{Weedbrook2004,Weedbrook2006} and
implemented~\cite{Lance2005}, known as the no-switching protocol, where homodyne detection is replaced by heterodyne detection. This enables the honest
parties to exploit both quadratures in the distribution of the secret key. It
is important to note that the coherent-state encoding introduced by
\textcite{Grosshans2002} is today at the core of the most promising
continuous-variable QKD implementations, thanks to the possibility of using
standard telecom components
\cite{Lodewyck2005,Lodewyck2007,Lodewyck2007b,Lodewyck2007c,Fossier2009}.

In order to reach significant transmission distances, i.e., corresponding to more than
$3$~dB of loss, two main techniques are commonly used: reverse reconciliation~\cite{Grosshans2003} and post-selection~\cite{Silberhorn2002}. Furthermore, the introduction of new
protocols using two-way quantum communication~\cite{Pirandola2008} and
discrete modulation~\cite{Leverrier2009b,Leverrier2010c} have shown the
possibility of further improvements in terms of transmission range. Recently, it was shown by \textcite{Weedbrook2010} that a secure key could, in principle, be generated over short distances at
wavelengths considerably longer than optical and down into the microwave regime, providing a potential platform for noise-tolerant short-range QKD.

The first security proof in continuous-variable QKD was given by
\textcite{Gottesman2001} using squeezed states.
The proof used techniques from discrete-variable quantum error correction and
worked for states with squeezing greater than $2.51$~dB. Subsequent proofs for
continuous-variable QKD followed, including a proof against individual attacks for coherent-state protocols~\cite{Grosshans2004} and an unconditional security proof which reduced coherent attacks to collective attacks~\cite{Renner20009}. Using the latter result, a
large family of QKD protocols can be analyzed against the simpler collective
Gaussian attacks~\cite{Navascues2006,GarciaPatron2006,Leverrier2010b} which
have been fully characterized by \textcite{Pirandola2008c}. More recently,
finite-size effects have begun to be studied~\cite{Leverrier2010a}, with the
aim of assessing unconditional security when only a finite number of quantum systems have been exchanged.

This section is structured as follows. In Sec.~\ref{sec:QKD1}
we present the various continuous-variable QKD protocols using Gaussian
states. This is followed by an analysis of their security in
Sec.~\ref{sec:QKD2} and finally, in Sec.~\ref{sec:QKD3}, we discuss the future
directions of the field.

\subsection{Continuous-variable QKD protocols\label{sec:QKD1}}

In this section we start by presenting a generic QKD protocol. Then we
continue by illustrating the most important families of continuous-variable
QKD protocols based on the use of Gaussian states. These protocols are
presented as \textit{prepare-and-measure} schemes,
where Alice prepares an ensemble of signal states using a random number generator. In
Sec.~\ref{sec:EBdescription}\ we also discuss the entanglement-based
representation, where Alice's preparation is realized by a suitable measurement
over an entangled source.

\subsubsection{A generic protocol}

Any QKD protocol, be it based on discrete or continuous variables, can be
divided into two steps: (1) quantum communication followed by (2) classical
post-processing. During the quantum communication, Alice and Bob exchange a
significant number of quantum states over a communication channel, which is
modeled as a quantum channel. In each round, Alice encodes a classical random
variable $a$ onto a quantum system which is sent to Bob. This system is
measured by Bob at the output of the channel, thus extracting a random variable
$b$ which is correlated to Alice's. Repeating this procedure many times, Alice and Bob
generate two sets of correlated data, known as the raw keys.

Quantum communication is followed by classical post-processing where the two
raw keys are mapped into a shared secret key (i.e., the final key used to
encode the secret message). The classical post-processing is divided into several
stages \cite{Gisin2002,Assche2006,Scarani2009}. The first stage is the \textit{sifting} of
the keys where Alice and Bob communicate which basis or quadrature they used to encode/decode the information, thus discarding incompatible data. We then have
\textit{parameter estimation}, where the two parties compare a randomly chosen
subset of their data. This step allows them to analyze the channel and
upper-bound the information stolen by Eve. Next, we have \textit{error correction}, where the two parties
communicate the syndromes of the errors affecting their data. As a result,
Alice's and Bob's raw keys are transformed into the same string of bits.
Finally, we have \textit{privacy amplification}. During this step, the two
parties generate a smaller but secret key, reducing Eve's knowledge of the key
to a negligible amount~\cite{Assche2006}. The amount of data to discard is
given by the upper-bound on Eve's information which has been computed during
the parameter estimation stage.

It is important to note that the classical post-processing stages of error correction and privacy amplification involves a public
channel that Alice and Bob use by means of either one-way or two-way classical
communication (the initial stages of sifting and parameter estimation always involve two-way communication). Two-way classical communication is allowed in the postselection protocol which we introduce later. When one-way
classical communication is used and is forward, i.e., from Alice to
Bob, we have direct reconciliation. In this case, Alice's data is the reference
which must be estimated by Bob (and Eve). By contrast, if one-way classical
communication is backward, i.e., from Bob to Alice, then we have reverse
reconciliation, where Bob's data must be estimated by Alice (and Eve). As discussed by \textcite{Pirandola2009b}, both direct and reverse reconciliation can be in
principle realized by using a \textit{single} classical communication. This
observation enables a simple definition of the most general protocols in
direct and reverse reconciliation (direct and reverse protocols). By using
these protocols we can define the direct and reverse secret-key capacities of
an arbitrary quantum channel (see Sec.~\ref{sec:escap}). Another important observation is that the public channel used for the
classical communication must be authenticated. This means that Alice and Bob
have to identify themselves by using a pre-shared secret
key~\cite{Renner2005c}. As a result, QKD does not create secret keys out of
nothing, but rather expands initial secret keys into longer ones.

\subsubsection{Coherent-state protocol (homodyne detection)}

A seminal result in QKD using continuous variables was the discovery that coherent
states are sufficient to distribute secret keys~\cite{Grosshans2002,Ralph2003}. Because
coherent states are much easier to generate in the lab than any other Gaussian state,
this result opened the door to experimental demonstrations and field
implementations. The first Gaussian modulated coherent state protocol utilized direct reconciliation~\cite{Grosshans2002}, followed shortly after by reverse reconciliation~\cite{Grosshans2003}. Since then nearly all proposals have used coherent states as its substrate. The security of coherent-state protocols is based on the fact that coherent states are non-orthogonal (cf. Eq.~\ref{eq: overlap coherent state}), which on its own is a sufficient condition
for QKD (i.e., the no-cloning theorem applies). The quantum communication starts by Alice generating two real variables,
$a_{q}$ and $a_{p}$, each drawn from a Gaussian distribution of variance
$V_{a}$ and zero mean. These variables are encoded onto a coherent state resulting in a mean of
$(a_{q},a_{p})$. By imposing $V_{a}=V-1$, we obtain an average output state
which is thermal of variance $V$. For each incoming state, Bob draws a random
bit $u^{\prime}$ and measures either the $\hat{q}$ or $\hat{p}$ quadrature
using homodyne detection based on the outcome of $u'$. After repeating these steps many times, Alice ends
up with a long string of data encoding the values $(a_{q},a_{p})$ which are
correlated with Bob's homodyne outcomes $b$. The post-processing starts by Bob
revealing his string of random bits $u^{\prime}$ and Alice keeping as the final
string of data $a$ the values ($a_{q}$ or $a_{p}$) matching Bob's quadratures.

\subsubsection{No-switching protocol (heterodyne detection)}

In the previous protocols, Alice generates two real random
variables but in the end only one is ultimately used for the key after the sifting stage. Thus,
one can modify the protocol in order to use both values for the generation
of the key, as shown by \textcite{Weedbrook2004}. The quantum communication
part of the protocol is equivalent to the previous protocols except for
Bob's measurement which is now replaced by heterodyne detection, and enables him to measure
$\hat{q}$ and $\hat{p}$ simultaneously (albeit with a noise penalty demanded by the uncertainty principle). Since there is no longer the random switching
between the two conjugated bases, the random number generator at Bob's side is
no longer needed. After repeating these steps many times, Alice ends up with two
strings of data $(a_{q},a_{p})$ correlated with Bob's data
$(b_{q},b_{p})$. Heterodyne detection allows for a simpler experimental setup, producing higher secret-key rates and can be used in conjunction with all known continuous-variable QKD protocols.

\subsubsection{Squeezed-state protocols}

The ability to use coherent states was a milestone in continuous-variable QKD and is currently, by far, the most popular state to use both theoretically and experimentally. However, the first protocol based on the Gaussian modulation of Gaussian states with Gaussian measurements was given by
\textcite{Cerf2001} and involved using squeezed states. Here Alice
generates a random bit $u$ and a real variable $a$ drawn from a Gaussian
distribution of variance $V_{a}$ and zero mean. Subsequently, she generates a
squeezed vacuum state and displaces it by an amount $a$. Before sending the state through the quantum channel, Alice applies a random phase of
$\theta=u\pi/2$. This is equivalent to randomly choosing to squeeze and
displace either the $\hat{q}$ or $\hat{p}$ quadrature. Averaging the output states over
the Gaussian distribution gives a thermal state whose variance $V$ is the same for $u=0$ and $u=1$, which prevents Eve
from extracting information on which quadrature was selected by Alice. This
imposes the constraint $V_{a}+1/V=V$ on Alice's modulation. Once the state has
reached Bob, he generates a random bit $u^{\prime}$ informing him which quadrature he should measure. Alice and Bob then publicly reveal their strings
of random bits keeping only the data which
corresponds to the same measured quadrature.

Another squeezed state protocol was developed by \textcite{GarciaPatron2009a} where Alice again randomly sends displaced squeezed states to Bob. However, this time Bob uses heterodyne detection rather than homodyne detection, but still disregards either one of his quadrature measurements, depending on Alice's quadrature choice. This reverse reconciliation protocol can be seen
as a noisy version of the protocol with squeezed states and homodyne
detection. Thanks
to this addition of noise, the protocol has an enhanced robustness versus the
noise of the channel which can be interpreted as the continuous-variable
counterpart of the effect described by \textcite{Renner2005} for qubit-based
protocols. Note that such an effect can also be seen in the work of \textcite{Navascues2005}, where the protocol with coherent states and
homodyne detection has a better performance than the protocol with squeezed states and
homodyne detection when using direct reconciliation. Further evidence that
noise can improve the performance of QKD is provided in the work of \textcite{Pirandola2009b}.

\subsubsection{Fully-Gaussian protocols and entanglement-based representation}\label{sec:EBdescription}

\begin{figure}[!ht]
\begin{center}
\includegraphics[width=8cm]{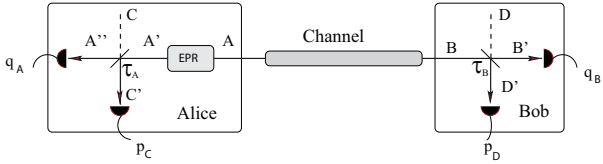}
\caption{Entanglement-based representation for the
fully-Gaussian protocols. Alice has an EPR state $\left\vert
V\right\rangle _{A' A}$ sending mode $A$ to Bob while keeping mode
$A'$. Alice (Bob) mixes her (his) mode $A'$ ($B$) with a vacuum mode in
a beam splitter of transmissivity $\tau_{A}$ ($\tau_{B}$) and subsequently
homodynes the output quadratures. Depending on the value of $\tau_{A}$, Alice
generates a source of squeezed states ($\tau_{A}=1$) or coherent states
($\tau_{A}=1/2$). Then, Bob applies homodyne ($\tau_{B}=1$) or heterodyne
($\tau_{B}=1/2$) detection.}\label{fig:EB-Gen}
\end{center}
\end{figure}

The previous protocols based on coherent states encoding and homodyne detection,
together with the no-switching protocols and the squeezed states protocols are
all based on the Gaussian modulation of Gaussian states followed by Gaussian
measurements. For this reason, we refer to these protocols as
\textit{fully-Gaussian protocols}. Because they can be implemented in direct or
reverse reconciliation, they represent a family of eight protocols. As we will
discuss afterwards, their unconditional security can be simply assessed
against collective Gaussian attacks. By adopting an entanglement-based representation
\cite{Bennett1992,Grosshans2003b}, these\ protocols can be described by a
unique scheme \cite{GarciaPatron2007} where Alice has an EPR state
$\left\vert V\right\rangle _{A' A}$ with noise variance $V$ and sends
one mode $A$ to Bob while keeping the other mode $A'$ for herself (see
Fig.~\ref{fig:EB-Gen}).
Then, Alice mixes her mode $A'$ with a vacuum mode $C$ into a
beam splitter of transmissivity $\tau_{A}$, followed by a measurement of the
output quadratures $\hat{q}_{A}$ and $\hat{p}_{C}$. As a result, she projects
the EPR mode $A$ into a Gaussian state with mean $\mathbf{d}=(\gamma
_{q}q_{A},\gamma_{p}p_{C})$ and covariance matrix $\mathbf{V}=\mathrm{diag}%
(x^{-1},x)$, where $x=(\mu V+1)/(V+\mu)$, $\mu=(1-\tau_{A})/\tau_{A}$ and
\begin{equation}
\gamma_{q}=\frac{\sqrt{\tau_{A}(V^{2}-1)}}{\tau_{A}V+(1-\tau_{A})},~\gamma
_{p}=\frac{\sqrt{(1-\tau_{A})(V^{2}-1)}}{(1-\tau_{A})V+\tau_{A}}.
\end{equation}
It is easy to check that Alice generates a source of squeezed (coherent)
states for $\tau_{A}=1$ ($\tau_{A}=1/2$). Then, Bob applies homodyne
($\tau_{B}=1$) or heterodyne ($\tau_{B}=1/2$) detection depending on which protocol they
want to implement. It is important to note that the entanglement-based representation is a
powerful tool to study many other QKD protocols, including discrete modulation
and two-way protocols. In general, any prepare-and-measure protocol admits an entanglement-based
representation. This is because any ensemble of states on a system $A$ can be
realized by applying a partial measurement on a larger bipartite system
$A+A^{\prime}$ \cite{Hughston1993}.

\subsubsection{Postselection}

Originally, it was believed that the range of continuous-variable QKD protocols could
not exceed the $3$~dB loss limit, as first encountered by direct reconciliation. Exceeding such a limit corresponds to having less than $50\%$ transmission which intuitively means that Eve is getting more information on Alice's data than what Bob is. However, two proposals showed that
such a limit can actually be surpassed, namely reverse reconciliation (as discussed previously) and postselection~\cite{Silberhorn2002}. The quantum communication part of the postselection
protocol is equivalent to the previously mentioned coherent-state protocols. However, the main difference occurs in the classical post-processing stage. In the sifting stage, once Bob has
revealed which quadrature he measured, Alice replies with the absolute value
of her corresponding quadrature ($|a_{q}|$ or $|a_{p}|$). Subsequently, Bob, depending on Alice's revealed value and the absolute value of his measurement outcome $|b|$, decides, following a
pre-established rule, whether they should discard or keep parts of their data. The main concept is
that, every pair of values ($|a|,|b|$) can be associated with a discrete
channel and a binary protocol, based on the signs of $a$ and $b$. A
theoretical secret-key rate $K(|a|,|b|)$ can be calculated for each channel
($|a|,|b|$) from the data obtained during the parameter estimation stage of
the post-processing. The postselection protocol discards those channels for
which $K(|a|,|b|)\leq 0$, keeping only those channels with a positive contribution. A variant of this protocol consists in Bob applying heterodyne instead of
homodyne detection, i.e., a no-switching postselection
protocol~\cite{Lance2005,Lorenz2004}. In such a case, Alice and Bob can extract information from
both quadratures thus increasing the secret-key rate. This version of the postselection protocol has also been experimentally demonstrated~\cite{Lance2005,Lorenz2006}.

\subsubsection{Discrete modulation of Gaussian states}

The very first continuous-variable QKD protocols were based on a discrete (and
hence, non-Gaussian) encoding of Gaussian
states~\cite{Ralph1999,Hillery2000,Reid2000}. However, after the discovery of Gaussian modulated coherent states as a viable resource, the discrete encoding took a back seat with only a small number of papers continuing with the idea~\cite{Namiki2003,Namiki2006,Heid2006}. In recent times though, there has been renewed interest in the discrete encoding of coherent states~\cite{Sych2009,Zhao2009,Leverrier2009b,Leverrier2010c,Leverrier2010d} due to it being experimentally easier to implement as well as its higher error correction efficiencies which promotes continuous-variable QKD over longer distances. A generalized protocol using a discrete modulation~\cite{Sych2009} consists of an alphabet of $N$ coherent states $\ket{\alpha_k}=\ket{a e^{i2\pi k/N}}$ with relative phase
$2\pi k/N$, where $k$ encodes the secret key. Bob uses either homodyne or heterodyne detection in order to estimate $k$. Such a multi-letter encoding scheme can achieve higher key rates under the assumption of a lossy channel. Of the proposals introduced thus far, the classical post-processing stage uses either postselection or reverse reconciliation. The current drawback with discrete modulation Gaussian protocols is the infancy of their security analysis, although promising advances have been made recently~\cite{Zhao2009,Leverrier2009b,Leverrier2010c}.

\subsubsection{Two-way quantum communication}\label{sec_2way}

\begin{figure}[h]
\vspace{-2.0cm}
\par
\begin{center}
\includegraphics[width=0.50\textwidth]{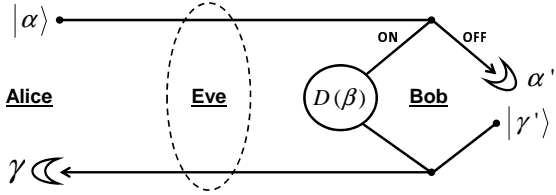}
\end{center}
\par
\vspace{-2.5cm}\caption{Two-way coherent-state protocol. Alice sends a random
coherent state to Bob, who selects between two configurations, ON or OFF.
In ON, Bob applies a random displacement $D(\beta)$.
In OFF, he heterodynes and prepares another
random coherent state $\ket{\gamma'}$. In both configurations,
the output state is sent back to Alice who
performs heterodyne detection. The figure also
displays a two-mode attack (discussed later).}%
\label{2wayPROT}%
\end{figure}

In standard QKD\ protocols the quantum communication is one-way, i.e., quantum
systems are sent from Alice to Bob. In two-way protocols, this process is
bidirectional, with the systems transformed by Bob and sent back to Alice
\cite{Bostroem2002,Bostroem2008}. Recently, \textcite{Pirandola2008}
introduced this idea in continuous-variable QKD, showing how the use of
two-way quantum communication can increase the robustness to noise of the key
distribution. As a result, bosonic channels which are too noisy for one-way
protocols may become secure for two-way protocols. This \textquotedblleft
security activation\textquotedblright\ can have non-trivial applications,
especially in realistic communication lines where the noise is high. For simplicity, we discuss only the two-way coherent-state protocol
depicted in Fig.~\ref{2wayPROT}, which is a two-way extension of the
no-switching protocol. 
Let Alice prepare a random coherent state $|\alpha\rangle$, whose amplitude
$\alpha$ is Gaussian modulated. This state is sent to Bob, who randomly
chooses between two configurations, ON\ and OFF. In ON, Bob applies a random
displacement $D(\beta)$ with $\beta$ Gaussian modulated. In OFF, Bob
heterodynes the incoming state with outcome $\alpha^{\prime}$ and prepares
another random coherent state $\left\vert \gamma^{\prime}\right\rangle $.\ In
both configurations, the state is finally sent back to Alice, who performs
heterodyne detection with outcome $\gamma$. During sifting, Bob declares the
configuration chosen in each round. In ON, Alice processes $\alpha$ and
$\gamma$ to estimate $\beta$. In OFF, Alice considers $\alpha\simeq
\alpha^{\prime}$ and $\gamma\simeq\gamma^{\prime}$. During parameter
estimation, Alice and Bob analyze the noise properties of the channel,
checking for the presence of memory between the forward and backward paths. If
memory is present, they select the OFF configuration only. In this way, they
can destroy the effect of the memory in the post-processing, by choosing only
one of the two paths and processing its data in direct or reverse
reconciliation. In this case the protocol is at least as robust as the underlying
one-way protocol. By contrast, if memory is absent, Alice and Bob can use both
the ON\ and OFF configurations. In this case, the key distribution is more
robust to the noise of the channel, with the enhancement provided by the use
of the ON configuration (see Sec.~\ref{sec_2waySEC} for details.)

\subsubsection{Thermal state QKD}

Generally, it is assumed in all the previous continuous-variable QKD protocols that Alice's initial states originate from encoding classical information onto pure vacuum states. However, in practice this is never possible with some level of impurity occurring due to experimental imperfections. Thermal state QKD therefore addresses this issue where the protocol is now analyzed with respect to Alice using noisy coherent states. This was first investigated by \textcite{Filip2008} and \textcite{Usenko2010} who showed that by using reverse reconciliation the distance over which QKD was secure, fell
rapidly as the states became significantly impure. Extending upon this initial work, \textcite{Weedbrook2010} showed that by using direct reconciliation, and provided that the channel transmission $\tau$ is greater than $50\%$, the security of quantum
cryptography is not dependent on the amount of preparation noise on Alice's states. This is a counterintuitive result as we might naturally expect that as Alice's states
become more and more thermalized, secure transmission
over any finite distance would become impossible. Consequently, the best strategy to deal with preparation noise is to use a combination of direct ($\tau>0.5$) and reverse reconciliation ($\tau \leq 0.5$). This motivated analysis into secure key generation at different wavelengths and was shown that secure regions exists from the optical and infrared all the way down into the microwave region~\cite{Weedbrook2010}.

\subsection{Security analysis\label{sec:QKD2}}

The strongest definition of security in a quantum scenario was given by
\textcite{Renner2005b}. A QKD protocol is said to be $\epsilon$-secure if
\begin{align}
D(\hat{\rho}_{abE},\hat{\sigma}_{ab}\otimes\hat{\rho}_{E})\leq\epsilon
\end{align}
where $D$ is the trace distance as defined in Eq.~(\ref{eq: trace distance}). Here $\hat{\rho}_{abE}$ is the final joint state of Alice, Bob and Eve and
$\hat{\sigma}_{ab} \otimes \hat{\rho}_E$ is the ideal secret-key state. Therefore, up to a probability $\epsilon$ Alice and Bob generate a shared secret key identical to an ideal key and with probability $1-\epsilon$ they abort. In the following we present the necessary tools
to calculate the secret-key rate $K$ for various continuous-variable QKD protocols.

\subsubsection{Main eavesdropping attacks}\label{types of attacks}

To prove the unconditional security of a QKD protocol, the following
assumptions on Eve have to be satisfied: (1) full access to the quantum
channel; (2) no computational (classical or quantum) limitation; (3)
capable of monitoring the public channel, without modifying the messages
(authenticated channel); (4) no access to Alice's and Bob's setups. Under these
assumptions, the most powerful attack that Eve can implement is known as a
\textit{coherent attack}. This consists in Eve preparing a global ancillary system and
making it interact with all the signals sent through the quantum channel, and then
storing the output ancillary system into a quantum memory\footnote{Quantum memory is a device that allows the storage and retrieval of quantum information. It plays a role in many continuous-variable quantum information protocols. For more theoretical details and the status of experimental demonstrations, see e.g., \textcite{Hammerer2010} and \textcite{Lvovsky2009}.}. Finally, after
having listened to all the classical communication over the public channel, Eve
applies an optimal joint measurement on the quantum memory. The security against coherent attacks is extremely complex to address.
Interestingly, by using the quantum de Finetti theorem, proven by
\textcite{Renner20007} for discrete variables and by \textcite{Renner20009}
for continuous variables, we can prove unconditional security in the
asymptotic regime by analyzing the simpler class of collective attacks. For an arbitrary QKD protocol in the entanglement-based
representation, if the multimode entangled state (shared between Alice
and Bob after many uses of the channel)
is permutationally invariant, then, for the quantum de
Finetti theorem, this state can be approximated (asymptotically)
by a mixture of independent and identically distributed
two-mode states. This corresponds to considering the simpler
case of a collective attack.

In a \textit{collective attack} Eve has a set of independent and identically prepared
systems (ancillas) each one interacting individually with a single signal sent
by Alice. In the entanglement-based representation, this implies that the
output state of Alice, Bob and Eve is in a tensor product of $n$ identical
states ($\hat{\rho}_{\mathbf{abE}}=\hat{\rho}_{abE}^{\otimes n}$). Eve's
ancillas are stored in a quantum memory and then, after listening to Alice and Bob's
classical communication, Eve applies an optimal
measurement on the quantum memory. In the asymptotic regime ($n\rightarrow
\infty$), the secret-key rate $K$ can be computed via the formula
\cite{Renner2005}:
\begin{equation}
K= \varphi \left[I(a\mathrm{:}b)-S(x\mathrm{:}E)\right]  , \label{eq:Kcoll}%
\end{equation}
where $I(a\mathrm{:}b)$ is the mutual information between the variables of
Alice ($a$) and Bob ($b$) and $S(x\mathrm{:}E)$ is the Holevo bound between
Alice's (Bob's) variable $x=a$ ($x=b$) and Eve's quantum memory, when direct
(reverse) reconciliation is used. For more on the mutual information and the Holevo bound see Sec.~\ref{sec:ccap}. The coefficient $\varphi \in [0,1]$ models the effect of the
sifting. For instance, we have $\varphi=1$\ for the no-switching protocol,
while $\varphi=1/2$ for the protocol with coherent states and homodyne
detection.

\subsubsection{Finite-size analysis}\label{finite size analysis}

Until now we have considered the asymptotic scenario where Alice and Bob
exchange infinitely many signals. This ideal situation is useful when we are
interested in comparing the optimal performances of different protocols.
However, in practice the number of signals is always finite. The formalism to
address this problem was recently developed for discrete-variable QKD~\cite{Scarani2008,Cai2009}. In what follows we explain the
most important features of finite-size analysis in the continuous-variable
scenario \cite{Leverrier2010a}. In such a situation, the secret-key rate reads
\begin{equation}
K=\frac{\varphi n}{N}\left[  \beta I(a\mathrm{:}b)-S_{\epsilon_{PE}%
}(a(b)\mathrm{:}E)-\Delta(n)-D(n)\right]  ,
\end{equation}
where $N$ is the total number of signals exchanged; $n$ is the numbers of
signals used for the establishment of the key ($N-n$ is used for parameter
estimation); $\beta$ is the \textit{reconciliation efficiency} (ranging from $0$ when no information is extracted to $1$ for perfect reconciliation); $S_{\epsilon_{PE}}(a(b)\mathrm{:}E)$ is the maximal value of Eve's
information compatible with the parameter estimation data; $\Delta(n)$ is
related to the security of the privacy amplification and the speed of
convergence of the smooth min-entropy towards the von Neumann entropy; $D(n)$
is the penalty due to considering collective attacks instead of coherent attacks~\cite{Renner20007,Renner20009,Christandl2009}. The principal
finite-size negative effect in discrete-variable QKD is due to the parameter
estimation \cite{Cai2009} which is expected to be also the case for
continuous-variable QKD~\cite{Leverrier2010a}.

Despite the fact that \textcite{Renner20009} have shown that collective attacks are as
powerful as coherent attacks in the asymptotic regime, the correction $D(n)$
provided for the finite regime leads to a result that could be improved. An alternative approach using the natural symmetries of bosonic channels, was suggested by \textcite{Leverrier2009a}, with only partial results
obtained so far~\cite{Leverrier2009c}. An ideal solution would be finding a
generalization of the \textcite{Leverrier2010b} result by showing that collective Gaussian
attacks are optimal in the finite regime, i.e., $D(n)=0$ (which is the case for the asymptotic scenario as discussed in the next section). The study of finite-size effects in
continuous-variable QKD is very recent and further investigations are needed.

\subsubsection{Optimality of collective Gaussian attacks}\label{optimality of collective Gaussian attacks}

The fully-Gaussian protocols, have the most developed security proofs due to their high symmetry. As we have discussed, \textcite{Renner20009} have shown that, assuming the permutation symmetry of the
classical post-processing, collective attacks are as efficient as coherent
attacks. Therefore, in order to guarantee the security against collective attacks we need
to know what type of collective attack is the most dangerous. A crucial step
in that direction was the discovery that the optimal attack Eve can implement
is one based on Gaussian operations~\cite{Navascues2006,GarciaPatron2006,Leverrier2010b}. This consequently makes the security analysis much easier. \textcite{GarciaPatron2006} showed that, for an entanglement-based
QKD protocol characterized by a tripartite state $\hat{\rho}_{abE}$, i.e.,
resulting from Alice's and Bob's measurements on the pure state
$\ket{\psi}_{ABE}$, of covariance matrix $\mathbf{V}_{ABE}$, the secret-key is
minimized by the Gaussian state $\hat{\rho}_{abE}^{G}$ of the same covariance
matrix, i.e.,
\begin{equation}
K(\hat{\rho}_{abE})\geq K(\hat{\rho}_{abE}^{G}). \label{eq:GaussMinimize}%
\end{equation}
As a result, collective Gaussian attacks represent the fundamental benchmark
to test the asymptotic security of continuous-variable QKD protocols based on
the Gaussian modulation of Gaussian states.

\subsubsection{Full characterization of collective Gaussian
attacks\label{sec:collectiveGauss}}

\begin{figure}[th]
\vspace{-0.6cm}
\begin{center}
\includegraphics[width=7.8cm]{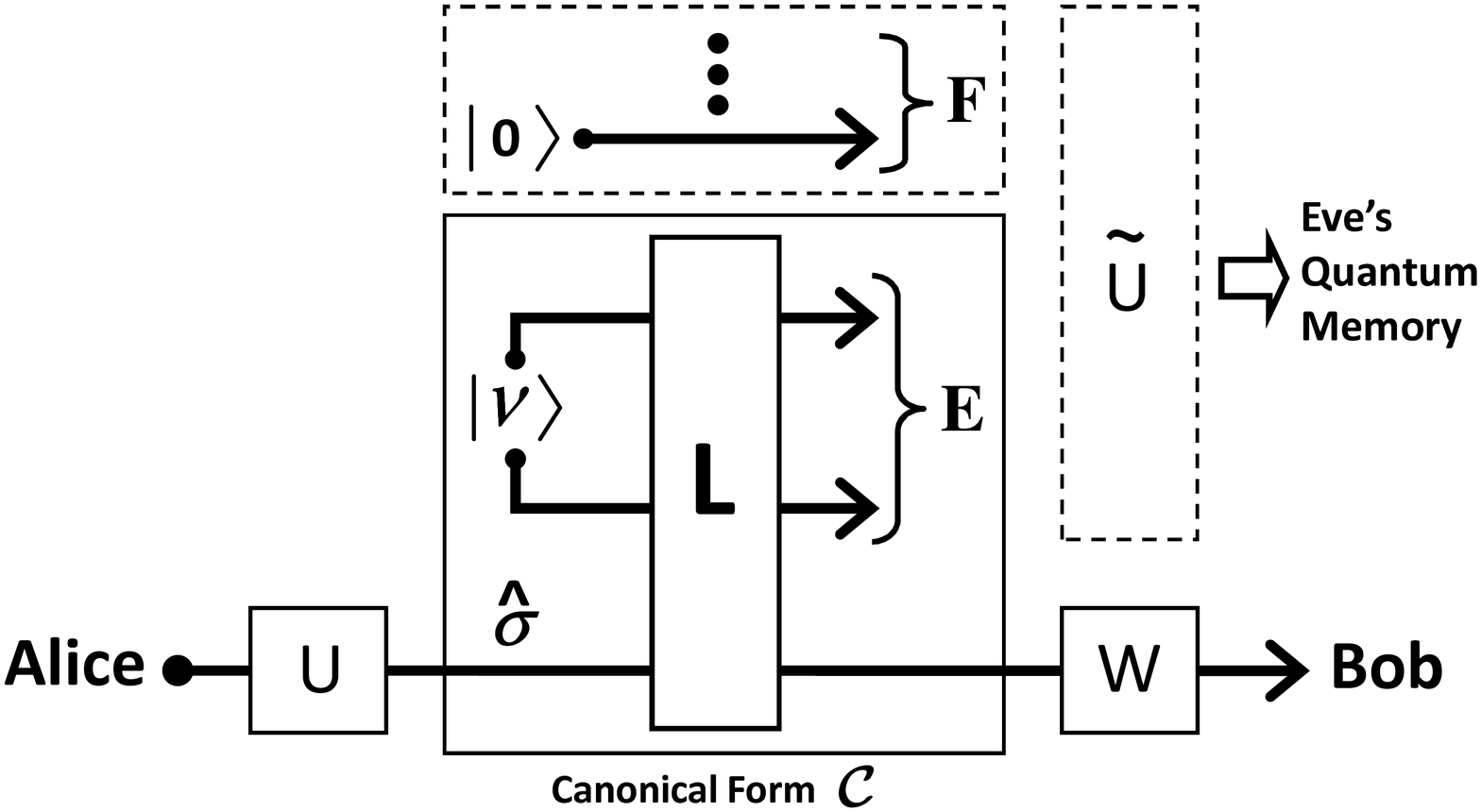}
\end{center}
\par
\vspace{-1.1cm}\caption{Construction of a collective Gaussian attack in four
steps. (1) Any one-mode\ Gaussian channel $\mathcal{G}$ can be reduced to a
canonical form $\mathcal{C}$\ via two Gaussian unitaries $U$ and $W$. (2) Form
$\mathcal{C}$ can be dilated into a symplectic transformation $\mathbf{L}$
mixing the input state $\hat{\sigma}$ with an EPR\ state $\left\vert \nu
\right\rangle $. (3) In a finite box, the dilation is unique up to a unitary
$\tilde{U}$\ combining the output EPR modes $\mathbf{E}$\ with a countable set
of vacuum modes $\mathbf{F}$. (4) After $\tilde{U}$ all of the output is stored in
a quantum memory that Eve measures at the end of the protocol.}%
\label{GCattack}%
\end{figure}

The most general description of a collective Gaussian attack is achieved by
dilating the most general one-mode Gaussian channel into an environment which
is controlled by Eve. As discussed in Sec.~\ref{sec:onechannels}, an arbitrary
one-mode Gaussian channel $\mathcal{G}$ is associated with three symplectic
invariants: transmissivity $\tau$, rank $r$, and thermal number $\bar{n}$.
These quantities identify a simpler channel, the canonical form $\mathcal{C}%
(\tau,r,\bar{n})$, which is equivalent to $\mathcal{G}$\ up to a pair of
Gaussian unitaries $U$ and $W$ (see Fig.~\ref{GCattack}). The canonical form
can be dilated into a symplectic transformation $\mathbf{L}(\tau,r)$ which
mixes the incoming state $\hat{\sigma}$ with an EPR state $\left\vert \nu\right\rangle $ of
variance $\nu=2\bar{n}+1$ (see Fig.~\ref{GCattack}). Now if we treat the
environment as a large but finite box, the dilation is unique up to a unitary
$\tilde{U}$ which transforms the output EPR modes\ $\mathbf{E}$ together with a
countable set of vacuum modes $\mathbf{F}$ (see Fig.~\ref{GCattack}). Thus,
for each use of the channel, Eve's modes $\{\mathbf{E},\mathbf{F}\}$ are
transformed by some $\tilde{U}$\ and then stored in a quantum memory. This
memory is detected at the end of the protocol by means of an optimal coherent
measurement $\mathcal{M}$ which estimates Alice's data (in direct
reconciliation) or Bob's data (in reverse reconciliation). This is the most
general description of a collective Gaussian attack \cite{Pirandola2008c}.

This scenario can be greatly simplified if we use the Holevo bound for Eve's
accessible information. For instance, this happens when we consider the
asymptotic regime, so that Eq.~(\ref{eq:Kcoll}) holds. In this case, we can
ignore the details of $\mathcal{M}$, the extra unitary $\tilde{U}$ and the
extra ancillas $\mathbf{F}$. As a result, the attack is simply described by
the canonical dilation \{$\mathbf{L}(\tau,r),\left\vert \nu\right\rangle $\}
and the one-mode Gaussian unitaries $\{U,W\}$ (see solid boxes in
Fig.~\ref{GCattack}). As discussed in Ch.~II, the Gaussian unitaries $\{U,W\}$
can be further decomposed in displacements, rotations and squeezings. By definition, we call `canonical' the
attacks with $U=W=I$. These attacks are fully described by the canonical
dilation \{$\mathbf{L}(\tau,r),\left\vert \nu\right\rangle $\}
\cite{Pirandola2008c}. The most important canonical attack is the (collective)
entangling-cloner attack~\cite{Grosshans2003}. In this attack, the symplectic
transformation $\mathbf{L}$ represents a beam splitter of transmissivity
$0<\tau<1$\ mixing the incoming signal mode with one mode only of the EPR state
$\left\vert \nu\right\rangle $. Thus, from the point of view of Alice and Bob,
we have a lossy channel with transmissivity $\tau$\ and thermal number
$\bar{n}=(\nu-1)/2$. This channel is the most common, representing the
standard description for communication lines such as optical fibers.

\subsubsection{Secret-key rates}

In this section, we discuss the secret-key rates of the continuous-variable
QKD protocols given in Sec.~\ref{sec:QKD1}. These rates are derived in the
presence of a collective entangling cloner attack which is the most important collective Gaussian attack
in the experimental sense. This attack can be identified by the parameters
of the corresponding lossy channel, i.e., transmission $\tau$\ and thermal
number $\bar{n}$. Equivalently, we can consider $\tau$\ and the
excess noise $\chi:=2\bar{n}(1-\tau
)\tau^{-1}$, i.e., the noise on Bob's side referred to the input (Alice). These parameters are inferred by Alice and Bob during the
parameter estimation stage. Given a specific protocol, the corresponding
secret-key rate can be expressed in terms of the two channel parameters as
$K=K(\tau,\chi)$. Furthermore, the equation $K=0$ defines the
security threshold of the protocol,
expressed in terms of tolerable excess noise $\bar{\chi}$ versus the transmissivity of the
channel, i.e., $\bar{\chi}=\bar{\chi}(\tau)$.

Note that we can derive more general expressions for the secret-key rates by
considering the most general form of a collective Gaussian attack (cf. Sec.~\ref{sec:collectiveGauss}). This
generalization can be found in \textcite{Pirandola2008c} for the no-switching
protocol of \textcite{Weedbrook2004}. The secret-key rates of the other protocols
could be generalized as well. This generalization involves not only the study
of other canonical attacks but also the analysis of phase-effects (mixing of
the quadratures) which derive from the Gaussian unitaries $U$ and $W$. These
effects can be taken into account by introducing suitable corrections in the
expressions of the rates. Another possibility is reducing an attack to a canonical attack ($U = W =1$)
by means of random transformations in the post-processing stage, which
sacrifices part of the secret data. This symmetrization has been recently
used by \textcite{Leverrier2009a} to delete phase-effects from lossy channels.

\paragraph{Fully-Gaussian protocols\newline}

Here we discuss the secret-key rates for the family of fully-Gaussian protocols. In the entanglement-based description, one mode of an EPR state of
variance $V$ is sent through the lossy channel with transmissivity $\tau$\ and
excess noise $\chi$. At the output of the channel, Alice and Bob's bipartite
state is Gaussian with covariance matrix
\begin{equation}
\mathbf{V}_{AB}=\left(
\begin{array}
[c]{cc}%
x\mathbf{I} & z\mathbf{Z}\\
z\mathbf{Z} & y\mathbf{I}%
\end{array}
\right)  , \label{eq:CMchannel}%
\end{equation}
$x=V$, $y=\tau(V+\chi)$ and $z=\sqrt{\tau(V^{2}-1)}$. Now the various
protocols differ for the measurements of both Alice and Bob and the kind of
reconciliation used. In order to estimate Eq.~(\ref{eq:Kcoll}) we first calculate
Alice and Bob's mutual information~\cite{GarciaPatron2007}
\begin{equation}
I(a\mathrm{:}b)=\frac{w}{2}\log\left[  (V+\chi)/(\chi+\lambda V^{-1})\right],
\end{equation}
where $w=1$, except for the no-switching protocol where $w=2$; then $\lambda=V$
($\lambda=1$) for protocols with coherent (squeezed) states. The calculation of Eve's information is more involved. As an example, we consider the calculation of $S(b\mathrm{:}E)=S(E)-S(E|b)$ for reverse reconciliation
using coherent states and homodyne detection. First we use the fact that
Eve's system $E$ purifies $AB$, i.e., $S(E)=S(AB)$, where
$S(AB)$ can be calculated from the symplectic eigenvalues of the matrix $\mathbf{V}_{AB}$ using Eq.~(\ref{symple}), which are then substituted into
Eq.~(\ref{VN_Gauss}). Next, to calculate the term $S(E|b)$ we find the symplectic eigenvalues of the covariance matrix $V_{E|b}$, computed using Eq.~(\ref{eq: CM under homo}), and then proceed as before.

\paragraph{Postselection\newline}

Determining the secret-key rates for postselection is more challenging than either direct or reverse reconciliation. Despite the fact that the postselection protocol involves Gaussian elements in the quantum
communication part, its description becomes non-Gaussian after the filtering
of data. Consequently, using Eq.~(\ref{eq:GaussMinimize}) to upper-bound Eve's information no longer applies. Therefore, a subtler analysis has to be carried out to
obtain tighter bounds. Here we present the basic security analysis for the postselection protocol against collective entangling cloner attacks~\cite{Heid2007,Symul2007}. When Bob performs homodyne detection, the mutual information between Alice
and Bob for a given pair of variables ($|a_{q(p)}|,|b|$) is given by Shannon's
formula for a binary channel~\cite{Shannon1948}
\begin{equation}
I(a:b)=1+p_{e}\log p_{e}+(1-p_{e})\log(1-p_{e})~. \label{eq:ShannonFormula}%
\end{equation}
Here $p_{e}$ is Bob's error in determining the value of Alice's sign and is given by
\begin{equation}
p_{e}=\left\{  1+\exp\left[  8\sqrt{\tau}|a_{q(p)}b|/(1+\chi)\right]
\right\}  ^{-1}.
\end{equation}
Now if Bob performs heterodyne detection, we have to consider the Shannon
formula for two parallel binary channels, one per quadrature, and is given by~\cite{Lance2005}
\begin{equation}
p_{e}=\left\{  1+\exp\left[  4\sqrt{2\tau}|a_{q(p)}b|/(1+\chi)\right]
\right\}^{-1}.
\end{equation}
Eve's information is calculated using the Holevo bound between Eve's system
and the information bits used as reference for the key ($a$ in direct
reconciliation and $b$ in reverse reconciliation). The key rate for a given
pair of values ($a,b$) reads
\begin{equation}
\Delta K=\varphi \hspace{1mm} \mathrm{max}\{\beta I(a:b)-S^{a,b}(E:x),0\},
\end{equation}
where again $\varphi$\ accounts for the sifting and $\beta$\ for the reconciliation
efficiency.
The postselection is then modeled by the \textit{maximum} function, imposing
a zero contribution of the effective binary channel when $\beta I(a:b)-S^{a,b}%
(E:x)<0$, as expected. Finally, the evaluation of the overall secret-key
rate needs to be calculated numerically and is given by
\begin{equation}
K=\int p(a,b)~\Delta K(a,b)~da~db,
\end{equation}
where $p(a,b)$ is a joint probability distribution. A detailed experimental analysis was
carried out by \textcite{Symul2007} using postselection with the no-switching protocol.

\paragraph{Discrete modulation of Gaussian states\newline}

One of the technical advantages of continuous-variable QKD is that it relies solely on standard high-speed optical telecom components. However to date, field
implementations have been restricted to short distances ($27$~km by
\textcite{Fossier2009}). The main reason is the low efficiency of the reconciliation stage for protocols using Gaussian modulation~\cite{Lodewyck2007b}. This is especially true at low signal-to-noise
ratios~\cite{Leverrier2008}, which is the working regime when distributing
secret keys over long distances. On the other hand, extremely good reconciliation protocols exist for discrete
modulations, as the error correction procedure is greatly simplified. In this
case, the problem can be mapped onto a binary channel with additive noise, for
which there exists very good codes, such as low-density-parity check
codes~\cite{Richardson2001}. Unfortunately, protocols based on discrete
modulation, even if using Gaussian states, have non-Gaussian
entanglement-based representations. As a result, the calculation of Eve's
information can no longer rely on the previous optimality proofs~\cite{Navascues2006,GarciaPatron2006,Leverrier2010b}.

However, the proof by \textcite{GarciaPatron2006} can still be used to provide
a (non-tight) Gaussian upper bound on Eve's information. This idea was used by
\textcite{Leverrier2009b}, where a protocol with four coherent states was
shown to outperform Gaussian-modulated protocols in the regime of low
signal-to-noise ratio. The crucial point was the observation that the
four-state modulation well approximates the Gaussian modulation for low
modulation variances. As a result, the Gaussian upper bound can still be used, being
nearly tight in the studied regime. The following year \textcite{Leverrier2010c} proposed another non-Gaussian (but continuous) modulation
protocol able to exploit the Gaussian upper bound. In this protocol, Alice
generates points centered on an eight-dimensional sphere to decide which
ensemble of four successive coherent states are to be sent. Then, Bob uses the no-switching protocol (heterodyne detection) to guess the point selected by Alice. The secret-key rate
reached by this new protocol is higher, by nearly an order of magnitude, for
realistic parameters, which enables the distribution of a secret keys over distances of the order of
$50$~km, even after taking all finite-size effects into account.

\paragraph{Two-way quantum communication\newline}\label{sec_2waySEC}

In general, the security analysis of two-way protocols is quite involved. For
simplicity, here we consider the two-way coherent-state protocol of
Fig.~\ref{2wayPROT}. Using symmetrization arguments
\cite{Renner2005b,Renner20009}, one can reduce an arbitrary coherent attack to
a two-mode attack, affecting each round-trip independently. This attack can
have a residual memory between the two uses of the quantum channel. If this
memory is present, Alice and Bob use the OFF\ configuration, thus collapsing
the protocol to one-way quantum communication. Correspondingly, the attack is
reduced to one-mode, i.e., collective, which can be bounded by assuming a
Gaussian interaction (collective Gaussian attack). Thus, in OFF, the security
threshold is given by the underlying one-way protocol (the no switching
protocol in this case). The advantage occurs when no memory is present, which
is the most practical situation. In particular, this happens when the original
attack is already collective. In this case, Alice and Bob can use both the ON
and OFF configurations to process their data. While the OFF configuration is
equivalent to two instances of one-way protocol (forward and backward), the ON
configuration is based on a coherent two-way quantum communication. Let us
consider the case of a collective entangling cloner attack, which results in a
one-mode lossy channel with transmissivity $\tau$\ and excess noise $\chi$.
For every $\tau\in(0,1)$, the key distribution is possible whenever the excess
noise $\chi$ is below a certain value $\bar{\chi}$ specified by the security
threshold $\bar{\chi}=\bar{\chi}(\tau)$. As shown by~\textcite{Pirandola2008},
the security threshold in ON configuration is higher than the one in OFF
configuration. For instance, if we consider reverse reconciliation, we have
$\bar{\chi}_{\text{ON}}(\tau)>\bar{\chi}_{\text{OFF}}(\tau)$ for every
$\tau\in(0,1)$. As a result, there are lossy channels whose excess noise
$\chi$ is intolerable in OFF but still tolerable in ON. Thanks to this
security activation, the two-way coherent-state protocol is able to distribute
secret keys in communication lines which are too noisy for the corresponding
one-way protocol. This result, which has been proven for large modulation and
many rounds (asymptotic regime), is also valid for other Gaussian modulation
protocols, extended to two-way quantum communication via the hybrid ON/OFF
formulation.

\subsection{Future directions}\label{sec:QKD3}

Continuous-variable QKD offers a promising alternative to the traditional discrete-variable QKD systems (for a state-of-the-art comparison between the various QKD platforms see \textcite{Scarani2009}). An important next step for continuous-variable QKD is to prove unconditional
security in a fully realistic scenario, for example, by improving the reconciliation
procedures and taking finite-size effects into account. This could potentially provide
extremely high secret-key rates over distances which are comparable to the
ones of discrete-variable protocols (about 100 km). Thus, additional research
efforts are focused to extend the range of continuous-variable QKD protocols. As opposed
to the single-photon detectors of discrete-variable QKD, the use of homodyne
detection in continuous-variable QKD provides an outcome even for the vacuum
input. Filtering out this vacuum noise is the main weakness in the
reconciliation procedures. From this point of view, postselection is the best
choice. Therefore, proving the unconditional security of the postselection
protocol would be of great interest\footnote{At the time of writing a paper by \textcite{Walk2011} has appeared which addresses this issue.}. Another possibility is the design of new protocols which are more robust to
excess noise, i.e., with higher security thresholds. This would enable the
reconciliation procedures to work much more efficiently. Such a possibility has been
already shown by the use of two-way protocols. Thus further directions include
the full security analysis of protocols based on multiple quantum
communication.

The further development of continuous-variable quantum
repeaters is also an important research direction. Quantum repeaters would allow one to distribute entanglement between
two end-points of a long communication line, which can later be used to
extract a secret key. This technique combines entanglement distillation,
entanglement swapping, and the use of quantum memories. Unfortunately,
Gaussian operations cannot distill Gaussian entanglement which poses a serious
limitation to this approach. However, there has been ongoing research effort in the direction of Gaussian preserving optical entanglement distillation employing non-Gaussian elements~\cite{Bro03,Eisert2004,Ral09,Menzies2007,Fiurasek2003a}.

\section{Continuous-Variable Quantum Computation using Gaussian Cluster States}\label{chapter computation}

Quantum computation using continuous variables was first considered by \textcite{Llo99} in the circuit model of quantum computing~\cite{Nielsen2000}. They showed that arbitrary quantum logic gates (i.e., simple unitaries) could
be created using Hamiltonians that are polynomial in the quadrature
operators $\hat{q}$ and $\hat{p}$ of the harmonic oscillator. Years later, a different but computationally equivalent model of continuous-variable quantum computation, known as cluster state quantum computation, was developed by \textcite{Zha06} and \textcite{Men06}. This measurement-based protocol
of quantum computation was originally developed by \textcite{Rau01} for discrete variables and forgoes actively implementing quantum gates. Instead, the computation is achieved via local measurements on a highly entangled multimode state, known as a cluster state. In the ideal case, the continuous-variable cluster state is created using infinitely squeezed states, but in practice, is approximated by a finitely squeezed Gaussian entangled state.

These two models of continuous-variable computation must be associated with a fault tolerant and error correctable system, where at some point
the continuous variables are discretized. With this in mind, there is a third type of continuous-variable quantum computer, known as the Gottesman-Kitaev-Preskill quantum computer~\cite{Got01}. This proposal shows how to encode finite-dimensional qubits into the infinite-dimensional harmonic oscillator, thus facilitating fault tolerance and quantum error correction. In this section, we focus primarily on cluster state quantum computation while still using important elements of the Gottesman-Kitaev-Preskill computer and the Lloyd-Braunstein model.

This section is structured as follows. We begin by introducing important continuous-variable gates as well as defining what constitutes a universal set of gates. In Sec.~\ref{sec: one-way QC using CVs} we introduce the notion of one-way quantum computation using continuous variables and how one can gain an understanding of it by considering a teleportation circuit. The important tools of graph states and nullifiers follows in Sec.~\ref{sec: graph states and nullifiers}, while the realistic case of Gaussian computational errors due to finite squeezing is discussed in Sec.~\ref{sec: Gaussian Errors from Finite Squeezing}. The various proposals for optically implementing Gaussian cluster states are revealed in Sec.~\ref{Chapter 8 subsection optical implementation}. In Secs.~\ref{Chapter 8 universal quantum computation} and \ref{chapter 8 error correction}, achieving universal quantum computation and quantum error correction for continuous variables, are discussed respectively. Two examples of algorithms for a continuous-variable quantum computer are given in Sec.~\ref{ch: QC algorithms}, before ending with future directions of the field in Sec.~\ref{ch: QC future directions}.

\subsection{Continuous-variable quantum gates}\label{section: CV gates}

Before introducing the quantum gates used in continuous-variable quantum computation we remind the reader that the displacement gate $D(\alpha)$, the beam splitter gate $B$, and the one and two-mode squeezing gates, $S$ and $S_2$, are important Gaussian gates which have already been introduced in Sec.~\ref{sec: examples of gaussian states}. To begin with, in Gaussian quantum information processing there are the Heisenberg-Weyl operators which comprise of the position and momentum phase-space displacement operators, given respectively as
\begin{align}
X(s) = {\rm exp}(-i s \hat{p}/2), \hspace{8mm} Z(t) = {\rm exp} (i t \hat{q}/2 ),
\end{align}
where $X(s)$ gives a shift by an amount $s$ in the $q$ direction and $Z(t)$ a momentum shift by an amount $t$, i.e., in terms of the displacement operator $D(\alpha)$ they can be rewritten as $X = D(s/2)$ and $Z = D(it/2)$. They are related via $X(s) Z(t) = e^{-ist/2} Z(t) X(s)$ and act on the position computational basis states $\ket{q}$ as
\begin{align}
X(s)\ket{q} = \ket{q+s}, &\hspace{8mm} Z(t)\ket{q} = e^{itq/2}\ket{q}.
\end{align}
and on the momentum basis states $\ket{p}$ as
\begin{align}
X(s)\ket{p} = e^{-isp/2}\ket{p}, &\hspace{8mm} Z(t)\ket{p} = \ket{p+t}.
\end{align}
The position and momentum basis states are related via a Fourier transform as defined in Eq.~(\ref{eq: basis states relation via fourier }). The Fourier gate $F$ is the Gaussian version of the qubit Hadamard gate and can be defined in terms of the annihilation and creation operators as well as the quadrature operators
\begin{align}
F = {\rm exp}\Big(\frac{i \pi}{4}\Big) {\rm exp}\Big[\frac{i\pi}{2} \hat{a}^{\dagger} \hat{a}\Big] = {\rm exp}\Big[\frac{i\pi}{8} \Big(\hat{q}^2 + \hat{p}^2 \Big)\Big].
\end{align}
In the phase-space picture, the Fourier gate is a $\pi/2$ rotation, e.g., from one quadrature to the other
\begin{align}
F^{\dagger} \hat{q} F = -\hat{p}, \hspace{1cm} F^{\dagger} \hat{p} F = \hat{q}.
\end{align}
The Fourier gate acts on the displacement gates as follows
\begin{align}
F^{\dagger} Z(t) F = X(t), \hspace{1cm} F X(s) F^{\dagger} = Z(s).
\end{align}
Finally, the Fourier gate acting on the quadrature eigenstates gives
\begin{align}
F \ket{x}_q &= \ket{x}_p, \hspace{1.5cm} F^{\dagger} \ket{x}_q = \ket{-x}_p,\\
F \ket{x}_p &= \ket{-x}_q, \hspace{1.2cm} F^{\dagger} \ket{x}_p = \ket{x}_q,
\end{align}
where the subscript is used to remind us whether we are in the computational $q$ basis or the conjugate $p$ basis. The phase gate $P(\eta)$ can be thought of as a type of shearing operation, i.e., a combination of rotations and squeezers. It is defined as
\begin{align}
P(\eta) = {\rm exp}\Big[\Big(\frac{i \eta}{4}\Big) \hat{q}^2 \Big].
\end{align}
where $\eta \in \mathbb{R}$. The phase gate acts on the $X(s)$ displacement gate as
\begin{align}
P^{\dagger} X(s) P &= e^{i \eta s^2/4} X(s) Z(\eta s),
\end{align}
while leaving $Z$ unaltered. The phase gate affects the quadratures as
\begin{align}
P^{\dagger} \hat{q} P = \hat{q}, \hspace{1cm} P^{\dagger} \hat{p} P = \hat{p} + \eta \hat{q}.
\end{align}
The controlled-phase gate, or \textit{\small CPHASE} for short, is a two-mode Gaussian gate defined as
\begin{align}\label{eq: CPHASE gate}
C_Z = {\rm exp}\Big[ \Big(\frac{i}{2}\Big)  \hat{q}_1 \otimes \hat{q}_2 \Big].
\end{align}
The effect of this two-mode gate on the computational basis states is given by
\begin{align}
C_Z \ket{q_1} \ket{q_2} = e^{iq_1 q_2/2} \ket{q_1} \ket{q_2},
\end{align}
In the Heisenberg picture, the {\small CPHASE} gate transforms the momentum quadratures according to
\begin{align}
\hat{p}_1 \rightarrow \hat{p}_1 + \hat{q}_2, \hspace{1cm} \hat{p}_2 \rightarrow \hat{p}_2 + \hat{q}_1,
\end{align}
while doing nothing to the position quadratures $\hat{q}_1 \rightarrow \hat{q}_1$ and $\hat{q}_2 \rightarrow \hat{q}_2$. The {\small CPHASE} gate and the phase gate both get their names from the analogous discrete-variable gates and their similar actions on the Pauli matrices~\cite{Nielsen2000}.

Finally, we note the graphical representation of the quantum gates in the circuit model of computation. The single-mode Gaussian gates: Heisenberg-Weyl displacement gates, single-mode squeeze gate, Fourier gate and the phase gate are given respectively as
\begin{align}
\Qcircuit @C=1em @R=.7em {
& \gate{X} & \qw & & \gate{Z} & \qw & & \gate{S} & \qw & & \gate{F} & \qw &  & \gate{P} & \qw
}
\end{align}
While the two-mode Gaussian gates: the \textit{\small CPHASE} gate, the beam splitter gate and the two-mode squeeze gate, are denoted respectively by
\begin{eqnarray}
    \Qcircuit @C=1.5em @R=1.8em @!R {
     & \ctrl{1} & \qw & & \multigate{1}{B} & \qw & & \multigate{1}{S} & \qw \\
     & \control \qw & \qw & & \ghost{B} & \qw & & \ghost{S} & \qw
     }
\end{eqnarray}

\subsubsection{Universal set of quantum gates}

A continuous-variable quantum computer is said to be universal if it can implement an arbitrary Hamiltonian with arbitrarily small error. So what are the necessary and sufficient conditions for a continuous-variable quantum computer to be universal? This is given by the Lloyd-Braunstein criterion~\cite{Llo99} which tells us which gates are needed to generate any unitary transformation to arbitrary accuracy. This consists of the two families of gates:
\begin{enumerate}
  \item $Z(t)$, $P(\eta)$, $F$ and $U_G$ (which is any multimode Gaussian gate, e.g., $C_Z$ or $B$), $\forall$ $t,\eta \in \mathbb{R}$. This first family generates all possible Gaussian operations.
  \item $\exp[it\hat{q}^{n\ge 3}]$ (for some value of $t$) which is a non-linear transformation of polynomial degree $3$ or higher and corresponds to a family of non-Gaussian gates.
\end{enumerate}
Note that if we were restricted to using only Gaussian gates we would not be able to synthesize an arbitrary Hamiltonian. In fact, the continuous-variable version~\cite{Bar02} of the Gottesman-Knill theorem~\cite{Got98} tells us that starting from an initial Gaussian state, Gaussian processing (which includes Gaussian measurements and Gaussian operations) can be efficiently simulated on a classical computer.

\subsection{One-way quantum computation using continuous variables}\label{sec: one-way QC using CVs}

One-way quantum computation~\cite{Rau01} using continuous variables~\cite{Zha06,Men06,Gu09} allows one to perform any computational algorithm by implementing a sequence of single-mode measurements on a specially entangled state known as a cluster state (note that we often begin our analysis using a perfectly entangled state but move to the more realistic case of a Gaussian cluster state as we progress). Here quantum gates are not required, as arbitrary Hamiltonians are simulated via measurements alone. After each measurement is performed the resulting measurement outcome is used to select the basis of the next measurement. In general, the order in which measurements are made does matter, a property known as \textit{adaptiveness}. However, when implementing only Gaussian gates, this condition is relaxed and the order no longer matters, a property known as \textit{parallelism}. The two basic steps of continuous-variable cluster state quantum computation can be summarized as follows:
\begin{enumerate}
   \item Cluster state preparation: All qumodes are initialized as highly squeezed vacuum states, approximating momentum eigenstates $\ket{0}_p$. The $C_Z$ gate is applied to the relevant qumodes in order to create the entangled cluster state.
   \item Measurements: single-mode measurements are made on the relevant qumodes where each result is used to select the subsequent measurement basis.
 \end{enumerate}
Here a quantum mode, or \textit{qumode} for short, is the continuous-variable analogue of the discrete-variable qubit and is simply a continuous-variable quantum state or mode. Note that, up until this point in the review, we have simply referred to such states as modes. However in line with the terminology used in the current research of continuous-variable cluster states, we will refer to such quantum states as qumodes.

\subsubsection{Understanding one-way computation via teleportation}\label{Chapter 8 Understanding One-Way Computation via Teleportation}

To get a feel of how measurements allow us to generate arbitrary evolutions in cluster state computation, it is helpful to look at quantum teleportation from the perspective of the quantum circuit model. The quantum circuit for the gate teleportation of a single-mode continuous-variable quantum state $\ket{\psi}$ is given by~\cite{Men06}
\begin{eqnarray}
    \Qcircuit @C=1em @R=1.8em @!R {
     \lstick{\ket \psi}  & \ctrl{1} & \meter & \rstick{\hat{p}=m_1} \qw \\
     \lstick{\ket 0_p} & \control \qw & \qw & \rstick{X(m_1) F \ket \psi} \qw
     }\\\nonumber
\end{eqnarray}
The above circuit can be understood in the following way. The input states consist of the arbitrary state $\ket{\psi}$ that we wish to teleport and a momentum eigenstate $\ket{0}_p$ (note that we will begin by considering the unphysical case of perfectly squeezed vacuum states with the realistic case of Gaussian squeezed states discussed later). They are entangled using a $C_Z$ gate. A $\hat{p}$ quadrature measurement is performed resulting in the outcome $m_1$. The state $\ket{\psi}$ is thus teleported from the top quantum wire to the bottom wire and can be fully restored by applying the corrections $F^{\dagger} X^{\dagger}(m_1)$ to the output state. We will now go through the above circuit in more detail. First, the two initial input states can be written as $\ket{\psi} \ket{0}_p$. Expanding them into the position basis gives $\ket{\psi} \ket{0}_p = (2\sqrt{\pi})^{-1} \int dq_1 dq_2 \psi(q_1) \ket{q_1} \ket{q_2}$
where $\ket{\psi} = \int dq_1 \psi(q_1) \ket{q_1}$ and $\ket{0}_p = (2\sqrt{\pi})^{-1} \int dq_2 \ket{q_2}$. Applying the {\small CPHASE} gate leads to
\begin{align}
C_Z \ket{\psi} \ket{0}_p = \frac{1}{2\sqrt{\pi}} \int dq_1 dq_2 \psi(q_1) e^{i q_1 q_2/2} \ket{q_1} \ket{q_2}.
\end{align}
After measuring $\hat{p}$ of the first mode, using the projector $\ket{m_1}\bra{m_1}_p$, and obtaining the result $m_1$ we get
\begin{align}\nonumber
\frac{1}{4 \pi} \int dq_1 dq_2 \psi(q_1) e^{i q_1 (q_2 - m_1)/2} \ket{q_2},
\end{align}
where we used $\langle m_1 \ket{q_1}  = (2 \sqrt{\pi})^{-1} {\rm exp}(-i q_1 m_1/2)$. The above state can be rewritten as: $\ket{\psi'} = X(m_1) F \ket{\psi}$. As mentioned before applying the corrections gives back the initial state: $F^{\dagger} X^{\dagger}(m_1)\ket{\psi'} = \ket{\psi}$.

Having considered the teleportation of an arbitrary state, we are now in a position to consider the teleportation of a quantum gate~\cite{Bartlett2003}, which is at the heart of the measurement-based model of computation. This requires only a slight alteration to the previous circuit as we will be considering teleporting gates that are diagonal in the computational basis and thus commute through the {\small CPHASE} gate. For example, the circuit below teleports the state $\ket{\psi'} = U \ket{\psi}$ where $U = \exp[i f(\hat{q})]$ is a gate diagonal in the computational basis
\begin{eqnarray}\label{circuit: cubic teleportation final222}
    \Qcircuit @C=1em @R=1.8em @!R {
     \lstick{U \ket \psi}  & \ctrl{1} & \meter & \rstick{\hat{p}=m_1} \qw \\
     \lstick{\ket 0_p} & \control \qw & \qw & \rstick{X(m_1) F U\ket \psi} \qw
     }\\\nonumber
\end{eqnarray}
The above circuit is equivalent to the circuit below, where $U$ can be absorbed into the measurement process
\begin{eqnarray}\label{circuit: cubic teleportation final333}
    \Qcircuit @C=1em @R=1.8em @!R {
     \lstick{\ket \psi}  & \ctrl{1}      & \meter & \rstick{U^{\dagger} \hspace{1mm} \hat{p} \hspace{1mm} U =m_1} \qw \\
     \lstick{\ket 0_p} & \control \qw & \qw & \rstick{X(m_1) F U\ket \psi} \qw
     }\\\nonumber
\end{eqnarray}
The above circuit forms the basis for our understanding of measurement-based quantum computation. Let us stop for a moment to consider why this is. Circuit~(\ref{circuit: cubic teleportation final222}) is the typical quantum circuit where an algorithm (in this case gate teleportation) is achieved by first implementing a quantum gate $U$ onto a quantum state $\ket{\psi}$. However, circuit~(\ref{circuit: cubic teleportation final333}) shows us that we no longer need to explicitly implement the quantum gate but we can simulate the effect of the gate using only measurements in a new basis. This effect is the building block of cluster state computation where we can concatenate a number of these circuits to form a larger cluster state.

\subsubsection{Implementing gates using measurements}

We have just shown that by performing a measurement in the basis $U^{\dagger} \hat{p} U$ we can simulate the effect of the $U$ gate on an arbitrary state. Using this result with the previously mentioned Lloyd-Braunstein criterion, we are able to implement the set of universal Hamiltonians $\hat{q},\hat{q}^2, \hat{q}^3$ using only measurements. We can forget about the two-mode Gaussian gate $U_G$ from the set as we have already used it in creating the cluster via the $C_Z$ gate. We also use the Hamiltonian $\hat{q}^3$, rather than any other higher order polynomial, because we know how to optically implement it (more on this in Sec.~\ref{Chapter 8 universal quantum computation}). This corresponds to the following three transformations
\begin{align}
U_{j}^{\dagger} \hat{p} \hspace{1mm} U_{j} = \hat{p} + t \hat{q}^{j-1}
\end{align}
for $j=1,2,3$ and where the gates diagonal in the computational basis are conveniently written as $U_{j} = \exp[(it/2j) \hat{q}^j]$. Notice that $U_{1}$ corresponds to the Heisenberg-Weyl displacement operator $Z(t)$, $U_{2}$ is the phase gate $P(t)$ and $U_{3}$ is known as the cubic phase gate, denoted as $V(t)$. So how are the above transformations optically implemented? Well, the first one is achieved by simply measuring $\hat{p}$ and adding $t$ to the measurement result. The second one is a homodyne measurement in a rotated quadrature basis: $(\hat{p}\hspace{1mm} {\rm cos} \theta - \hat{q} \hspace{1mm} {\rm sin}\theta)/{\rm cos \theta}$. However, the cubic Hamiltonian is a little more difficult to implement than the previous two and will be discussed in more detail in Sec.~\ref{Chapter 8 universal quantum computation}.

\subsection{Graph states and nullifiers}\label{sec: graph states and nullifiers}

\subsubsection{Graph states}

A common and convenient way of depicting cluster states is by using graphs. The continuous-variable version of graph states was
defined by \textcite{Zha06} where every continuous-variable cluster
state can be represented by a graph~\cite{Gu09} known as a graph state\footnote{From now on we will use the terms ``cluster states" and ``graph states" interchangeably. Note that some authors technically refer to a cluster state as one which has a graph that is universal for measurement-based computation (e.g., a square lattice); while a graph state could be any arbitrary graph. However, in the continuous-variable literature~\cite{Men06,Men08,Fla09,Men10, Men10a,Men10b} it is common to use them synonymously with context providing clarity.}. Specifically, a graph $G = (V,E)$ consists of a set of vertices (or nodes) $V$ and a set of edges $E$. The following recipe allows us to construct a corresponding graph state
\begin{enumerate}
  \item Each squeezed momentum eigenstate becomes a vertex in the graph.
  \item Each $C_Z$ operation applied between two qumodes is an edge in the graph.
\end{enumerate}
To illustrate, we give a simple example of a two-mode cluster state. Below we have the first step of initializing the two squeezed momentum eigenstates (represented by vertices and labeled $1$ and $2$). In the second step the $C_Z$ gate is applied, indicated by the edge joining vertices $1$ and $2$. The final step illustrates how measurements are indicated on a graph. Here a $\hat{p}$ quadrature measurement on the first node is implemented
\begin{align}
 \Qcircuit[2em] @R=1em @C=1em {
    \node{\text{1}} & \node{\text{2}}
    }
     \hspace{2mm}
     \rightarrow
     \hspace{2mm}
 \Qcircuit[2em] @R=1em @C=1em {
    \node{\text{1}} & \node{\text{2}} \link{0}{-1}
    }
 \hspace{2mm}
     \rightarrow
     \hspace{2mm}
 \Qcircuit[2em] @R=1em @C=1em {
    \node{\text{$\hat{p}$}} & \node{\text{2}} \link{0}{-1}
    }
\end{align}
Previously we introduced the quantum circuit formalism to understand measurement-based computation. Therefore we give below an equivalence between the teleportation circuit (on the left) and the graph state formalism (on the right)
\begin{align}
    \Qcircuit @C=1em @R=1.8em @!R {
     \lstick{\ket 0_p}  & \ctrl{1} & \meter & \rstick{\hat{p}=m_1} \qw \\
     \lstick{\ket 0_p} & \control \qw & \qw & \qw
 }
 \hspace{2.2cm}
 \Qcircuit[2em] @R=1em @C=1em {
    \node{\text{$\hat{p}$}} & \node{\text{2}} \link{0}{-1}
    }
\end{align}

The concept of continuous-variable graph states has been developed in a number of papers by Zhang~\cite{Zha08,Zha08a,Zha10} and independently by others~\cite{Zhang2009,Aolita2011,Pfi07,Men07,Zai08,Men08,Fla09,Fla09a}. Recently, a more general approach was introduced by \textcite{Men10a} allowing the graphical calculus formalism to be applied to the practical case where continuous-variable cluster states are created using finitely squeezed Gaussian states rather than ideal perfectly squeezed eigenstates.

\subsubsection{Stabilizers and nullifiers}

The stabilizer formalism~\cite{Got97} for continuous variables~\cite{Got01,Bar04,van07} is a useful way of both defining and analyzing cluster states (or graph states). An operator $\hat{O}$ is a stabilizer of a state $\ket{\psi}$ if $\hat{O}\ket{\psi} = \ket{\psi}$, i.e., it has an eigenvalue of $+1$. For example, a zero momentum eigenstate $\ket{0}_p$ is stabilized by the displacement operator $X(s)$, i.e., $X(s)\ket{0}_p = \ket{0}_p$ for all values of $s$. For an arbitrary continuous-variable graph state $\ket{\phi}$ with graph $G=(V,E)$ on $n$ qumodes, the stabilizers are defined as
\begin{align}\label{Chapter 8 stabilizers}
K_i(s) = X_i(s) \prod_{j \in N(i)} Z_j(s),
\end{align}
for $i=1,...,n$ and for all $s \in \mathbb{R}$. Here $N(i)$ means the set of vertices in the neighborhood of $v_i$, i.e., $N(i) = \{j|(v_j,v_i) \in E\}$. A variation of these stabilizers $K_i$ involves using what is known as \textit{nullifiers} $H_i$. Here  every stabilizer is the exponential of a nullifier, i.e., $K_i(s) = e^{-is H_i}$ for all $s \in \mathbb{R}$. This results in $H_i \ket{\phi} = 0$ where the set of nullifiers are given by
\begin{align}\label{Chapter 8 nullifier}
H_i = \hat{p}_i - \sum_{j \in N(i)} \hat{q}_j.
\end{align}\
Therefore the graph state $\ket{\phi}$ is a zero eigenstate of the above nullifiers where any linear superposition satisfies $H_i \ket{\psi} = 0$ and $[H_i,H_j]=0$. An example might be helpful here. Suppose we have a simple three-node linear cluster where the nodes are labeled $1,2$ and $3$. Then, according to Eq.~(\ref{Chapter 8 nullifier}), the nullifiers are given by: $\hat{p}_1 - \hat{q}_2, \hat{p}_2 - \hat{q}_1- \hat{q}_3, \hat{p}_3 - \hat{q}_2$. Furthermore, according to Eq.~(\ref{Chapter 8 stabilizers}), the set of stabilizers can be written as: $X_1(s) Z_2(s), Z_1(s) X_2(s) Z_3(s), Z_2(s) X_3(s)$, for all $s$. Therefore, by simply looking at a given graph we can write down the nullifiers and stabilizers of that particular graph. We note another useful way of analyzing graph states, other than the nullifier formalism, is by using the Wigner representation~\cite{Gu09}.

\subsubsection{Shaping clusters: removing nodes and shortening wires}\label{Chapter 8 Shaping Clusters: Removing Nodes and Shortening Wires}

The nullifier formalism provides a useful way of understanding how graph states are transformed by quadrature measurements. It can be shown~\cite{Gu09} that $\hat{q}$ (computational) measurements remove a given node (modulo some known displacement), while $\hat{p}$ quadrature measurements also remove the node but preserve the correlations between neighboring nodes (modulo a displacement and Fourier transform). For example, in the graph state picture, a position quadrature measurement on the second node has the following effect
\begin{align}\nonumber
\Qcircuit[2em] @R=1em @C=1em {
    \node{\text{1}} & \node{\text{$\hat{q}$}} \link{0}{-1} & \node{\text{3}} \link{0}{-1}
    }
 \hspace{0.5cm}
 &\longrightarrow
 \hspace{0.5cm}
 \Qcircuit[2em] @R=1em @C=1em {
    \node{\text{1}} &   \node{\text{3}}
    }
\end{align}
and a momentum measurement on the second node we have
\begin{align}\nonumber
    \Qcircuit[2em] @R=1em @C=1em {
    \node{\text{1}} & \node{\text{$\hat{p}$}} \link{0}{-1} & \node{\text{3}} \link{0}{-1}
    }
 \hspace{0.5cm}
 &\longrightarrow
 \hspace{0.5cm}
 \Qcircuit[2em] @R=1em @C=1em {
    \node{\text{1}} &  \node{\text{3}} \link{0}{-1}
    }
\end{align}
Using the above techniques we can shape a Gaussian cluster in order to put it into the required topology to perform a specific algorithm. For example, below we can create the graph state on the right by first performing a sequence of quadrature measurements on an initial $4 \times 5$ cluster given on the left
\begin{align}
    \Qcircuit[1.5em] @R=1em @C=1em {
     & \node{} \link{1}{0} \link{0}{1} & \node{} \link{1}{0} \link{0}{1} & \node{} \link{1}{0} \link{0}{1} & \node{} \link{1}{0}  & \\
     & \node{\text{$\hat{q}$}} \link{1}{0} \link{0}{1} & \node{\text{$\hat{q}$}} \link{1}{0} \link{0}{1} & \node{\text{$\hat{p}$}} \link{1}{0} \link{0}{1} & \node{\text{$\hat{q}$}} \link{1}{0} &  \\
     & \node{} \link{1}{0} \link{0}{1} & \node{} \link{1}{0} \link{0}{1} & \node{} \link{1}{0} \link{0}{1} & \node{} \link{1}{0} & \longrightarrow \\
     & \node{\text{$\hat{q}$}} \link{1}{0} \link{0}{1} & \node{\text{$\hat{p}$}} \link{1}{0} \link{0}{1} & \node{\text{$\hat{q}$}} \link{1}{0} \link{0}{1} & \node{\text{$\hat{q}$}} \link{1}{0} &  \\
     & \node{} \link{0}{1} & \node{} \link{0}{1}  & \node{} \link{0}{1}  & \node{} &
     }
     %
     %
     %
     \Qcircuit[1.5em] @R=1em @C=0.3em {\\
     & \node{}  \link{0}{1} & \node{}  \link{0}{1} & \node{} \link{1}{0} \link{0}{1} & \node{}  &\\
     & \node{}  \link{0}{1} & \node{} \link{1}{0} \link{0}{1} & \node{}  \link{0}{1} & \node{} & \\
     %
     %
     %
     %
     & \node{} \link{0}{1} & \node{} \link{0}{1}  & \node{} \link{0}{1}  & \node{} &
    }
    \end{align}
Recently, an experimental demonstration of continuous-variable cluster state shaping was performed using a four-mode linear cluster using homodyne detection and feedforward~\cite{Miw10}.

\subsection{Gaussian errors from finite squeezing}\label{sec: Gaussian Errors from Finite Squeezing}

In the next section, we will look at ways in which Gaussian cluster states can be implemented optically. As soon as we start discussing practical implementations we have to consider using finitely squeezed Gaussian states in our analysis which inevitably introduces errors into our computations. To illustrate the effect of finite squeezing we show what happens to the propagation of quantum information in a simple teleportation protocol. We now go back to the teleportation circuits from Sec.~\ref{Chapter 8 Understanding One-Way Computation via Teleportation} where we showed the effect of teleporting, first a qumode and then a gate diagonal in the computational basis, from one quantum wire to another. In that particular scenario, the nodes of the cluster were momentum eigenstates. In the calculations that follow they will be replaced by Gaussian squeezed states, i.e., $\ket{0}_p \longrightarrow \ket{0,V_S}_p$ where $V_S<1$ is the variance of the input squeezing. Suppose we initially start off with the two input states $\ket{\psi}$ and $\ket{0,V_S}_p$ where both can again be expanded in terms of arbitrary position bases, i.e., $\ket{\psi} = \int dq_1 \psi(q_1) \ket{q_1}$ and
\begin{align}
\ket{0,V_S}_p = (\pi V_S)^{-1/4} \int dq_2 \hspace{1mm} e^{-q_2^2/2V_S} \ket{q_2},
\end{align}
Applying the {\small CPHASE} gate to these two states gives
\begin{align}
(\pi V_S)^{-1/4} \int dq_1 dq_2 \psi(q_1) e^{-q_2^2/2V_S} e^{i q_1 q_2/2} \ket{q_1} \ket{q_2}.
\end{align}
After performing a measurement on the first mode in the momentum basis we end up with: $\ket{\psi'} = \mathcal{M} X(m_1)F \ket{\psi}$, where $\mathcal{M}$ is a Gaussian distortion~\cite{Men06}
\begin{align}
\mathcal{M} \ket{\psi} \propto \int dq \hspace{1mm} e^{q^2 V_S/2} \ket{q} \bra{q} \psi \rangle.
\end{align}
So effectively what we have is a Gaussian distortion, with zero mean and variance $1/V_S$, applied to the original input state as a result of propagating through a cluster created using finite squeezing (equivalently, this can be thought of as a convolution in momentum space by a Gaussian of variance $V_S$). The same reasoning holds when we consider the gate teleportation situation where the output is given by $\mathcal{M} X(m_1)F U \ket{\psi}$. We note that the above distortions due to finite squeezing errors of both state propagation and universal gate teleportation have also been analyzed by \textcite{Gu09} from the point of view of the Wigner representation.

\subsection{Optical implementations of Gaussian cluster states}\label{Chapter 8 subsection optical implementation}

Here we look at the various methods to optically implement continuous-variable cluster states using Gaussian elements. The advantage of the continuous-variable optical approach, compared to the discrete-variable approach, is that the generation of continuous-variable cluster states is completely deterministic. Furthermore, once the cluster is setup only homodyne detection is needed to implement any multimode Gaussian transformation~\cite{Uka10a}. However, the errors introduced into the computations, due to the finitely squeezed resources, are a downside to this unconditionality. The five methods for cluster state production are outlined below.

\subsubsection{Canonical method}

The canonical method was first introduced in 2006~\cite{Zha06,Men06} and proposed a literal interpretation of how to implement an optical Gaussian cluster state. By that we mean each mode is first prepared as a momentum squeezed vacuum state and then an appropriate number of $C_Z$ gates are applied to create the required cluster. The $C_Z$ gate is optically implemented using two beam splitters and two online\footnote{An online squeezer (also known as inline squeezer) is the squeezing of an arbitrary, possibly unknown, state. An offline squeezer is the squeezing of a known state, typically the vacuum.} squeezers~\cite{Yur85,Walls1995}. One of the advantages of this method is that the $C_Z$ gates commute with one another (i.e., the order in which they are applied does not matter) and thus facilitates theoretical analysis. On the other hand, the implementation of the $C_Z$ gate is experimentally challenging~\cite{Ukai2011} due to the difficulty of online squeezing~\cite{Yur85,Por89,Yos07} and therefore is not very efficient as $C_Z$ gates are needed for every link in the cluster. Note that the demonstration of another type of quantum non-demolition (QND) gate (the SUM gate) has also been achieved~\cite{Yos08}.

\subsubsection{Linear-optics method}

The linear-optics method was conceived by van Loock~\cite{van07} and provided a way of greatly simplifying the optical implementation of the canonical method. Put simply, the linear-optics method allows the creation of a cluster state using only offline squeezed states and a beam splitter network. Experimentally this represented an important advancement in the building of continuous-variable cluster states as the difficult part of online squeezing was now moved offline~\cite{Gu09}. In this work~\cite{van07} two algorithms were developed using squeezed vacuum states and linear optics to create two varieties of cluster states: (1) canonical Gaussian cluster states and (2) generalized cluster-type states. The first algorithm, known as the decompositional algorithm, used the Bloch-Messiah reduction~(cf. Sec.~\ref{Euler Decomposition of Canonical Unitaries}) to show that the canonical method can be decomposed into offline squeezed states and beam splitters to create the original canonical cluster. The second algorithm (this time independent of the Bloch-Messiah reduction) showed how to create a more general class of Gaussian cluster states, known as cluster-type states (from which the canonical cluster states are a special case). This was shown by requiring that the final cluster state (again created from squeeze vacuum states and carefully configured passive linear optics) satisfies the nullifier relation of Eq.~(\ref{Chapter 8 nullifier}) in the limit of infinite squeezing. However, when the finite squeezing case is also considered a larger family of non-canonical cluster states are created.

One of the benefits of the second algorithm is that the antisqueezing components are suppressed thus making it experimentally more appealing~\cite{Yuk08}. Also smaller levels of input squeezing are required to create cluster-type states compared to using the canonical method to create canonical states with the same kind of correlations. A number of experiments using the linear-optics method have been demonstrated from setting up an initial four-mode Gaussian cluster~\cite{Su07,Yuk08} (including linear, T-shape and square clusters~\cite{Yuk08}) to simple continuous-variable one-way quantum computations on a four-mode linear cluster~\cite{Uka10,Miw10}.

\subsubsection{Single-OPO method}

The single-OPO-method~\cite{Men07,Men08,Fla09} was developed around the same time as the
linear-optics method and shows how to create an ultra-compact and scalable, universal $N$-mode cluster state using only a single optical parametric oscillator\footnote{A simple OPO consists of an optical cavity (e.g., two facing mirrors) with a crystal inside. Typically this crystal is nonlinear (e.g., second-order $\chi^{(2)}$) and is pumped by a laser beam which can lead to the down-conversion or the up-conversion of the initial frequencies.}(OPO). Effectively this means that the cluster state can be created in just a single step using a top-down approach and requires the same amount of resources as the linear-optics method (i.e., $\mathcal{O}(N^2)$). However, unlike the linear-optics method it does not require an interferometer which can be cumbersome for large $N$ thus removing the beam splitter network altogether. It therefore holds great promise of scalability for universal continuous-variable cluster states.

The initial proposal~\cite{Men07} showed that, by using an appropriately constructed multi-frequency pump beam, the single OPO could generate any continuous-variable cluster state with a bipartite graph. Mathematically this result relied on showing a connection between continuous-variable cluster states and Hamiltonian graphs or $\mathcal{H}$-graphs\ for
short~\cite{Men07,Zai08,Men08,Fla09,Fla09a}. These $\mathcal{H}$-graphs correspond to those states produced by an OPO. With this result we have effectively gone from requiring $N$ single-mode squeezers (OPOs) to a single multimode OPO which is pumped by an $\mathcal{O}(N^2)$-mode beam. Further progress showed~\cite{Men08,Fla09} that this method can in fact produce a whole family of universal continuous-variable cluster states where the encoding scheme of the single-OPO involves using the optical frequency comb. Here each independent qumode corresponds to a different frequency in the optical frequency comb (which derives its name from the equal spacings between each qumode). The main advantage of this method is that the number of pump frequencies is not $\mathcal{O}(N^2)$ but in fact constant. Recently, the first experimental demonstration of cluster state generation in the optical frequency was performed~\cite{Pysher2011}. Here $15$ quadripartite cluster states were created over $60$ cavity qumodes, exhibiting its potential for scalable quantum computation.

\subsubsection{Single-QND-gate method}

In the four years since the canonical method, and its reliance on $C_Z$ gates, all of the previous methods have purposely shied away from using this non-demolition gate due to its difficulty in being experimentally implemented~\cite{Ukai2011}. However, in a novel approach~\cite{Men10b}, the $C_Z$ gate once again makes an appearance in a compact scheme devised to generate arbitrarily large cluster states. In the canonical method, $O(N^2)$ low-noise $C_Z$ gates are needed to set up the initial $N$-mode cluster. However, in this new approach, all that is required is a single $C_Z$ gate. In fact, for universal quantum computation, only a single copy of the following key optical ingredients are needed: a single-mode vacuum squeezer, a $C_Z$ gate, a homodyne detector and a photon counter. The basic premise of the single-QND-gate method involves building the cluster on the go where the cluster is extended and measured as needed, according to the particular algorithm to be executed.

The specific design of this method can be understood from considering a simple linear cluster. In this case momentum squeezed vacuum states are generated at regular intervals and repeatedly fed into a single $C_Z$ gate. One output of the gate is directed towards a detector while the other is fed back into the $C_Z$ gate. Because in general all qumodes travel the same optical path, but importantly at different times, the encoding scheme of the qumodes is temporal. This process of creating and measuring is repeated over and over during the duration of the algorithm and can be extended in much the same way to create universal cluster states. One advantage that the single-QND-gate method offers over the previous approaches is that maintaining the coherence of a large cluster becomes less of an issue. This is because we are only concerned with the coherence of a small instance of the cluster at any one time.

\subsubsection{Temporal-mode-linear-optics method}

The latest approach was developed by \textcite{Men10} and combines the essential features and benefits of the previous three methods into one. This method, known as the temporal-mode-linear optics method, offers an improvement over the single-QND-gate approach in that it uses the techniques from the linear-optics scheme to move the experimentally challenging online squeezing, offline. This new temporal-mode encoding, where again the input squeezed vacuum states are repeatedly sent through the same optical hardware but at different times, still maintains the finite coherence and scalability features of the previous model. This implementation is achieved by recognizing that the output states of the single-OPO method are in fact Gaussian projected entangled pair states~\cite{Ohl10}. Gaussian projected entangled pair states are pairs of Gaussian two-mode squeezed states that are locally projected down to a lower-dimensional subspace. For example, in the cluster state formalism, this corresponds to having two two-node graph states where a measurement projects the ends of both nodes down to a single node and in doing so creates a linear three-node graph~\cite{Ohl10}. The graphical formalism developed by \textcite{Men10a} is used to describe and formulate this Gaussian projected entangled pair states construction and can be optically implemented using single-mode offline squeezers and linear optics.

\subsection{Universal quantum computation}\label{Chapter 8 universal quantum computation}

As previously noted, the Lloyd-Braunstein criterion tells us that in order to achieve universal quantum computation, i.e., the ability to generate an arbitrary Hamiltonian, we need the addition of a non-Gaussian element, such as a non-Gaussian operation or a non-Gaussian state, to our tool box. In continuous-variable cluster state quantum computation the cubic phase gate and the cubic phase state, respectively, are examples of such elements. The cubic phase gate is defined as
\begin{align}
V = {\rm exp}(i \gamma \hat{q}^3),
\end{align}
where $\gamma \in \mathbb{R}$. The action of the cubic phase gate on a zero momentum eigenstate $\ket{0}_p$ creates what is known as the cubic phase state $\ket{\gamma}$ which is an unnormalizable non-Gaussian state
\begin{align}
\ket{\gamma} =  V\ket{0}_p = \frac{1}{2 \sqrt{\pi}} \int dq \hspace{1mm} e^{i \gamma q^3}\ket{q}.
\end{align}
Ultimately the cubic phase state will be used as a resource to implement the cubic phase gate onto an arbitrary state in the cluster. We begin by first showing how to create such a state and then explore two methods of implementing the cubic phase gate.

\subsubsection{Creating the cubic phase state}

It was shown by \textcite{Got01} that the cubic phase state could be created by implementing the following quantum circuit~\cite{Gu09}
\begin{eqnarray}
\label{eq:cubicphasestateGKP2}
    \Qcircuit @C=1em @R=1.8em @!R {
    \lstick{S(r) \ket 0}  & \ctrl{1}  & \meter & \rstick{{X^{\dagger} \hat{n} X}} \qw \\
     \lstick{S(r) \ket 0} &  \control \qw &  \qw & \rstick{\approx V' \ket{0}_p} \qw
     }
\end{eqnarray}
where $V'=\exp(i\gamma'(n)\hat q^3)$ and the strength of the gate depends on the probabilistic measurement result: $\gamma'(n) = (6 \sqrt{2n +1})^{-1}$. Therefore the above circuit corresponds to a simple two-mode graph state with a displaced photon counting measurement on the first node
\begin{eqnarray}
\label{graph: cubic phase}
 \Qcircuit[3.5em] @R=1em @C=1em {
    \node{ \text{$X^{\dagger} \hat{n} X$}} & \node{\text{$\gamma'$}} \link{0}{-1}
    }
\end{eqnarray}
Notice that the output state above is not quite in the form we would like, i.e., it is $\gamma'$ and not $\gamma$. To correct this note that $V'$ can be decomposed into two squeeze gates and the cubic phase gate~\cite{Got01}, i.e.,
\begin{align}\label{eq: cubic phase squeezing}
V' = S(f) V S^{\dagger} (f),
\end{align}
where $f:=f(n) = [\gamma/\gamma'(n)]^{1/3}$. Once these squeezing corrections are implemented the cubic phase state can be synthesized.

\subsubsection{Implementing the cubic phase gate}

Now that we have a way of creating the cubic phase state we look at two possible approaches to inducing the action of the cubic phase gate onto an arbitrary state using the cubic phase state as a resource. The first approach deviates from the typical measurement-based scheme by performing Gaussian measurements on a non-Gaussian cluster state. The second approach revisits the typical setup of a Gaussian cluster but requires that the measurement tool box now consists of both Gaussian and non-Gaussian measurements.

\paragraph{Non-Gaussian cluster and Gaussian measurements\\}

In the standard cluster-state model of computation all nodes are initialized as zero momentum eigenstates before becoming entangled via the {\small CPHASE} gate. However, one way to implement the cubic phase gate is to first embed the cubic phase state into the original cluster. In the regime of finite squeezing the initial Gaussian cluster state now becomes non-Gaussian. A computation is performed as before using only Gaussian (homodyne) measurements. Once the cubic phase state $\ket{\gamma}$ is part of the initial cluster a variation of gate teleportation can be used to teleport $V$ onto an arbitrary state $\ket{\phi}$ of the cluster (where for instance $\ket{\phi}$ is the state of a node in the cluster at a particular point in time). The following circuit achieves this~\cite{Gu09}
\begin{align}\label{eqn:resource_circuit2}
\Qcircuit @C=1em @R=1em {
    \lstick{\ket{\phi}} & \ctrl{1} & \qw & \meter & \rstick{\hat{p}=m_1} \qw
     \\
    \lstick{\ket{\gamma} = V \ket{0}_p} & \ctrl{0} & \ctrl{1} & \meter & \rstick{\hat{p}=m_2} \qw
    \\
    \lstick{\ket{0}_p} & \qw & \ctrl{0} & \qw & \rstick{\approx \ket{\phi'} = e^{i \gamma \hat{q}^3} \ket{\phi}} \qw\\
}
\end{align}
\\
modulo known Gaussian corrections~\cite{Wee09} (hence the $\approx$). In the graph state formalism the above circuit is depicted as
\begin{eqnarray}\nonumber
\Qcircuit[2.5em] @R=1em @C=1em {
    \node{\text{$\phi$}} & \node{\text{$\gamma$}} \link{0}{-1} & \node{\text{3}} \link{0}{-1}
    }
 \hspace{0.4cm}
 \longrightarrow
 \hspace{0.4cm}
\Qcircuit[2.5em] @R=1em @C=1em {
    \node{\text{$\hat{p}$}} & \node{\text{$\hat{p}$}} \link{0}{-1} & \node{\text{$\phi'$}} \link{0}{-1}
    }
\end{eqnarray}
Note that the above graph corresponds to a subgraph of a much larger cluster. One way to think about what is happening is that we have the cubic phase gate $V$ acting on a zero momentum eigenstate on the second node and by performing momentum quadrature measurements on the first and second nodes we are effectively teleporting $V$ onto $\ket{\phi}$ (modulo corrections) with the resulting state appearing at the third node. The beauty of using a non-Gaussian cluster from the beginning is that once it is created only quadrature measurements are needed to perform any algorithm. However, creating such an improved resource is experimentally challenging.

\paragraph{Gaussian cluster and non-Gaussian measurements\\}

Now the previous approach is tailored to creating the cubic phase state offline, i.e., as one of the initial resource states prior to the entangling gate. However, if we wanted to emulate the original cluster state formalism by beginning with a universal Gaussian cluster, our set of measurements would need to be both Gaussian and non-Gaussian (e.g., photon counting). One consequence of this is the need to perform the squeezing corrections of Eq.~(\ref{eq: cubic phase squeezing}) online. With this in mind circuit~(\ref{eqn:resource_circuit2}) now becomes
\begin{align}
\Qcircuit @C=.8em @R=1em {
    \lstick{\ket{\phi}} & \gate{S(f)} & \ctrl{1} & \qw & \meter & \rstick{\hat{p}=m_1} \qw
     \\
    \lstick{V'  \ket{0}_p} & \qw &\ctrl{0} & \ctrl{1} &  \meter & \rstick{\hat{p}=m_2} \qw
    \\
    \lstick{\ket{0}_p} & \qw & \qw & \ctrl{0}  & \gate{S^\dag(f,m_1)}  & \rstick{\ket{\phi'} \approx V \ket{\phi}} \qw
}
\end{align}
where the second squeezing correction is dependent on the first measurement result $m_1$ because it is itself dependent on the photon counting result $n$~\cite{Wee09}. Translating the above circuit into the graph state formalism and using graph~(\ref{graph: cubic phase}) gives the following
\begin{align}\nonumber
\Qcircuit[3.5em] @R=1em @C=1em {
    & & \node{\text{$X^{\dagger}\hat{n} X$}} \link{1}{0} & & & \\
    \node{\text{$\hat{p}$}} & \gate{S(f)} & \node{\text{$\hat{p}$}} \link{0}{-1} & \gate{S^{\dagger}(f,m_1)} & \node{\text{$\phi'$}} \link{0}{-1}
    }
\end{align}
where the input state $\ket{\phi}$ is the first node from the left. Here the square boxes represent subgraphs through which the squeezing corrections, $S(f)$ and $S^{\dagger}(f,m_1)$, are implemented via homodyne measurements~\cite{van07a}. Another consequence of the non-Gaussian measurement is that the concept of parallelism is not longer valid as the center top node needs to be measured first due to the probabilistic nature of the measurement outcome. After which the amount of squeezing required on the first node is dependent on the result $n$. Hence the time ordering of measurements now matters and adaptiveness plays a role (depending on the specific algorithm performed, the value of $\gamma$ might also depend on previous measurement results as well). From the above graph one can notice that ideally the top center node is only attached to the node below it. This is where the shaping tools from Sec.~\ref{Chapter 8 Shaping Clusters: Removing Nodes and Shortening Wires} play a part. For example, if there is a point in the computation on the Gaussian cluster where the cubic phase state needs to be created then removing or deleting nodes from the cluster would allow one to have it in the required form.

\subsection{Quantum error correction}\label{chapter 8 error correction}

To argue that a particular physical system is capable of universal quantum computation it is not sufficient to show that the system in question can implement arbitrary unitary evolutions. In any physical implementation there will be imperfections in the system that will inevitably lead to random errors being introduced. Even if these errors are small, when large scale quantum processing is considered, we have to worry about their propagation during gate operations. If uncorrected, such errors will grow uncontrollably and make the computation useless. The answer to this problem is fault tolerant error correction~\cite{Sho95,Ste96}. Thus to make a necessary and sufficient argument that a particular physical system is capable of continuous-variable quantum computation, strictly one must also show that fault tolerant error correction is possible.

The idea of error correction is self-explanatory, though the description of its application to quantum systems requires some care. Classically we might consider using a redundancy code such that, for example, $0 \to 000$ and $1 \to 111$. If a bit flip occurs on one of the bits we might end up with $010$ or $101$ but we can recover the original bit value by taking a majority vote. An example of a quantum redundant encoding for qubits is $\alpha |0 \rangle + \beta |1 \rangle \to \alpha |000 \rangle + \beta |111 \rangle$ where we have created an entangled state rather than copies. It is then possible to identify an error without collapsing the superposition, by reading out the parity of pairs of qubits. For example a bit-flip error might result in the state $\alpha |001 \rangle + \beta |110 \rangle$. The parity of the first two qubits will be zero whilst the parity of the second two qubits will be one, thus unambiguously identifying that an error has occurred on the last qubit. Because we are measuring the parity, not the qubit value, the superposition is not collapsed. Such codes can be expanded to cope with the possibility of more than one error occurring between correction attempts and to cope with multiple types of errors. Of course the gates being used to detect and correct the errors may themselves be faulty. An error correction code is said to be fault tolerant if error propagation can be prevented even if the components used to do the error correction introduce errors themselves. Typically this is only possible if the error rate per operation is below some level known as the fault tolerant threshold.

The first error correction protocol for continuous variables~\cite{Llo98,Bra98a} was developed as a direct generalization of the qubit redundancy codes~\cite{Sho95}. In Braunstein's simplified version~\cite{Bra98b}, eight ancilla squeezed states are mixed on beam splitters with the signal state to create a nine-mode encoded state. Decoding is similarly achieved with beam splitters, with homodyne detection on eight of the modes providing information about errors on the remaining signal mode. This protocol has recently been demonstrated experimentally~\cite{Aok09}. On any particular run the code can correct any error that occurs on any single
mode of the encoded state~\cite{Walker2010}. This was shown to extend to multiple errors
provided they occur in a stochastic way~\cite{vanLoock2008}. Unfortunately,
error models of this kind are non-Gaussian error models and do not
correspond to the Gaussian errors that typically occur in experiments
due to loss and thermal noise. Other protocols for correcting more specific types of non-Gaussian noise imposed on Gaussian states have been proposed and experimentally demonstrated~\cite{Nis98,Las10}. It has been proven that error correction of Gaussian noise, imposed on Gaussian states, using only Gaussian operations is impossible~\cite{Nis09}.

This no-go theorem does not apply if the initial states are non-Gaussian. An example of such a protocol is that developed by \textcite{Got01}. Here, the information is discretized by encoding qubit states as non-Gaussian continuous-variable states. Such states could be generated in optical modes by means of cross-Kerr interactions~\cite{Pirandola2004}. Error correction against Gaussian errors can then be achieved using Gaussian operations. This protocol is known to be fault tolerant, although the threshold requirements are quite extreme~\cite{Gla06}. A simpler encoding of qubits into continuous-variable states is the coherent state encoding~\cite{Ral03}. Fault tolerant error correction against Gaussian errors can also be achieved with this system, with a better threshold behavior~\cite{Lun08}. The price paid for this improvement  though, is that non-Gaussian operations are also required.

A third possibility that is not explicitly forbidden by the no-go theorem is to error correct Gaussian states against Gaussian noise using non-Gaussian operations, that none-the-less result in a Gaussian output state. Such protocols have been proposed~\cite{Bro03} and demonstrated~\cite{Xia10} for continuous-variable entanglement distillation. Entanglement distillation~\cite{Ben96} is a non-deterministic error detection protocol, useful for extending the reach of quantum communication systems. Continuous-variable protocols have also been developed to distill entanglement against non-Gaussian noise~\cite{Hag08,Dong08} and for non-Gaussian states against Gaussian noise~\cite{Our09}. In principle, continuous-variable teleportation and continuous-variable distillation protocols based on noiseless amplification can be combined to error correct Gaussian states against Gaussian noise~\cite{Ralph2011}. However, it is not currently known if such protocols can be made fault tolerant.

Error correction cannot be directly introduced into the continuous-variable cluster state model by simply simulating a circuit model error correction protocol with the cluster~\cite{Ohl10,Cable2010}. This is a generalization of the result of \textcite{Nie05} that similarly restricts error correction for discrete-variable cluster states. These authors showed that error correction could only be incorporated into the cluster state computation model provided the construction and measurement of the cluster occurred concurrently, and the off-line, non-deterministic production of special states was allowed. In continuous-variable cluster state computation, without fully-fletched continuous-variable fault tolerance, continuous-variable cluster states based on any finite squeezing are strictly speaking not resources for continuous-variable cluster state quantum computing~\cite{Ohl10}. However, it has been argued that in principle, combining
the techniques of \textcite{Nie05} with the oscillator encoding scheme of \textcite{Got01} would allow
fault tolerant continuous-variable cluster state computation to
be carried out~\cite{Gu09}, though this has not been shown explicitly.

\subsection{Continuous-variable algorithms}\label{ch: QC algorithms}

Finally, before discussing future directions, we briefly mention two algorithms that have been developed for a continuous-variable quantum computer: Grover's search algorithm~\cite{Grover1997} and the Deutsch-Jozsa algorithm~\cite{Deutsch1992}.  These algorithms were originally developed for discrete-variable systems~\cite{Nielsen2000} and later analogs were found for continuous-variable systems in terms of the quantum circuit model formalism. Grover's search algorithm using continuous variables was presented by \textcite{Pati2000} and showed that a square-root speed-up in searching an unsorted database could be achieved in analogy with the qubit case. A continuous-variable version of one of the earliest quantum algorithms, the Deutsch-Jozsa algorithm, was first developed by \textcite{Pati2003}. Here the goal of determining whether a function is constant or balanced was constructed in the ideal case of perfectly squeezed qumodes. Later, this algorithm was analyzed in more detail by \textcite{Adcock2009} and reformulated using Gaussian states by \textcite{Zwierz2010}.

\subsection{Future directions}\label{ch: QC future directions}

Research interest in the field of continuous-variable quantum computation has increased significantly in the last few years. This is particularly true in the case of cluster state quantum computation. Therefore it is worth making a brief comparison between the continuous-variable cluster states discussed here and
the discrete-variable approach based on single-photon qubits and linear optics techniques~\cite{Knill2001,Nielsen2004}. The key trade-off is between construction of the cluster
and its measurements. In the continuous-variable approach construction is deterministic, whilst in the
single-photon approach it is non-deterministic and requires a very large overhead in terms
of photon sources and memory in order to make it near deterministic. On the other hand,
all required measurements are trivial in the single-photon approach, whilst non-Gaussian
measurements pose a major challenge in the continuous-variable approach. At this point it is difficult to
say which of these problems represents the biggest impediment to building a large scale system.

There are a number of important avenues for future research in continuous-variable quantum computation. Perhaps the most important at this stage is the development of continuous-variable fault tolerance for cluster state quantum computation. Another avenue would be to incorporate continuous-variable quantum algorithms, such as Grover's algorithm and the Deutsch-Jozsa algorithm, into the cluster state model. Additionally, the development of further algorithms for a continuous-variable quantum computer, e.g., an optical version of Shor's factoring algorithm~\cite{Shor1997}, would also be interesting, especially for future experimental demonstrations.

\section{Conclusion and Perspectives}\label{chapter conclusion}

This review examined the power of continuous-variable quantum information from a Gaussian
perspective. The processing of Gaussian quantum information involves the use of any combination of
Gaussian states, Gaussian operations, and Gaussian measurements. The ability to characterize Gaussian
states and operations via their first and second-order statistical moments offers a major simplification
in the mathematical analysis of quantum information protocols. Over the last decade, optical and atomic Gaussian states and operations have being recognized as key resources for quantum
information processing. For example,
continuous-variable quantum teleportation only requires Gaussian entanglement and Gaussian operations, while
it can be used to teleport arbitrary, even highly non-Gaussian, quantum states. Similarly, continuous-variable quantum key distribution works with coherent states, while it achieves the unconditional
security once believed to reside only with highly non-classical resources. Yet another unexpected property
is that a Gaussian cluster state is provably a universal resource for quantum computation. All of these
findings have put forward the idea that Gaussian protocols deserve a front row position in quantum
information science.

Beyond a comprehensive description of Gaussian quantum information protocols, this review also
examined bounds on the distinguishability of Gaussian states, and features of Gaussian bosonic quantum communication channels such as their capacity
and statistical discrimination. Future directions in quantum information sciences include the
exploration of more complex scenarios of quantum communication, involving different protocols such
as quantum cloning or teleportation networks. In this context, the Gaussian approach is particularly
promising, allowing us to explore these future directions with powerful mathematical tools and
standard optical components.

From a purely Gaussian perspective, i.e., when one is restricted to using states, operations
and measurements that are all Gaussian, certain protocols are not possible. For example, universal quantum computation,
entanglement distillation, and error correction all require that the protocol be supplemented with either a non-Gaussian state, operation, or measurement. For some tasks, hybrid systems, which combine elements
from continuous and discrete-variable quantum information processing, are then favored as they may
outperform purely discrete-variable systems. Interestingly, the powerful mathematical tools of Gaussian analysis can sometimes be used even when
non-Gaussian processing or non-Gaussian states are involved. For example, in certain quantum key
distribution protocols, even if Alice and Bob use non-Gaussian distributions or Eve makes a non-Gaussian attack, the security can be ensured by considering the worst case of a Gaussian attack against
a Gaussian protocol. Such an analysis would still hold if quantum repeaters based on non-Gaussian
processing were used by Alice and Bob. Similarly, universal
quantum computation achieved by a Gaussian cluster state and non-Gaussian detection is an example
of the power brought by the application of Gaussian analysis tools to hybrid quantum systems.

In conclusion, we anticipate that Gaussian quantum information will play a key role in future
developments of quantum information sciences, both theoretical and experimental. This is due to the simplicity
and versatility of the involved protocols as well as the availability of the required technologies. We
hope that this review will help encourage these developments.

\section*{Acknowledgments}

We would like to thank Sam Braunstein, Jens Eisert, Chris Fuchs, Saikat Guha, Alexander Holevo, Anthony Leverrier, Stefano Mancini, Christoph Marquardt, Nick Menicucci, Matteo Paris, Eugene Polzik, Joachim Sch\"afer, Denis Sych, Graeme Smith, David Vitali, Nathan Walk, Mark Wilde, Howard Wiseman, and Jing Zhang for providing valuable feedback on the manuscript. C.~W. would also like to thank Hoi-Kwong Lo and Shakti Sivakumar. C.~W. acknowledges support from the Ontario postdoctoral fellowship program,
CQIQC postdoctoral fellowship program, CIFAR, Canada Research Chair program,
NSERC, QuantumWorks and the W.~M.~Keck Foundation. S.~P. acknowledges support from the W.~M.~Keck Foundation and the European Community (Marie Curie Action). R.~G.~P acknowledges support from the W.~M.~Keck Foundation and the Alexander von Humboldt foundation. N.~J.~C acknowledges support of the Future and Emerging Technologies programme
of the European Commission under the FP7 FET-Open grant no. 212008 (COMPAS) and
by the Interuniversity Attraction Poles program of the Belgian Science Policy Office under grant IAP P6-10 (photonics@be). T.~C.~R acknowledges support from the Australian Research
Council. J.~H.~S acknowledges support from the W.~M.~Keck Foundation, the ONR, and DARPA. S.~L. acknowledges the support of NSF, DARPA, NEC, Lockheed Martin, and ENI
via the MIT energy initiative.

\bibliographystyle{apsrmp}

\begin{thebibliography}{42}
\expandafter\ifx\csname natexlab\endcsname\relax\def\natexlab#1{#1}\fi
\expandafter\ifx\csname bibnamefont\endcsname\relax
    \def\bibnamefont#1{#1}\fi
\expandafter\ifx\csname bibfnamefont\endcsname\relax
    \def\bibfnamefont#1{#1}\fi
\expandafter\ifx\csname citenamefont\endcsname\relax
    \def\citenamefont#1{#1}\fi
\expandafter\ifx\csname url\endcsname\relax
    \def\url#1{\texttt{#1}}\fi
\expandafter\ifx\csname urlprefix\endcsname\relax\def\urlprefix{URL }\fi
\providecommand{\bibinfo}[2]{#2}
\providecommand{\eprint}[2][]{\url{#2}}

\bibitem[{\citenamefont{Ac\'{\i}n}(2001)}]{Acin2001}
\bibinfo{author}{\bibnamefont{Ac\'{\i}n, A.}},
\bibinfo{year}{2001},
\bibinfo{journal}{Phys. Rev. Lett.} \textbf{\bibinfo{volume}{87}},
\bibinfo{pages}{177901}.

\bibitem[{\citenamefont{Adcock {\em et al.}}(2009)}]{Adcock2009}
\bibinfo{author}{\bibnamefont{Adcock, M.~R.~A, P.~H\o{}yer, and B.~C.~Sanders}},
\bibinfo{year}{2009},
\bibinfo{journal}{New J. Phys.} \textbf{\bibinfo{volume}{11}},
\bibinfo{pages}{103035}.

\bibitem[{\citenamefont{Adesso and Datta}(2010)}]{Adesso2010}
\bibinfo{author}{\bibnamefont{Adesso, G., and A.~Datta}},
\bibinfo{year}{2010},
\bibinfo{journal}{Phys. Rev. Lett.} \textbf{\bibinfo{volume}{105}},
\bibinfo{pages}{030501}.

\bibitem[{\citenamefont{Adesso and Illuminati}(2005)}]{Adesso2005}
\bibinfo{author}{\bibnamefont{Adesso, G., and F.~Illuminati}},
\bibinfo{year}{2005},
\bibinfo{journal}{Phys. Rev. Lett.} \textbf{\bibinfo{volume}{95}},
\bibinfo{pages}{150503}.

\bibitem[{\citenamefont{Adesso and Illuminati}(2007)}]{Adesso2007}
\bibinfo{author}{\bibnamefont{Adesso, G., and F.~Illuminati}},
\bibinfo{year}{2007},
\bibinfo{journal}{J. Phys. A: Math. Theor.} \textbf{\bibinfo{volume}{40}},
\bibinfo{pages}{7821}.

\bibitem[{\citenamefont{Adesso {\em et al.}}(2004)}]{Adesso2004}
\bibinfo{author}{\bibnamefont{Adesso, G., A.~Serafini, and F.~Illuminati}},
\bibinfo{year}{2004},
\bibinfo{journal}{Phys. Rev. A} \textbf{\bibinfo{volume}{70}},
\bibinfo{pages}{022318}.

\bibitem[{\citenamefont{Aoki {\em et al.}}(2009)}]{Aok09}
\bibinfo{author}{\bibnamefont{Aoki, T., G.~Takahashi, T.~Kajiya, J.~Yoshikawa, S.~L.~Braunstein, P.~van Loock, and A.~Furusawa}},
\bibinfo{year}{2009},
\bibinfo{journal}{Nature Physics} \textbf{\bibinfo{volume}{5}},
\bibinfo{pages}{541}.


\bibitem[{\citenamefont{Aolita {\em et al.}}(2011)}]{Aolita2011}
\bibinfo{author}{\bibnamefont{Aolita, L., A.~J.~Roncaglia, A.~Ferraro, and A.Ac\'{\i}n}},
\bibinfo{year}{2011},
\bibinfo{journal}{Phys. Rev. Lett.} \textbf{\bibinfo{volume}{106}},
\bibinfo{pages}{090501}.


\bibitem[{\citenamefont{Andersen {\em et al.}}(2005)}]{Andersen2005}
\bibinfo{author}{\bibnamefont{Andersen, U.~L., V.~Josse, and G.~Leuchs}},
\bibinfo{year}{2005},
\bibinfo{journal}{Phys. Rev. Lett.} \textbf{\bibinfo{volume}{94}},
\bibinfo{pages}{240503}.

\bibitem[{\citenamefont{Andersen {\em et al.}}(2010)}]{Andersen2010}
\bibinfo{author}{\bibnamefont{Andersen, U.~L., G.~Leuchs, C.~Silberhorn}},
\bibinfo{year}{2010},
\bibinfo{journal}{Laser \& Photonics Reviews} \textbf{\bibinfo{volume}{4}},
\bibinfo{pages}{337}.


\bibitem[{\citenamefont{Arthurs and Kelly, Jr.}(1965)}]{Arthurs1965}
\bibinfo{author}{\bibnamefont{Arthurs, E., and J.~L.~Kelly, Jr.}},
\bibinfo{year}{1965},
\bibinfo{journal}{Bell~Syst.~Tech.~J.} \textbf{\bibinfo{volume}{44}},
\bibinfo{pages}{725}.

\bibitem[{\citenamefont{Arvind {\em et al.}}(1995)}]{Arvind1995}
\bibinfo{author}{\bibnamefont{Arvind, B.~Dutta, N.~Mukunda, and R.~Simon}},
\bibinfo{year}{1995},
\bibinfo{journal}{Pramana J. Phys.} \textbf{\bibinfo{volume}{45}},
\bibinfo{pages}{471}.

\bibitem[{\citenamefont{Audenaert {\em et al.}}(2007)}]{Audenaert2007}
\bibinfo{author}{\bibnamefont{Audenaert, K.~M.~R., J.~Calsamiglia, L.~Masanes, R.~Munoz-Tapia, A.~Ac\'{\i}n, E.~Bagan, and F.~Verstraete}},
\bibinfo{year}{2007},
\bibinfo{journal}{Phys. Rev. Lett.} \textbf{\bibinfo{volume}{98}},
\bibinfo{pages}{160501}.

\bibitem[{\citenamefont{Audenaert {\em et al.}}(2008)}]{Audenaert2008}
\bibinfo{author}{\bibnamefont{Audenaert, K.~M.~R., M.~Nussbaum, A.~Szkola, and F.~Verstraete}},
\bibinfo{year}{2008},
\bibinfo{journal}{Commun.~Math.~Phys.} \textbf{\bibinfo{volume}{279}},
\bibinfo{pages}{251}.

\bibitem[{\citenamefont{Bachor and Ralph}(2004)}]{Bachor2004}
\bibinfo{author}{\bibnamefont{Bachor, H.~-A., and T.~C.~Ralph}},
  \bibinfo{year}{2004}, \emph{\bibinfo{booktitle}{A Guide to Experiments in Quantum Optics}} (\bibinfo{publisher}{Wiley-VCH, Weinheim}).

\bibitem[{\citenamefont{Ban}(1999)}]{Ban1999}
\bibinfo{author}{\bibnamefont{Ban, M.}},
\bibinfo{year}{1999},
\bibinfo{journal}{J.~Opt.~B: Quantum Semiclass. Opt.} \textbf{\bibinfo{volume}{1}},
\bibinfo{pages}{L9}.

\bibitem[{\citenamefont{Barnes}(2004)}]{Bar04}
\bibinfo{author}{\bibnamefont{Barnes, R.~L.}},
\bibinfo{year}{2004},
\bibinfo{booktitle}{``Stabilizer Codes for Continuous-variable Quantum Error Correction"},
\bibinfo{journal}{arXiv:quant-ph/0405064}.


\bibitem[{\citenamefont{Barnum {\em et al.}}(2000)}]{Barnum00}
\bibinfo{author}{\bibnamefont{Barnum, H., E.~Knill, and M.~A.~Nielsen}},
\bibinfo{year}{2000},
\bibinfo{journal}{IEEE Trans.~Info.~Theor.} \textbf{\bibinfo{volume}{46}},
\bibinfo{pages}{1317}.

\bibitem[{\citenamefont{Bartlett and Munro}(2003)}]{Bartlett2003}
\bibinfo{author}{\bibnamefont{Bartlett, S.~D., and W.~J.~Munro}},
\bibinfo{year}{2003},
\bibinfo{journal}{Phys. Rev. Lett.} \textbf{\bibinfo{volume}{90}},
\bibinfo{pages}{117901}.

\bibitem[{\citenamefont{Bartlett {\em et al.}}(2002)}]{Bar02}
\bibinfo{author}{\bibnamefont{Bartlett, S.~D., B.~C.~Sanders, S.~L.~Braunstein, and K.~Nemoto}},
\bibinfo{year}{2002},
\bibinfo{journal}{Phys. Rev. Lett.} \textbf{\bibinfo{volume}{88}},
\bibinfo{pages}{097904}.



\bibitem[{\citenamefont{Bennett, Bernstein, {\em et al.}}(1996)}]{Bennett1996}
\bibinfo{author}{\bibnamefont{Bennett, C.~H., H.~J.~Bernstein, S.~Popescu, and B.~Schumacher}},
\bibinfo{year}{1996},
\bibinfo{journal}{Phys. Rev. A} \textbf{\bibinfo{volume}{53}},
\bibinfo{pages}{2046}.

\bibitem[{\citenamefont{Bennett and Brassard}(1984)}]{Bennett1984}
\bibinfo{author}{\bibnamefont{Bennett, C.~H., and G.~Brassard}},
\bibinfo{year}{1984}, in \emph{\bibinfo{booktitle}{Proceedings of the IEEE International Conference on Computers, Systems
and Signal Processing}},
(\bibinfo{publisher}{Bangalore, India, (IEEE, New York)}),
\bibinfo{pages}{175}.


\bibitem[{\citenamefont{Bennett {\em et al.}}(1993)}]{Bennett1993}
\bibinfo{author}{\bibnamefont{Bennett, C.~H., G.~Brassard, C.~Crepeau, R.~Jozsa, A.~Peres, and W.~K.~Wootters}},
\bibinfo{year}{1993},
\bibinfo{journal}{Phys. Rev. Lett.} \textbf{\bibinfo{volume}{70}},
\bibinfo{pages}{1895}.

\bibitem[{\citenamefont{Bennett {\em et al.}}(1992)}]{Bennett1992}%
\bibinfo{author}{\bibnamefont{Bennett, C.~H., G.~Brassard, and N.~D.~Mermin}},
\bibinfo{year}{1992}, \bibinfo{journal}{Phys. Rev. Lett.}
\textbf{\ \bibinfo{volume}{68}}, \bibinfo{pages}{557}.


\bibitem[{\citenamefont{Bennett, Brassard, {\em et al.}}(1996)}]{Ben96}
\bibinfo{author}{\bibnamefont{Bennett, C.~H., G.~Brassard, S.~Popescu, B.~Schumacher,
J.~A.~Smolin, and W.~K.~Wootters}},
\bibinfo{year}{1996},
\bibinfo{journal}{Phys. Rev. Lett.} \textbf{\bibinfo{volume}{76}},
\bibinfo{pages}{722}.

\bibitem[{\citenamefont{Bennett, DiVincenzo, {\em et al.}}(1996)}]{Bennett1996b}
\bibinfo{author}{\bibnamefont{Bennett C.~H., D.~P.~DiVincenzo, J.~A.~Smolin, and W.~K.~Wootters}},
\bibinfo{year}{1996},
\bibinfo{journal}{Phys. Rev. A} \textbf{\bibinfo{volume}{54}},
\bibinfo{pages}{3824}.


\bibitem[{\citenamefont{Bennett {\em et al.}}(2002)}]{Bennett2002}
\bibinfo{author}{\bibnamefont{Bennett, C.~H., P.~W.~Shor, J.~A.~Smolin, and A.~V.~Thapliyal}},
\bibinfo{year}{2002},
\bibinfo{journal}{IEEE Trans.~Inf.~Theor.} \textbf{\bibinfo{volume}{48}},
\bibinfo{pages}{2637}.

\bibitem[{\citenamefont{Bennett and Wiesner}(1992)}]{Ben92}
\bibinfo{author}{\bibnamefont{Bennett, C.~H., and S.~J.~Wiesner}},
\bibinfo{year}{1992},
\bibinfo{journal}{Phys. Rev. Lett.} \textbf{\bibinfo{volume}{69}},
\bibinfo{pages}{2881}.


\bibitem[{\citenamefont{Bergou {\em et al.}}(2004)}]{Bergou2004}
\bibinfo{author}{\bibnamefont{Bergou, J., U.~Herzog, and M.~Hillery}},
\bibinfo{year}{2004}, in \emph{\bibinfo{booktitle}{Lecture Notes in Physics}}, edited by \bibinfo{editor}{\bibnamefont{M.~Paris and J.~Rehacek}},
(\bibinfo{publisher}{Springer, Berlin}), \textbf{\bibinfo{volume}{649}},
\bibinfo{pages}{417}.

\bibitem[{\citenamefont{Bisio {\em et al.}}(2011)}]{Bisio2011}
\bibinfo{author}{\bibnamefont{Bisio A., M.~Dall'Arno, and G.~M.~D'Ariano}},
\bibinfo{year}{2011},
\bibinfo{journal}{Phys. Rev. A} \textbf{\bibinfo{volume}{84}},
\bibinfo{pages}{012310}.


\bibitem[{\citenamefont{Bondurant}(1993)}]{Bon93}
\bibinfo{author}{\bibnamefont{Bondurant, R.~S.}},
\bibinfo{year}{1993},
\bibinfo{journal}{Opt. Lett.} \textbf{\bibinfo{volume}{18}},
\bibinfo{pages}{1896}.

\bibitem[{\citenamefont{Bostr\"{o}m and Felbinger}(2002)}]{Bostroem2002}
\bibinfo{author}{\bibnamefont{Bostr\"{o}m, K., and T.~Felbinger}},
\bibinfo{year}{2002},
\bibinfo{journal}{Phys. Rev. Lett.} \textbf{\bibinfo{volume}{89}},
\bibinfo{pages}{187902}.

\bibitem[{\citenamefont{Bostr\"{o}m and Felbinger}(2008)}]{Bostroem2008}
\bibinfo{author}{\bibnamefont{Bostr\"{o}m, K., and T.~Felbinger}},
\bibinfo{year}{2008},
\bibinfo{journal}{Phys. Lett. A} \textbf{\bibinfo{volume}{372}},
\bibinfo{pages}{3953}.

\bibitem[{\citenamefont{Botero and Reznik}(2003)}]{Botero2003}
\bibinfo{author}{\bibnamefont{Botero, A., and B.~Reznik}},
\bibinfo{year}{2003},
\bibinfo{journal}{Phys. Rev. A} \textbf{\bibinfo{volume}{67}},
\bibinfo{pages}{052311}.


\bibitem[{\citenamefont{Bowen {\em et al.}}(2003)}]{Bowen2003}
\bibinfo{author}{\bibnamefont{Bowen, W.~P., N.~Treps, B.~C.~Buchler, R.~Schnabel, T.~C.~Ralph, H.~-A.~Bachor, T.~Symul, and P.~K.~Lam}},
\bibinfo{year}{2003},
\bibinfo{journal}{Phys. Rev. A} \textbf{\bibinfo{volume}{67}},
\bibinfo{pages}{032302}.



\bibitem[{\citenamefont{Braunstein}(1990)}]{Braunstein1990}
\bibinfo{author}{\bibnamefont{Braunstein, S.~L.}},
\bibinfo{year}{1990},
\bibinfo{journal}{Phys. Rev. A} \textbf{\bibinfo{volume}{42}},
\bibinfo{pages}{474}.

\bibitem[{\citenamefont{Braunstein}(1998a)}]{Bra98a}
\bibinfo{author}{\bibnamefont{Braunstein, S.~L.}},
\bibinfo{year}{1998a},
\bibinfo{journal}{Phys. Rev. Lett.} \textbf{\bibinfo{volume}{80}},
\bibinfo{pages}{4084}.

\bibitem[{\citenamefont{Braunstein}(1998b)}]{Bra98b}
\bibinfo{author}{\bibnamefont{Braunstein, S.~L.}},
\bibinfo{year}{1998b},
\bibinfo{journal}{Nature} \textbf{\bibinfo{volume}{394}},
\bibinfo{pages}{47}.


\bibitem[{\citenamefont{Braunstein}(2005)}]{Bra05}
\bibinfo{author}{\bibnamefont{Braunstein, S.~L.}},
\bibinfo{year}{2005},
\bibinfo{journal}{Phys. Rev. A} \textbf{\bibinfo{volume}{71}},
\bibinfo{pages}{055801}.

\bibitem[{\citenamefont{Braunstein, Cerf {\em et al.}}(2001)}]{Braunstein2001b}
\bibinfo{author}{\bibnamefont{Braunstein, S.~L., N.~J.~Cerf, S.~Iblisdir, P.~van~Loock, and S.~Massar}},
\bibinfo{year}{2001},
\bibinfo{journal}{Phys. Rev. Lett.} \textbf{\bibinfo{volume}{86}},
\bibinfo{pages}{4938}.


\bibitem[{\citenamefont{Braunstein and Crouch}(1991)}]{Braunstein1991}
\bibinfo{author}{\bibnamefont{Braunstein, S.~L., and D.~D.~Crouch}},
\bibinfo{year}{1991},
\bibinfo{journal}{Phys. Rev. A} \textbf{\bibinfo{volume}{43}},
\bibinfo{pages}{330}.

\bibitem[{\citenamefont{Braunstein {\em et al.}}(2000)}]{Braunstein2000}
\bibinfo{author}{\bibnamefont{Braunstein, S.~L, C.~A.~Fuchs, and H.~J.~Kimble}},
\bibinfo{year}{2000},
\bibinfo{journal}{J.~Mod.~Opt.} \textbf{\bibinfo{volume}{47}},
\bibinfo{pages}{267}.

\bibitem[{\citenamefont{Braunstein, Fuchs {\em et al.}}(2001)}]{Braunstein2001}
\bibinfo{author}{\bibnamefont{Braunstein, S.~L, C.~A.~Fuchs, H.~J.~Kimble, and P.~van Loock}},
\bibinfo{year}{2001},
\bibinfo{journal}{Phys. Rev. A} \textbf{\bibinfo{volume}{64}},
\bibinfo{pages}{022321}.

\bibitem[{\citenamefont{Braunstein and Kimble}(1998)}]{Braunstein1998}
\bibinfo{author}{\bibnamefont{Braunstein, S.~L., and H.~J.~Kimble}},
\bibinfo{year}{1998},
\bibinfo{journal}{Phys. Rev. Lett.} \textbf{\bibinfo{volume}{80}},
\bibinfo{pages}{869}.

\bibitem[{\citenamefont{Braunstein and Kimble}(1999)}]{Bra99}
\bibinfo{author}{\bibnamefont{Braunstein, S.~L., and H.~J.~Kimble}},
\bibinfo{year}{1999},
\bibinfo{journal}{Phys. Rev. A} \textbf{\bibinfo{volume}{61}},
\bibinfo{pages}{042302}.

\bibitem[{\citenamefont{Braunstein and Pati}(2003)}]{Braunstein2003}
\bibinfo{author}{\bibnamefont{Braunstein, S.~L., and A.~K.~Pati}},
  \bibinfo{year}{2003}, \emph{\bibinfo{booktitle}{Quantum information with continuous variables}} (\bibinfo{publisher}{Kluwer
Academic, Dordrecht}).

\bibitem[{\citenamefont{Braunstein and van Loock}(2005)}]{Bra04}
\bibinfo{author}{\bibnamefont{Braunstein, S.~L., and P.~van Loock}},
\bibinfo{year}{2005},
\bibinfo{journal}{Rev. Mod. Phys.} \textbf{\bibinfo{volume}{77}},
\bibinfo{pages}{513}.

\bibitem[{\citenamefont{Bravyi}(2005)}]{Bravyi2005}
\bibinfo{author}{\bibnamefont{Bravyi, S.}},
\bibinfo{year}{2005},
\bibinfo{journal}{Quantum Inf. and Comp.} \textbf{\bibinfo{volume}{5}},
\bibinfo{pages}{216}.

\bibitem[{\citenamefont{Browne {\em et al.}}(2003)}]{Bro03}
\bibinfo{author}{\bibnamefont{Browne, D.~E., J.~Eisert, S.~Scheel, and M.~B.~Plenio}},
\bibinfo{year}{2003},
\bibinfo{journal}{Phys. Rev. A} \textbf{\bibinfo{volume}{67}},
\bibinfo{pages}{062320}.

\bibitem[{\citenamefont{Buono {\em et al.}}(2010)}]{Buono2010}
\bibinfo{author}{\bibnamefont{Buono, D., G.~Nocerino, V.~D'Auria, A.~Porzio, S.~Olivares, and M.~G.~A.~Paris}},
\bibinfo{year}{2010},
\bibinfo{journal}{J.~Opt.~Soc.~Am.~B} \textbf{\bibinfo{volume}{27}},
\bibinfo{pages}{A110}.

\bibitem[{\citenamefont{Bu\v{z}ek and Hillery}(1996)}]{Buzek96}
\bibinfo{author}{\bibnamefont{Bu\v{z}ek, V., and M.~Hillery}},
\bibinfo{year}{1996},
\bibinfo{journal}{Phys. Rev. A} \textbf{\bibinfo{volume}{54}},
\bibinfo{pages}{1844}.

\bibitem[{\citenamefont{Cable and Browne}(2010)}]{Cable2010}
\bibinfo{author}{\bibnamefont{Cable, H., and D.~E.~Browne}},
\bibinfo{year}{2010},
\bibinfo{journal}{New J. Phys.} \textbf{\bibinfo{volume}{12}},
\bibinfo{pages}{113046}.



\bibitem[{\citenamefont{Cai and Scarani}(2009)}]{Cai2009}
\bibinfo{author}{\bibnamefont{Cai, R.~Y.~Q., and V.~Scarani}},
\bibinfo{year}{2009},
\bibinfo{journal}{New J. Phys.} \textbf{\bibinfo{volume}{11}},
\bibinfo{pages}{045024}.

\bibitem[{\citenamefont{Calsamiglia {\em et al.}}(2008)}]{Calsamiglia2008}
\bibinfo{author}{\bibnamefont{Calsamiglia, J., R. Munoz-Tapia, L. Masanes, A.
Ac\'{\i}n, and E. Bagan}},
\bibinfo{year}{2008},
\bibinfo{journal}{Phys. Rev. A} \textbf{\bibinfo{volume}{77}},
\bibinfo{pages}{032311}.

\bibitem[{\citenamefont{Caruso {\em et al.}}(2006)}]{Caruso2006}
\bibinfo{author}{\bibnamefont{Caruso, F., V.~Giovannetti, and A.~S.~Holevo}},
\bibinfo{year}{2006},
\bibinfo{journal}{New
Journal of Physics} \textbf{\bibinfo{volume}{8}},
\bibinfo{pages}{310}.

\bibitem[{\citenamefont{Caruso and Giovannetti}(2006)}]{Caruso2006b}
\bibinfo{author}{\bibnamefont{Caruso, F., and V.~Giovannetti}},
\bibinfo{year}{2006},
\bibinfo{journal}{Phys. Rev. A} \textbf{\bibinfo{volume}{74}},
\bibinfo{pages}{062307}.

\bibitem[{\citenamefont{Caruso {\em et al.}}(2008)}]{Caruso2008}
\bibinfo{author}{\bibnamefont{Caruso, F., J.~Eisert, V.~Giovannetti, and A.~S.~Holevo}},
\bibinfo{year}{2008},
\bibinfo{journal}{New J. Phys.} \textbf{\bibinfo{volume}{10}},
\bibinfo{pages}{083030}.

\bibitem[{\citenamefont{Caruso {\em et al.}}(2011)}]{Caruso2010}
\bibinfo{author}{\bibnamefont{Caruso, F., J. Eisert, V. Giovannetti, and A. S. Holevo}},
\bibinfo{year}{2011},
\bibinfo{journal}{Phys. Rev. A} \textbf{\bibinfo{volume}{84}},
\bibinfo{pages}{022306}.

\bibitem[{\citenamefont{Caves and Drummond}(1994)}]{Cav94}
\bibinfo{author}{\bibnamefont{Caves, C.~M., and P.~D.~Drummond}},
\bibinfo{year}{1994},
\bibinfo{journal}{Rev.~Mod.~Phys.} \textbf{\bibinfo{volume}{66}},
\bibinfo{pages}{481}.

\bibitem[{\citenamefont{Cerf}(2003)}]{Cerf-chapter}
\bibinfo{author}{\bibnamefont{Cerf, N.~J.}},
\bibinfo{year}{2003}, in \bibinfo{booktitle}{{\it Quantum Information with Continuous Variables}}, edited by
\bibinfo{editor}{\bibnamefont{S.~L.~Braunstein and A.~K.~Pati}},
(\bibinfo{publisher}{Kluwer Academic, Dordrecht}),
\bibinfo{pages}{277-293}.


\bibitem[{\citenamefont{Cerf and Fiur\`{a}\v{s}ek}(2006)}]{CerfCloning06}
\bibinfo{author}{\bibnamefont{Cerf, N.~J., and J.~Fiur\`{a}\v{s}ek}},
\bibinfo{year}{2006}, in \bibinfo{booktitle}{{\it Progress in Optics}}, edited by
\bibinfo{editor}{\bibnamefont{E.~Wolf}},
(\bibinfo{publisher}{Elsevier, Amsterdam}),
\bibinfo{pages}{455-545}.


\bibitem[{\citenamefont{Cerf and Grangier}(2007)}]{Cerf2007}
\bibinfo{author}{\bibnamefont{Cerf, N.~J., and P.~Grangier}},
\bibinfo{year}{2007},
\bibinfo{journal}{J.~Opt.~Soc.~Am.~B} \textbf{\bibinfo{volume}{24}},
\bibinfo{pages}{324}.


\bibitem[{\citenamefont{Cerf and Iblisdir}(2000)}]{Cerf2000a}
\bibinfo{author}{\bibnamefont{Cerf, N.~J., and S.~Iblisdir}},
\bibinfo{year}{2000},
\bibinfo{journal}{Phys. Rev. A} \textbf{\bibinfo{volume}{62}},
\bibinfo{pages}{040301(R)}.

\bibitem[{\citenamefont{Cerf and Iblisdir}(2001a)}]{Cerf2001a}
\bibinfo{author}{\bibnamefont{Cerf, N.~J., and S.~Iblisdir}},
\bibinfo{year}{2001a},
\bibinfo{journal}{Phys. Rev. Lett.} \textbf{\bibinfo{volume}{87}},
\bibinfo{pages}{247903}.

\bibitem[{\citenamefont{Cerf and Iblisdir}(2001b)}]{Cerf2001b}
\bibinfo{author}{\bibnamefont{Cerf, N.~J., and S.~Iblisdir}},
\bibinfo{year}{2001b},
\bibinfo{journal}{Phys. Rev. A} \textbf{\bibinfo{volume}{64}},
\bibinfo{pages}{032307}.

\bibitem[{\citenamefont{Cerf {\em et al.}}(2002)}]{Cer02}
\bibinfo{author}{\bibnamefont{Cerf, N.~J., S.~Iblisdir and G.~van.~Assche}},
\bibinfo{year}{2002},
\bibinfo{journal}{Eur.~Phys.~J.~D} \textbf{\bibinfo{volume}{18}},
\bibinfo{pages}{211}.

\bibitem[{\citenamefont{Cerf {\em et al.}}(2000)}]{Cerf2000}
\bibinfo{author}{\bibnamefont{Cerf, N. J., A.~Ipe, and X.~Rottenberg}},
\bibinfo{year}{2000},
\bibinfo{journal}{Phys. Rev. Lett.} \textbf{\bibinfo{volume}{85}},
\bibinfo{pages}{1754}.

\bibitem[{\citenamefont{Cerf {\em et al.}}(2005)}]{Cerf2005b}
\bibinfo{author}{\bibnamefont{Cerf, N.~J., O.~Kr\"uger, P.~Navez, R.~F.~Werner, and M.~M.~Wolf}},
\bibinfo{year}{2005},
\bibinfo{journal}{Phys. Rev. Lett.} \textbf{\bibinfo{volume}{95}},
\bibinfo{pages}{070501}.

\bibitem[{\citenamefont{Cerf {\em et al.}}(2001)}]{Cerf2001}
\bibinfo{author}{\bibnamefont{Cerf, N.~J., M.~Levy, and G.~van Assche}},
\bibinfo{year}{2001},
\bibinfo{journal}{Phys. Rev. A} \textbf{\bibinfo{volume}{63}},
\bibinfo{pages}{052311}.


\bibitem[{\citenamefont{Cerf {\em et al.}}(2007)}]{Cerf2007a}
\bibinfo{author}{\bibnamefont{Cerf, N.~J., G.~Leuchs, and E.~S.~Polzik}},
\bibinfo{year}{2007}, \emph{\bibinfo{booktitle}{Quantum Information with Continuous Variables of Atoms and Light}} (\bibinfo{publisher}{Imperial College Press, London}).

\bibitem[{\citenamefont{Chefles}(2000)}]{A.Chefles2000}
\bibinfo{author}{\bibnamefont{Chefles, A.}},
\bibinfo{year}{2000},
\bibinfo{journal}{Contemp. Phys.} \textbf{\bibinfo{volume}{41}},
\bibinfo{pages}{401}.

\bibitem[{\citenamefont{Chefles and Barnett}(1998a)}]{Che98}
\bibinfo{author}{\bibnamefont{Chefles, A., and S.~M.~Barnett}},
\bibinfo{year}{1998a},
\bibinfo{journal}{J.~Mod.~Opt.} \textbf{\bibinfo{volume}{45}},
\bibinfo{pages}{1295}.

\bibitem[{\citenamefont{Chefles and Barnett}(1998b)}]{A.Chefles1998}
\bibinfo{author}{\bibnamefont{Chefles, A., and S.~M.~Barnett}},
\bibinfo{year}{1998b},
\bibinfo{journal}{Phys. Lett. A} \textbf{\bibinfo{volume}{250}},
\bibinfo{pages}{223}.

\bibitem[{\citenamefont{Chen and Zhang}(2007)}]{Chen2007}
\bibinfo{author}{\bibnamefont{Chen, H., and J.~Zhang}},
\bibinfo{year}{2007},
\bibinfo{journal}{Phys. Rev. A} \textbf{\bibinfo{volume}{75}},
\bibinfo{pages}{022306}.


\bibitem[{\citenamefont{Childs {\em et al.}}(2000)}]{Childs2000}
\bibinfo{author}{\bibnamefont{Childs, A., J.~Preskill, and J.~Renes}},
\bibinfo{year}{2000},
\bibinfo{journal}{J.~Mod.~Opt.} \textbf{\bibinfo{volume}{47}},
\bibinfo{pages}{155}.

\bibitem[{\citenamefont{Chiribella {\em et al.}}(2008)}]{Chiribella2008}
\bibinfo{author}{\bibnamefont{Chiribella, G., G.~D'Ariano, and P.~Perinotti}},
\bibinfo{year}{2008},
\bibinfo{journal}{Phys. Rev. Lett.} \textbf{\bibinfo{volume}{101}},
\bibinfo{pages}{180501}.

\bibitem[{\citenamefont{Chizhov {\em et al.}}(2002)}]{Chizhov2002}
\bibinfo{author}{\bibnamefont{Chizhov, A.~V., L.~Kn\"{o}ll, and D.~G.~Welsch}},
\bibinfo{year}{2002},
\bibinfo{journal}{Phys. Rev. A} \textbf{\bibinfo{volume}{65}},
\bibinfo{pages}{022310}.

\bibitem[{\citenamefont{Christandl {\em et al.}}(2009)}]{Christandl2009}
\bibinfo{author}{\bibnamefont{Christandl, M., R.~K\"onig, and R.~Renner}},
\bibinfo{year}{2009},
\bibinfo{journal}{Phys. Rev. Lett.} \textbf{\bibinfo{volume}{102}},
\bibinfo{pages}{020504}.

\bibitem[{\citenamefont{Cochrane {\em et al.}}(2004)}]{Cochrane2004}
\bibinfo{author}{\bibnamefont{Cochrane, P.~T., T.~C.~Ralph, and A.~Dolinska}},
\bibinfo{year}{2004},
\bibinfo{journal}{Phys. Rev. A} \textbf{\bibinfo{volume}{69}},
\bibinfo{pages}{042313}.

\bibitem[{\citenamefont{Cook {\em et al.}}(2007)}]{Coo07}
\bibinfo{author}{\bibnamefont{Cook, R.~L., P.~J.~Martin, and J.~M.~Geremia}},
\bibinfo{year}{2007},
\bibinfo{journal}{Nature} \textbf{\bibinfo{volume}{446}},
\bibinfo{pages}{774}.


\bibitem[{\citenamefont{Cover and Thomas}(2006)}]{Cover2006}
\bibinfo{author}{\bibnamefont{Cover, T.~M., J.~A.~Thomas}},
\bibinfo{year}{2006}, \emph{\bibinfo{booktitle}{Elements of
Information Theory}} (\bibinfo{publisher}{Wiley, Hoboken}).



\bibitem[{\citenamefont{D'Auria {\em et al.}}(2009)}]{Auria2009}
\bibinfo{author}{\bibnamefont{D'Auria, V., S.~Fornaro, A.~Porzio, S.~Solimeno, S.~Olivares, and M.~G.~A.~Paris}},
\bibinfo{year}{2009},
\bibinfo{journal}{Phys. Rev. Lett.} \textbf{\bibinfo{volume}{102}},
\bibinfo{pages}{020502}.

\bibitem[{\citenamefont{Demoen {\em et al.}}(1977)}]{Demoen1977}
\bibinfo{author}{\bibnamefont{Demoen, B., P.~Vanheuverzwijn, and A.~Verbeure}},
\bibinfo{year}{1977},
\bibinfo{journal}{Lett. Math. Phys.} \textbf{\bibinfo{volume}{2}},
\bibinfo{pages}{161}.

\bibitem[{\citenamefont{Deutsch and Jozsa}(1992)}]{Deutsch1992}
\bibinfo{author}{\bibnamefont{Deutsch, D., and R.~Jozsa}},
\bibinfo{year}{1992},
\bibinfo{journal}{Proc. R. Soc. London A} \textbf{\bibinfo{volume}{439}},
\bibinfo{pages}{553}.


\bibitem[{\citenamefont{Devetak}(2005)}]{Devetak2005a}
\bibinfo{author}{\bibnamefont{Devetak, I.}},
\bibinfo{year}{2005},
\bibinfo{journal}{IEEE Trans. Inf. Theory} \textbf{\bibinfo{volume}{51}},
\bibinfo{pages}{44}.


\bibitem[{\citenamefont{Devetak and Shor}(2005)}]{Devetak2005b}
\bibinfo{author}{\bibnamefont{Devetak, I., and P.~W.~Shor}},
\bibinfo{year}{2005},
\bibinfo{journal}{Commun. Math. Phys.} \textbf{\bibinfo{volume}{256}},
\bibinfo{pages}{287}.

\bibitem[{\citenamefont{Devetak and Winter}(2004)}]{Devetak2004}
\bibinfo{author}{\bibnamefont{Devetak, I., and A.~Winter}},
\bibinfo{year}{2004},
\bibinfo{journal}{Phys. Rev. Lett.} \textbf{\bibinfo{volume}{93}},
\bibinfo{pages}{080501}.

\bibitem[{\citenamefont{Dieks}(1982)}]{Dieks82}
\bibinfo{author}{\bibnamefont{Dieks, D.}},
\bibinfo{year}{1982},
\bibinfo{journal}{Phys. Lett.} \textbf{\bibinfo{volume}{92A}},
\bibinfo{pages}{271}.

\bibitem[{\citenamefont{Di Vincenzo {\em et al.}}(1998)}]{DiVicenzo1998}
\bibinfo{author}{\bibnamefont{Di Vincenzo, D.~P., P.~W.~Shor, and J.~A.~Smolin}},
\bibinfo{year}{1998},
\bibinfo{journal}{Phys. Rev. A} \textbf{\bibinfo{volume}{57}},
\bibinfo{pages}{830}.

\bibitem[{\citenamefont{Di Vincenzo and Terhal}(2005)}]{DiVicenzo2005}
\bibinfo{author}{\bibnamefont{Di Vincenzo, D.~P., and B.~M.~Terhal}},
\bibinfo{year}{2005},
\bibinfo{journal}{Found. Phys.} \textbf{\bibinfo{volume}{35}},
\bibinfo{pages}{1967}.

\bibitem[{\citenamefont{Dolinar}(1973)}]{Dol73}
\bibinfo{author}{\bibnamefont{Dolinar, S.}},
\bibinfo{year}{1973},
\bibinfo{journal}{Research Laboratory of Electronics, MIT,
Quarterly Progress Report} \textbf{\bibinfo{volume}{111}},
\bibinfo{pages}{115}.

\bibitem[{\citenamefont{Dong {\em et al.}}(2008)}]{Dong08}
\bibinfo{author}{\bibnamefont{Dong, R., M.~Lassen, J.~Heersink, C.~Marquardt, R.~Filip, G.~Leuchs, and U.~L.~Andersen}},
\bibinfo{year}{2008},
\bibinfo{journal}{Nature Physics} \textbf{\bibinfo{volume}{4}},
\bibinfo{pages}{919}.

\bibitem[{\citenamefont{Duan {\em et al.}}(2000)}]{Duan2000}
\bibinfo{author}{\bibnamefont{Duan, L.~-M., G.~Giedke, J.~I.~Cirac, and P.~Zoller}},
\bibinfo{year}{2000},
\bibinfo{journal}{Phys. Rev. Lett.} \textbf{\bibinfo{volume}{84}},
\bibinfo{pages}{2722}.

\bibitem[{\citenamefont{Duan {\em et al.}}(2009)}]{Duan2009}
\bibinfo{author}{\bibnamefont{Duan, R., Y.~Feng, and M.~Ying}},
\bibinfo{year}{2009},
\bibinfo{journal}{Phys. Rev. Lett.} \textbf{\bibinfo{volume}{103}},
\bibinfo{pages}{210501}.

\bibitem[{\citenamefont{Eisert}(2001)}]{Eisert2001}
\bibinfo{author}{\bibnamefont{Eisert, J.}},
\bibinfo{year}{2001},
\bibinfo{journal}{Ph.D. thesis (Potsdam)}.

\bibitem[{\citenamefont{Eisert {\em et al.}}(2004)}]{Eisert2004}
\bibinfo{author}{\bibnamefont{Eisert, J., D.~E.~Browne,
S.~Scheel, and M.~B.~Plenio}},
\bibinfo{year}{2004},
\bibinfo{journal}{Annals of Physics} \textbf{\bibinfo{volume}{311}},
\bibinfo{pages}{431}.


\bibitem[{\citenamefont{Eisert {\em et al.}}(2010)}]{Eisert2010}
\bibinfo{author}{\bibnamefont{Eisert, J., M.~Cramer, and M.~B.~Plenio}},
\bibinfo{year}{2010},
\bibinfo{journal}{Rev. Mod. Phys.} \textbf{\bibinfo{volume}{82}},
\bibinfo{pages}{277}.


\bibitem[{\citenamefont{Eisert and Plenio}(2003)}]{Eisert2003}
\bibinfo{author}{\bibnamefont{Eisert, J., and M.~B.~Plenio}},
\bibinfo{year}{2003},
\bibinfo{journal}{Int. J. Quant. Inf.} \textbf{\bibinfo{volume}{1}},
\bibinfo{pages}{479}.

\bibitem[{\citenamefont{Eisert {\em et al.}}(2002)}]{Eisert2002}
\bibinfo{author}{\bibnamefont{Eisert, J., S.~Scheel, and M.~B.~Plenio}},
\bibinfo{year}{2002},
\bibinfo{journal}{Phys. Rev. Lett.} \textbf{\bibinfo{volume}{89}},
\bibinfo{pages}{137903}.

\bibitem[{\citenamefont{Eisert and Wolf}(2007)}]{Eisert2007}
\bibinfo{author}{\bibnamefont{Eisert, J., and M.~M.~Wolf}},
\bibinfo{year}{2007}, in \emph{\bibinfo{booktitle}{Quantum Information with Continuous Variables of Atoms and Light}}, edited by \bibinfo{editor}{\bibnamefont{E.~Polzik, N.~Cerf, and G. ~Leuchs}},
(\bibinfo{publisher}{Imperial College Press, London}).

\bibitem[{\citenamefont{Enk}(2002)}]{Elk2002}
\bibinfo{author}{\bibnamefont{Enk, S. V.}},
\bibinfo{year}{2002},
\bibinfo{journal}{Phys. Rev. A} \textbf{\bibinfo{volume}{66}},
\bibinfo{pages}{042313}.


\bibitem[{\citenamefont{Fernholz {\em et al.}}(2008)}]{Fernholz2008}
\bibinfo{author}{\bibnamefont{Fernholz, T., H.~Krauter, K.~Jensen, J.~F.~Sherson, A.~S.~Sorensen, and E.~S.~Polzik}},
\bibinfo{year}{2008},
\bibinfo{journal}{Phys. Rev. Lett.} \textbf{\bibinfo{volume}{101}},
\bibinfo{pages}{073601}.

\bibitem[{\citenamefont{Ferraro {\em et al.}}(2005)}]{Ferraro2005}
\bibinfo{author}{\bibnamefont{Ferraro, A., S.~Olivares, and M.~G.~A.~Paris}},
\bibinfo{year}{2005}, \emph{\bibinfo{booktitle}{Gaussian states in quantum information}} (\bibinfo{publisher}{Biliopolis, Napoli}).



\bibitem[{\citenamefont{Filip}(2008)}]{Filip2008}
\bibinfo{author}{\bibnamefont{Filip, R.}},
\bibinfo{year}{2008},
\bibinfo{journal}{Phys. Rev. A} \textbf{\bibinfo{volume}{77}},
\bibinfo{pages}{022310}.

\bibitem[{\citenamefont{Fiur\`{a}\v{s}ek}(2001)}]{Fiurasek2001}
\bibinfo{author}{\bibnamefont{Fiur\`{a}\v{s}ek, J.}},
\bibinfo{year}{2001},
\bibinfo{journal}{Phys. Rev. Lett.} \textbf{\bibinfo{volume}{86}},
\bibinfo{pages}{4942}.

\bibitem[{\citenamefont{Fiur\`{a}\v{s}ek}(2002a)}]{Fiurasek2002}
\bibinfo{author}{\bibnamefont{Fiur\`{a}\v{s}ek, J.}},
\bibinfo{year}{2002a},
\bibinfo{journal}{Phys. Rev. Lett.} \textbf{\bibinfo{volume}{89}},
\bibinfo{pages}{137904}.

\bibitem[{\citenamefont{Fiur\`{a}\v{s}ek}(2002b)}]{Fiurasek2002b}
\bibinfo{author}{\bibnamefont{Fiur\`{a}\v{s}ek, J.}},
\bibinfo{year}{2002b},
\bibinfo{journal}{Phys. Rev. A} \textbf{\bibinfo{volume}{66}},
\bibinfo{pages}{012304}.

\bibitem[{\citenamefont{Fiur\`{a}\v{s}ek}(2003)}]{Fiurasek2003}
\bibinfo{author}{\bibnamefont{Fiur\`{a}\v{s}ek, J.}},
\bibinfo{year}{2003},
\bibinfo{journal}{Phys. Rev. A} \textbf{\bibinfo{volume}{67}},
\bibinfo{pages}{012321}.

\bibitem[{\citenamefont{Fiur\`{a}\v{s}ek and Cerf}(2007)}]{Fiurasek2007}
\bibinfo{author}{\bibnamefont{Fiur\`{a}\v{s}ek, J., and N.~J.~Cerf}},
\bibinfo{year}{2007},
\bibinfo{journal}{Phys. Rev. A} \textbf{\bibinfo{volume}{75}},
\bibinfo{pages}{052335}.

\bibitem[{\citenamefont{Fiur\`{a}\v{s}ek {\em et al.}}(2004)}]{Fiurasek2004}
\bibinfo{author}{\bibnamefont{Fiur\`{a}\v{s}ek, J., N.~J.~Cerf, and E.~S.~Polzik}},
\bibinfo{year}{2004},
\bibinfo{journal}{Phys. Rev. Lett.} \textbf{\bibinfo{volume}{93}},
\bibinfo{pages}{180501}.

\bibitem[{\citenamefont{Fiur\`{a}\v{s}ek {\em et al.}}(2003)}]{Fiurasek2003a}
\bibinfo{author}{\bibnamefont{Fiur\`{a}\v{s}ek, J., L.~Mi\v{s}ta, Jr., and R.~Filip}},
\bibinfo{year}{2003},
\bibinfo{journal}{Phys. Rev. A} \textbf{\bibinfo{volume}{67}},
\bibinfo{pages}{022304}.


\bibitem[{\citenamefont{Flammia {\em et al.}}(2009)}]{Fla09}
\bibinfo{author}{\bibnamefont{Flammia, S.~T., N.~C.~Menicucci, and O.~Pfister}},
\bibinfo{year}{2009},
\bibinfo{journal}{J. Phys. B} \textbf{\bibinfo{volume}{42}},
\bibinfo{pages}{114009}.

\bibitem[{\citenamefont{Flammia and Severini}(2009)}]{Fla09a}
\bibinfo{author}{\bibnamefont{Flammia, S.~T, and S.~Severini}},
\bibinfo{year}{2009},
\bibinfo{journal}{J. Phys. A} \textbf{\bibinfo{volume}{42}},
\bibinfo{pages}{065302}.

\bibitem[{\citenamefont{Fossier {\em et al.}}(2009)}]{Fossier2009}
\bibinfo{author}{\bibnamefont{Fossier, S., E.~Diamanti, T.~Debuisschert, A.~Villing,
R.~Tualle-Brouri, and P.~Grangier}},
\bibinfo{year}{2009},
\bibinfo{journal}{New J. Phys.} \textbf{\bibinfo{volume}{11}},
\bibinfo{pages}{045023}.

\bibitem[{\citenamefont{Fuchs}(2000)}]{Fuchs2000}
\bibinfo{author}{\bibnamefont{Fuchs, C.~A.}},
\bibinfo{year}{2000}, in \emph{\bibinfo{booktitle}{Quantum Communication, Computing, and Measurement}}, edited by \bibinfo{editor}{\bibnamefont{P.~Kumar, G.~M.~D'Ariano, and O.~Hirota}},
(\bibinfo{publisher}{Kluwer, Dordrecht}), \textbf{\bibinfo{volume}{649}},
\bibinfo{pages}{11}.

\bibitem[{\citenamefont{Fuchs and de Graaf}(1999)}]{Fuchs99a}
\bibinfo{author}{\bibnamefont{Fuchs, C.~A., and J.~V.~de Graaf}},
\bibinfo{year}{1999},
\bibinfo{journal}{IEEE Trans. Inf. Theory} \textbf{\bibinfo{volume}{45}},
\bibinfo{pages}{1216}.

\bibitem[{\citenamefont{Furusawa {\em et al.}}(1998)}]{Furusawa1998}
\bibinfo{author}{\bibnamefont{Furusawa, A., J.~L.~S\o rensen, S.~L.~Braunstein, C.~A.~Fuchs, H.~J.~Kimble, and E.~S.~Polzik}},
\bibinfo{year}{1998},
\bibinfo{journal}{Science} \textbf{\bibinfo{volume}{282}},
\bibinfo{pages}{706}.

\bibitem[{\citenamefont{Furusawa and van Loock}(2011)}]{Furusawa2011}
\bibinfo{author}{\bibnamefont{Furusawa, A., and P.~van Loock}},
  \bibinfo{year}{2011}, \emph{\bibinfo{booktitle}{Quantum Teleportation and Entanglement: A Hybrid Approach to Optical Quantum Information Processing}} (\bibinfo{publisher}{Wiley-VCH, Weinheim}).

\bibitem[{\citenamefont{Gabriel {\em et al.}}(2010)}]{Gabriel2010}
\bibinfo{author}{\bibnamefont{Gabriel, C., C.~Wittmann, D.~Sych, R.~Dong, W.~Mauerer,	 U.~L.~Andersen,	C.~Marquardt, and G.~Leuchs}},
\bibinfo{year}{2010},
\bibinfo{journal}{Nature Photonics} \textbf{\bibinfo{volume}{4}},
\bibinfo{pages}{711}.

\bibitem[{\citenamefont{Garc\'{\i}a-Patr\'{o}n}(2007)}]{GarciaPatron2007}
\bibinfo{author}{\bibnamefont{Garc\'{\i}a-Patr\'{o}n, R.}},
\bibinfo{year}{2007},
\bibinfo{journal}{Ph.D. thesis (Universit\'{e} Libre de Bruxelles)}.

\bibitem[{\citenamefont{Garc\'{\i}a-Patr\'{o}n and Cerf}(2006)}]{GarciaPatron2006}
\bibinfo{author}{\bibnamefont{Garc\'{\i}a-Patr\'{o}n, R., and N. J. Cerf}},
\bibinfo{year}{2006},
\bibinfo{journal}{Phys. Rev. Lett.} \textbf{\bibinfo{volume}{97}},
\bibinfo{pages}{190503}.

\bibitem[{\citenamefont{Garc\'{\i}a-Patr\'{o}n and Cerf}(2009)}]{GarciaPatron2009a}
\bibinfo{author}{\bibnamefont{Garc\'{\i}a-Patr\'{o}n, R., and N.~J.~Cerf}},
\bibinfo{year}{2009},
\bibinfo{journal}{Phys. Rev. Lett.} \textbf{\bibinfo{volume}{102}},
\bibinfo{pages}{130501}.

\bibitem[{\citenamefont{Garc\'{\i}a-Patr\'{o}n {\em et al.}}(2009)}]{GarciaPatron2009}
\bibinfo{author}{\bibnamefont{Garc\'{\i}a-Patr\'{o}n, R., S.~Pirandola, S.~Lloyd,
J.~H.~Shapiro}},
\bibinfo{year}{2009},
\bibinfo{journal}{Phys. Rev. Lett.} \textbf{\bibinfo{volume}{102}},
\bibinfo{pages}{210501}.

\bibitem[{\citenamefont{Geremia}(2004)}]{Ger04}
\bibinfo{author}{\bibnamefont{Geremia, J.~M.}},
\bibinfo{year}{2004},
\bibinfo{journal}{Phys. Rev. A} \textbf{\bibinfo{volume}{70}},
\bibinfo{pages}{062303}.

\bibitem[{\citenamefont{Gerry and Knight}(2005)}]{Gerry2005}
\bibinfo{author}{\bibnamefont{Gerry, C.~C., and P.~L.~Knight}},
\bibinfo{year}{2000}, \emph{\bibinfo{booktitle}{Introductory Quantum Optics}} (\bibinfo{publisher}{Cambridge University Press, Cambridge}).

\bibitem[{\citenamefont{Giedke and Cirac}(2002)}]{Giedke2002}
\bibinfo{author}{\bibnamefont{Giedke, G., and J.~I.~Cirac}},
\bibinfo{year}{2002},
\bibinfo{journal}{Phys. Rev. A} \textbf{\bibinfo{volume}{66}},
\bibinfo{pages}{032316}.

\bibitem[{\citenamefont{Giedke, Duan, {\em et al.}}(2001)}]{Giedke2001}
\bibinfo{author}{\bibnamefont{Giedke, G., L.~-M.~Duan, J.~I.~Cirac, and P.~Zoller}},
\bibinfo{year}{2001},
\bibinfo{journal}{Int. J. Quant. Inf.} \textbf{\bibinfo{volume}{3}},
\bibinfo{pages}{79}.

\bibitem[{\citenamefont{Giedke, Kraus, {\em et al.}}(2001)}]{Giedke2001b}
\bibinfo{author}{\bibnamefont{Giedke, G., B.~Kraus, M.~Lewenstein, and J.~I.~Cirac}},
\bibinfo{year}{2001},
\bibinfo{journal}{Phys. Rev. Lett.} \textbf{\bibinfo{volume}{87}},
\bibinfo{pages}{167904}.



\bibitem[{\citenamefont{Giedke {\em et al.}}(2003)}]{Giedke2003b}
\bibinfo{author}{\bibnamefont{Giedke, G., M.~M.~Wolf, O.~Kruger, R.~F.~Werner, and J.~I.~Cirac}},
\bibinfo{year}{2003},
\bibinfo{journal}{Phys. Rev. Lett.} \textbf{\bibinfo{volume}{91}},
\bibinfo{pages}{107901}.

\bibitem[{\citenamefont{Giorda and Paris}(2010)}]{Giorda2010}
\bibinfo{author}{\bibnamefont{Giorda, P., and M.~G.~A.~Paris}},
\bibinfo{year}{2010},
\bibinfo{journal}{Phys. Rev. Lett.} \textbf{\bibinfo{volume}{105}},
\bibinfo{pages}{020503}.

\bibitem[{\citenamefont{Giovannetti {\em et al.}}(2004a)}]{Giovannetti2004a}
\bibinfo{author}{\bibnamefont{Giovannetti, V., S.~Guha, S.~Lloyd, L.~Maccone, J.~H.~Shapiro, and H.~P.~Yuen}},
\bibinfo{year}{2004a},
\bibinfo{journal}{Phys. Rev. Lett.} \textbf{\bibinfo{volume}{16}},
\bibinfo{pages}{027902}.


\bibitem[{\citenamefont{Giovannetti {\em et al.}}(2004b)}]{Giovannetti2004b}
\bibinfo{author}{\bibnamefont{Giovannetti, V., S.~Guha, S.~Lloyd, L.~Maccone, and J.~H.~Shapiro}},
\bibinfo{year}{2004b},
\bibinfo{journal}{Phys. Rev. A} \textbf{\bibinfo{volume}{70}},
\bibinfo{pages}{032315}.


\bibitem[{\citenamefont{Giovannetti and Lloyd}(2004)}]{Giovannetti2004d}
\bibinfo{author}{\bibnamefont{Giovannetti, V., and S.~Lloyd}},
\bibinfo{year}{2004},
\bibinfo{journal}{Phys. Rev. A} \textbf{\bibinfo{volume}{70}},
\bibinfo{pages}{062307}.

\bibitem[{\citenamefont{Giovannetti, Lloyd, {\em et al.}}(2004)}]{Giovannetti2004c}
\bibinfo{author}{\bibnamefont{Giovannetti, V., S.~Lloyd, L.~Maccone, J.~H.~Shapiro, and B.~J.~Yen}},
\bibinfo{year}{2004},
\bibinfo{journal}{Phys. Rev. A} \textbf{\bibinfo{volume}{70}},
\bibinfo{pages}{022328}.

\bibitem[{\citenamefont{Giovannetti {\em et al.}}(2003)}]{Giovannetti2003}
\bibinfo{author}{\bibnamefont{Giovannetti, V., S.~Lloyd, L.~Maccone, and P.~W.~Shor}},
\bibinfo{year}{2003},
\bibinfo{journal}{Phys. Rev. A} \textbf{\bibinfo{volume}{68}},
\bibinfo{pages}{062323}.


\bibitem[{\citenamefont{Gisin {\em et al.}}(2006)}]{Gisin2006}
\bibinfo{author}{\bibnamefont{Gisin, N., S.~Fasel, B.~Kraus, H.~Zbinden, and G.~Ribordy}},
\bibinfo{year}{2006},
\bibinfo{journal}{Phys. Rev. A} \textbf{\bibinfo{volume}{73}},
\bibinfo{pages}{022320}.

\bibitem[{\citenamefont{Gisin {\em et al.}}(2002)}]{Gisin2002}
\bibinfo{author}{\bibnamefont{Gisin, N., G.~Ribordy, W.~Tittel, and H.~Zbinden}},
\bibinfo{year}{2002},
\bibinfo{journal}{Rev. Mod. Phys.} \textbf{\bibinfo{volume}{74}},
\bibinfo{pages}{145}.

\bibitem[{\citenamefont{Glancy and Knill}(2006)}]{Gla06}
\bibinfo{author}{\bibnamefont{Glancy, S., and E. Knill}},
\bibinfo{year}{2006},
\bibinfo{journal}{Phys. Rev. A} \textbf{\bibinfo{volume}{73}},
\bibinfo{pages}{012325}.

\bibitem[{\citenamefont{Gottesman}(1997)}]{Got97}
\bibinfo{author}{\bibnamefont{Gottesman, D.}},
\bibinfo{year}{1997},
\bibinfo{journal}{Ph.D. thesis (California Institute of Technology)}.

\bibitem[{\citenamefont{Gottesman}(1998)}]{Got98}
\bibinfo{author}{\bibnamefont{Gottesman, D.}},
\bibinfo{year}{1998},
\bibinfo{booktitle}{``The Heisenberg Representation of Quantum Computers"},
\bibinfo{journal}{arXiv:quant-ph/9807006v1}.

\bibitem[{\citenamefont{Gottesman {\em et al.}}(2001)}]{Got01}
\bibinfo{author}{\bibnamefont{Gottesman, D., A.~Kitaev, and J.~Preskill}},
\bibinfo{year}{2001},
\bibinfo{journal}{Phys. Rev. A} \textbf{\bibinfo{volume}{64}},
\bibinfo{pages}{012310}.


\bibitem[{\citenamefont{Gottesman and Preskill}(2001)}]{Gottesman2001}
\bibinfo{author}{\bibnamefont{Gottesman, D., and J.~Preskill}},
\bibinfo{year}{2001},
\bibinfo{journal}{Phys. Rev. A} \textbf{\bibinfo{volume}{63}},
\bibinfo{pages}{022309}.

\bibitem[{\citenamefont{Grosshans and Cerf}(2004)}]{Grosshans2004}
\bibinfo{author}{\bibnamefont{Grosshans, F., and N.~J.~Cerf}},
\bibinfo{year}{2004},
\bibinfo{journal}{Phys. Rev. Lett.} \textbf{\bibinfo{volume}{92}},
\bibinfo{pages}{047905}.

\bibitem[{\citenamefont{Grosshans, Cerf, {\em et al.}}(2003)}]{Grosshans2003b}
\bibinfo{author}{\bibnamefont{Grosshans, F., N.~J.~Cerf, J.~Wenger, R.~Tualle-Brouri, and P.~Grangier}},
\bibinfo{year}{2003},
\bibinfo{journal}{Quantum Inf. Comput.} \textbf{\bibinfo{volume}{3}},
\bibinfo{pages}{535}.

\bibitem[{\citenamefont{Grosshans and Grangier}(2001)}]{Grosshans2001}
\bibinfo{author}{\bibnamefont{Grosshans, F., and P.~Grangier}},
\bibinfo{year}{2001},
\bibinfo{journal}{Phys. Rev. A} \textbf{\bibinfo{volume}{64}},
\bibinfo{pages}{010301 (R)}.


\bibitem[{\citenamefont{Grosshans and Grangier}(2002)}]{Grosshans2002}
\bibinfo{author}{\bibnamefont{Grosshans, F., and P.~Grangier}},
\bibinfo{year}{2002},
\bibinfo{journal}{Phys. Rev. Lett.} \textbf{\bibinfo{volume}{88}},
\bibinfo{pages}{057902}.

\bibitem[{\citenamefont{Grosshans, van~Assche, {\em et al.}}(2003)}]{Grosshans2003}
\bibinfo{author}{\bibnamefont{Grosshans, F., G.~van~Assche, J.~Wenger, R.~Tualle-Brouri, N.~J.~Cerf, and P.~Grangier}},
\bibinfo{year}{2003},
\bibinfo{journal}{Nature} \textbf{\bibinfo{volume}{421}},
\bibinfo{pages}{238}.

\bibitem[{\citenamefont{Grover}(1997)}]{Grover1997}
\bibinfo{author}{\bibnamefont{Grover, L.~K.}},
\bibinfo{year}{1997},
\bibinfo{journal}{Phys. Rev. Lett.} \textbf{\bibinfo{volume}{79}},
\bibinfo{pages}{325}.

\bibitem[{\citenamefont{Gu {\em et al.}}(2009)}]{Gu09}
\bibinfo{author}{\bibnamefont{Gu, M., C.~Weedbrook, N.~C.~Menicucci, T.~C.~Ralph, and P.~van Loock}},
\bibinfo{year}{2009},
\bibinfo{journal}{Phys. Rev. A} \textbf{\bibinfo{volume}{79}},
\bibinfo{pages}{062318}.

\bibitem[{\citenamefont{Guha}(2008)}]{Guha2008}
\bibinfo{author}{\bibnamefont{Guha, S.}},
\bibinfo{year}{2008},
\bibinfo{journal}{Ph.D. thesis (Massachusetts Institute of Technology, Cambridge)}.

\bibitem[{\citenamefont{Guha and Erkmen}(2009)}]{Guha2009}
\bibinfo{author}{\bibnamefont{Guha, S., and B.~Erkmen}},
\bibinfo{year}{2009},
\bibinfo{journal}{Phys. Rev. A} \textbf{\bibinfo{volume}{80}},
\bibinfo{pages}{052310}.


\bibitem[{\citenamefont{Hage {\em et al.}}(2008)}]{Hag08}
\bibinfo{author}{\bibnamefont{Hage, B., A.~Samblowski, J.~DiGuglielmo, A.~Franzen, J.~Fiur\`{a}\v{s}ek, and R.~Schnabel}},
\bibinfo{year}{2008},
\bibinfo{journal}{Nature Physics} \textbf{\bibinfo{volume}{4}},
\bibinfo{pages}{915}.

\bibitem[{\citenamefont{Hall}(1994)}]{Hall1994}
\bibinfo{author}{\bibnamefont{Hall, M.~J.~W.}},
\bibinfo{year}{1994},
\bibinfo{journal}{Phys. Rev. A} \textbf{\bibinfo{volume}{50}},
\bibinfo{pages}{3295}.

\bibitem[{\citenamefont{Hammerer {\em et al.}}(2005)}]{Hammerer2005}
\bibinfo{author}{\bibnamefont{Hammerer, K., M.~M.~Wolf, E.~S.~Polzik, and J.~I.~Cirac}},
\bibinfo{year}{2005},
\bibinfo{journal}{Phys. Rev. Lett.} \textbf{\bibinfo{volume}{94}},
\bibinfo{pages}{150503}.

\bibitem[{\citenamefont{Hammerer {\em et al.}}(2010)}]{Hammerer2010}
\bibinfo{author}{\bibnamefont{Hammerer, K., A.~S.~Sorensen, and E.~S.~Polzik}},
\bibinfo{year}{2010},
\bibinfo{journal}{Rev. Mod. Phys.} \textbf{\bibinfo{volume}{82}},
\bibinfo{pages}{1041}.

\bibitem[{\citenamefont{Harrington and Preskill}(2001)}]{Harrington2001}
\bibinfo{author}{\bibnamefont{Harrington, J., and J.~Preskill}},
\bibinfo{year}{2001},
\bibinfo{journal}{Phys. Rev. A} \textbf{\bibinfo{volume}{64}},
\bibinfo{pages}{062301}.

\bibitem[{\citenamefont{Harrow {\em et al.}}(2010)}]{Harrow2010}
\bibinfo{author}{\bibnamefont{Harrow, A. W, A.~Hassidim, D.~W.~Leung, and J.~Watrous}},
\bibinfo{year}{2010},
\bibinfo{journal}{Phys. Rev. A} \textbf{\bibinfo{volume}{81}},
\bibinfo{pages}{032339}.

\bibitem[{\citenamefont{Hastings}(2009)}]{Hastings2009}
\bibinfo{author}{\bibnamefont{Hastings, M.~B.}},
\bibinfo{year}{2009},
\bibinfo{journal}{Nature Physics} \textbf{\bibinfo{volume}{5}},
\bibinfo{pages}{255}.

\bibitem[{\citenamefont{Hayashi}(2009)}]{Hayashi2009}
\bibinfo{author}{\bibnamefont{Hayashi, M.}},
\bibinfo{year}{2009},
\bibinfo{journal}{IEEE Trans. Inf. Theory} \textbf{\bibinfo{volume}{55}},
\bibinfo{pages}{3807}.

\bibitem[{\citenamefont{Heid and L\"utkenhaus}(2006)}]{Heid2006}
\bibinfo{author}{\bibnamefont{Heid, M., and N.~L\"utkenhaus}},
\bibinfo{year}{2006},
\bibinfo{journal}{Phys. Rev. A} \textbf{\bibinfo{volume}{73}},
\bibinfo{pages}{052316}.

\bibitem[{\citenamefont{Heid and L\"utkenhaus}(2007)}]{Heid2007}
\bibinfo{author}{\bibnamefont{Heid, M., and N.~L\"utkenhaus}},
\bibinfo{year}{2007},
\bibinfo{journal}{Phys. Rev. A} \textbf{\bibinfo{volume}{76}},
\bibinfo{pages}{022313}.

\bibitem[{\citenamefont{Helstrom}(1976)}]{Helstrom1976}
\bibinfo{author}{\bibnamefont{Helstrom, C. W.}},
\bibinfo{year}{1976}, \emph{\bibinfo{booktitle}{Quantum Detection and
Estimation Theory}}, Mathematics in Science and Engineering, \textbf{\bibinfo{volume}{123}},
(\bibinfo{publisher}{Academic Press, New York}).

\bibitem[{\citenamefont{Hillery}(2000)}]{Hillery2000}
\bibinfo{author}{\bibnamefont{Hillery, M.}},
\bibinfo{year}{2000},
\bibinfo{journal}{Phys. Rev. A} \textbf{\bibinfo{volume}{61}},
\bibinfo{pages}{022309}.

\bibitem[{\citenamefont{Hiroshima}(2006)}]{Hiroshima2006}
\bibinfo{author}{\bibnamefont{Hiroshima, T.}},
\bibinfo{year}{2006},
\bibinfo{journal}{Phys. Rev. A} \textbf{\bibinfo{volume}{73}},
\bibinfo{pages}{012330}.

\bibitem[{\citenamefont{Hirota}(2011)}]{Hirota2011}
\bibinfo{author}{\bibnamefont{Hirota, O.}},
\bibinfo{year}{2011},
\bibinfo{booktitle}{``Error free quantum reading by quasi bell state of entangled coherent
states"},
\bibinfo{journal}{arXiv:1108.4163}.

\bibitem[{\citenamefont{Holevo}(1972)}]{Holevo72}
\bibinfo{author}{\bibnamefont{Holevo, A.~S.}},
\bibinfo{year}{1972},
\bibinfo{journal}{Probl. Inform. Transm.} \textbf{\bibinfo{volume}{8}},
\bibinfo{pages}{63}.

\bibitem[{\citenamefont{Holevo}(1973)}]{Holevo1973}
\bibinfo{author}{\bibnamefont{Holevo, A.~S.}},
\bibinfo{year}{1973},
\bibinfo{journal}{Probl. Inform. Transm.} \textbf{\bibinfo{volume}{9}},
\bibinfo{pages}{177}.

\bibitem[{\citenamefont{Holevo}(1975)}]{Holevo1975}
\bibinfo{author}{\bibnamefont{Holevo, A.~S.}},
\bibinfo{year}{1975},
\bibinfo{journal}{IEEE Trans.~Inform.~Theory} \textbf{\bibinfo{volume}{21}},
\bibinfo{pages}{533}.

\bibitem[{\citenamefont{Holevo}(1998)}]{Holevo1998}
\bibinfo{author}{\bibnamefont{Holevo, A.~S.}},
\bibinfo{year}{1998},
\bibinfo{journal}{IEEE Trans. Inf. Theory} \textbf{\bibinfo{volume}{44}},
\bibinfo{pages}{269}.

\bibitem[{\citenamefont{Holevo}(2005)}]{Holevo2005}
\bibinfo{author}{\bibnamefont{Holevo, A.~S.}},
\bibinfo{year}{2005},
\bibinfo{journal}{Theory of Probability and its
Applications} \textbf{\bibinfo{volume}{50}},
\bibinfo{pages}{86}.

\bibitem[{\citenamefont{Holevo}(2007)}]{Holevo2007}
\bibinfo{author}{\bibnamefont{Holevo, A.~S.}},
\bibinfo{year}{2007},
\bibinfo{journal}{Prob. of Inf. Transm.} \textbf{\bibinfo{volume}{43}},
\bibinfo{pages}{1}.

\bibitem[{\citenamefont{Holevo}(2011)}]{Holevo2011}
\bibinfo{author}{\bibnamefont{Holevo, A.~S.}},
  \bibinfo{year}{2011}, \emph{\bibinfo{booktitle}{Probabilistic and Statistical Aspects of Quantum Theory}} (\bibinfo{publisher}{Edizioni della
Normale, Pisa}).

\bibitem[{\citenamefont{Holevo {\em et al.}}(1999)}]{Holevo1999}
\bibinfo{author}{\bibnamefont{Holevo, A.~S., M.~Sohma, and O.~Hirota}},
\bibinfo{year}{1999},
\bibinfo{journal}{Phys. Rev. A} \textbf{\bibinfo{volume}{59}},
\bibinfo{pages}{1820}.

\bibitem[{\citenamefont{Holevo and Werner}(2001)}]{Holevo2001}
\bibinfo{author}{\bibnamefont{Holevo, A.~S., and R.~F.~Werner}},
\bibinfo{year}{2001},
\bibinfo{journal}{Phys. Rev. A} \textbf{\bibinfo{volume}{63}},
\bibinfo{pages}{032312}.


\bibitem[{\citenamefont{Horn and Johnson}(1985)}]{Horn1985}
\bibinfo{author}{\bibnamefont{Horn, R.~A., and C.~R.~Johnson}},
\bibinfo{year}{1985}, \emph{\bibinfo{booktitle}{Matrix Analysis}} (\bibinfo{publisher}{Cambridge University Press, Cambridge}).


\bibitem[{\citenamefont{Horodecki {\em et al.}}(1996)}]{Horodecki1996}
\bibinfo{author}{\bibnamefont{Horodecki, M., P.~Horodecki, and R.~Horodecki}},
\bibinfo{year}{1996},
\bibinfo{journal}{Phys. Lett. A} \textbf{\bibinfo{volume}{223}},
\bibinfo{pages}{1}.

\bibitem[{\citenamefont{Horodecki {\em et al.}}(1998)}]{Horodecki1998}
\bibinfo{author}{\bibnamefont{Horodecki, M., P.~Horodecki, and R.~Horodecki}},
\bibinfo{year}{1998},
\bibinfo{journal}{Phys. Rev. Lett.} \textbf{\bibinfo{volume}{80}},
\bibinfo{pages}{5239}.

\bibitem[{\citenamefont{Horodecki {\em et al.}}(2000)}]{Horodecki2000}
\bibinfo{author}{\bibnamefont{Horodecki, M., P.~Horodecki, and R.~Horodecki}},
\bibinfo{year}{2000},
\bibinfo{journal}{Phys. Rev. Lett.} \textbf{\bibinfo{volume}{84}},
\bibinfo{pages}{2014}.

\bibitem[{\citenamefont{Horodecki {\em et al.}}(2009)}]{Horodecki2009}
\bibinfo{author}{\bibnamefont{Horodecki, M., P.~Horodecki, and R.~Horodecki}},
\bibinfo{year}{2009},
\bibinfo{journal}{Rev. Mod. Phys.} \textbf{\bibinfo{volume}{81}},
\bibinfo{pages}{865}.


\bibitem[{\citenamefont{Hudson}(1974)}]{Hudson1974}
\bibinfo{author}{\bibnamefont{Hudson, R.~L.}},
\bibinfo{year}{1974},
\bibinfo{journal}{Rep.~Math.~Phys.} \textbf{\bibinfo{volume}{6}},
\bibinfo{pages}{249}.

\bibitem[{\citenamefont{Hughston {\em et al.}}(1993)}]{Hughston1993}
\bibinfo{author}{\bibnamefont{Hughston, L.~P., R.~Jozsa, and W.~K.~Wootters}},
\bibinfo{year}{1993},
\bibinfo{journal}{Phys. Lett. A } \textbf{\bibinfo{volume}{183}},
\bibinfo{pages}{14}.

\bibitem[{\citenamefont{Hyllus and Eisert}(2006)}]{Hyllus2006}
\bibinfo{author}{\bibnamefont{Hyllus, P., and J.~Eisert}},
\bibinfo{year}{2006},
\bibinfo{journal}{New J. Phys.} \textbf{\bibinfo{volume}{8}},
\bibinfo{pages}{51}.



\bibitem[{\citenamefont{Ide {\em et al.}}(2001)}]{Ide2001}
\bibinfo{author}{\bibnamefont{Ide, T., H.~F.~Hofmann, T.~Kobayashi, and A.~Furusawa}},
\bibinfo{year}{2001},
\bibinfo{journal}{Phys. Rev. A} \textbf{\bibinfo{volume}{65}},
\bibinfo{pages}{012313}.

\bibitem[{\citenamefont{Invernizzi {\em et al.}}(2011)}]{Invernizzi2010}
\bibinfo{author}{\bibnamefont{Invernizzi, C., M.~G.~A.~Paris, and S.~Pirandola}},
\bibinfo{year}{2011},
\bibinfo{journal}{Phys. Rev. A} \textbf{\bibinfo{volume}{84}},
\bibinfo{pages}{022334}.


\bibitem[{\citenamefont{Jeong and Ralph}(2007)}]{Jeo05}
\bibinfo{author}{\bibnamefont{Jeong, H., and T.~C.~Ralph}},
\bibinfo{year}{2007}, in \emph{\bibinfo{booktitle}{Quantum Information with Continuous Variables of Atoms and Light}}, edited by \bibinfo{editor}{\bibnamefont{E.~Polzik, N.~Cerf, and G. ~Leuchs}},
(\bibinfo{publisher}{Imperial College Press, London}).

\bibitem[{\citenamefont{Jia {\em et al.}}(2004)}]{Jia2004}
\bibinfo{author}{\bibnamefont{Jia, X., X.~Su, Q.~Pan, J.~Gao, C.~Xie, and K.~Peng}},
\bibinfo{year}{2004},
\bibinfo{journal}{Phys. Rev. Lett.} \textbf{\bibinfo{volume}{93}},
\bibinfo{pages}{250503}.

\bibitem[{\citenamefont{Josse {\em et al.}}(2006)}]{Josse2006}
\bibinfo{author}{\bibnamefont{Josse, V., M.~Sabuncu, N.~J.~Cerf, G.~Leuchs, and U.~L.~Andersen}},
\bibinfo{year}{2006},
\bibinfo{journal}{Phys. Rev. Lett.} \textbf{\bibinfo{volume}{96}},
\bibinfo{pages}{163602}.



\bibitem[{\citenamefont{Jozsa}(1994)}]{Jozsa1994}
\bibinfo{author}{\bibnamefont{Jozsa, R.}},
\bibinfo{year}{1994},
\bibinfo{journal}{J.~Mod.~Opt.} \textbf{\bibinfo{volume}{41}},
\bibinfo{pages}{2315}.

\bibitem[{\citenamefont{Julsgaard {\em et al.}}(2001)}]{Julsgaard2001}
\bibinfo{author}{\bibnamefont{Julsgaard, B., A.~Kozhekin, and E.~S.~Polzik}},
\bibinfo{year}{2001},
\bibinfo{journal}{Nature} \textbf{\bibinfo{volume}{413}},
\bibinfo{pages}{400}.


\bibitem[{\citenamefont{Kennedy}(1973)}]{Ken73}
\bibinfo{author}{\bibnamefont{Kennedy, R.~S.}},
\bibinfo{year}{1973},
\bibinfo{journal}{Research Laboratory of Electronics, MIT,
Quarterly Progress Report} \textbf{\bibinfo{volume}{108}},
\bibinfo{pages}{219}.

\bibitem[{\citenamefont{Knill {\em et al.}}(2001)}]{Knill2001}
\bibinfo{author}{\bibnamefont{Knill, E., R.~Laflamme, and G.~J.~Milburn}},
\bibinfo{year}{2001},
\bibinfo{journal}{Nature} \textbf{\bibinfo{volume}{409}},
\bibinfo{pages}{46}.

\bibitem[{\citenamefont{Kok and Lovett}(2010)}]{Kok10}
\bibinfo{author}{\bibnamefont{Kok, P., and B.~Lovett}},
  \bibinfo{year}{2010}, \emph{\bibinfo{booktitle}{Introduction to optical quantum
information processing}} (\bibinfo{publisher}{Cambridge University Press, Cambridge}).


\bibitem[{\citenamefont{Lam {\em et al.}}(1997)}]{Lam1997}
\bibinfo{author}{\bibnamefont{Lam, P.~K., T.~C.~Ralph, E.~H.~Huntington, and H.~-A.~Bachor}},
\bibinfo{year}{1997},
\bibinfo{journal}{Phys. Rev. Lett.} \textbf{\bibinfo{volume}{79}},
\bibinfo{pages}{1471}.

\bibitem[{\citenamefont{Lance {\em et al.}}(2004)}]{Lance2004}
\bibinfo{author}{\bibnamefont{Lance, A.~M., T.~Symul, W.~P.~Bowen, B.~C.~Sanders, and P.~K.~Lam}},
\bibinfo{year}{2004},
\bibinfo{journal}{Phys. Rev. Lett.} \textbf{\bibinfo{volume}{92}},
\bibinfo{pages}{177903}.

\bibitem[{\citenamefont{Lance {\em et al.}}(2005)}]{Lance2005}
\bibinfo{author}{\bibnamefont{Lance, A.~M., T.~Symul, V.~Sharma, C.~Weedbrook, T.~C.~Ralph, and P.~K.~Lam}},
\bibinfo{year}{2005},
\bibinfo{journal}{Phys. Rev. Lett.} \textbf{\bibinfo{volume}{95}},
\bibinfo{pages}{180503}.

\bibitem[{\citenamefont{La Porta {\em et al.}}(1989)}]{Por89}
\bibinfo{author}{\bibnamefont{La Porta, A., R.~E.~Slusher, and B.~Yurke}},
\bibinfo{year}{1989},
\bibinfo{journal}{Phys. Rev. Lett.} \textbf{\bibinfo{volume}{62}},
\bibinfo{pages}{28}.

\bibitem[{\citenamefont{Lassen {\em et al.}}(2010)}]{Las10}
\bibinfo{author}{\bibnamefont{Lassen, M., M.~Sabuncu, A.~Huck, J.~Niset, G.~Leuchs, N.~J.~Cerf, and U.~L.~Andersen}},
\bibinfo{year}{2010},
\bibinfo{journal}{Nature Physics} \textbf{\bibinfo{volume}{4}},
\bibinfo{pages}{700}.

\bibitem[{\citenamefont{Leonhardt}(2010)}]{Leonhardt2010}
\bibinfo{author}{\bibnamefont{Leonhardt, U.}},
  \bibinfo{year}{2010}, \emph{\bibinfo{booktitle}{Essential Quantum Optics}} (\bibinfo{publisher}{Cambridge University Press, Cambridge}).

\bibitem[{\citenamefont{Leverrier {\em et al.}}(2008)}]{Leverrier2008}
\bibinfo{author}{\bibnamefont{Leverrier, A., R.~All\'{a}ume, J.~Boutros, G.~Z\'{e}mor, and P.~Grangier}},
\bibinfo{year}{2008},
\bibinfo{journal}{Phys. Rev. A} \textbf{\bibinfo{volume}{77}},
\bibinfo{pages}{042325}.

\bibitem[{\citenamefont{Leverrier and Cerf}(2009)}]{Leverrier2009c}
\bibinfo{author}{\bibnamefont{Leverrier, A., and N.~J.~Cerf}},
\bibinfo{year}{2009},
\bibinfo{journal}{Phys. Rev. A} \textbf{\bibinfo{volume}{80}},
\bibinfo{pages}{010102}.

\bibitem[{\citenamefont{Leverrier and Grangier}(2009)}]{Leverrier2009b}
\bibinfo{author}{\bibnamefont{Leverrier, A., and P.~Grangier}},
\bibinfo{year}{2009},
\bibinfo{journal}{Phys. Rev. Lett.} \textbf{\bibinfo{volume}{102}},
\bibinfo{pages}{180504}.

\bibitem[{\citenamefont{Leverrier and Grangier}(2010a)}]{Leverrier2010b}
\bibinfo{author}{\bibnamefont{Leverrier, A., and P.~Grangier}},
\bibinfo{year}{2010a},
\bibinfo{journal}{Phys. Rev. A} \textbf{\bibinfo{volume}{81}},
\bibinfo{pages}{062314}.

\bibitem[{\citenamefont{Leverrier and Grangier}(2010b)}]{Leverrier2010c}
\bibinfo{author}{\bibnamefont{Leverrier, A., and P.~Grangier}},
\bibinfo{year}{2010b},
\bibinfo{booktitle}{``Long distance quantum key distribution with continuous variables"},
\bibinfo{journal}{arXiv:1005.0328}.

\bibitem[{\citenamefont{Leverrier and Grangier}(2010c)}]{Leverrier2010d}
\bibinfo{author}{\bibnamefont{Leverrier, A., and P.~Grangier}},
\bibinfo{year}{2010c},
\bibinfo{booktitle}{``Continuous-variable quantum key distribution protocols with a discrete modulation"},
\bibinfo{journal}{arXiv:1002.4083}.

\bibitem[{\citenamefont{Leverrier {\em et al.}}(2010)}]{Leverrier2010a}
\bibinfo{author}{\bibnamefont{Leverrier, A., F.~Grosshans, and P.~Grangier}},
\bibinfo{year}{2010},
\bibinfo{journal}{Phys. Rev. A} \textbf{\bibinfo{volume}{81}},
\bibinfo{pages}{062343}.

\bibitem[{\citenamefont{Leverrier {\em et al.}}(2009)}]{Leverrier2009a}
\bibinfo{author}{\bibnamefont{Leverrier, A., E.~Karpov, P.~Grangier, and N.~J.~Cerf}},
\bibinfo{year}{2009},
\bibinfo{journal}{New J. Phys.} \textbf{\bibinfo{volume}{11}},
\bibinfo{pages}{115009}.

\bibitem[{\citenamefont{Li {\em et al.}}(2002)}]{Li02}
\bibinfo{author}{\bibnamefont{Li, X., Q.~Pan, J.~Jing, J.~Zhang, C.~Xie, and K.~Peng}},
\bibinfo{year}{2002},
\bibinfo{journal}{Phys. Rev. Lett.} \textbf{\bibinfo{volume}{88}},
\bibinfo{pages}{047904}.

\bibitem[{\citenamefont{Lindblad}(2000)}]{Lindblad2000}
\bibinfo{author}{\bibnamefont{Lindblad G.}},
\bibinfo{year}{2000},
\bibinfo{journal}{J. Phys. A} \textbf{\bibinfo{volume}{33}},
\bibinfo{pages}{5059}.


\bibitem[{\citenamefont{Lita {\em et al.}}(2008)}]{Lita2008}
\bibinfo{author}{\bibnamefont{Lita, A.~E., A.~J.~Miller, and S.~W.~Nam}},
\bibinfo{year}{2008},
\bibinfo{journal}{Optics Express} \textbf{\bibinfo{volume}{16}},
\bibinfo{pages}{3032}.



\bibitem[{\citenamefont{Lloyd}(1997)}]{Lloyd1997}
\bibinfo{author}{\bibnamefont{Lloyd, S.}},
\bibinfo{year}{1997},
\bibinfo{journal}{Phys. Rev. A} \textbf{\bibinfo{volume}{55}},
\bibinfo{pages}{1613}.

\bibitem[{\citenamefont{Lloyd}(2008)}]{Lloyd2008}
\bibinfo{author}{\bibnamefont{Lloyd, S.}},
\bibinfo{year}{2008},
\bibinfo{journal}{Science} \textbf{\bibinfo{volume}{321}},
\bibinfo{pages}{1463}.

\bibitem[{\citenamefont{Lloyd and Braunstein}(1999)}]{Llo99}
\bibinfo{author}{\bibnamefont{Lloyd, S., and S.~L.~Braunstein}},
\bibinfo{year}{1999},
\bibinfo{journal}{Phys. Rev. Lett.} \textbf{\bibinfo{volume}{82}},
\bibinfo{pages}{1784}.

\bibitem[{\citenamefont{Lloyd {\em et al.}}(2009)}]{Lloyd2009}
\bibinfo{author}{\bibnamefont{Lloyd, S., V.~Giovannetti, L.~Maccone, N.~J.~Cerf, S.~Guha,
R.~Garcia-Patron, S.~Mitter, S.~Pirandola, M.~B.~Ruskai, J.~H.~Shapiro, and H.~Yuan}},
\bibinfo{year}{2009},
\bibinfo{booktitle}{``The bosonic minimum output entropy conjecture and Lagrangian minimization"},
\bibinfo{journal}{arXiv:0906.2758}.

\bibitem[{\citenamefont{Lloyd and Slotine}(1998)}]{Llo98}
\bibinfo{author}{\bibnamefont{Lloyd, S., and J.-J. E. Slotine}},
\bibinfo{year}{1998},
\bibinfo{journal}{Phys. Rev. Lett.} \textbf{\bibinfo{volume}{80}},
\bibinfo{pages}{4088}.

\bibitem[{\citenamefont{Lodewyck {\em et al.}}(2007)}]{Lodewyck2007b}
\bibinfo{author}{\bibnamefont{Lodewyck, J., M.~Bloch, R.~Garc\'{\i}a-Patr\'{o}n, S.~Fossier, E.~Karpov, E.~Diamanti, T.~Debuisschert, N.~J.~Cerf, R.~Tualle-Brouri, S.~W.~McLaughlin, and P.~Grangier}},
\bibinfo{year}{2007},
\bibinfo{journal}{Phys. Rev. A} \textbf{\bibinfo{volume}{76}},
\bibinfo{pages}{042305}.

\bibitem[{\citenamefont{Lodewyck {\em et al.}}(2005)}]{Lodewyck2005}
\bibinfo{author}{\bibnamefont{Lodewyck, J., T.~Debuisschert, R.~Tualle-Brouri, and P.~Grangier}},
\bibinfo{year}{2005},
\bibinfo{journal}{Phys. Rev. A} \textbf{\bibinfo{volume}{72}},
\bibinfo{pages}{050303}.

\bibitem[{\citenamefont{Lodewyck {\em et al.}}(2007)}]{Lodewyck2007}
\bibinfo{author}{\bibnamefont{Lodewyck, J., T.~Debuisschert, R.~Garc\'{\i}a-Patr\'{o}n, R.~Tualle-~Brouri, N.~J.~Cerf, and P.~Grangier}},
\bibinfo{year}{2007},
\bibinfo{journal}{Phys. Rev. Lett.} \textbf{\bibinfo{volume}{98}},
\bibinfo{pages}{030503}.

\bibitem[{\citenamefont{Lodewyck and Grangier}(2007)}]{Lodewyck2007c}
\bibinfo{author}{\bibnamefont{Lodewyck, J., and P.~Grangier}},
\bibinfo{year}{2007},
\bibinfo{journal}{Phys. Rev. A} \textbf{\bibinfo{volume}{76}},
\bibinfo{pages}{022332}.

\bibitem[{\citenamefont{Lorenz {\em et al.}}(2004)}]{Lorenz2004}
\bibinfo{author}{\bibnamefont{Lorenz, S., N.~Korolkova, and G.~Leuchs}},
\bibinfo{year}{2004},
\bibinfo{journal}{Appl.~Phys.~B: Lasers Opt.} \textbf{\bibinfo{volume}{79}},
\bibinfo{pages}{273}.

\bibitem[{\citenamefont{Lorenz {\em et al.}}(2006)}]{Lorenz2006}
\bibinfo{author}{\bibnamefont{Lorenz, S., J.~Rigas, M.~Heid, U.~L.~Andersen, N.~L\"{u}tkenhaus,  and G.~Leuchs}},
\bibinfo{year}{2006},
\bibinfo{journal}{Phys. Rev. A} \textbf{\bibinfo{volume}{74}},
\bibinfo{pages}{042326}.



\bibitem[{\citenamefont{Lund {\em et al.}}(2008)}]{Lun08}
\bibinfo{author}{\bibnamefont{Lund, A.~P., T.~C.~Ralph, and H.~L.~Haselgrove}},
\bibinfo{year}{2008},
\bibinfo{journal}{Phys. Rev. Lett.} \textbf{\bibinfo{volume}{100}},
\bibinfo{pages}{030503}.




\bibitem[{\citenamefont{Lupo {\em et al.}}(2011)}]{Lupo2011}
\bibinfo{author}{\bibnamefont{Lupo, C., V.~Giovannetti, S.~Pirandola, S.~Mancini, and S.~Lloyd}},
\bibinfo{year}{2011},
\bibinfo{journal}{Phys. Rev. A} \textbf{\bibinfo{volume}{84}},
\bibinfo{pages}{010303 (R)}.

\bibitem[{\citenamefont{Lupo {\em et al.}}(2011)}]{Lupo2011a}
\bibinfo{author}{\bibnamefont{Lupo, C., S.~Pirandola, P.~Aniello, and S.~Mancini}},
\bibinfo{year}{2011},
\bibinfo{journal}{Phys. Scr.} \textbf{\bibinfo{volume}{T143}},
\bibinfo{pages}{014016}.



\bibitem[{\citenamefont{Lvovsky and Raymer}(2009)}]{Lvovsky2009a}
\bibinfo{author}{\bibnamefont{Lvovsky, A.~I., and M.~G.~Raymer}},
\bibinfo{year}{2009},
\bibinfo{journal}{Rev. Mod. Phys.} \textbf{\bibinfo{volume}{81}},
\bibinfo{pages}{299}.


\bibitem[{\citenamefont{Lvovsky {\em et al.}}(2009)}]{Lvovsky2009}
\bibinfo{author}{\bibnamefont{Lvovsky, A.~I., B.~C.~Sanders, and W.~Tittel}},
\bibinfo{year}{2009},
\bibinfo{journal}{Nature Photonics} \textbf{\bibinfo{volume}{3}},
\bibinfo{pages}{706}.

\bibitem[{\citenamefont{Magnin {\em et al.}}(2010)}]{Magnin2010}
\bibinfo{author}{\bibnamefont{Magnin, L., F.~Magniez, A.~Leverrier, and N.~J.~Cerf}},
\bibinfo{year}{2010},
\bibinfo{journal}{Phys. Rev. A} \textbf{\bibinfo{volume}{81}},
\bibinfo{pages}{010302 (R)}.

\bibitem[{\citenamefont{Mancini {\em et al.}}(2003)}]{Mancini2003}
\bibinfo{author}{\bibnamefont{Mancini, S., D.~Vitali, and P.~Tombesi}},
\bibinfo{year}{2003},
\bibinfo{journal}{Phys. Rev. Lett.} \textbf{\bibinfo{volume}{90}},
\bibinfo{pages}{137901}.


\bibitem[{\citenamefont{Mandilara and Cerf}(2011)}]{Mandilara2011}
\bibinfo{author}{\bibnamefont{Mandilara, A., and N.~J.~Cerf}},
\bibinfo{year}{2011},
\bibinfo{booktitle}{``Quantum bit commitment under Gaussian constraints"},
\bibinfo{journal}{arXiv:1105.2140v1}.


\bibitem[{\citenamefont{Mandilara {\em et al.}}(2009)}]{Mandilara2009}
\bibinfo{author}{\bibnamefont{Mandilara, A., E.~Karpov, and N.~J.~Cerf}},
\bibinfo{year}{2009},
\bibinfo{journal}{Phys. Rev. A} \textbf{\bibinfo{volume}{79}},
\bibinfo{pages}{062302}.

\bibitem[{\citenamefont{Mari and Vitali}(2008)}]{Mari2008}
\bibinfo{author}{\bibnamefont{Mari, A., and D.~Vitali}},
\bibinfo{year}{2008},
\bibinfo{journal}{Phys. Rev. A} \textbf{\bibinfo{volume}{78}},
\bibinfo{pages}{062340}.

\bibitem[{\citenamefont{Mattle {\em et al.}}(1996)}]{Mat96}
\bibinfo{author}{\bibnamefont{Mattle, K., H.~Weinfurter, P.~G.~Kwiat,
and A.~Zeilinger}},
\bibinfo{year}{1996},
\bibinfo{journal}{Phys. Rev. Lett.} \textbf{\bibinfo{volume}{76}},
\bibinfo{pages}{4656}.

\bibitem[{\citenamefont{Menezes {\em et al.}}(1997)}]{Menezes1997}
\bibinfo{author}{\bibnamefont{Menezes, A., P.~van~Oorschot, S.~Vanstone}},
  \bibinfo{year}{1997}, \emph{\bibinfo{booktitle}{Handbook of Applied Cryptography}} (\bibinfo{publisher}{CRC Press, Boca Raton}).

\bibitem[{\citenamefont{Menicucci}(2011)}]{Men10}
\bibinfo{author}{\bibnamefont{Menicucci, N.~C.}},
\bibinfo{year}{2011},
\bibinfo{journal}{Phys. Rev. A} \textbf{\bibinfo{volume}{83}},
\bibinfo{pages}{062314}.

\bibitem[{\citenamefont{Menicucci {\em et al.}}(2008)}]{Men08}
\bibinfo{author}{\bibnamefont{Menicucci, N.~C., S.~T.~Flammia, and O.~Pfister}},
\bibinfo{year}{2008},
\bibinfo{journal}{Phys. Rev. Lett.} \textbf{\bibinfo{volume}{101}},
\bibinfo{pages}{130501}.

\bibitem[{\citenamefont{Menicucci {\em et al.}}(2011)}]{Men10a}
\bibinfo{author}{\bibnamefont{Menicucci, N.~C., S.~T.~Flammia, and P.~van Loock}},
\bibinfo{year}{2011},
\bibinfo{journal}{Phys. Rev. A} \textbf{\bibinfo{volume}{83}},
\bibinfo{pages}{042335}.

\bibitem[{\citenamefont{Menicucci {\em et al.}}(2007)}]{Men07}
\bibinfo{author}{\bibnamefont{Menicucci, N.~C., S.~T.~Flammia, H.~Zaidi, and O.~Pfister}},
\bibinfo{year}{2007},
\bibinfo{journal}{Phys. Rev. A} \textbf{\bibinfo{volume}{76}},
\bibinfo{pages}{010302(R)}.

\bibitem[{\citenamefont{Menicucci {\em et al.}}(2010)}]{Men10b}
\bibinfo{author}{\bibnamefont{Menicucci, N.~C., X.~Ma, and T.~C.~Ralph}},
\bibinfo{year}{2010},
\bibinfo{journal}{Phys. Rev. Lett.} \textbf{\bibinfo{volume}{104}},
\bibinfo{pages}{250503}.

\bibitem[{\citenamefont{Menicucci {\em et al.}}(2006)}]{Men06}
\bibinfo{author}{\bibnamefont{Menicucci, N.~C., P.~van Loock, M.~Gu, C.~Weedbrook, T.~C.~Ralph, and M.~A.~Nielsen}},
\bibinfo{year}{2006},
\bibinfo{journal}{Phys. Rev. Lett.} \textbf{\bibinfo{volume}{97}},
\bibinfo{pages}{110501}.

\bibitem[{\citenamefont{Menzies and Korolkova}(2007)}]{Menzies2007}
\bibinfo{author}{\bibnamefont{Menzies, D., and N.~Korolkova}},
\bibinfo{year}{2007},
\bibinfo{journal}{Phys. Rev. A} \textbf{\bibinfo{volume}{76}},
\bibinfo{pages}{062310}.


\bibitem[{\citenamefont{Milburn and Braunstein}(1999)}]{Milburn1999}
\bibinfo{author}{\bibnamefont{Milburn, G.~J., and S.~L.~Braunstein}},
\bibinfo{year}{1999},
\bibinfo{journal}{Phys. Rev. A} \textbf{\bibinfo{volume}{60}},
\bibinfo{pages}{937}.

\bibitem[{\citenamefont{Miwa, {\em et al.}}(2010)}]{Miw10}
\bibinfo{author}{\bibnamefont{Miwa, Y., R.~Ukai, J.~Yoshikawa, R.~Filip, P.~van Loock, and A.~Furusawa}},
\bibinfo{year}{2010},
\bibinfo{journal}{Phys. Rev. A} \textbf{\bibinfo{volume}{82}},
\bibinfo{pages}{032305}.

\bibitem[{\citenamefont{Mizuno {\em et al.}}(2005)}]{Miz05}
\bibinfo{author}{\bibnamefont{Mizuno, J., K.~Wakui, A.~Furusawa, M.~Sasaki}},
\bibinfo{year}{2005},
\bibinfo{journal}{Phys. Rev. A} \textbf{\bibinfo{volume}{71}},
\bibinfo{pages}{012304}.

\bibitem[{\citenamefont{Nair}(2011)}]{Nair2011}
\bibinfo{author}{\bibnamefont{Nair, R.}},
\bibinfo{year}{2011},
\bibinfo{journal}{Phys. Rev. A} \textbf{\bibinfo{volume}{84}},
\bibinfo{pages}{032312}.

\bibitem[{\citenamefont{Namiki and Hirano}(2003)}]{Namiki2003}
\bibinfo{author}{\bibnamefont{Namiki R., and T.~Hirano}},
\bibinfo{year}{2003},
\bibinfo{journal}{Phys. Rev. A} \textbf{\bibinfo{volume}{67}},
\bibinfo{pages}{022308}.


\bibitem[{\citenamefont{Namiki and Hirano}(2006)}]{Namiki2006}
\bibinfo{author}{\bibnamefont{Namiki R., and T.~Hirano}},
\bibinfo{year}{2006},
\bibinfo{journal}{Phys. Rev. A} \textbf{\bibinfo{volume}{74}},
\bibinfo{pages}{032302}.

\bibitem[{\citenamefont{Navascu\'es and Ac\'in}(2005)}]{Navascues2005}
\bibinfo{author}{\bibnamefont{Navascu\'es, M., and A.~Ac\'in}},
\bibinfo{year}{2005},
\bibinfo{journal}{Phys. Rev. Lett.} \textbf{\bibinfo{volume}{94}},
\bibinfo{pages}{020505}.

\bibitem[{\citenamefont{Navascu\'es {\em et al.}}(2006)}]{Navascues2006}
\bibinfo{author}{\bibnamefont{Navascu\'es, M., F.~Grosshans, and A.~Ac\'in}},
\bibinfo{year}{2006},
\bibinfo{journal}{Phys. Rev. Lett.} \textbf{\bibinfo{volume}{97}},
\bibinfo{pages}{190502}.


\bibitem[{\citenamefont{Nha and Carmichael}(2005)}]{Nha2005}
\bibinfo{author}{\bibnamefont{Nha, H., and H.~J.~Carmichael}},
\bibinfo{year}{2005},
\bibinfo{journal}{Phys. Rev. A} \textbf{\bibinfo{volume}{71}},
\bibinfo{pages}{032336}.

\bibitem[{\citenamefont{Nielsen}(2004)}]{Nielsen2004}
\bibinfo{author}{\bibnamefont{Nielsen, M.~A.}},
\bibinfo{year}{2004},
\bibinfo{journal}{Phys. Rev. Lett.} \textbf{\bibinfo{volume}{93}},
\bibinfo{pages}{040503}.

\bibitem[{\citenamefont{Nielsen and Chuang}(2000)}]{Nielsen2000}
\bibinfo{author}{\bibnamefont{Nielsen, M.~A., and I.~L.~Chuang}},
  \bibinfo{year}{2000}, \emph{\bibinfo{booktitle}{Quantum
Computation and Quantum Information}} (\bibinfo{publisher}{Cambridge University Press, Cambridge}).

\bibitem[{\citenamefont{Nielsen and Dawson}(2005)}]{Nie05}
\bibinfo{author}{\bibnamefont{Nielsen, M.~A., and C.~M.~Dawson}},
\bibinfo{year}{2005},
\bibinfo{journal}{Phys. Rev. A} \textbf{\bibinfo{volume}{71}},
\bibinfo{pages}{042323}.


\bibitem[{\citenamefont{Niset {\em et al.}}(2007)}]{Niset2007}
\bibinfo{author}{\bibnamefont{Niset, J., A.~Ac\'{\i}n, U.~L.~Andersen, N.~J.~Cerf, R.~Garc\'{\i}a-Patr\'{o}n, M.~Navascues, and M.~Sabuncu}},
\bibinfo{year}{2007},
\bibinfo{journal}{Phys. Rev. Lett.} \textbf{\bibinfo{volume}{98}},
\bibinfo{pages}{260404}.

\bibitem[{\citenamefont{Niset {\em et al.}}(2008)}]{Nis98}
\bibinfo{author}{\bibnamefont{Niset, J., U.L. Andersen, and N.J. Cerf}},
\bibinfo{year}{2008},
\bibinfo{journal}{Phys. Rev. Lett.} \textbf{\bibinfo{volume}{101}},
\bibinfo{pages}{130503}.

\bibitem[{\citenamefont{Niset {\em et al.}}(2009)}]{Nis09}
\bibinfo{author}{\bibnamefont{Niset, J., J.~Fiur\`{a}\v{s}ek, and N~J.~Cerf}},
\bibinfo{year}{2009},
\bibinfo{journal}{Phys. Rev. Lett.} \textbf{\bibinfo{volume}{102}},
\bibinfo{pages}{120501}.

\bibitem[{\citenamefont{Nussbaum and Szkola}(2009)}]{Nussbaum2009}
\bibinfo{author}{\bibnamefont{Nussbaum, M., and A.~Szkola}},
\bibinfo{year}{2009},
\bibinfo{journal}{Ann. Stat.} \textbf{\bibinfo{volume}{37}},
\bibinfo{pages}{1040}.


\bibitem[{\citenamefont{Ohliger {\em et al.}}(2010)}]{Ohl10}
\bibinfo{author}{\bibnamefont{Ohliger, M., K.~Kieling, and J.~Eisert}},
\bibinfo{year}{2010},
\bibinfo{journal}{Phys. Rev. A} \textbf{\bibinfo{volume}{82}},
\bibinfo{pages}{042336}.


\bibitem[{\citenamefont{Olivares and Paris}(2004)}]{Oli04}
\bibinfo{author}{\bibnamefont{Olivares, S., and M.~G.~A.~Paris}},
\bibinfo{year}{2004},
\bibinfo{journal}{J.Opt.B: Quantum Semiclass. Opt.} \textbf{\bibinfo{volume}{6}},
\bibinfo{pages}{69}.

\bibitem[{\citenamefont{Olivares {\em et al.}}(2006)}]{Olivares2006}
\bibinfo{author}{\bibnamefont{Olivares, S., M.~G.~A.~Paris, and U.~L.~Andersen}},
\bibinfo{year}{2006},
\bibinfo{journal}{Phys. Rev. A} \textbf{\bibinfo{volume}{73}},
\bibinfo{pages}{062330}.

\bibitem[{\citenamefont{Osaki {\em et al.}}(1996)}]{Osa96}
\bibinfo{author}{\bibnamefont{Osaki M., M.~Ban, and O.~Hirota}},
\bibinfo{year}{1996},
\bibinfo{journal}{Phys. Rev. A} \textbf{\bibinfo{volume}{54}},
\bibinfo{pages}{1691}.

\bibitem[{\citenamefont{Ourjoumtsev {\em et al.}}(2009)}]{Our09}
\bibinfo{author}{\bibnamefont{Ourjoumtsev, A., F.~Ferreyrol, R.~Tualle-Brouri, and Ph.~Grangier}},
\bibinfo{year}{2009},
\bibinfo{journal}{Nature Physics} \textbf{\bibinfo{volume}{5}},
\bibinfo{pages}{189}.

\bibitem[{\citenamefont{Ourjoumtsev {\em et al.}}(2007)}]{Ourjoumtsev2007}
\bibinfo{author}{\bibnamefont{Ourjoumtsev, A., H.~Jeong, R.~Tualle-Brouri, and P.~Grangier}},
\bibinfo{year}{2007},
\bibinfo{journal}{Nature} \textbf{\bibinfo{volume}{448}},
\bibinfo{pages}{784}.

\bibitem[{\citenamefont{Parker {\em et al.}}(2000)}]{Parker2000}
\bibinfo{author}{\bibnamefont{Parker, S., S.~Bose, and M.~B.~Plenio}},
\bibinfo{year}{2000},
\bibinfo{journal}{Phys. Rev. A} \textbf{\bibinfo{volume}{61}},
\bibinfo{pages}{032305}.

\bibitem[{\citenamefont{Pati and Braunstein}(2003)}]{Pati2003}
\bibinfo{author}{\bibnamefont{Pati, A.~K., and S.~L.~Braunstein}},
\bibinfo{year}{2003}, in \emph{\bibinfo{booktitle}{Quantum information with continuous variables}}, edited by \bibinfo{editor}{\bibnamefont{S.~L.~Braunstein, and A.~K.~Pati}},
(\bibinfo{publisher}{Kluwer Academic, Dordrecht}),
\bibinfo{pages}{31-36}.

\bibitem[{\citenamefont{Pati {\em et al.}}(2000)}]{Pati2000}
\bibinfo{author}{\bibnamefont{Pati, A.~K., S.~L.~Braunstein, and S.~Lloyd}},
\bibinfo{year}{2000},
\bibinfo{booktitle}{``Quantum searching with continuous variables"},
\bibinfo{journal}{arXiv:quant-ph/0002082v2}.

\bibitem[{\citenamefont{Paulsen}(2002)}]{Paulsen2002}
\bibinfo{author}{\bibnamefont{Paulsen, V.}},
  \bibinfo{year}{2002}, \emph{\bibinfo{booktitle}{Completely bounded maps and operator algebras}} (\bibinfo{publisher}{Cambridge University Press, Cambridge}).


\bibitem[{\citenamefont{Pereira {\em et al.}}(2000)}]{Per00}
\bibinfo{author}{\bibnamefont{Pereira, S.~F, Z.~Y.~Ou, and H.~J.~Kimble}},
\bibinfo{year}{2000},
\bibinfo{journal}{Phys. Rev. A} \textbf{\bibinfo{volume}{62}},
\bibinfo{pages}{042311}.

\bibitem[{\citenamefont{Peres}(1996)}]{Peres1996}
\bibinfo{author}{\bibnamefont{Peres, A.}},
\bibinfo{year}{1996},
\bibinfo{journal}{Phys. Rev. Lett.} \textbf{\bibinfo{volume}{77}},
\bibinfo{pages}{1413}.

\bibitem[{\citenamefont{Pfister}(2007)}]{Pfi07}
\bibinfo{author}{\bibnamefont{Pfister, O.}},
\bibinfo{year}{2007},
\bibinfo{booktitle}{``Graph states and carrier-envelope phase squeezing"},
\bibinfo{journal}{arXiv:quant-ph/0701104}.

\bibitem[{\citenamefont{Pirandola}(2005)}]{Pirandola2005b}
\bibinfo{author}{\bibnamefont{Pirandola, S.}},
\bibinfo{year}{2005},
\bibinfo{journal}{Int. J. Quant. Inf.} \textbf{\bibinfo{volume}{3}},
\bibinfo{pages}{239}.

\bibitem[{\citenamefont{Pirandola}(2011)}]{Pirandola2010}
\bibinfo{author}{\bibnamefont{Pirandola, S.}},
\bibinfo{year}{2011},
\bibinfo{journal}{Phys. Rev. Lett.} \textbf{\bibinfo{volume}{106}},
\bibinfo{pages}{090504}.

\bibitem[{\citenamefont{Pirandola, Braunstein, and Lloyd}(2008)}]{Pirandola2008c}
\bibinfo{author}{\bibnamefont{Pirandola, S., S.~L.~Braunstein, and S.~Lloyd}},
\bibinfo{year}{2008},
\bibinfo{journal}{Phys. Rev. Lett.} \textbf{\bibinfo{volume}{101}},
\bibinfo{pages}{200504}.

\bibitem[{\citenamefont{Pirandola, Garc\'{\i}a-Patr\'{o}n, {\em et al.}}(2009)}]{Pirandola2009b}
\bibinfo{author}{\bibnamefont{Pirandola, S., R.~Garc\'{\i}a-Patr\'{o}n, S.~L.~Braunstein, and S.~Lloyd}},
\bibinfo{year}{2009},
\bibinfo{journal}{Phys. Rev. Lett.} \textbf{\bibinfo{volume}{102}},
\bibinfo{pages}{050503}.

\bibitem[{\citenamefont{Pirandola and Lloyd}(2008)}]{Pirandola2008b}
\bibinfo{author}{\bibnamefont{Pirandola, S., and S.~Lloyd}},
\bibinfo{year}{2008},
\bibinfo{journal}{Phys. Rev. A} \textbf{\bibinfo{volume}{78}},
\bibinfo{pages}{012331}.

\bibitem[{\citenamefont{Pirandola {\em et al.}}(2011)}]{Pirandola2011}
\bibinfo{author}{\bibnamefont{Pirandola, S., C.~Lupo, V.~Giovannetti,
S.~Mancini, and S.~L.~Braunstein}},
\bibinfo{year}{2011},
\bibinfo{booktitle}{``Quantum reading capacity"},
\bibinfo{journal}{arXiv:1107.3500}.

\bibitem[{\citenamefont{Pirandola and Mancini}(2006)}]{Pirandola2006b}
\bibinfo{author}{\bibnamefont{Pirandola, S., and S.~Mancini}},
\bibinfo{year}{2006},
\bibinfo{journal}{Laser Physics} \textbf{\bibinfo{volume}{16}},
\bibinfo{pages}{1418}.

\bibitem[{\citenamefont{Pirandola, Mancini, {\em et al.}}(2008)}]{Pirandola2008}
\bibinfo{author}{\bibnamefont{Pirandola, S., S.~Mancini, S.~Lloyd, and S.~L.~Braunstein}},
\bibinfo{year}{2008},
\bibinfo{journal}{Nature Physics} \textbf{\bibinfo{volume}{4}},
\bibinfo{pages}{726}.



\bibitem[{\citenamefont{Pirandola {\em et al.}}(2005)}]{Pirandola2005}
\bibinfo{author}{\bibnamefont{Pirandola, S., S.~Mancini, and D.~Vitali}},
\bibinfo{year}{2005},
\bibinfo{journal}{Phys. Rev. A} \textbf{\bibinfo{volume}{71}},
\bibinfo{pages}{042326}.

\bibitem[{\citenamefont{Pirandola {\em et al.}}(2003)}]{Pirandola2003}
\bibinfo{author}{\bibnamefont{Pirandola, S., S.~Mancini, D.~Vitali, and P.~Tombesi}},
\bibinfo{year}{2003},
\bibinfo{journal}{Phys. Rev. A} \textbf{\bibinfo{volume}{68}},
\bibinfo{pages}{062317}.

\bibitem[{\citenamefont{Pirandola {\em et al.}}(2004)}]{Pirandola2004}
\bibinfo{author}{\bibnamefont{Pirandola, S., S.~Mancini, D.~Vitali, and P.~Tombesi}},
\bibinfo{year}{2004},
\bibinfo{journal}{Europhys. Lett.} \textbf{\bibinfo{volume}{68}},
\bibinfo{pages}{323}.

\bibitem[{\citenamefont{Pirandola, Serafini, and Lloyd}(2009)}]{Pirandola2009}
\bibinfo{author}{\bibnamefont{Pirandola, S., A.~Serafini, and S.~Lloyd}},
\bibinfo{year}{2009},
\bibinfo{journal}{Phys. Rev. A} \textbf{\bibinfo{volume}{79}},
\bibinfo{pages}{052327}.

\bibitem[{\citenamefont{Pirandola {\em et al.}}(2006)}]{Pirandola2006}
\bibinfo{author}{\bibnamefont{Pirandola, S., D.~Vitali, P.~Tombesi, and S.~Lloyd}},
\bibinfo{year}{2006},
\bibinfo{journal}{Phys. Rev. Lett.} \textbf{\bibinfo{volume}{97}},
\bibinfo{pages}{150403}.


\bibitem[{\citenamefont{Plenio}(2005)}]{Plenio2005}
\bibinfo{author}{\bibnamefont{Plenio, M.~B.}},
\bibinfo{year}{2005},
\bibinfo{journal}{Phys. Rev. Lett.} \textbf{\bibinfo{volume}{95}},
\bibinfo{pages}{090503}.

\bibitem[{\citenamefont{Polkinghorne and Ralph}(1999)}]{Polkinghorne1999}
\bibinfo{author}{\bibnamefont{Polkinghorne, R.~E.~S, and T.~C.~Ralph}},
\bibinfo{year}{1999},
\bibinfo{journal}{Phys. Rev. Lett.} \textbf{\bibinfo{volume}{83}},
\bibinfo{pages}{2095}.

\bibitem[{\citenamefont{Pysher}(2011)}]{Pysher2011}
\bibinfo{author}{\bibnamefont{Pysher, M., Y.~Miwa, R.~Shahrokhshahi, R.~Bloomer, and O.~Pfister}},
\bibinfo{year}{2011},
\bibinfo{journal}{Phys. Rev. Lett.} \textbf{\bibinfo{volume}{107}},
\bibinfo{pages}{030505}.

\bibitem[{\citenamefont{Ralph}(1999a)}]{Ralph1999}
\bibinfo{author}{\bibnamefont{Ralph, T.~C.}},
\bibinfo{year}{1999a},
\bibinfo{journal}{Phys. Rev. A} \textbf{\bibinfo{volume}{61}},
\bibinfo{pages}{010303}.

\bibitem[{\citenamefont{Ralph}(1999b)}]{Ralph1999a}
\bibinfo{author}{\bibnamefont{Ralph, T.~C}},
\bibinfo{year}{1999b},
\bibinfo{journal}{Optics Letters} \textbf{\bibinfo{volume}{24}},
\bibinfo{pages}{348}.

\bibitem[{\citenamefont{Ralph}(2001)}]{Ralph2001}
\bibinfo{author}{\bibnamefont{Ralph, T.~C.}},
\bibinfo{year}{2001},
\bibinfo{journal}{Phys. Rev. A} \textbf{\bibinfo{volume}{65}},
\bibinfo{pages}{012319}.

\bibitem[{\citenamefont{Ralph}(2003)}]{Ralph2003}
\bibinfo{author}{\bibnamefont{Ralph, T.~C.}},
\bibinfo{year}{2003}, in \emph{\bibinfo{booktitle}{Quantum Information with Continuous Variables}}, edited by \bibinfo{editor}{\bibnamefont{S.~L.~Braunstein and A.~K.~Pati}},
(\bibinfo{publisher}{Kluwer Academic Publishers, The Netherlands}).

\bibitem[{\citenamefont{Ralph}(2011)}]{Ralph2011}
\bibinfo{author}{\bibnamefont{Ralph, T.~C}},
\bibinfo{year}{2011},
\bibinfo{journal}{Phys. Rev. A} \textbf{\bibinfo{volume}{84}},
\bibinfo{pages}{022339}.


\bibitem[{\citenamefont{Ralph {\em et al.}}(2003)}]{Ral03}
\bibinfo{author}{\bibnamefont{Ralph, T.~C., A.~Gilchrist, G.~J.~Milburn, W.~J.~Munro, and S.Glancy}},
\bibinfo{year}{2003},
\bibinfo{journal}{Phys. Rev. A} \textbf{\bibinfo{volume}{68}},
\bibinfo{pages}{042319}.

\bibitem[{\citenamefont{Ralph and Huntington}(2002)}]{Ral02}
\bibinfo{author}{\bibnamefont{Ralph, T.~C., and E.~H.~Huntington}},
\bibinfo{year}{2002},
\bibinfo{journal}{Phys. Rev. A} \textbf{\bibinfo{volume}{66}},
\bibinfo{pages}{042321}.


\bibitem[{\citenamefont{Ralph and Lam}(1998)}]{Ral98}
\bibinfo{author}{\bibnamefont{Ralph, T.~C., and P.~K.~Lam}},
\bibinfo{year}{1998},
\bibinfo{journal}{Phys.~Rev.~Lett.} \textbf{\bibinfo{volume}{81}},
\bibinfo{pages}{5668}.



\bibitem[{\citenamefont{Ralph and Lund}(2009)}]{Ral09}
\bibinfo{author}{\bibnamefont{Ralph, T.~C., and Lund, A.~P.}},
\bibinfo{year}{2009}, in \emph{\bibinfo{booktitle}{Proceedings of 9th International Conference on Quantum Communication Measurement and Computing}}, edited by \bibinfo{editor}{\bibnamefont{A.~Lvovsky}},
(\bibinfo{publisher}{AIP}),
\bibinfo{pages}{155160}.

\bibitem[{\citenamefont{Raussendorf and Briegel}(2001)}]{Rau01}
\bibinfo{author}{\bibnamefont{Raussendorf, R., and H.~J.~Briegel}},
\bibinfo{year}{2001},
\bibinfo{journal}{Phys. Rev. Lett.} \textbf{\bibinfo{volume}{86}},
\bibinfo{pages}{5188}.

\bibitem[{\citenamefont{Reid}(2000)}]{Reid2000}
\bibinfo{author}{\bibnamefont{Reid, M.~D.}},
\bibinfo{year}{2000},
\bibinfo{journal}{Phys. Rev. A} \textbf{\bibinfo{volume}{62}},
\bibinfo{pages}{062308}.


\bibitem[{\citenamefont{Renner}(2005)}]{Renner2005b}
\bibinfo{author}{\bibnamefont{Renner, R.}},
\bibinfo{year}{2005},
\bibinfo{journal}{Ph.D. thesis (ETH, Zurich)}.

\bibitem[{\citenamefont{Renner}(2007)}]{Renner20007}
\bibinfo{author}{\bibnamefont{Renner, R.}},
\bibinfo{year}{2007},
\bibinfo{journal}{Nature Physics} \textbf{\bibinfo{volume}{9}},
\bibinfo{pages}{645}.

\bibitem[{\citenamefont{Renner and Cirac}(2009)}]{Renner20009}
\bibinfo{author}{\bibnamefont{Renner, R., and J.~I.~Cirac}},
\bibinfo{year}{2009},
\bibinfo{journal}{Phys. Rev. Lett.} \textbf{\bibinfo{volume}{102}},
\bibinfo{pages}{110504}.



\bibitem[{\citenamefont{Renner {\em et al.}}(2005)}]{Renner2005}
\bibinfo{author}{\bibnamefont{Renner, R., N.~Gisin, and B.~Kraus}},
\bibinfo{year}{2005},
\bibinfo{journal}{Phys. Rev. A} \textbf{\bibinfo{volume}{72}},
\bibinfo{pages}{012332}.

\bibitem[{\citenamefont{Renner and Wolf}(2005)}]{Renner2005c}
\bibinfo{author}{\bibnamefont{Renner, R., and S.~Wolf}},
\bibinfo{year}{2005}, \emph{\bibinfo{booktitle}{Advances in
Cryptology: CRYPTO 2003}},
\bibinfo{journal}{Lecture Notes in Computer Science} \textbf{\bibinfo{volume}{2729}},
\bibinfo{pages}{78}.



\bibitem[{\citenamefont{Richardson {\em et al.}}(2001)}]{Richardson2001}
\bibinfo{author}{\bibnamefont{Richardson, T.~J., M.~A.~Shokrollahi, and R.~L.~Urbanke}},
\bibinfo{year}{2001},
\bibinfo{journal}{IEEE Trans. Inf. Theory} \textbf{\bibinfo{volume}{47}},
\bibinfo{pages}{619}.

\bibitem[{\citenamefont{Rivest {\em et al.}}(1978)}]{Rivest1978}
\bibinfo{author}{\bibnamefont{Rivest, R., A.~Shamir, and L.~Adleman}},
\bibinfo{year}{1978},
\bibinfo{journal}{Commun. ACM} \textbf{\bibinfo{volume}{21}},
\bibinfo{pages}{120}.


\bibitem[{\citenamefont{Sacchi}(2005)}]{Sacchi2005}
\bibinfo{author}{\bibnamefont{Sacchi, M.}},
\bibinfo{year}{2005},
\bibinfo{journal}{Phys. Rev. A} \textbf{\bibinfo{volume}{72}},
\bibinfo{pages}{014305}.

\bibitem[{\citenamefont{Sasaki and Hirota}(1996)}]{Sas96}
\bibinfo{author}{\bibnamefont{Sasaki, M., and O.~Hirota}},
\bibinfo{year}{1996},
\bibinfo{journal}{Phys. Rev. A} \textbf{\bibinfo{volume}{54}},
\bibinfo{pages}{2728}.

\bibitem[{\citenamefont{Sasaki and Hirota}(1997)}]{Sas97}
\bibinfo{author}{\bibnamefont{Sasaki, M., and O.~Hirota}},
\bibinfo{year}{1997},
\bibinfo{journal}{Phys. Lett. A} \textbf{\bibinfo{volume}{224}},
\bibinfo{pages}{213}.



\bibitem[{\citenamefont{Scarani {\em et al.}}(2009)}]{Scarani2009}
\bibinfo{author}{\bibnamefont{Scarani, V., H.~Bechmann-Pasquinucci, N.~J.~Cerf, M.~Du\v{s}ek, N.~L\"{u}tkenhaus, and M.~Peev}},
\bibinfo{year}{2009},
\bibinfo{journal}{Rev. Mod. Phys.} \textbf{\bibinfo{volume}{81}},
\bibinfo{pages}{1301}.

\bibitem[{\citenamefont{Scarani {\em et al.}}(2005)}]{GisinCloning05}
\bibinfo{author}{\bibnamefont{Scarani, V., S.~Iblisdir, N.~Gisin, and A.~Ac\'{\i}n}},
\bibinfo{year}{2005},
\bibinfo{journal}{Rev. Mod. Phys.} \textbf{\bibinfo{volume}{77}},
\bibinfo{pages}{1225}.

\bibitem[{\citenamefont{Scarani and Renner}(2008)}]{Scarani2008}
\bibinfo{author}{\bibnamefont{Scarani, V., and R.~Renner}},
\bibinfo{year}{2008},
\bibinfo{journal}{Phys. Rev. Lett.} \textbf{\bibinfo{volume}{100}},
\bibinfo{pages}{200501}.


\bibitem[{\citenamefont{Schneier}(1995)}]{Schneier1995}
\bibinfo{author}{\bibnamefont{Schneier, B.}},
  \bibinfo{year}{2000}, \emph{\bibinfo{booktitle}{Applied Cryptography}} (\bibinfo{publisher}{Wiley, New York}).

\bibitem[{\citenamefont{Schumacher and Nielsen}(1996)}]{Schumacher1996}
\bibinfo{author}{\bibnamefont{Schumacher B., and M.~A.~Nielsen}},
\bibinfo{year}{1996},
\bibinfo{journal}{Phys. Rev. A} \textbf{\bibinfo{volume}{54}},
\bibinfo{pages}{2629}.

\bibitem[{\citenamefont{Schumacher and Westmoreland}(1997)}]{Schumacher1997}
\bibinfo{author}{\bibnamefont{Schumacher, B., and M.~D.~Westmoreland}},
\bibinfo{year}{1997},
\bibinfo{journal}{Phys. Rev. A} \textbf{\bibinfo{volume}{56}},
\bibinfo{pages}{131}.

\bibitem[{\citenamefont{Scutaru}(1998)}]{Scutaru1998}
\bibinfo{author}{\bibnamefont{Scutaru, H.}},
\bibinfo{year}{1998},
\bibinfo{journal}{J. Phys. A} \textbf{\bibinfo{volume}{31}},
\bibinfo{pages}{3659}.

\bibitem[{\citenamefont{SECOQC}(2007)}]{SECOQC2007}
\bibinfo{author}{\bibnamefont{SECOQC}},
\bibinfo{year}{2007},
\bibinfo{journal}{http://www.secoqc.net/}

\bibitem[{\citenamefont{Serafini}(2006)}]{Serafini2006}
\bibinfo{author}{\bibnamefont{Serafini, A.}},
\bibinfo{year}{2006},
\bibinfo{journal}{Phys. Rev. Lett.} \textbf{\bibinfo{volume}{96}},
\bibinfo{pages}{110402}.

\bibitem[{\citenamefont{Serafini and Adesso}(2007)}]{Serafini2007}
\bibinfo{author}{\bibnamefont{Serafini, A., and G.~Adesso}},
\bibinfo{year}{2007},
\bibinfo{journal}{J. Phys. A: Math. Theor.} \textbf{\bibinfo{volume}{40}},
\bibinfo{pages}{8041}.

\bibitem[{\citenamefont{Serafini, Adesso, and Illuminati}(2005)}]{Serafini2005}
\bibinfo{author}{\bibnamefont{Serafini A., G.~Adesso, and F.~Illuminati}},
\bibinfo{year}{2005},
\bibinfo{journal}{Phys. Rev. A} \textbf{\bibinfo{volume}{71}},
\bibinfo{pages}{032349}.

\bibitem[{\citenamefont{Serafini, Eisert, and Wolf}(2005)}]{Serafini2005b}
\bibinfo{author}{\bibnamefont{Serafini, A., J.~Eisert, and M.~M.~Wolf}},
\bibinfo{year}{2005},
\bibinfo{journal}{Phys. Rev. A} \textbf{\bibinfo{volume}{71}},
\bibinfo{pages}{012320}.

\bibitem[{\citenamefont{Serafini {\em et al.}}(2004)}]{Serafini2004}
\bibinfo{author}{\bibnamefont{Serafini, A., F.~Illuminati, and S.~De Siena}},
\bibinfo{year}{2004},
\bibinfo{journal}{J. Phys. B: At. Mol. Opt. Phys.} \textbf{\bibinfo{volume}{37}},
\bibinfo{pages}{L21}.


\bibitem[{\citenamefont{Shannon}(1948)}]{Shannon1948}
\bibinfo{author}{\bibnamefont{Shannon, C.~E.}},
\bibinfo{year}{1948},
\bibinfo{journal}{Bell Syst. Tech. J.} \textbf{\bibinfo{volume}{27}},
\bibinfo{pages}{379}.

\bibitem[{\citenamefont{Shannon}(1949)}]{Shannon1949}
\bibinfo{author}{\bibnamefont{Shannon, C.~E.}},
\bibinfo{year}{1949},
\bibinfo{journal}{Bell Syst. Tech. J.} \textbf{\bibinfo{volume}{28}},
\bibinfo{pages}{656}.

\bibitem[{\citenamefont{Shapiro}(1984)}]{Shapiro1984}
\bibinfo{author}{\bibnamefont{Shapiro, J.~H.}},
\bibinfo{year}{1984},
\bibinfo{journal}{IEEE J. Quantum Electron.} \textbf{\bibinfo{volume}{20}},
\bibinfo{pages}{803}.
%


\bibitem[{\citenamefont{Shapiro {\em et al.}}(2005)}]{Shapiro2005}
\bibinfo{author}{\bibnamefont{Shapiro, J. H., S.~Guha, B.~I.~Erkmen}},
\bibinfo{year}{2005},
\bibinfo{journal}{Journal of Optical Networking: Special Issue} \textbf{\bibinfo{volume}{4}},
\bibinfo{pages}{501}.

\bibitem[{\citenamefont{Shapiro and Lloyd}(2009)}]{Shapiro2009}
\bibinfo{author}{\bibnamefont{Shapiro, J. H., and Seth Lloyd}},
\bibinfo{year}{2009},
\bibinfo{journal}{New J. Phys.} \textbf{\bibinfo{volume}{11}},
\bibinfo{pages}{063045}.

\bibitem[{\citenamefont{Sherson {\em et al.}}(2006)}]{Sherson2006}
\bibinfo{author}{\bibnamefont{Sherson, J., H.~Krauter, R.~K.~Olsson, B.~Julsgaard, K.~Hammerer, I.~Cirac, and E.~S.~Polzik}},
\bibinfo{year}{2006},
\bibinfo{journal}{Nature} \textbf{\bibinfo{volume}{443}},
\bibinfo{pages}{557}.

\bibitem[{\citenamefont{Shor}(1995)}]{Sho95}
\bibinfo{author}{\bibnamefont{Shor, P.}},
\bibinfo{year}{1995},
\bibinfo{journal}{Phys. Rev. A} \textbf{\bibinfo{volume}{52}},
\bibinfo{pages}{2493}.

\bibitem[{\citenamefont{Shor}(1997)}]{Shor1997}
\bibinfo{author}{\bibnamefont{Shor, P.}},
\bibinfo{year}{1997},
\bibinfo{journal}{SIAM J. Comput.} \textbf{\bibinfo{volume}{26}},
\bibinfo{pages}{1484}.

\bibitem[{\citenamefont{Shor}(2002)}]{Shor2002}
\bibinfo{author}{\bibnamefont{Shor, P.}},
\bibinfo{year}{2002},
\emph{\bibinfo{booktitle}{The quantum channel capacity and coherent information}},
\bibinfo{journal}{Lecture Notes, MSRI Workshop on Quantum Computation}.

\bibitem[{\citenamefont{Shor}(2004)}]{Shor2004}
\bibinfo{author}{\bibnamefont{Shor, P.}},
\bibinfo{year}{2004},
\bibinfo{journal}{Comm. Math. Phys.} \textbf{\bibinfo{volume}{246}},
\bibinfo{pages}{453}.

\bibitem[{\citenamefont{Silberhorn {\em et al.}}(2002)}]{Silberhorn2002}
\bibinfo{author}{\bibnamefont{Silberhorn, Ch., T.~C.~Ralph, N.~L\"{u}tkenhaus, and G.~Leuchs}},
\bibinfo{year}{2002},
\bibinfo{journal}{Phys. Rev. Lett.} \textbf{\bibinfo{volume}{89}},
\bibinfo{pages}{167901}.

\bibitem[{\citenamefont{Simon}(2000)}]{Simon2000}
\bibinfo{author}{\bibnamefont{Simon, R.}},
\bibinfo{year}{2000},
\bibinfo{journal}{Phys. Rev. Lett.} \textbf{\bibinfo{volume}{84}},
\bibinfo{pages}{2726}.

\bibitem[{\citenamefont{Simon {\em et al.}}(1999)}]{Simon1999}
\bibinfo{author}{\bibnamefont{Simon, R., S.~Chaturvedi, and V.~Srinivasan}},
\bibinfo{year}{1999},
\bibinfo{journal}{J.~Math.~Phys.} \textbf{\bibinfo{volume}{40}},
\bibinfo{pages}{3632}.

\bibitem[{\citenamefont{Simon {\em et al.}}(1994)}]{Simon1994}
\bibinfo{author}{\bibnamefont{Simon, R., N.~Mukunda, and B.~Dutta}},
\bibinfo{year}{1994},
\bibinfo{journal}{Phys. Rev. A} \textbf{\bibinfo{volume}{49}},
\bibinfo{pages}{1567}.



\bibitem[{\citenamefont{Smith and Smolin}(2007)}]{Smith2007}
\bibinfo{author}{\bibnamefont{Smith, G., and J.~A.~Smolin}},
\bibinfo{year}{2007},
\bibinfo{journal}{Phys. Rev. Lett.} \textbf{\bibinfo{volume}{98}},
\bibinfo{pages}{030501}.

\bibitem[{\citenamefont{Smith {\em et al.}}(2011)}]{Smith2011}
\bibinfo{author}{\bibnamefont{Smith, G., J.~A.~Smolin, and J.~Yard}},
\bibinfo{year}{2011},
\bibinfo{journal}{Nature Photonics} \textbf{\bibinfo{volume}{5}},
\bibinfo{pages}{624}.

\bibitem[{\citenamefont{Smith and Yard}(2008)}]{Smith2008}
\bibinfo{author}{\bibnamefont{Smith, G., and J.~Yard}},
\bibinfo{year}{2008},
\bibinfo{journal}{Science} \textbf{\bibinfo{volume}{321}},
\bibinfo{pages}{1812}.

\bibitem[{\citenamefont{Soto {\em et al.}}(1983)}]{Soto1983}
\bibinfo{author}{\bibnamefont{Soto, F., and P.~Claverie}},
\bibinfo{year}{1983},
\bibinfo{journal}{J.~Math.~Phys.} \textbf{\bibinfo{volume}{24}},
\bibinfo{pages}{97}.

\bibitem[{\citenamefont{Steane}(1996)}]{Ste96}
\bibinfo{author}{\bibnamefont{Steane, A.~M.}},
\bibinfo{year}{1996},
\bibinfo{journal}{Phys. Rev. Lett.} \textbf{\bibinfo{volume}{77}},
\bibinfo{pages}{793}.

\bibitem[{\citenamefont{Stinespring}(1955)}]{Stinespring1955}
\bibinfo{author}{\bibnamefont{Stinespring, W.~F.}},
\bibinfo{year}{1955},
\bibinfo{journal}{Proc. Am. Math. Soc.} \textbf{\bibinfo{volume}{6}},
\bibinfo{pages}{211}.

\bibitem[{\citenamefont{Su {\em et al.}}(2007)}]{Su07}
\bibinfo{author}{\bibnamefont{Su, X., A.~Tan, X.~Jia, J.~Zhang, C.~Xie, and K.~Peng}},
\bibinfo{year}{2007},
\bibinfo{journal}{Phys. Rev. Lett.} \textbf{\bibinfo{volume}{98}},
\bibinfo{pages}{070502}.

\bibitem[{\citenamefont{Sabuncu {\em et al.}}(2007)}]{Sabuncu2007}
\bibinfo{author}{\bibnamefont{Sabuncu, M., U.~L.~Andersen, and G.~Leuchs}},
\bibinfo{year}{2007},
\bibinfo{journal}{Phys. Rev. Lett.} \textbf{\bibinfo{volume}{98}},
\bibinfo{pages}{170503}.

\bibitem[{\citenamefont{Sych and Leuchs}(2009)}]{Sych2009}
\bibinfo{author}{\bibnamefont{Sych, D., and G.~Leuchs}},
\bibinfo{year}{2009},
\bibinfo{journal}{New J. Phys.} \textbf{\bibinfo{volume}{12}},
\bibinfo{pages}{053019}.

\bibitem[{\citenamefont{Symul {\em et al.}}(2007)}]{Symul2007}
\bibinfo{author}{\bibnamefont{Symul, T, D.~J.~Alton, S.~M.~Assad, A.~M.~Lance, C.~Weedbrook, T.~C.~Ralph, and P.~K.~Lam}},
\bibinfo{year}{2007},
\bibinfo{journal}{Phys. Rev. A} \textbf{\bibinfo{volume}{76}},
\bibinfo{pages}{030303 (R)}.

\bibitem[{\citenamefont{Takahashi {\em et al.}}(2010)}]{Takahashi2010}
\bibinfo{author}{\bibnamefont{Takahashi, H., J.~S.~Neergaard-Nielsen, M.~Takeuchi,
M.~Takeoka, K.~Hayasaka, A.~Furusawa, and M.~Sasaki}},
\bibinfo{year}{2010},
\bibinfo{journal}{Nature Photonics} \textbf{\bibinfo{volume}{4}},
\bibinfo{pages}{178}.

\bibitem[{\citenamefont{Takei {\em et al.}}(2005)}]{Takei2005}
\bibinfo{author}{\bibnamefont{Takei, N., H.~Yonezawa, T.~Aoki, and A.~Furusawa}},
\bibinfo{year}{2005},
\bibinfo{journal}{Phys. Rev. Lett.} \textbf{\bibinfo{volume}{94}},
\bibinfo{pages}{220502}.

\bibitem[{\citenamefont{Takeoka {\em et al.}}(2005)}]{Tak05}
\bibinfo{author}{\bibnamefont{Takeoka, M., M.~Sasaki, P.~van Loock, and N.~L\"{u}tkenhaus}},
\bibinfo{year}{2005},
\bibinfo{journal}{Phys. Rev. A} \textbf{\bibinfo{volume}{71}},
\bibinfo{pages}{022318}.

\bibitem[{\citenamefont{Takeoka and Sasaki}(2008)}]{Tak08}
\bibinfo{author}{\bibnamefont{Takeoka, M., and M.~Sasaki}},
\bibinfo{year}{2008},
\bibinfo{journal}{Phys. Rev. A} \textbf{\bibinfo{volume}{78}},
\bibinfo{pages}{022320}.

\bibitem[{\citenamefont{Tan {\em et al.}}(2008)}]{Tan2008}
\bibinfo{author}{\bibnamefont{Tan, S.-H., B.~I.~Erkmen, V.~Giovannetti, S.~Guha, S.~Lloyd, L.~Maccone, S.~Pirandola, and J.~H.~Shapiro}},
\bibinfo{year}{2008},
\bibinfo{journal}{Phys. Rev. Lett.} \textbf{\bibinfo{volume}{101}},
\bibinfo{pages}{253601}.

\bibitem[{\citenamefont{Tsujino {\em et al.}}(2011)}]{Tsujino2011}
\bibinfo{author}{\bibnamefont{Tsujino, K., D.~Fukuda, G.~Fujii, S.~Inoue, M.~Fujiwara, M.~Takeoka, and M.~Sasaki}},
\bibinfo{year}{2011},
\bibinfo{journal}{Phys. Rev. Lett.} \textbf{\bibinfo{volume}{106}},
\bibinfo{pages}{250503}.

\bibitem[{\citenamefont{Tyc and Sanders}(2002)}]{Tyc2002}
\bibinfo{author}{\bibnamefont{Tyc, T., and B.~C.~Sanders}},
\bibinfo{year}{2002},
\bibinfo{journal}{Phys. Rev. A} \textbf{\bibinfo{volume}{65}},
\bibinfo{pages}{042310}.

\bibitem[{\citenamefont{Ukai {\em et al.}}(2011)}]{Ukai2011}
\bibinfo{author}{\bibnamefont{Ukai, R., S.~Yokoyama, J.~Yoshikawa, P.~van Loock, and A.~Furusawa}},
\bibinfo{year}{2011},
\bibinfo{booktitle}{``Demonstration of a controlled-phase gate
for continuous-variable one-way quantum computation"},
\bibinfo{journal}{arXiv:1107.0514}.

\bibitem[{\citenamefont{Ukai, Yoshikawa, {\em et al.}}(2010)}]{Uka10a}
\bibinfo{author}{\bibnamefont{Ukai, R., J.~Yoshikawa, N.~Iwata, P.~van Loock, and A.~Furusawa}},
\bibinfo{year}{2010},
\bibinfo{journal}{Phys. Rev. A} \textbf{\bibinfo{volume}{81}},
\bibinfo{pages}{032315}.

\bibitem[{\citenamefont{Ukai, Iwata, {\em et al.}}(2011)}]{Uka10}
\bibinfo{author}{\bibnamefont{Ukai, R., N.~Iwata, Y.~Shimokawa, S,~C.~Armstrong, A.~Politi, J.~Yoshikawa, P.~van Loock, and A.~Furusawa}},
\bibinfo{year}{2011},
\bibinfo{journal}{Phys. Rev. Lett.} \textbf{\bibinfo{volume}{106}},
\bibinfo{pages}{240504}.

\bibitem[{\citenamefont{Uhlmann}(1976)}]{Uhlmann1976}
\bibinfo{author}{\bibnamefont{Uhlmann, A.}},
\bibinfo{year}{1976},
\bibinfo{journal}{Rep. Math. Phys.} \textbf{\bibinfo{volume}{9}},
\bibinfo{pages}{273}.

\bibitem[{\citenamefont{Usenko and Filip}(2010)}]{Usenko2010}
\bibinfo{author}{\bibnamefont{Usenko, V.~C., and R.~Filip}},
\bibinfo{year}{2010},
\bibinfo{journal}{Phys. Rev. A} \textbf{\bibinfo{volume}{81}},
\bibinfo{pages}{022318}.

\bibitem[{\citenamefont{Usha Devi and Rajagopal}(2009)}]{Usha2009}
\bibinfo{author}{\bibnamefont{Usha Devi, A.~R., and A.~K.~Rajagopal}},
\bibinfo{year}{2009},
\bibinfo{journal}{Phys. Rev. A} \textbf{\bibinfo{volume}{79}},
\bibinfo{pages}{062320}.

\bibitem[{\citenamefont{Vahlbruch {\em et al.}}(2008)}]{Vah08}
\bibinfo{author}{\bibnamefont{Vahlbruch, H., M.~Mehmet, S.~Chelkowski, B.~Hage, A.~Franzen, N.~Lastzka, S.~Goszler, K.~Danzmann, and R.~Schnabel}},
\bibinfo{year}{2008},
\bibinfo{journal}{Phys. Rev. Lett.} \textbf{\bibinfo{volume}{100}},
\bibinfo{pages}{033602}.

\bibitem[{\citenamefont{Vaidman}(1994)}]{Vaidman1994}
\bibinfo{author}{\bibnamefont{Vaidman, L.}},
\bibinfo{year}{1994},
\bibinfo{journal}{Phys. Rev. A} \textbf{\bibinfo{volume}{49}},
\bibinfo{pages}{1473}.

\bibitem[{\citenamefont{van Assche}(2006)}]{Assche2006}
\bibinfo{author}{\bibnamefont{van Assche, G.}},
  \bibinfo{year}{2006}, \emph{\bibinfo{booktitle}{Quantum Cryptography and
Secret-Key Distillation}} (\bibinfo{publisher}{Cambridge University Press, Cambridge}).

\bibitem[{\citenamefont{van Loock}(2007)}]{van07a}
\bibinfo{author}{\bibnamefont{van Loock, P.}},
\bibinfo{year}{2007},
\bibinfo{journal}{J.~Opt.~Soc.~Am.~B} \textbf{\bibinfo{volume}{24}},
\bibinfo{pages}{2}.

\bibitem[{\citenamefont{van Loock}(2008)}]{vanLoock2008}
\bibinfo{author}{\bibnamefont{van Loock, P.}},
\bibinfo{year}{2008},
\bibinfo{booktitle}{``A note on quantum error correction with continuous variables"},
\bibinfo{journal}{arXiv:0811.3616}.


\bibitem[{\citenamefont{van Loock and Braunstein}(1999)}]{Loock1999}
\bibinfo{author}{\bibnamefont{van Loock, P., and S.~L.~Braunstein}},
\bibinfo{year}{1999},
\bibinfo{journal}{Phys. Rev. A} \textbf{\bibinfo{volume}{61}},
\bibinfo{pages}{010302(R)}.


\bibitem[{\citenamefont{van Loock and Braunstein}(2000)}]{Loock2000b}
\bibinfo{author}{\bibnamefont{van Loock, and S.~L.~Braunstein}},
\bibinfo{year}{2000},
\bibinfo{journal}{Phys. Rev. Lett.} \textbf{\bibinfo{volume}{84}},
\bibinfo{pages}{3482}.

\bibitem[{\citenamefont{van Loock and Braunstein}(2001)}]{Loock2001}
\bibinfo{author}{\bibnamefont{van Loock, and S.~L.~Braunstein}},
\bibinfo{year}{2001},
\bibinfo{journal}{Phys. Rev. Lett.} \textbf{\bibinfo{volume}{87}},
\bibinfo{pages}{247901}.

\bibitem[{\citenamefont{van Loock {\em et al.}}(2007)}]{van07}
\bibinfo{author}{\bibnamefont{van Loock, P., C.~Weedbrook, and M.~Gu}},
\bibinfo{year}{2007},
\bibinfo{journal}{Phys. Rev. A} \textbf{\bibinfo{volume}{76}},
\bibinfo{pages}{032321}.

\bibitem[{\citenamefont{Vedral {\em et al.}}(1997)}]{Vedral1997}
\bibinfo{author}{\bibnamefont{Vedral, V., M.~B.~Plenio, M.~A.~Rippin, and P.~L.~Knight}},
\bibinfo{year}{1997},
\bibinfo{journal}{Phys. Rev. Lett.} \textbf{\bibinfo{volume}{78}},
\bibinfo{pages}{2275}.

\bibitem[{\citenamefont{Vidal}(2000)}]{Vidal2000}
\bibinfo{author}{\bibnamefont{Vidal, G.}},
\bibinfo{year}{2000},
\bibinfo{journal}{J. Mod. Opt.} \textbf{\bibinfo{volume}{47}},
\bibinfo{pages}{355}.

\bibitem[{\citenamefont{Vernam}(1926)}]{Ver26}
\bibinfo{author}{\bibnamefont{Vernam, G.~S.}},
\bibinfo{year}{1926},
\bibinfo{journal}{J.~Am.~Inst.~Electr.~Eng.} \textbf{\bibinfo{volume}{45}},
\bibinfo{pages}{109}.

\bibitem[{\citenamefont{Vidal and Werner}(2002)}]{Vidal2002}
\bibinfo{author}{\bibnamefont{Vidal, G., and R.~F.~Werner}},
\bibinfo{year}{2002},
\bibinfo{journal}{Phys. Rev. A} \textbf{\bibinfo{volume}{65}},
\bibinfo{pages}{032314}.


\bibitem[{\citenamefont{Walk {\em et al.}}(2011)}]{Walk2011}
\bibinfo{author}{\bibnamefont{Walk, N., T.~Symul, T.~C.~Ralph, P.~K.~Lam}},
\bibinfo{year}{2011},
\bibinfo{booktitle}{``Security of post-selection based continuous-variable quantum key distribution against arbitrary attacks"},
\bibinfo{journal}{arXiv:1106.0825}.

\bibitem[{\citenamefont{Walker and Braunstein}(2010)}]{Walker2010}
\bibinfo{author}{\bibnamefont{Walker, T.~A., and S.~L.~Braunstein}},
\bibinfo{year}{2010},
\bibinfo{journal}{Phys. Rev. A} \textbf{\bibinfo{volume}{81}},
\bibinfo{pages}{062305}.

\bibitem[{\citenamefont{Walls and Milburn}(2008)}]{Walls1995}
\bibinfo{author}{\bibnamefont{Walls, D.~F., and G.~Milburn}},
  \bibinfo{year}{2008}, \emph{\bibinfo{booktitle}{Quantum
Optics}} (\bibinfo{publisher}{Springer, Berlin}).

\bibitem[{\citenamefont{Wang {\em et al.}}(2007)}]{Wang2007}
\bibinfo{author}{\bibnamefont{Wang, X.~-B., T.~Hiroshima, A.Tomita, and M.~Hayashi}},
\bibinfo{year}{2007},
\bibinfo{journal}{Physics Reports} \textbf{\bibinfo{volume}{448}},
\bibinfo{pages}{1-111}.

\bibitem[{\citenamefont{Wang and Ying}(2006)}]{Wang2006}
\bibinfo{author}{\bibnamefont{Wang, G., and M.~Ying}},
\bibinfo{year}{2006},
\bibinfo{journal}{Phys. Rev. A} \textbf{\bibinfo{volume}{73}},
\bibinfo{pages}{042301}.

\bibitem[{\citenamefont{Weedbrook}(2009)}]{Wee09}
\bibinfo{author}{\bibnamefont{Weedbrook, C.}},
\bibinfo{year}{2009},
\bibinfo{journal}{Ph.D. thesis (University of Queensland, Brisbane)}.


\bibitem[{\citenamefont{Weedbrook {\em et al.}}(2008)}]{Weedbrook2008}
\bibinfo{author}{\bibnamefont{Weedbrook, C., N.~B.~Grosse, T.~Symul, P.~K.~Lam, and T.~C.~Ralph}},
\bibinfo{year}{2008},
\bibinfo{journal}{Phys. Rev. A} \textbf{\bibinfo{volume}{77}},
\bibinfo{pages}{052313}.

\bibitem[{\citenamefont{Weedbrook {\em et al.}}(2004)}]{Weedbrook2004}
\bibinfo{author}{\bibnamefont{Weedbrook, C., A.~M.~Lance, W.~P.~Bowen, T.~Symul, T.~C.~Ralph, and P.~K.~Lam}},
\bibinfo{year}{2004},
\bibinfo{journal}{Phys. Rev. Lett.} \textbf{\bibinfo{volume}{93}},
\bibinfo{pages}{170504}.

\bibitem[{\citenamefont{Weedbrook {\em et al.}}(2006)}]{Weedbrook2006}
\bibinfo{author}{\bibnamefont{Weedbrook, C., A.~M.~Lance, W.~P.~Bowen, T.~Symul, T.~C.~Ralph, and P.~K.~Lam}},
\bibinfo{year}{2006},
\bibinfo{journal}{Phys. Rev. A} \textbf{\bibinfo{volume}{73}},
\bibinfo{pages}{022316}.

\bibitem[{\citenamefont{Weedbrook {\em et al.}}(2010)}]{Weedbrook2010}
\bibinfo{author}{\bibnamefont{Weedbrook,C., S.~Pirandola, S.~Lloyd, and T.~C.~Ralph}},
\bibinfo{year}{2010},
\bibinfo{journal}{Phys. Rev. Lett.} \textbf{\bibinfo{volume}{105}},
\bibinfo{pages}{110501}.

\bibitem[{\citenamefont{Werner and Wolf}(2001)}]{Werner2001}
\bibinfo{author}{\bibnamefont{Werner, R.~F., and M.~M.~Wolf}},
\bibinfo{year}{2001},
\bibinfo{journal}{Phys. Rev. Lett.} \textbf{\bibinfo{volume}{86}},
\bibinfo{pages}{3658}.





\bibitem[{\citenamefont{Williamson}(1936)}]{Williamson1936}
\bibinfo{author}{\bibnamefont{Williamson, J.}},
\bibinfo{year}{1936},
\bibinfo{journal}{Am. J. Math.} \textbf{\bibinfo{volume}{58}},
\bibinfo{pages}{141}.


\bibitem[{\citenamefont{Wittmann {\em et al.}}(2010a)}]{Wit10}
\bibinfo{author}{\bibnamefont{Wittmann, C., U.~L.~Andersen, M.~Takeoka, D.~Sych, and
G.~Leuchs}},
\bibinfo{year}{2010a},
\bibinfo{journal}{Phys. Rev. Lett.} \textbf{\bibinfo{volume}{104}},
\bibinfo{pages}{100505}.




\bibitem[{\citenamefont{Wittmann {\em et al.}}(2010b)}]{Wit10b}
\bibinfo{author}{\bibnamefont{Wittmann, C., U.~L.~Andersen, M.~Takeoka, D.~Sych, and
G.~Leuchs}},
\bibinfo{year}{2010b},
\bibinfo{journal}{Phys. Rev. A} \textbf{\bibinfo{volume}{81}},
\bibinfo{pages}{062338}.

\bibitem[{\citenamefont{Wittmann {\em et al.}}(2008)}]{Wit08}
\bibinfo{author}{\bibnamefont{Wittmann, C., M.~Takeoka, K.~N.~Cassemiro, M.~Sasaki, G.~Leuchs, and U.~L.~Andersen}},
\bibinfo{year}{2008},
\bibinfo{journal}{Phys. Rev. Lett.} \textbf{\bibinfo{volume}{101}},
\bibinfo{pages}{210501}.

\bibitem[{\citenamefont{Wolf {\em et al.}}(2003)}]{Wolf2003}
\bibinfo{author}{\bibnamefont{Wolf, M.~M., J.~Eisert, and M.~B.~Plenio}},
\bibinfo{year}{2003},
\bibinfo{journal}{Phys. Rev. Lett.} \textbf{\bibinfo{volume}{90}},
\bibinfo{pages}{047904}.

\bibitem[{\citenamefont{Wolf {\em et al.}}(2006)}]{Wolf2006}
\bibinfo{author}{\bibnamefont{Wolf, M.~M., G.~Giedke, and J.~I.~Cirac}},
\bibinfo{year}{2006},
\bibinfo{journal}{Phys. Rev. Lett.} \textbf{\bibinfo{volume}{96}},
\bibinfo{pages}{080502}.



\bibitem[{\citenamefont{Wolf {\em et al.}}(2004)}]{Wolf2004}
\bibinfo{author}{\bibnamefont{Wolf, M.~M., G.~Giedke, O.~Kr\"uger, R.~F.~Werner, and J.~I.~Cirac}},
\bibinfo{year}{2004},
\bibinfo{journal}{Phys. Rev. A} \textbf{\bibinfo{volume}{69}},
\bibinfo{pages}{052320}.

\bibitem[{\citenamefont{Wolf {\em et al.}}(2007)}]{Wolf2007b}
\bibinfo{author}{\bibnamefont{Wolf, M.~M., D.~P\'{e}rez-Garc\'{\i}a, and G.~Giedke}},
\bibinfo{year}{2007},
\bibinfo{journal}{Phys. Rev. Lett.} \textbf{\bibinfo{volume}{98}},
\bibinfo{pages}{130501}.


\bibitem[{\citenamefont{Wootters and Zurek}(1982)}]{Woo82}
\bibinfo{author}{\bibnamefont{Wootters, W.~K., and W.~.H.~Zurek}},
\bibinfo{year}{1982},
\bibinfo{journal}{Nature} \textbf{\bibinfo{volume}{299}},
\bibinfo{pages}{802}.


\bibitem[{\citenamefont{Xiang {\em et al.}}(2010)}]{Xia10}
\bibinfo{author}{\bibnamefont{Xiang, G.~Y., T.~C.~Ralph, A.~P.~Lund, N.~Walk, and G.~J.~Pryde}},
\bibinfo{year}{2010},
\bibinfo{journal}{Nature Photonics} \textbf{\bibinfo{volume}{4}},
\bibinfo{pages}{316}.

\bibitem[{\citenamefont{Yamamoto and Haus}(1986)}]{Yam86}
\bibinfo{author}{\bibnamefont{Yamamoto, Y., and H.~A.~Haus}},
\bibinfo{year}{1986},
\bibinfo{journal}{Rev.~Mod.~Phys.} \textbf{\bibinfo{volume}{58}},
\bibinfo{pages}{1001}.



\bibitem[{\citenamefont{Yeoman and Barnett}(1993)}]{Yeo93}
\bibinfo{author}{\bibnamefont{Yeoman, G., and S.~M.~Barnett}},
\bibinfo{year}{1993},
\bibinfo{journal}{J.~Mod.~Opt} \textbf{\bibinfo{volume}{40}},
\bibinfo{pages}{1497}.

\bibitem[{\citenamefont{Yonezawa {\em et al.}}(2004)}]{Yonezawa2004}
\bibinfo{author}{\bibnamefont{Yonezawa, H., T.~Aoki, and A.~Furusawa}},
\bibinfo{year}{2004},
\bibinfo{journal}{Nature} \textbf{\bibinfo{volume}{431}},
\bibinfo{pages}{430}.




\bibitem[{\citenamefont{Yoshikawa {\em et al.}}(2007)}]{Yos07}
\bibinfo{author}{\bibnamefont{Yoshikawa, J., T.~Hayashi, T.~Akiyama, N.~Takei, A.~Huck, U.~L.~Andersen, and A.~Furusawa}},
\bibinfo{year}{2007},
\bibinfo{journal}{Phys. Rev. A} \textbf{\bibinfo{volume}{76}},
\bibinfo{pages}{060301}.

\bibitem[{\citenamefont{Yoshikawa {\em et al.}}(2008)}]{Yos08}
\bibinfo{author}{\bibnamefont{Yoshikawa, J., Y.~Miwa, A.~Huck, U.~L.~Andersen, P.~van Loock, and A.~Furusawa}},
\bibinfo{year}{2008},
\bibinfo{journal}{Phys. Rev. Lett.} \textbf{\bibinfo{volume}{101}},
\bibinfo{pages}{250501}.

\bibitem[{\citenamefont{Yuen}(1976)}]{Yuen1976}
\bibinfo{author}{\bibnamefont{Yuen, H.~P.}},
\bibinfo{year}{1976},
\bibinfo{journal}{Phys. Rev. A} \textbf{\bibinfo{volume}{13}},
\bibinfo{pages}{2226}.

\bibitem[{\citenamefont{Yuen and Nair}(2009)}]{Yuen2009}
\bibinfo{author}{\bibnamefont{Yuen, H.~P., and R.~Nair}},
\bibinfo{year}{2009},
\bibinfo{journal}{Phys. Rev. A} \textbf{\bibinfo{volume}{80}},
\bibinfo{pages}{023816}.

\bibitem[{\citenamefont{Yuen and Ozawa}(1993)}]{Yuen1993}
\bibinfo{author}{\bibnamefont{Yuen, H.~P., and M.~Ozawa}},
\bibinfo{year}{1993},
\bibinfo{journal}{Phys. Rev. Lett.} \textbf{\bibinfo{volume}{70}},
\bibinfo{pages}{363}.

\bibitem[{\citenamefont{Yuen and Shapiro}(1978)}]{Yuen1978}
\bibinfo{author}{\bibnamefont{Yuen, H.~P., and J.~H.~Shapiro}},
\bibinfo{year}{1978},
\bibinfo{journal}{IEEE Trans. Inf. Theory} \textbf{\bibinfo{volume}{24}},
\bibinfo{pages}{657}.

\bibitem[{\citenamefont{Yuen and Shapiro}(1980)}]{Yuen1980}
\bibinfo{author}{\bibnamefont{Yuen, H.~P., and J.~H.~Shapiro}},
\bibinfo{year}{1980},
\bibinfo{journal}{IEEE Trans. Inf. Theory} \textbf{\bibinfo{volume}{26}},
\bibinfo{pages}{78}.


\bibitem[{\citenamefont{Yukawa {\em et al.}}(2008)}]{Yuk08}
\bibinfo{author}{\bibnamefont{Yukawa, M., R.~Ukai, P.~van Loock, and A.~Furusawa}},
\bibinfo{year}{2008},
\bibinfo{journal}{Phys. Rev. A} \textbf{\bibinfo{volume}{78}},
\bibinfo{pages}{012301}.

\bibitem[{\citenamefont{Yurke}(1985)}]{Yur85}
\bibinfo{author}{\bibnamefont{Yurke, B.}},
\bibinfo{year}{1985},
\bibinfo{journal}{J. Opt. Soc. Am. B} \textbf{\bibinfo{volume}{2}},
\bibinfo{pages}{732}.

\bibitem[{\citenamefont{Zaidi {\em et al.}}(2008)}]{Zai08}
\bibinfo{author}{\bibnamefont{Zaidi, H., N.~C.~Menicucci, S.~T.~Flammia, R.~Bloomer, M.~Pysher, and O.~Pfister}},
\bibinfo{year}{2008},
\bibinfo{journal}{Laser Phys.} \textbf{\bibinfo{volume}{18}},
\bibinfo{pages}{659}.


\bibitem[{\citenamefont{Zhang}(2008a)}]{Zha08}
\bibinfo{author}{\bibnamefont{Zhang, J.}},
\bibinfo{year}{2008a},
\bibinfo{journal}{Phys. Rev. A} \textbf{\bibinfo{volume}{78}},
\bibinfo{pages}{034301}.

\bibitem[{\citenamefont{Zhang}(2008b)}]{Zha08a}
\bibinfo{author}{\bibnamefont{Zhang, J.}},
\bibinfo{year}{2008b},
\bibinfo{journal}{Phys. Rev. A} \textbf{\bibinfo{volume}{78}},
\bibinfo{pages}{052307}.

\bibitem[{\citenamefont{Zhang}(2010)}]{Zha10}
\bibinfo{author}{\bibnamefont{Zhang, J.}},
\bibinfo{year}{2010},
\bibinfo{journal}{Phys. Rev. A} \textbf{\bibinfo{volume}{82}},
\bibinfo{pages}{034303}.

\bibitem[{\citenamefont{Zhang {\em et al.}}(2009)}]{Zhang2009}
\bibinfo{author}{\bibnamefont{Zhang, J., G.~Adesso, C.~Xie, and K.~Peng}},
\bibinfo{year}{2009},
\bibinfo{journal}{Phys. Rev. Lett.} \textbf{\bibinfo{volume}{103}},
\bibinfo{pages}{0708501}.

\bibitem[{\citenamefont{Zhang and Braunstein}(2006)}]{Zha06}
\bibinfo{author}{\bibnamefont{Zhang, J., and S.~L.~Braunstein}},
\bibinfo{year}{2006},
\bibinfo{journal}{Phys. Rev. A} \textbf{\bibinfo{volume}{73}},
\bibinfo{pages}{032318}.

\bibitem[{\citenamefont{Zhang {\em et al.}}(2003)}]{Zhang2003}
\bibinfo{author}{\bibnamefont{Zhang, T.~C., K.~W.~Goh, C.~W.~Chou, P.~Lodahl, and H.~J.~Kimble}},
\bibinfo{year}{2003},
\bibinfo{journal}{Phys. Rev. A} \textbf{\bibinfo{volume}{67}},
\bibinfo{pages}{033802}.

\bibitem[{\citenamefont{Zhao {\em et al.}}(2009)}]{Zhao2009}
\bibinfo{author}{\bibnamefont{Zhao Y.~-B., M.~Heid, J.~Rigas, and N.~L\"{u}tkenhaus}},
\bibinfo{year}{2009},
\bibinfo{journal}{Phys. Rev. A} \textbf{\bibinfo{volume}{79}},
\bibinfo{pages}{012307}.

\bibitem[{\citenamefont{Zwierz {\em et al.}}(2010)}]{Zwierz2010}
\bibinfo{author}{\bibnamefont{Zwierz, M., C.~A.~P\'erez-Delgado, and P.~Kok}},
\bibinfo{year}{2010},
\bibinfo{journal}{Phys. Rev. A} \textbf{\bibinfo{volume}{82}},
\bibinfo{pages}{042320}.

\end{thebibliography}

\end{document}